\documentclass[twocolumn,epjc3]{svjour3}  
\smartqed  
\RequirePackage{graphicx}
\RequirePackage{mathptmx}      
\journalname{Eur. Phys. J. C}
  \usepackage[english]{babel}
  \usepackage[dvips]{color}
  \usepackage{latexsym}
  \usepackage{cite}
  \usepackage{amsmath,bm}
  \usepackage{amssymb}  
  \usepackage{afterpage}
  \usepackage{url}
  \usepackage{multirow}
  \usepackage{tabularx}
  \usepackage{longtable}
  \usepackage{ogonek}
  \usepackage{balance}
  \usepackage[T1]{fontenc}
  \usepackage{filecontents}
  \usepackage{hyphenat}
  
   \usepackage{hyperref}
   \usepackage{breakurl}

  \newcommand{\MR}[2]{\multirow{#1}*{#2}}
  \newcommand{\MC}[3]{\multicolumn{#1}{#2}{#3}}
  \newcommand{\ndf}{\text{ndf}}
  \newcommand{\rat}{\varrho}  

  \def\MArun{$M_A^{\text{run}}$}
  \def\SMRFG_MArun{SM\,RFG\,+\,$M_A^{\text{run}}$}
  \def\SMRFGpMArun{SM  RFG $+$ $M_A^{\text{run}}$}
  \def\MARFG{$M_A^{\text{RFG}}$}
  \def\hA2018{hA\,2018}
  \def\hN2018{hN\,2018}
  \def\hAN2018{\hA2018\ \& \hN2018}

  \def\MINERvA{MINER$\nu$A}
  \def\GENIEG{{\sc GENIE}}
  \def\GENIE2{{\sc GENIE\,2}}
  \def\GENIE3{{\sc GENIE\,3}}
  \def\GENIEa{{\tt G18\_10a\_02\_11a}}
  \def\GENIEb{{\tt G18\_10b\_02\_11a}}
  \def\GENIEab{{\tt G18\_10a/b\_02\_11a}}
  \def\G18{{\tt G18}}
  \def\NUANCE{{\sc NUANCE}}
  \def\GiBUU{{\sc GiBUU}}
  \def\NuWro{{\sc NuWro}}
  
  \def\NEUT5{{\sc NEUT\,5.4.0}}
  
  \def\GEANT4{{\sc GEANT\,4}}
  \def\NEUGEN{{\sc NEUGEN}}
  \def\SuSAv2MEC{SuSAv2-MEC}
  \def\SuSAM{SuSAM$^*$}
  \def\T{\mathcal{T}}           

  \def\CN#1{$\dfrac{\chi_{#1}^2}{\text{ndf}}$}
  \def\CNN{$\dfrac{\chi_{\mathcal{N}}^2}{\text{ndf-1}}$}
  \def\CNNt{$\dfrac{\chi_{\mathcal{N}}^2}{\text{ndf-2}}$}
  \def\mc#1#2#3{\noalign{\medskip}\multicolumn{#1}{#2}{#3} \\ \noalign{\medskip}}
  \def\Nbestfit#1{$\mathcal{N}$}                                
  \def\hspaceFS{\hspace*{12pt}  forward scattering}             
  \def\hspaceBS{\hspace*{12pt} backward scattering}             
  \def\WithMono{\hspace*{12pt} with monoenergetic point}        

  \def\Vspace{3.0mm}   
  \def\Bel{$\left\{\begin{matrix} \nu: \\ \overline\nu: \end{matrix}\right.$}

  \renewcommand{\_}{{\fontfamily{ptm}\selectfont\textunderscore}}  

  \allowdisplaybreaks

  \numberwithin{table}{section}       

\begin{document}

\title{Running axial mass of the nucleon as a phenomenological tool for calculating
       quasielastic neutrino--nucleus cross sections
      }

\titlerunning{Running axial mass of the nucleon as a phenomenological tool for calculating
              QE neutrino--nucleus cross sections}        

\author{Igor~D.~Kakorin\thanksref{e1,addr1}
        \and
        Konstantin~S.~Kuzmin\thanksref{e2,addr1,addr2}
        \and
        Vadim~A.~Naumov\thanksref{e3,addr1}
       }
\thankstext{e1}{e-mail: Kakorin@jinr.ru}
\thankstext{e2}{e-mail: KKuzmin@theor.jinr.ru}
\thankstext{e3}{e-mail: VNaumov@theor.jinr.ru}


\institute{Bogoliubov Laboratory of Theoretical Physics, Joint Institute for Nuclear Research,
           RU-141980 Dubna,  Russia \label{addr1}
           \and
           Institute for Theoretical and Experimental Physics,
           RU-117218 Moscow, Russia \label{addr2}
}

\date{Received: date / Accepted: date}

\maketitle

\begin{abstract}
  We suggest an empirical rule-of-thumb for calculating the cross sections of charged-current quasielastic (CCQE) and CCQE-like
  interactions of neutrinos and antineutrinos with nuclei.
  The approach is based on the standard relativistic Fermi-gas model and on the notion of neutrino energy dependent axial-vector mass
  of the nucleon, governed by a couple of adjustable parameters, one of which is the conventional charged-current axial-vector mass.
  The inelastic background contributions and final-state interactions are therewith simulated using \GENIE3\ neutrino event generator.
  An extensive comparison of our calculations with earlier and current
  accelerator CCQE and CCQE-like data for different nuclear targets
  shows good or at least qualitative overall agreement over a wide energy range.
  We also discuss some problematical issues common to several competing contemporary models of the CCQE
  (anti)neutrino--nucleus scattering and to the current neutrino interaction generators.
\end{abstract}

\section{Introduction}
\label{sec:Introduction}
  

  An accurate calculation of the charged-current quasielastic (CCQE) neutrino-nucleus scattering cross sections
  remains an important issue to ensure the reliability and confidence level of extraction of neutrino oscillation
  parameters from atmospheric and accelerator neutrino experiments 
  \cite{Benhar:2015wva,Mosel:2019vhx}.
  This problem is closely related to a large experimental uncertainty in the determination of the weak axial-vector and,
  to a lesser degree, pseudoscalar form factors of the nucleon
  and usually reduces to the experimental uncertainty in the nucleon axial mass, $M_A$, which governs the $Q^2$ evolution
  of the axial-vector form factor in the conventional dipole parametrization,
 \[F_A\left(Q^2\right)= F_A(0)\left(1+\frac{Q^2}{M^2_A}\right)^{-2},\]
  where $Q^2$ is the modulus of the squared four-momentum transfer carried by the $W$-boson.

  Efforts were made in recent years to extract the value of the parameter $M_A$ from $\nu_{\mu}$D,
  $\overline{\nu}_\mu$H, and $\pi^\pm$ electroproduction experiments \cite{Bodek:2007wzPrep,Bodek:2007ym,Bodek:2007vi},
  and from all available at that time data on $\nu/\overline{\nu}$ scattering processes off light, intermediate and heavy nuclei
  \cite{Kuzmin:2006dt_PAN,Kuzmin:2006dh_APPB,Kuzmin:2007kr}.
  In the latter studies, the nuclear effects were accounted for 
  by using the closure over the dinucleon states and one-pion exchange currents 
  \cite{Singh:1971md, Singh:1974df, Singh:1986xh} for deuterium targets 
  and by applying the standard Smith-Moniz relativistic Fermi-gas (SM RFG) model \cite{Smith:1972xhWithErratum}
  (with the parameters extracted from electron-nucleus scattering measurements) for all other (heavier) nuclear targets.
  The most accurate models for the nucleon electromagnetic form factors were used in these calculations.
  It has been inferred from these studies that most of the then-existing CCQE and
  pion electroproduction data could be satisfactorily described with $M_A=M_A^{\text{RFG}}\simeq1$~GeV to within a few percent accuracy.
  This conclusion has been made before the modern high statistics measurements of the CCQE and CCQE-like scattering
  cross sections on carbon reach targets, performed in the FNAL experiments
  MiniBooNE \cite{Aguilar-Arevalo:2010zc,Aguilar-Arevalo:2013dva},
  SciBooNE \cite{AlcarazAunion:2009ku,Aunion:2010zz}, 
  \MINERvA\ \cite{Fiorentini:2013ezn,Fields:2013zhk,Walton:2014esl}, and
  MINOS \cite{Adamson:2014pgc},
  and also in the T2K experiments with two near detectors -- ND280 (off-axis) \cite{Abe:2013jth,Hadley:2014hoa,Abe:2014iza,Abe:2016tmq,Abe:2018pwo}
  and INGRID (on-axis) \cite{Abe:2015oar}.
  According to the RFG based calculations, the CCQE double-differential cross sections measured in MiniBooNE
  \cite{Aguilar-Arevalo:2010zc,Aguilar-Arevalo:2013dva} are well described with 
  $M_A^{\text{RFG}}=1.36\pm0.06$ GeV ($\nu_\mu$) and $1.31\pm0.03$ GeV ($\overline{\nu}_\mu$)
  (cf.\ also Refs.~\cite{Nieves:2011sc,Nieves:2011yp,Juszczak:2010ve,Golan:2013jtj,Butkevich:2013vva,Wilkinson:2016wmz}).
  These values are in reasonable agreement with other recent low-energy data but are incompatible
  with the formal world-average value of $M_A^{\text{RFG}}=1.026\pm0.021$~GeV \cite{Bernard:2001rs}
  as well as with the values of $1.07\pm0.11$~GeV ($\nu_\mu$) and $1.08\pm0.19$~GeV ($\overline{\nu}_\mu$),
  extracted from the total CCQE cross sections measured at higher energies in the NOMAD experiment \cite{Lyubushkin:2008pe}.

\begin{figure}[htb]
  \centering
  \includegraphics[width=\linewidth]
  {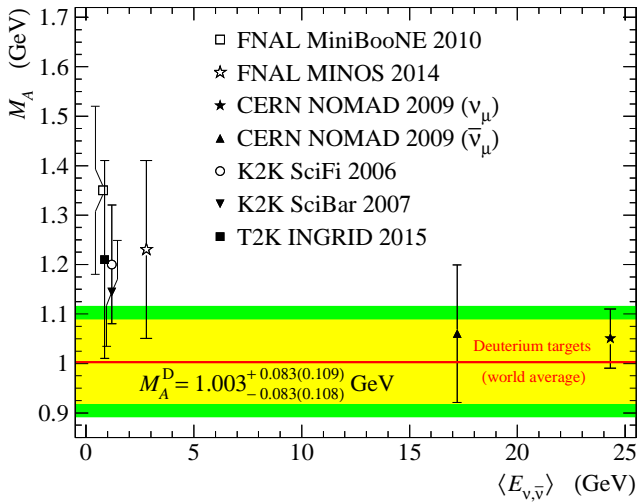}
  \caption{(Color online)
           The nucleon axial mass values vs.\ mean $\nu_\mu/\overline{\nu}_\mu$ energy,
           obtained in the experiments
           MiniBooNE \cite{Aguilar-Arevalo:2010zc},
           MINOS     \cite{Adamson:2014pgc},
           NOMAD     \cite{Lyubushkin:2008pe},
           SciFi     \cite{Gran:2006jn},
           SciBar    \cite{Espinal:2007zz}, and 
           INGRID    \cite{Abe:2015oar}.
           The straight line and surrounding shaded double band represent
           the $M_A$ value and its $1\sigma$ and $2\sigma$ uncertainties,
           as extracted from available deuterium data.}
  \label{Fig:MA_QES_REVIEW_shortlist}
  \end{figure}

  Figure~\ref{Fig:MA_QES_REVIEW_shortlist} shows the values of the nucleon axial mass obtained in the experiments
  \cite{Gran:2006jn,              
        Espinal:2007zz,           
        Lyubushkin:2008pe,        
        Aguilar-Arevalo:2010zc,   
        Adamson:2014pgc,          
        Abe:2015oar}              
  with composite (mainly carbon-rich) nuclear targets; the data are plotted as
  a function of the mean energy of the $\nu_\mu/\overline{\nu}_\mu$ beams. 
  Also shown is the result of our statistical analysis to the ``golden'' (consistent, non-overlapping)
  $\nu_{\mu}\rm{D}$ and $\overline{\nu}_{\mu}\text{H}$ data (see details below). 
  It is amply clear that the values of $M_A^{\text{RFG}}$ extracted using one or another version of the
  global RFG model from the low-energy data on heavy nuclear targets are in conflict with the deuterium data
  and also with the higher-energy data from NOMAD \cite{Lyubushkin:2008pe} and preceding experiments.
  Moreover, essentially all new low-energy data provide a hint that $M_A^{\text{RFG}}$
  increases with decreasing the mean $\nu_\mu/\overline{\nu}_\mu$ energy.


  Modern explanations of the recent experiments include the effects beyond the scope of the naive RFG and impulse approximation.
  Among these are various extensions of the standard (global) RFG model, such as
  local Fermi gas (LFG) model \cite{SajjadAthar:2016gnz},
  local density approximation (LDA) \cite{Graczyk:2004vv}, and 
  spectral function (SF) approach \cite{Benhar:2005dj,Benhar:2006nr,Ankowski:2014yfa,Vagnoni:2017hll,Rocco:2018mwt,Ivanov:2013saa,Bodek:2014pka,Sobczyk:2017mts,Ivanov:2018nlm,Ivanov:2019jqp}; 
  relativistic mean field and relativistic Green's function models \cite{Barbaro:2011iy,Meucci:2014bva}; 
  charged meson-exchange currents (MEC), intermediate $\Delta$ isobar or
  multi-nucleon excitations \cite{Ericson:2015cva,Gran:2013kda},
  short-range and long-range correlations (SRC and LRC) within random phase approximation
  (RPA) \cite{VanCuyck:2016fab,Nieves:2013fr,Pandey:2013cca};
  quantum-kinetic transport equations (implemented in the \GiBUU\ code) \cite{Lalakulich:2012hs,Mosel:2017ssx};
  parametrization of the observed enhancement in the transverse electron quasielastic response function
  (presumably because of MEC) \cite{Bodek:2011ps,Sobczyk:2012ah,Bodek:2012cm,Bodek:2014pka};
  a variety of so-called superscaling models, e.g.,
  SuSA \cite{Megias:2014kiaWithErratum,Barbaro:2021psv},
  SuSAv2 \cite{Gonzalez-Jimenez:2014eqa,Megias:2017cuh},
  SuSAv2-MEC \cite{Megias:2016fjk, Megias:2018ujz, Amaro:2019zos}, and
  \SuSAM\ \cite{RuizSimo:2018kdl}.
  The most comprehensive microscopic and phenomenological models
  usually increase the CCQE cross sections at low energies, thus providing better data explanation without increasing $M_A^{\text{RFG}}$
  (see Refs. \cite{Boyd:2009zz, Gallagher:2011zza, Garvey:2014exa, Alvarez-Ruso:2014bla, Katori:2016yel,Betancourt:2018bpu}
  for reviews and further references). 

  The main purposes of this study are to clarify the experiential state-of-the-art with the nucleon axial mass
  and provide a simple phenomenological method for an accurate description of the CCQE $\nu$ and $\overline{\nu}$
  interactions with nuclei at energies of interest for neutrino oscillation experiments, within the frameworks
  of conventional RFG model, but at the expense of having two adjustable parameters (instead of the only one, $M_A$)
  in the nuclear axial-vector structure function.
  The suggested recipe should never be considered as an alternative or competitor to the detailed microscopic models.
  Rather, it can serve as a complementary empirical tool which can easily be implemented in any Monte Carlo neutrino event generator
  \footnote{The model is implemented into the \GENIEG\ neutrino generator (version 2.11.0 and higher) as an option.}
  and used in the analyses of the experiments with accelerator and atmospheric neutrino and antineutrino fluxes.
\section{Running axial mass}
\label{sec:Method}

  The idea of the prescribed method is to calculate the cross sections for the CCQE $\nu/\overline{\nu}$ interactions
  with nuclei other than hydrogen and deuterium by using the neutrino energy dependent \emph{running axial-vector mass},
  \MArun, in the charged weak hadronic current, instead of the conventional constant axial-vector (dipole) mass $M_A$;
  below the latter will be referred to as the \emph{current axial mass}.

  Some motivation is required for the suggested ``trick'' in which the axial form factor $F_A$ -- the function of $Q^2$ only --
  is replaced by a function dependent on $Q^2$ and neutrino energy $E_\nu$ (through the function $M_A^\text{run}(E_\nu)$).
  According to the above-listed microscopic models, the multi-nucleon excitation mechanisms (such as the RPA long-range
  correlations or two particle -- two hole ($2p2h$) channels contributions caused by meson exchange currents)
  lead to an enhancement of the neutrino-nucleus flux-folded cross sections at low neutrino energies (see, e.g., Ref.~\cite{Nieves:2013fr}).
  The experimental hints shown in Fig.~\ref{Fig:MA_QES_REVIEW_shortlist} suggest that this enhancement can be phenomenologically
  reproduced only by adjusting the nucleon axial mass parameter, $M_A$, and remaining within the framework of
  the simple RFG approach that is without accounting for the nontrivial nuclear effects.
  The most straightforward way to do this would be to construct a suitable function $M_A^{\text{run}}(Q^2)$ and fine-tune 
  it on the appropriate datasets. 
  
  A more simple empirical solution is to use the well-known (close to linear)
  correlation between the mean $Q^2$ value, $\langle Q^2 \rangle$, and neutrino energy, $E_\nu$
  (see, e.g., Refs. \cite{Erriquez:1979,Eichten:1973cs,Barish:1976bj,Asratyan:1978rt,Fanourakis:1980si,Makeev:1981em}),
  which allows a two-step approximation in the calculating the flux-folded cross sections: 
  considering that $M_A^{\text{run}}(Q^2)$ is a relatively weakly dependent function of $Q^2$, one can approximately replace
  variable $Q^2$ in this function with its mean value, $\langle Q^2 \rangle$, and then use the mentioned correlation
  between $\langle Q^2 \rangle$ and $E_\nu$,
  \[M_A^{\text{run}}(Q^2) \longmapsto M_A^{\text{run}}(\langle Q^2 \rangle) \longmapsto M_A^{\text{run}}(E_\nu),\]
  thus arriving at the notion of running axial mass of the nucleon, $M_A^\text{run}(E_\nu)$.
%
%
  The energy dependence of \MArun\ can be then retrieved from available CCQE data.
  The outlined customization of the hadronic current must be treated as a purely empirical prescription to account for the experimental
  evidence of nuclear effects beyond the RFG model.

  The function \MArun\ must be parametrized in such a way that it asymptotically approaches the current (constant) $M_A$ at high energies
  and describes the lower-energy data and in general, it will be different for different modifications of the RFG models and other inputs.
  In the present analysis, we adopt the very simple parametrization,
  \begin{equation}
  \label{M_A_run}
  M_A^\text{run} = M_0\left(1+\frac{E_0}{E_\nu}\right),
  \end{equation}
  in which $E_\nu$ is the lab-frame neutrino energy which can be treated as Lorentz invariant
  ($E_\nu=(s-M^2)/2M$, where $M$ is the mass of the target nucleus) and the constant
  parameters of $M_0$ and $E_0$ are obtained from the global fit to available accelerator
  data on the CCQE $\nu_\mu$ and $\overline{\nu}_\mu$ interactions with nuclei.
  Hence the modified hadronic current formally retains its Lorentz-transformation property albeit
  loses the fundamental meaning.
  It turns out that the parametrization \eqref{M_A_run} is universal in the sense that it works rather well
  for all medium-to-heavy nuclear targets and at all available $\nu/\overline{\nu}$ energies.

  In our fit to \MArun, the $\nu_{\mu}$D and $\overline\nu_{\mu}$H cross sections are exploited for adjusting the parameter $M_0$ only,
  inasmuch as the function $M_A^\text{run}(E_\nu)$ can be applied to the $\nu/\overline\nu$ CCQE scattering from the heavier nuclear targets.
  Since $M^\text{run}_A \to M_0$ at high energies, $E_\nu \gg E_0$, where the RFG model works rather well, the constant $M_0$ can be
  treated as the current axial-vector mass $M_A$.
  On the other hand, the value of $M_A$ can be independently extracted from a fit to the $\nu_\mu$D data for which the nontrivial
  (beyond RFG) nuclear effects are relatively small and better understood.
  Such approach will provide the predictive power of the formalism over a rather wide kinematic region.

  \subsection{Parameters of the RFG model}

  \begin{figure*}[htb]
  \centering
  \includegraphics[width=\textwidth]
  {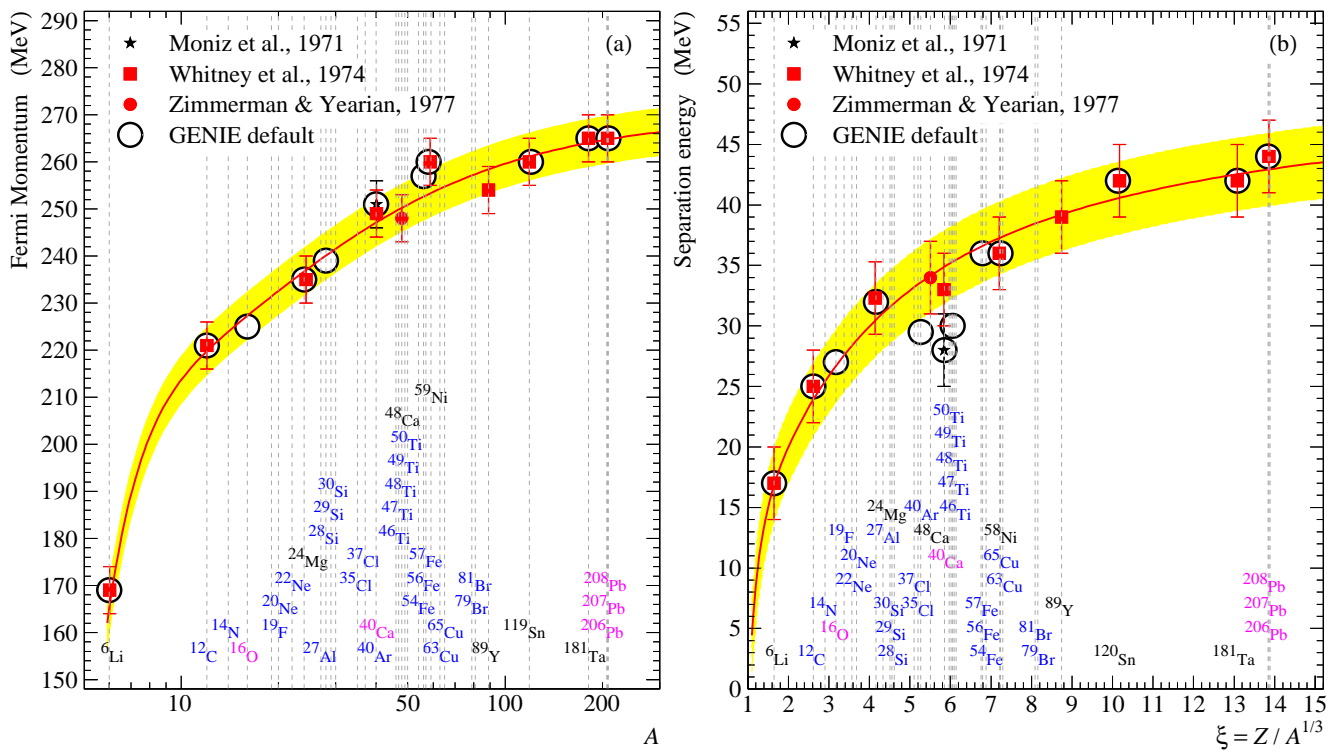}
  \caption{(Color online)
           The parametrizations of the Fermi momentum vs.\ mass number (a) and
           separation energy vs.\ parameter $\xi=Z/A^{1/3}$ (b)  according to Eqs.~\eqref{pFEb}.
           The data points are taken from Refs.~\cite{Moniz:1971mt,Whitney:1974hr,Zimmerman:1976}.
           The filled bands indicate the $1\sigma$ uncertainties of the interpolations. The dashed vertical lines represent
           the values of $A$ and $\xi$ for the isotopes studied in the experiments \cite{Moniz:1971mt,Whitney:1974hr,Zimmerman:1976}
           and  for those noticeably contained in the detector materials employed in the neutrino experiments whose data are
           involved into the present analysis. Current \GENIE3\ default inputs are also indicated.
          }
  \label{Fig:RFG_parameters}
  \end{figure*}
  The nuclear Fermi momenta, $p_F$, and binding energies (usually identified with the separation energies),
  $E_b$, are rather uncertain and values of these parameters used in the literature vary greatly
  (see Refs.~\cite{Bodek:2018lmc,Cai:2019hpx} for a more sophisticated treatment of these matters).
  Both parameters are usually subject of adjustment in each specific experiment.
  For a certain unification, in the ensuing calculations we will use the following interpolation formulas:
  \begin{equation}
  \label{pFEb}
  \begin{aligned}
  p_F = &\ p_F^0\left[1-\frac{4.2}{A}+\left(\frac{6.0}{A}\right)^2-\left(\frac{5.3}{A}\right)^3\right], \\
  E_b = &\ E_b^0\left[1-\frac{2.26}{\xi}+\left(\frac{1.73}{\xi}\right)^2-\left(\frac{1.21}{\xi}\right)^3\right],
  \end{aligned}
  \end{equation}
  where $p_F^0=270$~MeV, $E_b^0=50.4$~MeV, $\xi=Z/A^{1/3}$, $Z$ is the atomic number, and $A$ is the mass number.
  These interpolations are obtained from the available data on electron-nucleus scattering \cite{Whitney:1974hr,Zimmerman:1976} 
  and are sufficiently accurate for all nuclei with $A\ge6$, see Fig.~\ref{Fig:RFG_parameters}.
  The interpolation \eqref{pFEb} for the Fermi momenta is numerically close to that suggested in Ref.~\cite{Bodek:1980ar}
  and to the default \GENIEG\ inputs.
  Note that the previously published data~\protect\cite{Moniz:1971mt} shown in Fig.~\ref{Fig:RFG_parameters} were partially updated
  in Ref.~\cite{Whitney:1974hr}. The default \GENIEG\ values of the binding energies for $\xi\sim5-6$ are based on the obsolete
  data of Ref.~\protect\cite{Moniz:1971mt} leading to a certain ``dip'' in the function $E_b(\xi)$, which however does not affect
  the following analysis.
  
  The proton and neutron Fermi momenta are calculated in the conventional way \cite{Moniz:1971mt,Bodek:1980ar} as, respectively,
  \[
  p_F^p = \left(\frac{2Z}{A}\right)^{1/3}p_F 
  \enskip\text{and}\enskip
  p_F^n = \left(\frac{2N}{A}\right)^{1/3}p_F,
  \]
  where $N=A-Z$. These relations are based on the simplest assumption that the density of nuclear matter
  is approximately constant irrespective of the proton-to-neutron ratio $Z/N$.

  \subsection{CCQE-like background due to pion production}
  \label{sec:MArunGENIEsetting}

  For calculations of the inelastic contributions to the CCQE-like background with the \SMRFG_MArun\ model,
  we made several modification from the standard \GENIE3\ configuration \cite{GENIE:2021npt}.
  Specifically, for simulation of the single-pion neutrinoproduction
  (an essential contribution to the FSI correction for the CCQE-like cross sections, see Sec.\ \ref{sec:FSI})
  we use the modified extended Rein-Sehgal model (ExRS or KLN) \cite{Kuzmin:2003ji}
  supplemented by the pion-pole contribution to the hadronic axial current derived in 
  Ref.~\cite{Berger:2007rqWithErratum}. 
  This model is referred to as Berger-Sehgal (BS) or KLN-BS model in \GENIE3.
  Pauli-blocking effect is taken into account for the nuclear targets.
  Next, we refused the renormalization of the Breit-Wigner factors suggested in the original Rein-Sehgal model \cite{Rein:1980wg}
  and used in \GENIE3\ in a slightly modified form.
  Among several physical and technical reasons of that refusal, we only mention here that the normalization integral for the
  $S$-wave resonances diverges, leading to an unreasonable ambiguity due to an unphysical cutoff in invariant hadronic mass~\cite{Kakorin:2021mqo}. 
  The KLN-BS model properly takes into account the interference between the amplitudes of the resonances
  which have the same spin and orbital angular momentum. However, the current \GENIEG\ release of the KLN-BS model
  neglects this effect and operates with an incoherent sum over 17 resonance families. The interference essentially affects
  both on the absolute values and on the shapes of the pion production cross sections, but integrally causes less than $\sim2$\% effect 
  in the CCQE-like cross section calculations. This is within the level of expected accuracy of the \GENIEG3\ simulation procedure
  (which employs somewhat rough simplifications) and is well within the systematic errors of the experimental data
  under subsequent consideration. All the mentioned features and flaws are subject of further improvements of the \GENIEG\ package.
 
  Instead of the \GENIEG\ default ``resonance'' axial-vector mass value $M^\text{RES}_A=1.12$~GeV \cite{Kuzmin:2006dh_APPB,GENIE:2021npt},
  in the present analysis we use the updated value of $1.18$~GeV, obtained from the new global fit to all available $\nu_\mu$D
  single-pion production data~\cite{Kakorin:2021mqo}, for which the nonresonant background (NRB) is small compared to the
  resonance contribution. Note that treatment of NRB in \GENIEG\ is also different from that in the KLN-BS model.
  All other KLN-BS model inputs (resonance masses, widths, decay mode fractions) are updated according
  to the most recent data, as suggested in Ref.~\cite{Zyla:2020zbs}.
  
  The parametrization of the vector CCQE form factors from the \GENIEG\ default ``BBA(05)'' model
  are replaced to a more accurate ``BBBA$_{25}$(07)'' one \cite{Bodek:2007ym,Bodek:2007vi}.
  In all our calculations performed with the models/tunes incorporated in \GENIEG, we thoroughly accounted for all essential
  experimental features (cuts, elemental compositions of the detectors, etc.).
  The results of all other (non-\GENIEG) models under consideration are reproduced exactly as provided by their authors. 
  \section{Statistical analysis}
  \label{sec:StatisticalAnalysis}

  The measurements of the CCQE $\nu_\mu/\overline{\nu}_\mu$ cross sections were carried out
  from the mid-60s to present day, in the experiments at
  ANL \cite{Kustom:1969dh,
            Mann:1972,
            Mann:1973pr,
            Barish:1976dz,
            Barish:1977qk,
            Singer:1977rs,
            Miller:1982qi
           },
  BNL \cite{Cnops:1978zi,
            Fanourakis:1980si,
            Baker:1981su,
            Ahrens:1984gp,
            Ahrens:1988rr,
            Kitagaki:1990vs,Furuno:2003ng
           },
 FNAL \cite{Kitagaki:1983px,
            Asratyan:1984gh,Asratyan:1984ir,
            Ammosov:1986jn,
            Suwonjandee:2004aw,
            Aguilar-Arevalo:2010zc,
            Aguilar-Arevalo:2013dva,
            AlcarazAunion:2009ku,Aunion:2010zz,
            Fiorentini:2013ezn,
            Fields:2013zhk,
            Walton:2014esl,
            Wolcott:2015hda,
            Lu:2015tcr,
            Betancourt:2017uso,
            Lu:2018stk,
            Ruterbories:2018gub,
            Patrick:2018gvi,                                                   
            Betancourt:2013mba,
            Diaz:2014qla
           }
  LANL (LSND) \cite{Auerbach:2002iy},                                          
  CERN \cite{Block:1964,
             Burmeister:1965,
             Franzinetti:1965,
             Young:1967ud,
             Orkin-Lecourtois:1967,
             Holder:68,
             Budagov:1969bg,
             Eichten:1973cs,
             Haguenauer:74,
             Rollier:1975qr,
             Bonetti:1977cs,
             Musset:1978gf,
             Musset:1978gf,
             Pohl:1979zm,
             Armenise:1979zg,Singh:1992dc,                                     
             Allasia:1990uy,
             MartinezdelaOssaRomero:2007oxj,
             Lyubushkin:2008pe
            },
  IHEP \cite{Makeev:1981em,
             Belikov:1981fq,
             Belikov:1981ut,
             Belikov:1983kg,Belikov:1985mw,
             Grabosch:1986nu,
             Grabosch:1988js,
             Brunner:1989kw
            },
  K2K \cite{Gran:2006jn},                                                      
  and
  T2K \cite{Abe:2013jth,
            Hadley:2014hoa,
            Abe:2014iza,
            Abe:2015oar,
            Abe:2016tmq,
            Abe:2019sah
           }.
  The detector targets employed in these experiments were 
  hydrogen                                  \cite{Fanourakis:1980si},
  deuterium                                 \cite{Mann:1972,%
                                                  Mann:1973pr,%
                                                  Barish:1976dz,%
                                                  Barish:1977qk,%
                                                  Singer:1977rs,%
                                                  Cnops:1978zi,%
                                                  Baker:1981su,%
                                                  Miller:1982qi,%
                                                  Kitagaki:1983px,%
                                                  Kitagaki:1990vs,%
                                                  Furuno:2003ng,%
                                                  Allasia:1990uy%
                                                 },
  water                                     \cite{Gran:2006jn,Abe:2017rfw,Abe:2019sah}
  mineral oil                               \cite{Aguilar-Arevalo:2010zc,%
                                                  Aguilar-Arevalo:2013dva,%
                                                  Fiorentini:2013ezn,%
                                                  Fields:2013zhk%
                                                 },
  aluminium                                 \cite{Belikov:1981fq,%
                                                  Belikov:1981ut,%
                                                  Belikov:1983kg,%
                                                  Belikov:1985mw%
                                                 },
  argon                                     \cite{MartinezdelaOssaRomero:2007oxj},
  steel                                     \cite{Kustom:1969dh},
  iron                                      \cite{Suwonjandee:2004aw},
  freon                                     \cite{Block:1964,%
                                                  Burmeister:1965,%
                                                  Franzinetti:1965,%
                                                  Young:1967ud,%
                                                  Orkin-Lecourtois:1967,%
                                                  Eichten:1973cs,%
                                                  Haguenauer:74,%
                                                  Rollier:1975qr,%
                                                  Bonetti:1977cs,%
                                                  Musset:1978gf,%
                                                  Makeev:1981em,%
                                                  Grabosch:1986nu,%
                                                  Grabosch:1988js,%
                                                  Brunner:1989kw%
                                                 },
  propane--freon mixtures                   \cite{Musset:1978gf,%
                                                  Armenise:1979zg,%
                                                  Pohl:1979zm,%
                                                  Singh:1992dc%
                                                },
  neon--hydrogen mixture                    \cite{Asratyan:1984gh,%
                                                  Asratyan:1984ir,%
                                                  Ammosov:1986jn%
                                                 },
  complex carbon-bearing media (hydrocarbon, propane, polystyrene, etc.)
                                            \cite{Budagov:1969bg,%
                                                  Holder:68,%
                                                  Auerbach:2002iy,%
                                                  Ahrens:1984gp,%
                                                  Ahrens:1988rr,%
                                                  AlcarazAunion:2009ku,Aunion:2010zz,%
                                                  Lyubushkin:2008pe,%
                                                  Aguilar-Arevalo:2010zc,%
                                                  Aguilar-Arevalo:2013dva,%
                                                  AlcarazAunion:2009ku,Aunion:2010zz,%
                                                  Fiorentini:2013ezn,%
                                                  Fields:2013zhk,%
                                                  Wolcott:2015hda,%
                                                  Lu:2015tcr,%
                                                  Betancourt:2017uso,%
                                                  Lu:2018stk,%
                                                  Ruterbories:2018gub,%
                                                  Patrick:2018gvi,%
                                                  Betancourt:2013mba,%
                                                  Diaz:2014qla%
                                                 },
  and                                                
  complex carbonaceous targets of the T2K near detectors
                                            \cite{Abe:2013jth,%
                                                  Hadley:2014hoa,%
                                                  Abe:2014iza,%
                                                  Abe:2015oar,%
                                                  Abe:2016tmq%
                                                 }.
  Additional information can be found in review articles and data compilations
  \cite{Perkins:1972zt,Perkins:1975bj,Derrick:1974ema, Cline:1977si, Ermolov:1978ij, Alekhin:1987nn, Sakuda:2002rx,%
         Ammosov:1992ak, Baltay:1994he, Sorel:2007iv, Gran:2007kn} for the earlier and
  \cite{Katori:2016yel,Betancourt:2018bpu} for the contemporary experiments.

  For our statistical analysis, we employ the CCQE data on the total, 
  flux-averaged single-, and double-differential cross sections, 
  as well as the flux-weighted $Q^2$ distributions, $\langle{dN/dQ^2}\rangle$, or $Q^2$ distributions
  specified by the mean $\nu_\mu/\overline{\nu}_\mu$ energy.
  The full data set is formed by the most statistically reliable and self-contained measurements which were not reexamined
  (as a result of enlarged statistics, revised data processing, and so on) in the later reports of the same experimental groups.
  We avoid using the data from the experiments with poorly known (anti)neutrino energy spectra and/or with
  non-active detector targets (see Ref. \cite{Kuzmin:2007kr} for the details of the selection criteria).
  Namely, we use the results of the following experiments operating with different nuclear targets:
  \begin{itemize}
  \item[$\bullet$] hydrogen:
        BNL 1980 \cite{Fanourakis:1980si} ($dN_{\overline{\nu}}/dQ^2$, $5$);
  \item[$\bullet$] deuterium:
        ANL 1977 \cite{Barish:1977qk} ($\sigma_\nu$, $8$),
        ANL 1982 \cite{Miller:1982qi} ($\langle{dN_\nu/dQ^2}\rangle$, $39$),
        BNL 1990 \cite{Kitagaki:1990vs,Furuno:2003ng} ($\langle{dN_\nu/dQ^2}\rangle$, $37$),
        FNAL 1983 \cite{Kitagaki:1983px} ($\langle{dN_\nu/dQ^2}\rangle$, $20$), and
        Big European Bubble Chamber (BEBC) at CERN 1990 \cite{Allasia:1990uy} ($\langle{d\sigma_\nu/dQ^2}\rangle$, $8$);
  \item[$\bullet$] Ne-H$_2$ mixture:
        FNAL 1984 \cite{Asratyan:1984gh} ($dN_{\overline{\nu}}/dQ^2$, $14$);
  \item[$\bullet$] aluminium:
                   IHEP--ITEP 1985 \cite{Belikov:1983kg, Belikov:1985mw} ($\sigma_\nu$, $\sigma_{\overline{\nu}}$,
                   and $\langle{d\sigma_{\nu+\overline{\nu}}/dQ^2}\rangle$, $8$ in each dataset);
  \item[$\bullet$] carbon-rich media:
                   CERN NOMAD 2009 \cite{Lyubushkin:2008pe} ($\sigma_\nu$, $\sigma_{\overline{\nu}}$, $10$ and $6$, respectively),
                   FNAL MiniBooNE 2010 \cite{Aguilar-Arevalo:2010zc} ($\langle{d^2\sigma_\nu/dE_{\mu}d\cos\theta_\mu}\rangle$, $137$),
                   and MiniBooNE 2013 \cite{Aguilar-Arevalo:2013dva} ($\langle{d^2\sigma_{\overline{\nu}}/dE_{\mu}d\cos\theta_\mu}\rangle$, $78$),
                   T2K INGRID 2015 \cite{Abe:2015oar} ($\sigma_\nu$, $2$), and 
		           T2K ND280 2014 \cite{Abe:2014iza} ($\sigma_\nu$, $5$).
  \item[$\bullet$] liquid-argon time projection chamber (LAr-TPC) 2007 \cite{MartinezdelaOssaRomero:2007oxj} ($\sigma_\nu$, $1$);
  \item[$\bullet$] freon (CF$_3$Br), propane (C$_3$H$_8$), and freon-propane compounds:
                   bubble chamber Gargamelle (GGM) at CERN 1979 \cite{Armenise:1979zg,Singh:1992dc} ($\langle{dN_{\overline{\nu}}/dQ^2}\rangle$, $13$);
                   IHEP babble chamber SKAT 1990 \cite{Brunner:1989kw} ($\langle{d\sigma_\nu/dQ^2}\rangle$,
                   $\langle{d\sigma_{\overline{\nu}}/dQ^2}\rangle$, $8$ and $7$, respectively).
  \end{itemize}
  In the brackets we show the data types and numbers of the experimental bins involved into the analysis.
  Hence the full data set for the our analysis consists of $422$ data points with $290$, $124$, and $8$ ones
  for, respectively, $\nu_\mu$ ($68.7$\% of the full data set), $\overline{\nu}_\mu$ ($29.3$\%), and cumulative
  $\nu_\mu+\overline{\nu}_\mu$ ($1.9$\%) cross sections and distributions.
  The data are presented as $215$, $31$, $128$, and $48$ experimental data points for, respectively,
  the flux-folded double-differential CCQE cross sections measured by MiniBooNE ($51$\% of the full data set),
  differential in $Q^2$ cross sections ($7.4$\%), unnormalized $Q^2$ distributions ($30.3$\%), and
  flux-unfolded total CCQE cross sections ($11.4$\%). The full data set covers a wide energy range -- from the
  CCQE reaction threshold to about $100$~GeV.
  The data subset used for extracting the current axial mass $M_A$ ($\equiv M_A^\text{D}$) contains $117$ 
  data points (that constitutes $27.7$\% of the full data set) and is composed of the results of the experiments
  ANL 1977       \cite{Barish:1977qk},
      1982       \cite{Miller:1982qi}, 
  BNL 1980       \cite{Fanourakis:1980si},
      1990       \cite{Kitagaki:1990vs,Furuno:2003ng},
  FNAL 1983      \cite{Kitagaki:1983px}, and
  CERN BEBC 1990 \cite{Allasia:1990uy}.

  In the present analysis, we do not utilize the most recent data from
  T2K \cite{Abe:2016tmq,Abe:2017rfw,Abe:2018pwo,Abe:2019sah,Abe:2019cpx,Abe:2020jbf,Abe:2020uub,Abe:2019iyx,Abe:2020iop}, and 
  \MINERvA\ \cite{Fiorentini:2013ezn,Fields:2013zhk,Walton:2014esl,Wolcott:2015hda,Lu:2015tcr,Betancourt:2017uso,Lu:2018stk,%
                  Ruterbories:2018gub,Patrick:2018gvi,Carneiro:2019jds},
  as well as the CCQE-like double-differential cross sections and both
  CCQE and CCQE-like single-differential and total cross sections from MiniBooNE \cite{Aguilar-Arevalo:2010zc,Aguilar-Arevalo:2013dva}.
  Instead, a limited but representative part of these data is used for an attentive and thorough verification of the \SMRFG_MArun\ model
  and for a comparative analysis of several competing models (see Sec.\ \ref{sec:Comparison}). 
  We plan to study the remaining and new data in a future dedicated work.
  
  We use the ordinary least-square statistical model:
  \begin{equation}
  \label{Blobel_fixed}
  \chi^2 = \sum_{i}\left\{\sum_{j \in G_i}\frac{\left[N_iT_{ij}(\boldsymbol{\lambda})-E_{ij}\right]^2}{\sigma_{ij}^2}
           +\frac{\left(N_i-1\right)^2}{\delta_i^2}\right\}.
  \end{equation}
  Here, the index $i$ labels the experiments or data groups $G_i$, 
  index $j \in G_i$ enumerates the bin-averaged experimental data $E_{ij}$ from the group $G_i$,
  $\sigma_{ij}$ is the error of $E_{ij}$, without normalization uncertainties (due to the $\nu/\overline{\nu}$ flux indetermination
  and other sources). The normalization factors, $N_i$ (individual for each data group $G_i$), are treated as fitting parameters
  and are included into the ordinary penalty term, $\left(N_i-1\right)^2/\delta_i^2$, where $\delta_i$ is the relative
  normalization error. The value $T_{ij}(\boldsymbol{\lambda})$ represents the associated (also bin-averaged) model
  prediction, which is a function of a set of fitting parameters $\boldsymbol{\lambda}=(\lambda_1,\lambda_2,\ldots)$; in our particular case,
  $\boldsymbol{\lambda}=(M_0)$ for hydrogen and deuterium targets and $\boldsymbol{\lambda}=(M_0,E_0)$ for all other nuclear targets.

  The minimization procedure is significantly simplified by substituting into Eq.~\eqref{Blobel_fixed}
  $N_i=\mathcal{N}_i(\boldsymbol{\lambda})$, where the numbers
  \begin{equation}
  \label{Normalization_Factor}
  \mathcal{N}_i(\boldsymbol{\lambda})
  = \dfrac{1+\delta_i^2\sum_{j \in G_i}\sigma_{ij}^{-2}T_{ij}(\boldsymbol{\lambda})E_{ij}}
          {1+\delta_i^2\sum_{j \in G_i}\sigma_{ij}^{-2}T_{ij}^2(\boldsymbol{\lambda})}
  \end{equation}
  are the solutions of the minimization equations $\partial\chi^2/\partial N_i=0$
  (see \ref{Method_adds} for a slightly more complicated case).
  The $\chi^2$ value for the final fit to all data includes the penalty term
  \[\left[\left(M_0-M_A^\text{D}\right)/\Delta M_A^\text{D}\right]^2,\]
  which provides a ``soft anchoring'' of the parameter $M_0$ to the current axial mass
  $M_A^\text{D} \pm \Delta M_A^\text{D}$ obtained from the fitting of the robust deuterium data only.

  For extracting the value of $M_A$ from the CCQE $\nu_\mu$D data, the authors of the experiments 
  (see, e.g., Refs. \cite{Baker:1981su,Kitagaki:1990vs,Kitagaki:1983px})
  usually take into account the Pauli exclusion principle.
  In our analysis we try to use whenever possible the raw, uncorrected $\nu_\mu$D data (mainly $Q^2$ distributions).
  To account for the nuclear effects besides the trivial effects of Fermi motion and deuteron binding
  we adopt the closure approximation over the dinucleon states following Ref. \cite{Singh:1986xh},
  where the MEC contributions were estimated using the single-pion exchange diagrams in the static limit.
  In our calculations, the Reid hard-core potential and Hulthen  wave function for the deuteron were adopted,
  as providing the best description of the $\nu_\mu$D data.

  All the fits are done with the CERN function minimization and error analysis package MINUIT 
  \cite{James:1975343,James:1994vla-1998}.
  The errors of the output parameters quoted below correspond to one and two standard deviation.
  As follows from the analysis, the deviation of the normalization factors $\mathcal{N}_i$ from unity for each data group $G_i$
  does not exceed the doubled normalization uncertainty.

  As a result of the analysis of the deuterium and hydrogen data, the best-fit value of $M_A^\text{D}$ is found to be
  \begin{equation}
  \label{MAD}
  M_A^\text{D}=1.003_{-0.083\,(0.108)}^{+0.083\,(0.109)}~\text{GeV}
  \end{equation}
  with the corresponding $\chi^2/\ndf$ value of $127.5/(117-7)\approx1.16$.
  Values in brackets in Eq.~\eqref{MAD} and below are the two-standard-deviation (95\% C.L.) errors.
  The best-fit values of the running axial mass parameters are found as follows:
  \begin{equation}
  \label{Eqn:ME_QES_general}
  \begin{aligned}
  M_0 = &\ 1.052_{-0.094\,(0.113)}^{+0.095\,(0.114)}~\text{GeV}, \\
  E_0 = &\  278_{-111\,(131)}^{+130\,(158)}~\text{MeV},
  \end{aligned}
  \end{equation}
  with $\chi^2/\text{ndf}=272.7/(422-19)\approx0.68$. The data for all nuclear targets were involved into this analysis.

  As the next step, by taking into account that the SM RFG calculations well describe
  the high-energy data on all nuclear targets with the unique value
  of $M_A$ ($\approx M_A^\text{D}\approx1$~GeV) \cite{Kuzmin:2007kr},
  we add to the sum \eqref{Blobel_fixed} the penalty term
  \[\left(M_0-1.003\right)^2/0.083^2\] to constrain the bias of $M_0$ from $M_A^\text{D}$.
  The final global fit performed with this constraint yields
  \begin{equation}
  \label{Eqn:ME_QES_default}
  \begin{aligned}
  M_0 = &\ 1.008 \pm 0.025\,(0.029)~\text{GeV}, \\
  E_0 = &\  331_{-54\,(64)}^{+57\,(69)}~\text{MeV}
  \end{aligned}
  \end{equation}
  with $\chi^2/\text{ndf}=277.6/(422-19)\approx0.69$.

\begin{figure*}[!htb]
  \includegraphics[width=\linewidth]
  {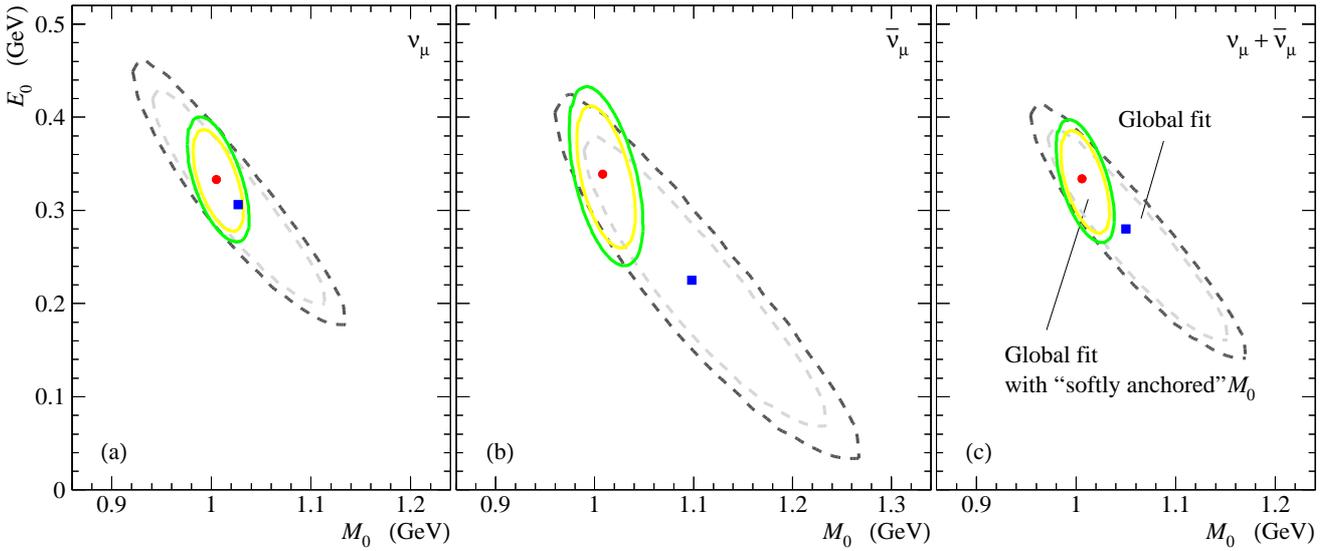}
  \caption{(Color online)
           Marginalized confidence contours 
           in the $(M_0,E_0)$ plane obtained in the global fits for
           $\nu_\mu$ (a), $\overline{\nu}_\mu$ (b), and $\nu_\mu+\overline{\nu}_\mu$ (c) CCQE datasets.
           The smaller solid contours are obtained from the fits in which the parameter $M_0$ has been softly
           anchored to the best-fit value extracted from the fit to deuterium data. 
           The inner and outer contours for both fits~\eqref{Eqn:ME_QES_general} and \eqref{Eqn:ME_QES_default}
           indicate, respectively, the $68$\% and $95$\% C.L.\ areas.
          }
  \label{Fig:ME_QES_M0E0_errors_BBBA25_shortlist}
  \end{figure*}

  Figure \ref{Fig:ME_QES_M0E0_errors_BBBA25_shortlist} shows the $1\sigma$ and $2\sigma$ confidence contours
  in the two-parameter plane, resulting from the three consecutive fits.
  It is in particular seen that the values of the parameters $M_0$ and $E_0$ obtained
  in the separate analyses used the $\nu_\mu$, $\overline{\nu}_\mu$, and $\nu_\mu+\overline{\nu}_\mu$ datasets
  are in agreement with each other within the $68$\% confidence contours; the agreement is worse for
  the best-fit values in the antineutrino case. It should be mentioned that the obtained values of
  $M_A^\text{D}$, $M_0$, and $E_0$ are strictly speaking valid only within the set of the inputs adopted
  in our analyses, such as the parameters of the RFG model (see Eq.~\eqref{pFEb}) and parametrization of
  the vector form factors of the nucleon (BBBA$_{25}$(07) model \cite{Bodek:2007ym}).
  However, as our study shows, these values are quite stable relative to small variations of the input parameters.

  \begin{figure}[ht!]
  \centering
  \includegraphics[width=\linewidth]
  {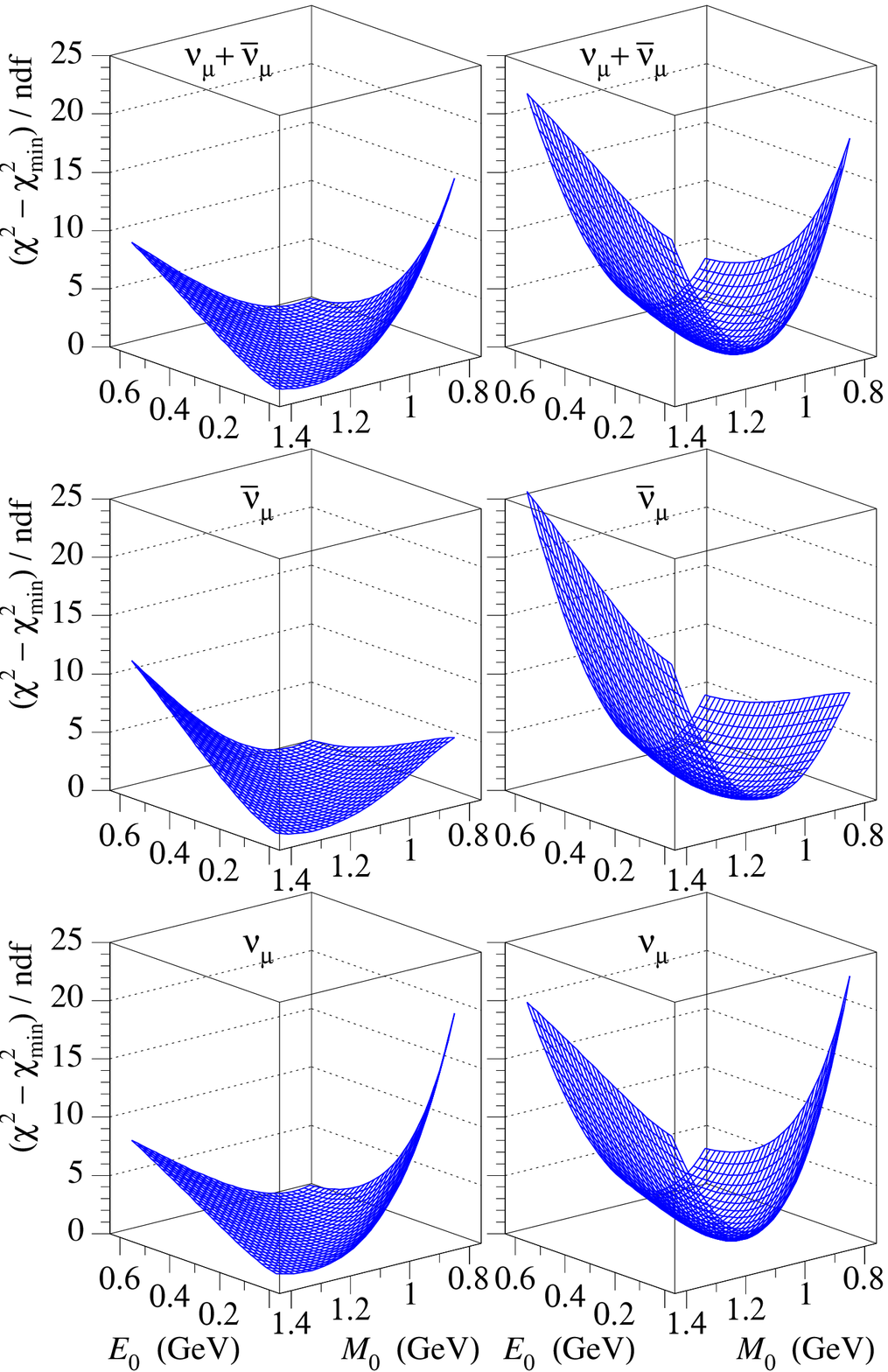}
  \caption{(Color online)
           Ratios $\left(\chi^2-\chi^2_{\text{min}}\right)/\text{ndf}$
           as functions of the free fitting parameters $E_0$ and $M_0$.
           The surfaces in left and right triplets of panels correspond to the global fits
           performed without and with ``softly anchored'' $M_0$, respectively.
          }
  \label{Fig:Chi2_M0E0_QES_5k_6k_3D}
  \end{figure}
  Figure~\ref{Fig:Chi2_M0E0_QES_5k_6k_3D} illustrates the impact of the anchoring of 
  $M_0$ on the $\chi^2$ values: the global minima of the $\chi^2$ for the standard
  fits without the anchoring (shown in three left panels) is not very distinct, while
  after the soft anchoring of $M_0$ to the deuterium value $M_A^\text{D}=1.003~\text{GeV}$
  the minima became distinctly visible.
 

  \begin{figure*}[!htb]
  \centering
  \includegraphics[width=\linewidth]
  {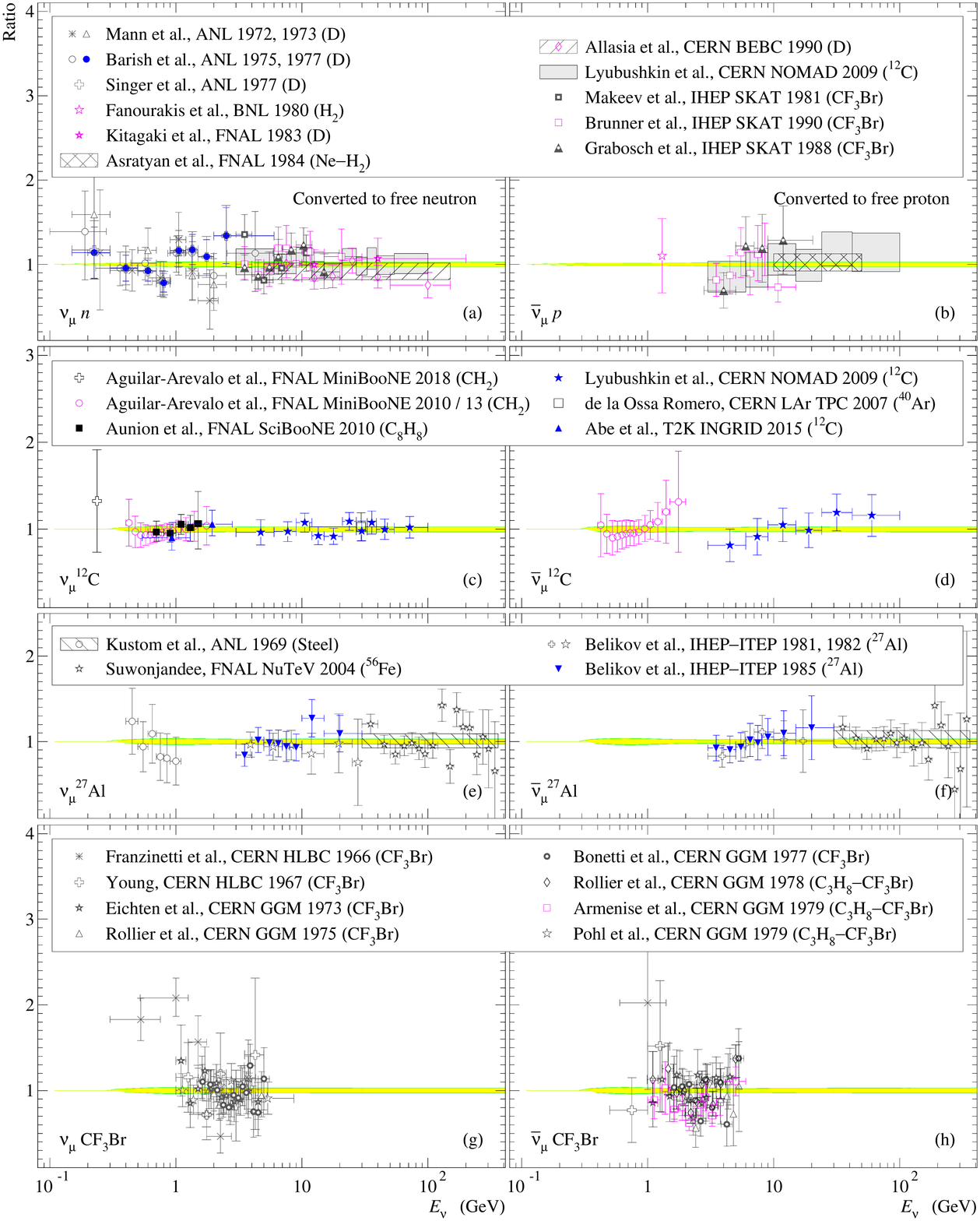}
  \caption{(Color online)
           Ratios of the total cross sections measured in different experiments
           to the corresponding predicted cross sections multiplied by the normalization factors $\mathcal{N}$
           listed in Table~\protect\ref{Tab:N_test}. See text for references.
          }
  \label{Fig:sQESCC_EXPvsTHE_101.3.31.301.6k_2_BBBA25_2}
  \end{figure*}

  \begin{figure*}[!htb]
  \centering
  \includegraphics[width=\linewidth]
  {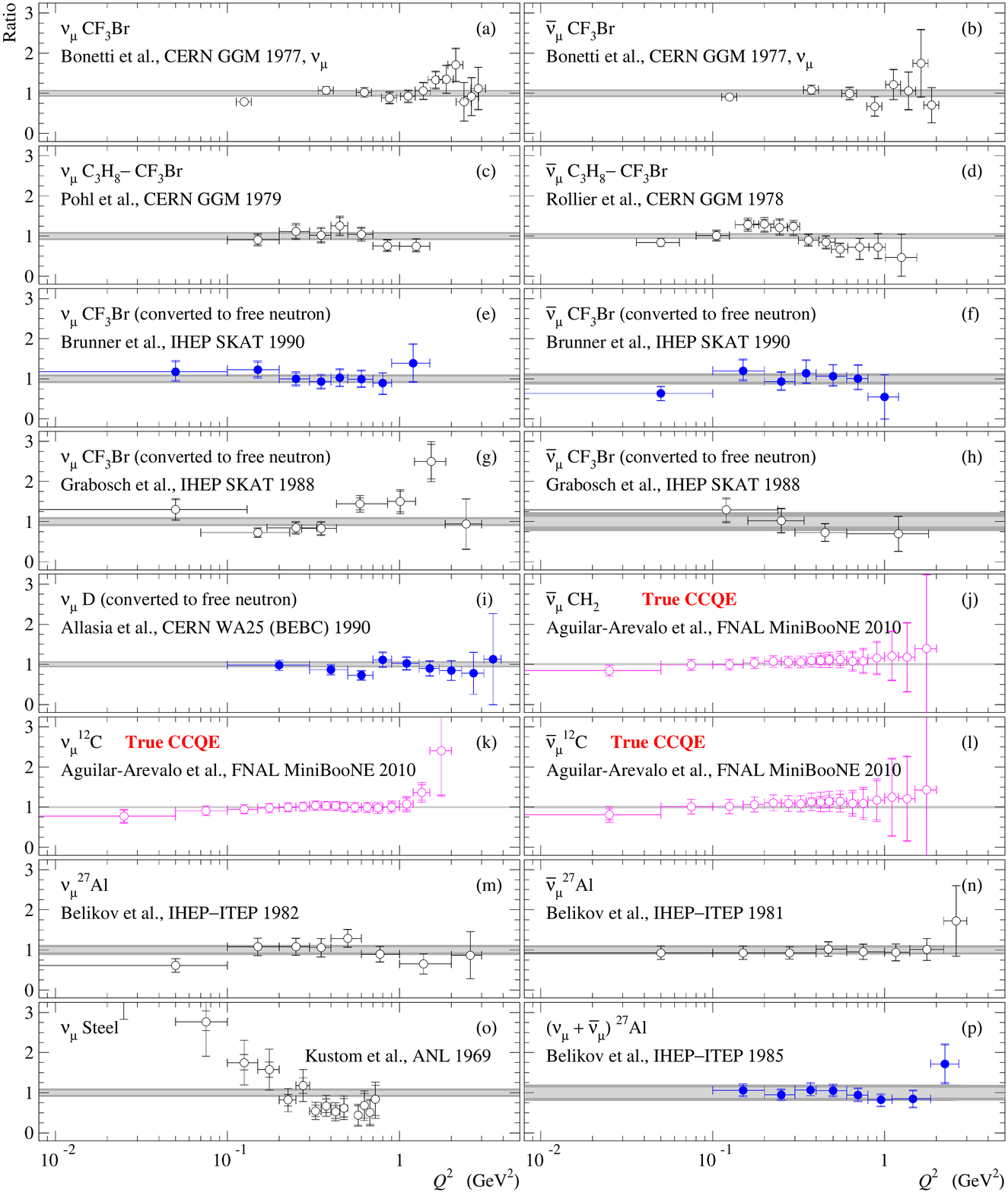}
  \caption{(Color online)
           Ratios of the differential cross sections $\langle{d\sigma/dQ^2}\rangle$ measured with
           ANL 1979 \cite{Kustom:1969dh},
           FNAL MiniBooNE 2010 \cite{Aguilar-Arevalo:2010zc}
                          2013 \cite{Aguilar-Arevalo:2013dva},
           CERN GGM 1977 \cite{Bonetti:1977cs},
                    1978 \cite{Rollier:1978kr},
                    1979 \cite{Pohl:1979zm},
                BEBC \cite{Allasia:1990uy},
           IHEP--ITEP 1981 \cite{Belikov:1981fq},
                      1982 \cite{Belikov:1981ut},
                      1985 \cite{Belikov:1983kg,Belikov:1985mw},
           IHEP SKAT 1988 \cite{Grabosch:1988js}, and
                     1990 \cite{Brunner:1989kw}
           to the corresponding predicted cross sections multiplied by the normalization factors $\mathcal{N}$
           listed in Table~\protect\ref{Tab:N_test}.
           The shaded double bands indicate the $1\sigma$ and $2\sigma$ uncertainties in the normalization factors.
           Only the data shown by filled symbols were included in the global fit. See text for more details.
          }
  \label{Fig:dsQESCC_dQ2_EXPvsTHE_2_101.1.30.301.8a_101.3.31.301.6k_BBBA25}
  \end{figure*}

  \begin{figure*}[htb]
  \centering
  \includegraphics[width=\linewidth]
  {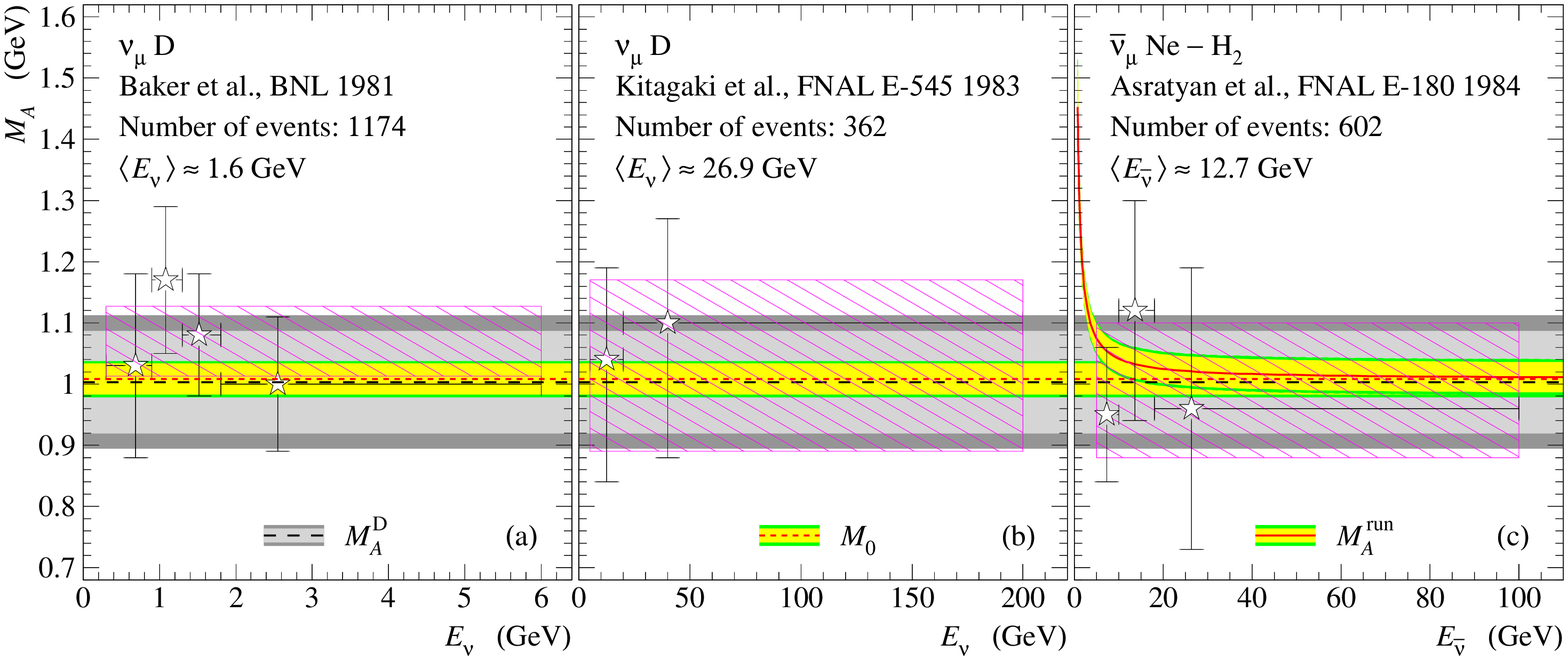}
  \caption{(Color online)
           Axial-vector mass, $M_A$, vs.\ (anti)neutrino energy extracted in the experiments
           BNL 1981 \cite{Baker:1981su} (a),
           FNAL E-545 (1983) \cite{Kitagaki:1983px} (b), and
           FNAL E-180 (1984) \cite{Asratyan:1984gh} (c).
           Number of events and estimated mean energies of the beams for each of the experiments
           are given in the legends.
           Hatched rectangles show the values of $M_A$ ($\pm1\sigma$) averaged over the full
           energy ranges (see text).
           Shaded bands around straight lines and around the curve in panel (c) indicate the
           $1\sigma$ and $2\sigma$ confidence intervals for the best-fitted values of $M^{\text{D}}_A$,
           $M_0$, and \MArun\ obtained in the global fits (see Eqs.~\eqref{MAD} and \eqref{Eqn:ME_QES_default}).
          }
  \label{Fig:MA_QES_Baker_BNL81_Kitagaki_FNAL83_Asratyan_FNAL84_101.3.31.301.6k_2_BBBA25}
  \end{figure*}

  A concentrated summary of comparison with the earlier data is presented in Figs.\ \ref{Fig:sQESCC_EXPvsTHE_101.3.31.301.6k_2_BBBA25_2}
  and \ref{Fig:dsQESCC_dQ2_EXPvsTHE_2_101.1.30.301.8a_101.3.31.301.6k_BBBA25}.
  Figure \ref{Fig:sQESCC_EXPvsTHE_101.3.31.301.6k_2_BBBA25_2} displays the ratios of the total CCQE cross sections measured
  in different experiments to the respective predicted cross sections renormalized to the data subsets.
  The following experiments are presented:
  ANL        1969 \cite{Kustom:1969dh},
             1972 \cite{Mann:1972},
             1973 \cite{Mann:1973pr},
             1975 \cite{Barish:1976dz} (see also \cite{Perkins:1975bj}), and
             1977 \cite{Singer:1977rs,Barish:1977qk};
  BNL        1980 \cite{Fanourakis:1980si},
  FNAL       1983 \cite{Kitagaki:1983px} and
             1984 \cite{Asratyan:1984gh},
  NuTeV      2004 \cite{Suwonjandee:2004aw};
  SciBooNE   2009 \cite{AlcarazAunion:2009ku,Aunion:2010zz},
  MiniBooNE  2010 \cite{Aguilar-Arevalo:2010zc},
             2013 \cite{Aguilar-Arevalo:2013dva}, and
			 2018 \cite{Aguilar-Arevalo:2018ylq},
  LAr-TPC    2007 \cite{MartinezdelaOssaRomero:2007oxj};
  CERN HLBC  1966 \cite{Franzinetti:1965},
       HLBC  1967 \cite{Young:1967ud},
       GGM   1973 \cite{Eichten:1973cs},
             1977 \cite{Bonetti:1977cs},
             1978 \cite{Rollier:1978kr}, and
             1979 \cite{Armenise:1979zg,Pohl:1979zm},
       BEBC  1990 \cite{Allasia:1990uy},
       NOMAD 2008 \cite{Lyubushkin:2008pe};
  IHEP--ITEP 1981 \cite{Belikov:1981fq},
             1982 \cite{Belikov:1981ut}, and
             1985 \cite{Belikov:1983kg,Belikov:1985mw};
  IHEP SKAT  1981 \cite{Makeev:1981em},
             1988 \cite{Grabosch:1988js}, and
             1990 \cite{Brunner:1989kw};
  T2K INGRID 2015 \cite{Abe:2015oar}. 
  The corresponding normalization factors, $\mathcal{N}$, are listed in Table~\ref{Tab:N_test}
  which is selfexplanatory.
  Notation used in panels (a), (b) and (c), (d) are the same as, respectively, in
  Figs.\ \ref{Fig:sQESCC_mn_n_C+ma_p_C_101.3.31.301.6k_2_BBBA25_1} and
         \ref{Fig:sQESCC_mn_n_D+ma_p_H_2_101.3.31.301.6k_2_BBBA25}.
  Here and in all figures below, the data points marked with filled symbols indicate the data included in the global fit.
  The vertical error bars and heights of the shaded rectangles in panels (a) and (b)
  represent the total errors which include the experimental normalization uncertainties.
  Narrow double bands represent the effect of $1\sigma$ and $2\sigma$ uncertainties in determination of the parameters $M_0$ and
  (for nuclei heavier than deuterium) $E_0$; the uncertainties in the normalization factors are not shown since they are
  in general different for the different data subsets shown in the same panel.
  It is seen that the ratios for deuterium and hydrogen are slowly sensitive to the uncertainties of $M_0$ at $E_\nu\lesssim200$~GeV
  and the ratios for the heavier nuclei are insensitive to the correlated uncertainties of $M_0$ and $E_0$ at $E_\nu\lesssim E_0\sim0.3$~GeV.
  The maximum sensitivity occurs at about $1$~GeV and does not exceeds $5$\%.
  The normalization factors for the total cross sections measured in the experiments 
  BNL 1980 \cite{Fanourakis:1980si},
  FNAL 1983 \cite{Kitagaki:1983px},
  MiniBooNE 2010 and 2013 \cite{Aguilar-Arevalo:2010zc,Aguilar-Arevalo:2013dva},
  CERN GGM  1979 \cite{Armenise:1979zg},
  CERN BEBC 1990 \cite{Allasia:1990uy}, and
  IHEP SKAT 1990 \cite{Brunner:1989kw}
  were obtained from the differential distributions included into the
  global fit since the total CCQE cross sections of the listed experiments were
  not included into the global fit.
  The respective absolute data and model predictions are shown
  in Figs.\ \ref{Fig:sQESCC_mn_n_C+ma_p_C_101.3.31.301.6k_2_BBBA25_1},
            \ref{Fig:sQESCC_mn_n_D+ma_p_H_2_101.3.31.301.6k_2_BBBA25}, and
            \ref{Fig:sQESCC_101.3.31.301.6k_2_BBBA25_NT} without renormalization.
  
  Similarly, Fig.\ \ref{Fig:dsQESCC_dQ2_EXPvsTHE_2_101.1.30.301.8a_101.3.31.301.6k_BBBA25}
  displays the ratios of the flux-weighted differential cross section $\langle{d\sigma/dQ^2}\rangle$
  measured in different experiments to the corresponding  predicted cross sections renormalized to the
  respective data subsets.
  The outer and inner vertical error bars represent the total errors of the experimental data
  and the total errors excluding the normalization uncertainties, respectively.
  In contrast to Fig.\ \ref{Fig:sQESCC_EXPvsTHE_101.3.31.301.6k_2_BBBA25_2},
  the shaded bands represent the effect of the uncertainties in the normalization factors and
  do not include the uncertainties of the parameters $M_0$ and $E_0$.
  The boundaries of the bands are calculated as
  $\pm \Delta{N}_{\nu, \overline{\nu}}/N_{\nu, \overline{\nu}}$,
  where $\Delta{N}_{\nu, \overline{\nu}}$ and $N_{\nu,\overline{\nu}}$ 
  are, respectively, the statistical uncertainty and best-fit value of the normalization factor.
  For the semi-sum of the $\nu_\mu$ and $\overline{\nu}_\mu$ cross sections measured in the IHEP-ITEP
  experiment \cite{Belikov:1983kg,Belikov:1985mw} (see panel (p)), the boundaries are calculated as 
 \[
  \pm\dfrac{\Delta{N_{\nu}}\;\langle{d\sigma_{\nu}/dQ^2}\rangle+\Delta{N_{\overline{\nu}}}\;\langle{d\sigma_{\overline\nu}/dQ^2}\rangle}
                  {N_{\nu} \;\langle{d\sigma_{\nu}/dQ^2}\rangle+       N_{\overline{\nu}} \;\langle{d\sigma_{\overline\nu}/dQ^2}\rangle}.
 \]
  Comparisons between the corresponding absolute data and the \SMRFG_MArun\ model predictions 
  are shown (without renormalization) in several figures in Sect.\ \ref{sec:MiniBooNE} (for the MiniBooNE CCQE data) 
  and in \ref{sec:GoldData} (for the earlier data).
  The ratios shown in Fig.\ \ref{Fig:dsQESCC_dQ2_EXPvsTHE_2_101.1.30.301.8a_101.3.31.301.6k_BBBA25}
  are obtained by dividing the experimental data in each bin by the predicted cross section integrated over
  the bin and multiplied by the normalization factor (see Table~\ref{Tab:N_test}) obtained for each
  experimental dataset from individual fits at fixed values of $M_0$ and $E_0$. 
  The slight difference in the normalization factors for the MiniBooNE data shown in Table~\ref{Tab:N_test} and
  in Tables \ref{Tab:MiniBooNE_Neutrino} and \ref{Tab:MiniBooNE_Antineutrino-CCQE}
  is explained below in Sec.~\ref{sec:MiniBooNE}.

  Figure \ref{Fig:MA_QES_Baker_BNL81_Kitagaki_FNAL83_Asratyan_FNAL84_101.3.31.301.6k_2_BBBA25}
  shows the current axial-vector mass, $M_A$, obtained in the three experiments
  at BNL \cite{Baker:1981su} and FNAL \cite{Kitagaki:1983px,Asratyan:1984gh} performed with
  bubble chambers filled with deuterium \cite{Baker:1981su,Kitagaki:1983px} and with a heavy neon-hydrogen
  mixture (64\% Ne) \cite{Asratyan:1984gh}. The data were extracted as functions of (anti)neutrino energy.
  Hatched regions indicate the average values of $M_A$ obtained for the full energy ranges:
  $\langle M_A \rangle=1.070\pm0.057$ GeV for $E_\nu=0.3-6$ GeV (a),
  $1.05_{-0.16}^{+0.12}$ GeV for $E_\nu=5-200$ GeV (b), and
  $0.99\pm0.11$ GeV for $E_{\overline\nu}=5-100$ GeV (c).
  The BNL result has been refined \cite{Kitagaki:1990vs} with increased statistics,
  $\langle M_A \rangle=1.070_{-0.045}^{+0.040}$ GeV for $E_\nu=0.5-6$ GeV.
  The cited results are consistent with each other and with the low-energy data points from the
  Argonne 12-foot bubble chamber filled with hydrogen and deuterium
  ($M_A=0.95\pm0.09$ GeV \cite{Barish:1977qk} and $1.00\pm0.05$ GeV \cite{Miller:1982qi};
  the neutrino energy spectrum peaks at $0.5$ GeV).
  It is important to keep in mind that the extractions in all these experiments were performed with different inputs.

  No significant variation of $M_A$ with energy is seen within rather wide energy range.
  The straight lines and curve with shaded bands represent the best-fit values of $M^{\text{D}}_A$, $M_0$, and \MArun\
  with corresponding $1\sigma$ and $2\sigma$ uncertainties drawn according to Eqs.\ \eqref{MAD} and \eqref{Eqn:ME_QES_default}. 
  As is seen, the straight lines (representing $M^{\text{D}}_A$ and $M_0$), are consistent with each other and with
  the deuterium data. They are also in very good agreement with the results of the former analyses
  \cite{Bodek:2007wzPrep,Bodek:2007ym,Bodek:2007vi,Kuzmin:2006dt_PAN,Kuzmin:2006dh_APPB,Kuzmin:2007kr}.
  The result of Ref. \cite{Asratyan:1984gh} (Ne-H$_2$ mixture) is formally consistent with both constant
  and running axial masses with about the same statistical significance.
  More comprehensive discussion of the earlier and current data on $M_A$ extractions is presented in Ref. \cite{Kakorin:2020atz}.

  \section{GENIE\,3 features}
  \label{sec:GENIE3}

  Let us discuss some features of the \GENIE3\ package that are essential for further consideration.
  The package provides a set of out-of-the-box comprehensive model configurations instead of a single ``default'' one,
  which has been a subject of customization by an user in the previous versions of \GENIEG.
  A set of models with their preset configurations is called ``tune''.
  \footnote{A full list of \GENIE3\ model configurations and tunes is maintained in URL: \protect\url{http://tunes.genie-mc.org}.}
  The advantages of this innovation is in that the user gets a self-consistent
  combination of the physical models in which a double-counting is excluded.
  Each tune is characterized by detailed MC comparisons with the data to which
  the input parameters were adjusted.
  Among the models presented in each tune, the most important are the models
  for elastic, quasielastic, and deep-inelastic scattering and resonance meson production.
  The tunes also include models for coherent and diffractive production of pions on nuclei,
  which make a small contribution to the CCQE-like background.
  Common to all tunes are the models for secondary interactions of hadrons inside nuclei.
  In addition to INTRANUKE -- the intranuclear transport simulation subpackage,
  which was the default in the \GENIEG\ releases prior to 3.0.0, its updated version
  (hereinafter referred to as INTRANUKE\,2018) is added.
  The INTRANUKE\,2018 consists of two models: \hA2018\ and \hN2018, whereas
  the former version of INTRANUKE included the only model called hA.
  The main distinctions between the models hA, \hA2018, and \hN2018\ are discussed below in some detail.

  \subsection{Final state interaction models}
  \label{sec:FSI}

  Neutrino interaction with a nucleon bound in a nucleus may produce secondary hadrons
  which then may interact (elastically or inelastically) on their way out of the nucleus.
  During the re-scattering, the secondaries can be absorbed, change their charge and 4-momentum, or produce new particles;
  besides multiple nucleons can be formed, causing spallation of the nucleus.
  The set of all these processes is commonly referred to as final state interaction (FSI) and its modeling is an important
  ingredient of any neutrino event generator \cite{Betancourt:2018bpu,Mosel:2019vhx}.
%
%
  It is amply clear that FSI change the distributions in outgoing nucleon variables, and, besides,
  may indirectly affect the distributions in terms of leptonic variables,
  because of experimental cuts imposed on the final-state nucleon variables affecting the event selection.
  One of the most important aspect of the FSI effects is in mimicking the CCQE topology
  by inelastic processes (e.g.\ pion production and absorption) inside the nucleus.


  The intranuclear cascade (INC) models treat the hadrons propagating in a nucleus 
  as free classical particles between two successive collisions
  on isolated nucleons bound in a potential well and undergoing Fermi motion.
  More sophisticated INC models (as, e.g., the model by Salcedo {\it et al.} Ref.~\cite{Salcedo:1987md})
  can account for the real part of hadron self-energy potential to simulate their elastic scattering on nucleons.
  In INC models, the probability per unit length of a certain reaction type
  (elastic scattering, charge exchange, pion absorption, pion production,
  etc.~\cite{Alam:2015nkk,Andreopoulos:2015wxa,GENIE_Manual_v3}) 
  is fundamental value which is determined by the corresponding cross sections and nuclear density.
%
%
  Although these models describe hadron scattering on nuclei quite well,
  the more accurate results can be reached by quantum-mechanical calculations
  (using a ``black disk'' approximation for quantum diffraction),
  when the interference between reaction types is accounted by summing of appropriate amplitudes.
%
%
  A pragmatic drawback of the quantum-mechanical calculations is the requirement of large
  computational resources, while the INC models are not so resource-intensive and in addition
  have the advantage that one can much easier trace propagation of hadrons in the nuclear medium.
  
  The intranuclear transport of hadrons and $\gamma$-quanta and their scattering are managed
  in \GENIE3\ \cite{GENIE_Manual_v3} by two INC simulation subpackages -- INTRANUKE (deprecated)
  and INTRANUKE\,2018.
%
  To determine the mean free paths of hadrons and $\gamma$s in nuclear medium, both INC subpackages use as input data
  the empirical cross sections for $p$, $\pi$, $K$, and $\gamma$ interactions with bare nucleons and nuclei; 
  the subpackage INTRANUKE\,2018 uses partially updated data and nuclear corrections of various kinds to the free nucleon cross sections.
  The nuclear densities are described by Gaussian or modified Gaussian shapes for nuclei with $A<20$, and
  for heavier nuclei -- by the Wood-Saxon density distribution dependent on surface thickness of the nucleus.

  The hA and \hA2018\ models are comparatively simple empirical models based on the total cross section data for different
  atomic nuclei, from helium to lead. Using these data, both models evaluate the probabilities of absorbing, generating,
  or recharging of nucleons, pions, and kaons with kinetic energies up to $1.2$~GeV.
  In the low-energy region of $(50-300)$ MeV, there are sufficient data for cascade modeling
  \cite{Ashery:1981tq, Navon:1983xj, Carroll:1976hj, Clough:1974qt, Bauhoff:1986gcb}
  but at higher energies, where the pion production probability becomes essential,
  the available data are rather fragmentary and thus the phenomenological cascade-exciton model CEM03
  \cite{Kerby:2014jra, Mashnik:2014eva, Mashnik:2016dmf}
  is applied. The cascade is modeled on an iron nucleus and the re-scaling factor $\propto A^{2/3}$
  is used to determine the cross sections on nuclei different from iron.
  The isospin symmetry is assumed in the models to recalculate the cross sections for $\pi^0$ from the charged pion data.
  The pion absorption is split into couple of different simulations: the absorption by two nucleons (by using the $\pi{d}\to{NN}$ data)
  and multi-nucleon absorption; the split probability is governed by empirical data.
  If there are two or more hadrons in the final state, the code distributes them evenly over the full phase space.


  The hA model can generate the hadron-nucleus elastic interactions when there is the probability
  that the scattered hadron gains energy due to the energy lost by the recoil nucleus.
  Since interaction of this type cannot be simulated in the INC spirit,
  it is added through an empirical model \cite{Dytman:2009zz}.
  Following recommendations of Ref.~\cite{Lu:2015tcr}
  the effect of energy gain by the scattered particle has been turned off in the \hA2018\ model because,
  it is assumed to be responsible for the discrepancy with the recent \MINERvA\ data~\cite{Lu:2018stk}.
  Other distinctions between the hA and \hA2018\ models consist in different values of the nucleon binding
  energies and in a 5~MeV correction applied for compound hadron cluster in the new model.
  For both models, in simulating the pion absorption, the binding energy is treated as a tunable parameter
  (for simplicity common to all nuclei) to fit the inclusive pion-nucleus scattering data from Ref.~\cite{Mckeown:1981pw}.
  The tuned value adopted in the hA (\hA2018) model is close to that of the SM RFG model for iron (carbon).

%

  The \hN2018\ model is fully INC model which allows to simulate all type of reactions for any nucleus and does not have the limitations
  peculiar to the \hA2018\ model. It utilizes the partial wave analysis of available data on the cross sections for $\pi N$, $KN$, and $NN$
  interactions as provided by the Scattering Analysis Interactive Dial-in (SAID) program
  \cite{Arndt:2003fj, Arndt:2006bf, GW_PWA}.
  For pions with kinetic energy below $350$~MeV the \hN2018\ uses the method of Ref.~\cite{Salcedo:1987md}
  which is based on ``$\Delta$ dominance'' (the pion-nucleus interactions are simulated through $\Delta(1232)$ excitation)
  and combines a microscopic field-theoretical calculation of the intrinsic probabilities for each reaction channels and
  a simulation procedure to follow the history of each pion.
  The ``$\Delta$ dominance'' model exploits the many-body techniques. 
  The calculations are performing in infinite nuclear matter with later correction for finite nuclei
  via the local density approximation and finite-range effects.
  The \hN2018\ model, unlike the \hA2018\ one, allows to simulate pre-equilibrium and compound nuclear states and
  the pion absorption is simulated as absorption on separate nucleons rather than on deuteron.
  The simulations of hadron transport through the nuclear medium have some common features in the \hN2018\ and \hA2018\ models.
  In particular, the interaction points in both models is estimated through a mean-field potential,
  the probability of interaction is determined in a similar way as value proportional
  to the mean free path, but the reaction channel is chosen on the basis of different input data. 

  \subsection{\G18\ tunes}
  \label{sec:GENIE3tune}

  For comparison purposes, we will deal with two of several comprehensive theory-driven \GENIE3\
  physics tunes designated as \GENIEa\ and \GENIEb\ (for short, these tunes will be sometimes abbreviated to \G18),
  incorporating, respectively, the \hA2018\ and \hN2018\ FSI models discussed in Sec.\ \ref{sec:FSI}.
  The remaining physical content of both tunes is quite the same.
  The CCQE sector is based on the model by Nieves \emph{et al.}~\cite{Nieves:2011pp} which is, in turn,
  an extension of the results of 
  Refs.~\cite{Gil:1997bm, Gil:1997jg, Nieves:2004wxWithErratum},
  where the quasielastic contribution to the inclusive electron and neutrino scattering on nuclei was studied.  
  The model is founded on a many-body expansion of the gauge boson absorption modes that includes $1p1h$, $2p2h$,
  and even $3p3h$ excitation mechanisms, as well as the excitation of $\Delta$ isobars.
  The $1p1h$ contribution is included within a local Fermi gas (LFG) picture incorporating several nuclear corrections,
  such as correct energy balance, long-range nuclear (RPA) correlations, Coulomb distortion, nuclear medium polarization,
  and dressing the nucleon propagators.
  
  The inelastic interactions (including coherent and diffractive meson production) in the ``\G18'' tunes are handled
  in an almost similar fashion as for the \MArun\ model. The main differences are in the Berger-Sehgal model
  \cite{Berger:2007rqWithErratum}
  for the resonance pion neutrinoproduction; in particular, the tunes use a bit refined version of the Rein-Sehgal
  normalization of the Breit-Wigner terms, and the vector and axial transition form factor model from 
  Ref.~\cite{Graczyk:2007bcWithErratum}.
  In contrast to the KLN-BS model, no Pauli blocking is included.
  
  All adjustable parameters, such as quasielastic and resonance axial-vector masses ($0.961$~GeV and $1.065$~GeV, respectively),
  cut in the invariant hadronic mass ($1.928$ GeV), specifications applied in the \NEUGEN\ neutrino generator \cite{Gallagher:2002sf}
  to the hadron multiplicity distributions
  (to avoid double counting of the DIS and resonance contributions), reweighing factors, etc.\ were tuned using all available
  data on the CCQE, CC$1\pi$, CC$2\pi$, and CC inclusive cross-sections and multiplicities.
  In our opinion, the most controversial point is the $12$\% overall reduction of the resonance production cross sections.
  Recall that the \MArun\ model does not use any reweighing factors and utilizes the common \GENIE3\ set of parameters
  excluding those mentioned in Sec.\ \ref{sec:Method}.
  
  Finally, it should be mentioned that all calculations with GENIE were done using the computational resources provided
  by the JINR cloud service
  \cite{Baranov:2016gvt, Balashov:2018}.


  \section{Comparison with recent data}
  \label{sec:Comparison}

  In this section, we compare in detail the contemporary comprehensive CCQE and CCQE-like data with the related predictions
  of the \SMRFG_MArun\ model (in conjunction with \hA2018\ and \hN2018\ FSI) and two \G18 physics tunes.
  In addition to that, we consider three more competing phenomenological models:
  \GiBUU~\cite{Gallmeister:2016dnq}, \SuSAv2MEC~\cite{Megias:2016fjk,Megias:2018ujz}, and
  \SuSAM~\cite{Amaro:2015zjaWithErratum,RuizSimo:2018kdl}.

  The \GiBUU\ model is a part of the \GiBUU\ (Giessen Boltzmann--Uehling--Uhlenbeck) project \cite{Buss:2011mx} and
  is an implementation of quantum-kinetic transport theory describing space-time evolution of a many-particle system 
  in a relativistic mean-field potential. The model is based on a refinement of the nuclear ground state and of $2p2h$
  interactions by using an empirical structure function that has separate momentum and energy dependence adjusted
  from electron scattering data. The $2p2h$ contribution is dependent on the target isospin, $\T$.
  Nucleon momenta come from an LFG distribution and the secondaries (also resonances) propagate through the residual nucleus
  in a nuclear potential that is consistent between initial and final state
  
  The \SuSAv2MEC\ model represents a fully relativistic approach, which exploits the scaling and superscaling \cite{Amaro:2004bs}
  features of inclusive electron scattering data \cite{Megias:2016lke} in order to predict neutrino-nucleus observables.
  The model is enhanced with the inclusion of relativistic mean-field theory effects and $2p2h$ axial and vector contributions
  to weak response functions in a relativistic Fermi gas. 
  Recently the \SuSAv2MEC\ model has been incorporated into the \GENIEG\ package \cite{Dolan:2019bxf},
  but in the present analysis we use only the original calculations from Ref. \cite{Megias:2016fjk}.

  The \SuSAM\ model is another representative of the superscaling approach, based on a scaling function extracted from
  a selection of electron-nucleus scattering data~\cite{Amaro:2018xdi}, and an effective nucleon mass inspired by the
  relativistic mean-field model; the effective mass phenomenologically incorporates the
  enhancement of the transverse current produced by the relativistic mean field and the
  scaling function merges nuclear effects beyond the impulse approximation, such as short-range correlations
  (responsible for tails in the scaling function).
  Note that the results of this model are usually presented as (rather wide) bands representing uncertainties in the implemented
  parameter set. In our estimations of the $\chi^2$ values (in exactly the same way as for other models) we use the central
  (rather than bin-averaged) values of the predicted cross sections and do not take into account the intrinsic model uncertainties.
  So the obtained $\chi^2$ values are significantly overestimated for the \SuSAM\ model and should be treated with care.
    A few more models will be discussed in subsections \ref{sec:T2K} and \ref{sec:MINERvA}.
  
  In order to provide the grounds for a quantitative comparison, in the subsequent discussion we employ the following definitions:
  \begin{align}
  \label{chi2_st}
  \chi^2_{\text{st}}  = &\ \left(\mathbf{E}-\mathbf{T}\right)^T\mathbf{W}^{-1}
                           \left(\mathbf{E}-\mathbf{T}\right),                                      \\
  \label{chi2_N}
  \chi^2_N            = &\ \left(\mathbf{E}-N\mathbf{T}\right)^T\widetilde{\mathbf{W}}^{-1}
                           \left(\mathbf{E}-N\mathbf{T}\right)+\frac{\left(N-1\right)^2}{\delta^2}, \\
  \label{chi2_log}
  \chi^2_{\text{log}} = &\ \left(\ln\mathbf{E}-\ln\mathbf{T}\right)^T\widetilde{\mathbf{V}}^{-1}
                           \left(\ln\mathbf{E}-\ln\mathbf{T}\right).
  \end{align}
  Here $\mathbf{E}$ and $\mathbf{T}$ are, respectively, the vectors of bin-averaged experimental data, $E_i$,
  and corresponding model predictions, $T_i$;
  $\ln\mathbf{E}$ and $\ln\mathbf{T}$ are, respectively, the vectors with the components $\ln(E_i)$ and $\ln(T_i)$;
  $\mathbf{W}=||W_{ij}||$ is the full covariance matrix, $\widetilde{\mathbf{W}}=\mathbf{W}-\mathbf{W}_{\text{norm}}$,
  where $\mathbf{W}_{\text{norm}}$ is the matrix of covariances due to the normalization uncertainties; 
  $\mathbf{V}=||W_{ij}/E_iE_j||$; $\delta_i$ is the (dimensionless) normalization uncertainty.
  Note that $\chi^2_{\text{log}}\approx\chi^2_{\text{st}}$ as $|T_i-E_i| \ll E_i$ for all $i$.
  We will mainly use the log-normal $\chi^2$s \eqref{chi2_log} for comparisons with the T2K and \MINERvA\ data.
  Very isolated singular contributions to $\chi^2_{\text{log}}$ are merely excluded from the analysis.
  
  The normalization factor in Eq.~\eqref{chi2_N} is defined by the minimization condition $\partial\chi^2_N/\partial N=0$:
  \begin{equation}
  \label{NormalizationWithCorrelations}
  N = \mathcal{N} = \frac{1+\delta^2\mathbf{T}^T\widetilde{\mathbf{W}}^{-1}\mathbf{E}}
                         {1+\delta^2\mathbf{T}^T\widetilde{\mathbf{W}}^{-1}\mathbf{T}}
  \end{equation}
  (cf.\ with Eq.~\eqref{Normalization_Factor}). From Eqs.~\eqref{chi2_N} and \eqref{NormalizationWithCorrelations}
  it follows that
  \begin{equation}
  \label{chi2_mathcalN}
  \chi^2_{\mathcal{N}} = \chi^2_1-\left(\mathcal{N}-1\right)^2
  \left(\frac{2}{\delta^2}+\mathbf{T}^T\widetilde{\mathbf{W}}^{-1}\mathbf{T}\right).
  \end{equation}
  It is clear that ``good'' values of $\chi^2_{\mathcal{N}}$ would indicate agreement
  in shape between the data and model prediction provided that $|\mathcal{N}-1|\lesssim\delta$.
%
%
  Unfortunately the formal solution \eqref{NormalizationWithCorrelations}
  (with some modifications due to the lack of the matrix $\mathbf{W}_{\text{norm}}$) as a rule appears
  to be unphysical for the T2K data on hydrocarbon and \MINERvA\ data 
  (see Sec.~\ref{sec:T2K_hydrocarbon}  and also Ref.~\cite{Bonus:2020yrd} where similar problem is discussed).
  Moreover, owing to strong correlations the standard $\chi^2$ values for the T2K C$_8$H$_8$ and \MINERvA\ data may
  sometimes look counterintuitive. So a contrasting of the standard and log-normal estimators would provide useful
  additional control.  

  Many data subsets (not necessarily the true CCQE ones) are reported in terms of reconstructed neutrino energy
  $E^{\text{QE}}_{\nu}$ and square of the four-momentum transferred $Q^2_{\text{QE}}$ for the simplest case when
  the target nucleon is at rest and there is no nuclear effects, except binding. 
  These ``quasielastic'' variables are defined through the muon kinematics which can be reconstructed squarely
  and without reference to an interaction model: 
  \begin{equation}
  \label{eq:QE-quantities}
  \begin{aligned} 
  E^{\text{QE}}_{\nu} = &\ \frac{M'^2-(M-E_b)^2-m_{\mu}^2+2(M-E_b)E_{\mu}}{2(M-E_b-E_{\mu}+P_{\mu}\cos\theta_{\mu})}, \\
  Q^2_{\text{QE}}     = &\ 2E^{\text{QE}}_{\nu}(E_{\mu}-P_{\mu}\cos\theta_{\mu}).
  \end{aligned}
  \end{equation}
  Here $m_{\mu}$, $M$, and $M'$ are the masses of muon, incoming, and outgoing nucleon, respectively;
  $E_b$ is the initial-state nucleon's removal energy
  (in fact an effective parameter tunable to reproduce the experimental data and thus in general different
  in different experiments, even those that work with targets of the same composition); 
  $E_{\mu}$, $P_{\mu}$, and $\theta_\mu$ are, respectively, the total energy, absolute value of the momentum, and
  scattering angle of the outgoing muon.
  Normally, the quantity $E^{\text{QE}}_{\nu}$ closely correlates with the true neutrino energy $E_\nu^{\text{True}}$ 
  (see, e.g., Fig.~12 in Ref.~\cite{Patrick:2018gvi}) and in any case the comparisons between the data and predictions
  expressed in terms of exactly the same well-defined variables do not introduce additional uncertainties or indeterminations.
 
  The comparisons of the model predictions with selected recent data are presented in Figs.\
  \ref{Fig:rRELCC_QELCC_Arevalo_MiniBooNE09_103.1.31.301.h4_103.1.31.101.k2_2_BSc_101.3.31.301.2k_2_BBBA25}--%
  \ref{Fig:MINERvA2018dQ2_E_QE} and in 
  Tables \ref{Tab:MiniBooNE_Neutrino}--\ref{Tab:MINERvA-1d_Q2_Neutrino-CCQE+CCQE-like-standard+lognorm},  
  which will be discussed in detail below in this section.
  The normalization factors, $\mathcal{N}$, shown in the Tables, are written with three digits in the mantissa.
  This certainly excessive precision is only needed in order to distinguish (usually small) differences between the two FSI models.
  Recall that the data of ND280 \cite{Abe:2017rfw,Abe:2018pwo} and \MINERvA\ \cite{Patrick:2018gvi,Ruterbories:2018gub},
  as well as an essential part of the MiniBooNE datasets~\cite{Aguilar-Arevalo:2010zc,Aguilar-Arevalo:2013dva}
  were not involved into the statistical analysis. These data are shown in the figures by open symbols.

  \subsection{MiniBooNE}
  \label{sec:MiniBooNE}

  In this section, we discuss a detailed comparison between the data from the Mini Booster Neutrino Experiment FNAL-E-0898
  (MiniBooNE) \cite{Aguilar-Arevalo:2010zc,Aguilar-Arevalo:2013dva} and respective predictions of the \SMRFG_MArun\ model,
  two \G18\ tunes and besides (for the true CCQE scattering data only) of the
  \SuSAv2MEC \cite{Megias:2016fjk}, \SuSAM \cite{RuizSimo:2018kdl,RuizSimo:2020}, and \GiBUU \cite{Gallmeister:2016dnq}.



  The ultra-pure mineral oil filling the MiniBooNE detector medium is composed of C$_n$H$_{2n+2}$ molecules
  with the length of the carbon chain $n=20-40$ ($\sim33$ on average, although the values differ in the literature).
  Since the cross sections only depend on the relative amount of each atomic species,
  in our simulations with \GENIE3, the $\nu/\overline\nu$ interaction is chosen to be an ``average'' single unit on the
  hydrocarbon chain $n(\text{CH}_{2.06})$. Hereafter, for brevity's sake, we will refer to the MiniBooNE oil as $\text{CH}_2$.
  For the \GiBUU\ and superscaling models under consideration, only the calculations for pure carbon target are currently available.
  
  The MiniBooNE event sample was selected by requiring a single well-reconstructed muon, and no final-state pions.
  The published dataset \cite{Aguilar-Arevalo:2010zc,Aguilar-Arevalo:2013dva,MiniBooNEdata} consists of both CCQE-like
  and CCQE-corrected cross sections.
  The former sample includes not only the FSI contributions but also complicated instrumental and methodical effects
  and the CCQE sample is cleared of it all; in particular, the contributions of single pion interactions in carbon is
  removed according to the Rein-Sehgal (RS) model \cite{Rein:1980wg} as it implemented into the \NUANCE\ MC neutrino event
  generator used in the MiniBooNE analyses.
  The $\overline\nu_\mu$ dataset \cite{Aguilar-Arevalo:2013dva} includes the CCQE and CCQE-like samples on both
  mineral oil and on carbon by subtracting the $\overline\nu_\mu$ CCQE events on hydrogen.
  
  \subsubsection{Treatment of CCQE-like backgrounds}
  \label{sec:MiniBooNE_Backgrounds}

\begin{figure}[!htb]
  \centering
  \includegraphics[width=\linewidth]
  {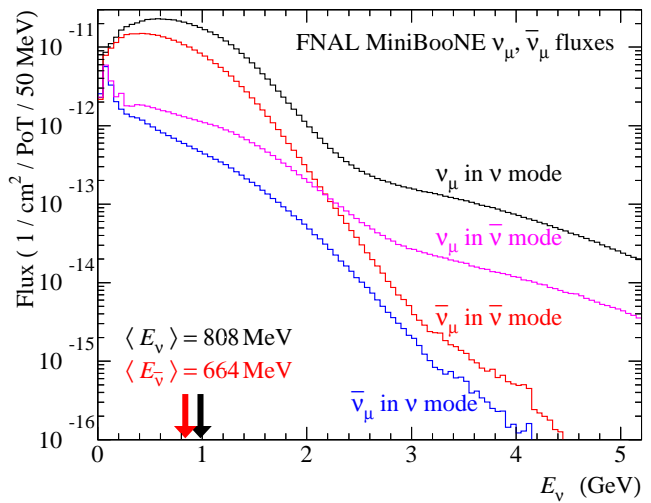}
  \caption{(Color online)
           Total $\nu_\mu$ and $\overline\nu_\mu$ flux predictions at the MiniBooNE detector with horn
           running in neutrino and antineutrino modes~\cite{Aguilar-Arevalo:2008yp,Aguilar-Arevalo:2011sz}.
          }
  \label{Fig:Fmn_Fma_MiniBooNE}
  \end{figure}

  Although the CCQE-like measurements are certainly less model-dependent than the CCQE measurements, their reliable
  modeling is complicated by indirect and very detector-dependent statistical analysis used in the MiniBooNE data
  processing for finding the CCQE-like backgrounds and by complex flavor composition of the NuMI beam in the $\nu$
  and $\overline\nu$ modes (see Fig.~\ref{Fig:Fmn_Fma_MiniBooNE}) and these complications are highly interconnected.
  The CCQE-like background in the $\nu$ mode is dominated by single positive pions.
  The procedure for selecting the CCQE sample and measuring the CC$1\pi^+$ background includes five
  steps \cite{Aguilar-Arevalo:2010zc}:
  (i)   selection of events with a clean CC signature;
  (ii)  separation of the selected CC sample into CCQE and CC$1\pi^+$ subsamples using a sequence of cuts;
  (iii) measurement of the CC$1\pi^+$ rate from the second subsample;
  (iv)  adjustment of the RS model used in the \NUANCE\ based CC$1\pi^+$ events simulation,
        to reproduce the measured CC$1\pi^+$ rate; and
  (v)   subtraction of the adjusted CC$1\pi^+$ background (and other predicted backgrounds) from the
        CCQE signal to produce the CCQE cross section.
 
  The background estimation for the $\overline\nu$ mode is more complicated.
  As is evident from Fig.~\ref{Fig:Fmn_Fma_MiniBooNE}, the $\nu_\mu$ contribution to the $\overline\nu$-mode flux
  is much more significant compared to the $\overline\nu_\mu$ component of the $\nu$-mode beam:
  the ratio of the integral $\nu_\mu$ and $\overline\nu_\mu$ fluxes in the $\nu$ mode is about $2.3$
  times larger than the inverse ratio in the $\overline\nu$ mode.
  This is mainly because of the leading-particle effect in hadronic interactions: proton-beryllium collisions,
  forming the $\nu_\mu$ and $\overline\nu_\mu$ beams, preferentially produce about twice as many $\pi^+$ as $\pi^-$
  \cite{Aguilar-Arevalo:2010zc}.
  Moreover, the overall contamination rate in MiniBooNE is much more significant in the $\overline\nu$-mode since
  the $\nu_\mu$ cross section is about thrice higher than the $\overline\nu_\mu$ ones in the $\sim1$~GeV energy range.
  The \NUANCE\ simulation predicts \cite{Aguilar-Arevalo:2010zc} that $\overline\nu_\mu$ events account for $\sim1$\%
  of the $\nu$-mode event sample, while $\nu_\mu$ events are the cause of about $30$\% of the full $\overline\nu$-mode sample.
  The fraction of all non-$\nu_\mu$ events in the final CCQE event sample in the $\nu$ mode is $1.4$\% \cite{Aguilar-Arevalo:2010zc},
  while in the $\overline\nu$ mode, the contribution of all $\nu_\mu$ events to the $\overline\nu_\mu$ CCQE sample
  is about $22$\% \cite{Grange:2015}. 
  Three independent methods were used to constrain, tune, and subtract the $\nu_\mu$ contamination prediction~\cite{Aguilar-Arevalo:2011sz}.


  The CCQE-like measurements exclude the $\overline\nu/\nu$ content of the subtracted data in the $\nu/\overline\nu$ mode.
  This, we repeat, is rather model- and detector-dependent procedure and it is difficult to reproduce this part of the
  MiniBooNE analysis in a third-party simulation. In order to avoid the unrealizable full reanalysis of the MiniBooNE data
  but without neglecting the above-mentioned features of the experiment, we use a simplified approach.
  Namely, we simulate the CCQE-like contributions in \GENIE3\ by using the RS model for the pion production with no cut on
  the hadronic invariant mass, $W$, and with all the input parameters chosen as close as possible to those adopted
  in the \NUANCE\ neutrino event generator used by the MiniBooNE experiment~\cite{Aguilar-Arevalo:2010zc,Aguilar-Arevalo:2013dva}.
  Then we form the ratios
  \begin{equation}
  \label{CCQE-likeCorrectionFactor}
  \frac{\text{CCQE-like contribution with \GENIEG\ setting}}
       {\text{CCQE-like contribution with \NUANCE\ setting}}
  \end{equation}
  for the cross sections of each type and for all model under consideration, and multiply it (bin by bin) to the corresponding
  full CCQE-like background contribution reported in Refs.~\cite{Aguilar-Arevalo:2010zc,Aguilar-Arevalo:2013dva}.
  This approach allows us to take into account all ``instrumental'' corrections and tunes, as well as the individual features
  of the investigated models for the inelastic interactions and FSI effects (see Sec.\ \ref{sec:GENIE3}).
  Although the \GENIE3\ generator cannot perfectly reproduce all the details of the hadronization model, models for shallow
  inelastic scattering (SIS) and other, less important but not negligible issues implemented into the \NUANCE\ generator
  \footnote{For example, in calculations of the denominator of Eq.\ \protect\eqref{CCQE-likeCorrectionFactor}, we use
            the \hA2018\ FSI model; \GENIE3\ treatment of the resonance to DIS transition region is also different
            from what used in the \NUANCE\ generator.},
  the expected overall impact of the differences to the CCQE-like cross sections is insignificant and can be neglected
  in the $\chi^2$ tests discussed below.
  
\begin{figure}[!hbt]
  \centering
  \includegraphics[width=\linewidth]
  {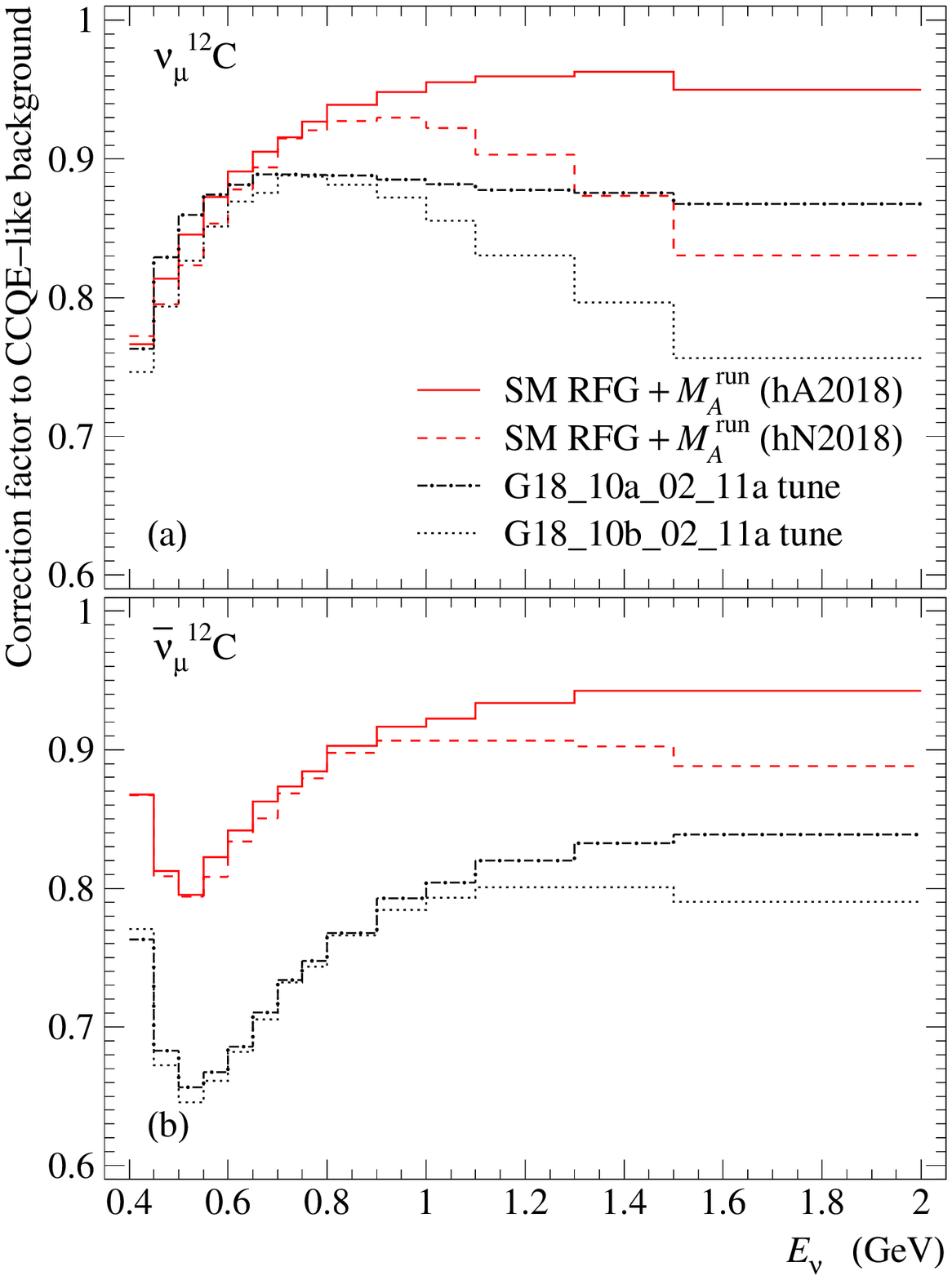}
  \caption{(Color online)
           The correction factors \eqref{CCQE-likeCorrectionFactor} for the $\nu_\mu{}^{12}$C and $\overline\nu_\mu{}^{12}$C
           total CCQE cross sections calculated using the settings for the \SMRFG_MArun\ model with \hA2018\ and \hN2018\
           FSI contributions, \GENIEa, and \GENIEb\ tunes.
          }
  \label{Fig:sQELCC_BSc_R+S_MiniBooNE10_13_101.3.31.301.6k_2_BBBA25}
  \end{figure}

\begin{figure}[!hbt]
  \centering
  \includegraphics[width=\linewidth]
  {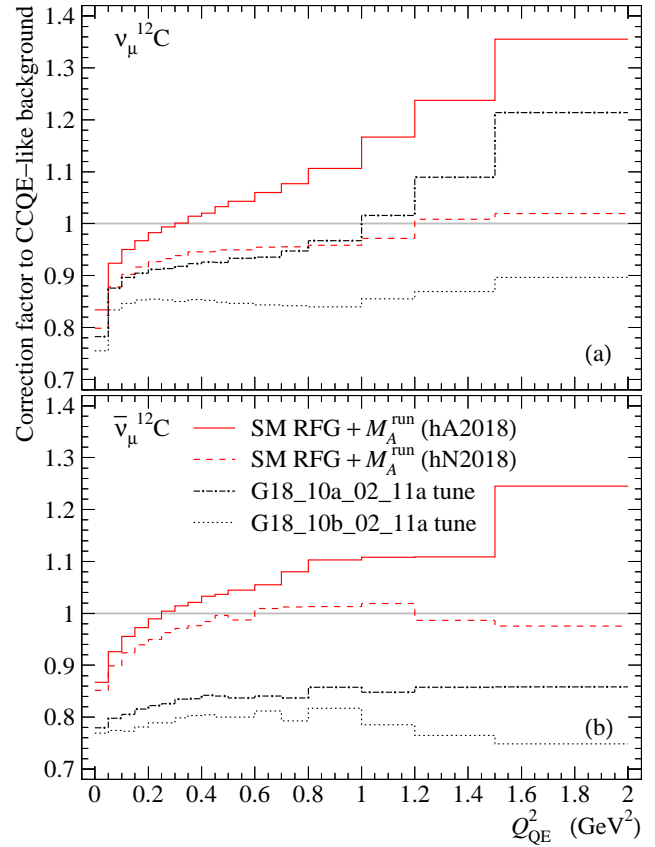}
  \caption{(Color online)
           The correction factors \eqref{CCQE-likeCorrectionFactor} for the $\nu_\mu{}^{12}$C and $\overline\nu_\mu{}^{12}$C
           differential CCQE cross sections, $d\sigma/dQ_{\text{QE}}^2$
           (see Fig.\ \protect\ref{Fig:dsQES_L_CC_dQ2_Arevalo_MiniBooNE10_13_101.3.31.301.6k_2_BBBA25_1}\ (b, f))
           calculated using the settings for the \SMRFG_MArun\ model with \hA2018\ and \hN2018\ FSI contributions,
           \GENIEa, and \GENIEb\ tunes.
          }
  \label{Fig:dsQELCC_dQ2_BSc_R+S_Arevalo_MiniBooNE10_13_101.3.31.301.6k_2_BBBA25}
  \end{figure}

\begin{figure}[!hbt]
  \centering
  \includegraphics[width=\linewidth]
  {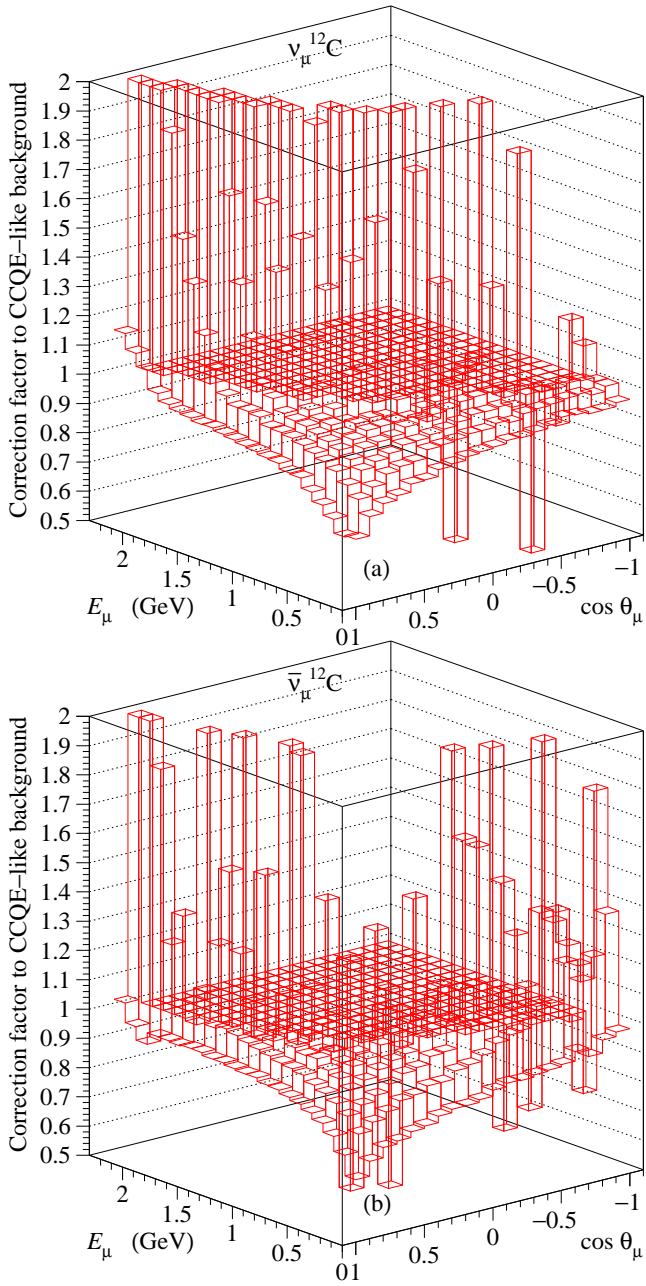}
  \caption{(Color online)
           The correction factors \eqref{CCQE-likeCorrectionFactor} for the $\nu_\mu{}^{12}$C and $\overline\nu_\mu{}^{12}$C
           double-differential CCQE cross sections, $d^2\sigma_{\nu}/dE_{\mu}d\cos\theta_{\mu}$ 
           (see Figs.\ \protect\ref{Fig:d2sQESCC_dEkdcosT_mn_n_CH_2.06} and
                       \protect\ref{Fig:d2sQESCC_dEkdcosT_ma_p_CH_2.06})
           calculated using the settings for the \SMRFG_MArun\ model with \hA2018\ FSI contributions.
          }
  \label{Fig:d2sQELCC_dEkdcosT_BSc_R+S_Arevalo_MiniBooNE10_13_101.3.31.301.6k_2_BBBA25_1_3D}
  \end{figure}

  Figures \ref{Fig:sQELCC_BSc_R+S_MiniBooNE10_13_101.3.31.301.6k_2_BBBA25},
          \ref{Fig:dsQELCC_dQ2_BSc_R+S_Arevalo_MiniBooNE10_13_101.3.31.301.6k_2_BBBA25}, and
          \ref{Fig:d2sQELCC_dEkdcosT_BSc_R+S_Arevalo_MiniBooNE10_13_101.3.31.301.6k_2_BBBA25_1_3D}
  show the ratios \eqref{CCQE-likeCorrectionFactor} representing the correction factors to the full CCQE-like backgrounds
  for the total, single-differential, and double-differential CCQE cross sections, respectively.
The 3D histograms in Fig.\ \ref{Fig:d2sQELCC_dEkdcosT_BSc_R+S_Arevalo_MiniBooNE10_13_101.3.31.301.6k_2_BBBA25_1_3D}
are cropped at the top to avoid showing the unphysical peaks that occur near the kinematic boundaries
(where the generated number of events is very small) due to statistical fluctuations in the Monte Carlo simulations.


It is seen that the correction factors for the backgrounds to the total CCQE cross sections are systematically less
than $1$ all our models (Fig.\ \ref{Fig:sQELCC_BSc_R+S_MiniBooNE10_13_101.3.31.301.6k_2_BBBA25});
at energies below $0.8-0.9$ GeV they slowly depend of the FSI model but at higher energies the differences become
more essential.
In case of the \G18\ model, the same is also true for the $\overline\nu_\mu{}^{12}$C cross sections of all kinds.
The correction factors for the single- and double-differential CCQE cross sections
(see Figs.\ \ref{Fig:dsQELCC_dQ2_BSc_R+S_Arevalo_MiniBooNE10_13_101.3.31.301.6k_2_BBBA25} and
            \ref{Fig:d2sQELCC_dEkdcosT_BSc_R+S_Arevalo_MiniBooNE10_13_101.3.31.301.6k_2_BBBA25_1_3D})
either decrease or increase the CCQE-like backgrounds in comparison to the \NUANCE\ predictions,
depending on the kinematic region, but the regions where the correction factors are $\gtrsim1$ make small contributions
to the cross sections and hence, integrally the corrections reduce the MiniBooNE estimations of the CCQE-like backgrounds.
Everywhere, the \hA2018\ FSI model provides larger corrections then the \hN2018\ model.
The \SMRFG_MArun\ inelastic contributions are either similar to (at very low energies in the $\nu$ mode) or
(everywhere in the $\overline\nu$ mode) larger than those estimated using the \G18\ inputs.

  \subsubsection{RES to QES Ratio}   
  \label{sec:MiniBooNE_RES2QESrat}
  
\begin{figure}[!htb]
  \centering
  \includegraphics[width=\linewidth]
  {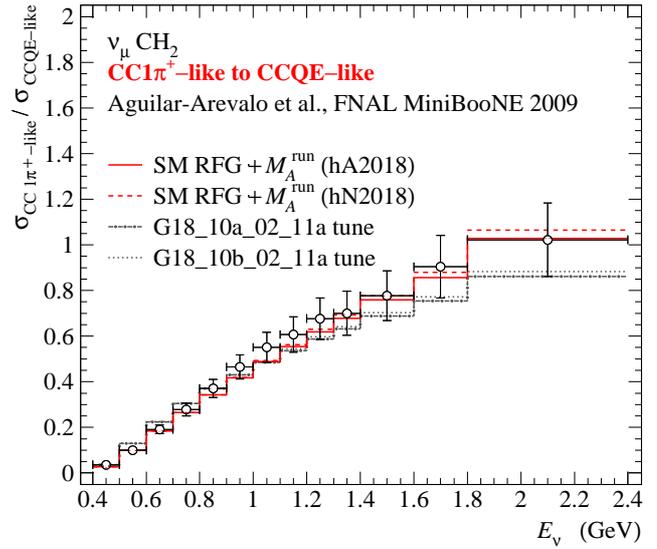}
  \caption{(Color online)
            Ratio of the total cross sections for CC singe $\pi^+$ neutrinoproduction and
            CCQE-like neutrino scattering on mineral oil
            measured with MiniBooNE~\cite{Aguilar-Arevalo:2009eb} vs.\ true neutrino energy.
            Vertical error bars include all sources of statistical and systematic uncertainty.
            Histograms show predictions of the \SMRFG_MArun\ model and \G18\ tunes with the two versions
            of the FSI effect simulation.
           }
  \label{Fig:rRELCC_QELCC_Arevalo_MiniBooNE09_103.1.31.301.h4_103.1.31.101.k2_2_BSc_101.3.31.301.2k_2_BBBA25}
  \end{figure}

  As an illustration of the accuracy of the CC1$\pi^+$ background simulation, we present in
  Fig.~\ref{Fig:rRELCC_QELCC_Arevalo_MiniBooNE09_103.1.31.301.h4_103.1.31.101.k2_2_BSc_101.3.31.301.2k_2_BBBA25}
  a comparison between the MiniBooNE data from Ref.~\cite{Aguilar-Arevalo:2009eb} and \GENIE3\ predictions
  (using the four models under discussion) for the ratio of the total CC$1\pi^+$-like and CCQE-like $\nu_{\mu}$CH$_2$
  cross sections. The ratios are plotted as functions of the true neutrino energy.
  This ratio is very convenient in that it is almost independent of the neutrino flux normalization and,
  to a lesser degree, contaminations.
  In all calculations with the \SMRFGpMArun\ model, we used the updated KLN-BS singe-pion production model
  shortly described in Sec.\ \ref{sec:Method}.
  It is seen that the predicted ratios are slowly sensible to the FSI models and for both of them
  are in good agreement with the data
  ($\chi_{\text{st}}^2/\ndf=0.55$ and $0.48$ for \hA2018\ and \hN2018, respectively).
  Agreement with the \G18\ tunes is worse
  ($\chi_{\text{st}}^2/\ndf=1.37$ and $1.30$ for \GENIEa\ and \GENIEb, respectively),
  but still quite acceptable. 
  Let us remind that the \GENIE3\ generator does not account for the interference between the resonances.
  We checked how this affects the ratio by applying a simple reweigting procedure and found that
  the interference effect is comparatively small but, unfortunately, slightly worsens
  the agreement of the \SMRFGpMArun\ model with the data; in terms of the standard $\chi^2$s we got:
  $\chi_{\text{st}}^2/\ndf=0.85$ and $0.76$ for, respectively, \hA2018\ and \hN2018\ FSI.
  The \G18\ tunes are less sensitive to the interference among the resonance amplitudes: after accounting
  for this effect, we obtained $\chi_{\text{st}}^2/\ndf=1.41$ and $1.33$ for, respectively, \GENIEa\ and \GENIEb.
  The neutrino flux uncertainties are largely canceled in the ratio, and hence
  the good accord with the data provides an additional evidence in favor of the CCQE, CC$1\pi^+$,
  and FSI models under consideration.

  The MiniBooNE Collaboration also reported the CC$1\pi^+$ to CCQE cross section ratio rescaled to an isoscalar target
  and the FSI corrected ratio.
  But the treatment of the FSI effects in Ref.~\cite{Aguilar-Arevalo:2009eb}
  is so different from that in \GENIE3\ that comparing these results with the \SMRFG_MArun\ models
  is not very interesting.
  For completeness, we add that we also tested the KLN-BS model using the more recent MiniBooNE measurements
  of the total CC1$\pi^+$ cross section~\cite{Aguilar-Arevalo:2010bm}. The comparison in particular demonstrates
  that (for both \hA2018\ and \hN2018\ FSI) the model slightly underestimates the data
  (with $\chi_{\text{st}}^2/\ndf=1.07$), but is in very good agreement with the cross-section shape in neutrino energy
  ($\chi_{\text{st}}^2/\ndf=0.40$).

  \subsubsection{Technical notes}
  \label{sec:MiniBooNE_TechNotes}
  
  A few more specialized remarks are necessary:
  
  (i) In our simulations, the (anti)neutrino flux averaging of the single- and double-differential cross sections
  is performed over the full energy range \cite{Aguilar-Arevalo:2008yp,Aguilar-Arevalo:2011sz}
  and thus our estimation of the mean $\nu_\mu$ and $\overline\nu_\mu$ beam energies (see legend in Fig.~\ref{Fig:Fmn_Fma_MiniBooNE})
  is slightly different from those given in Refs.~\cite{Aguilar-Arevalo:2010zc,Aguilar-Arevalo:2013dva} for the energy range
  $0<E_{\nu,\overline\nu}<3$~GeV. The spectral ``tails'' above $3$~GeV marginally affect the flux-folded CCQE
  cross sections but noticeably affect the CCQE-like background estimations, especially at high $Q^2$. 

  (ii) According to parameterization \eqref{pFEb} the separation energy for carbon is $E_b^{\text{C}}\approx25$~MeV.
  This value corresponds to one extracted from electron-nucleus scattering data \cite{Moniz:1971mt}
  within the RFG model and hence we use it in all our simulations for the targets containing carbon.
  At the same time, for more accurate comparison of our calculations with the MiniBooNE data reported in terms
  of $E_\nu^{\text{QE}}$ or $Q_{\text{QE}}^2$, the value of $E_b^{\text{C}}$ in Eqs.~\eqref{eq:QE-quantities} is set to
  $34$~MeV ($\pm9$~MeV) -- the value used in the MiniBooNE analyses.
 
  (iii) The global fit of the running axial mass described in Sec.\ \ref{sec:StatisticalAnalysis} is rather computer time consuming.
  To reduce the computing time, we used accurate analytical parametrizations for the $\nu_\mu$ and $\overline{\nu}_\mu$ energy spectra.
  The smooth (rational) parametrizations of the spectra, as opposed to the step-like dependences, significantly facilitate
  on-the-fly numerical integration. 
  In contrast to this, in all our calculations with \GENIE3, we used the energy spectra presented as histograms.
  The difference between the total cross sections computed using the analytical parametrizations
  and histogrammic representation of the spectra is about $1$\%.
  This is one of the reasons why the normalization factors obtained for the flux-averaged CCQE double-differential
  cross sections presented in Tables \ref{Tab:MiniBooNE_Neutrino} and \ref{Tab:MiniBooNE_Antineutrino-CCQE}
  and those listed in Table \ref{Tab:N_test} are somewhat different.

  
  Another reason of the marginal ($\lesssim 1\%$) differences is that in the global and individual fits, from which the
  normalization factors appeared in Table \ref{Tab:N_test} were extracted, we used the CCQE data, which were singled out by
  the experimental methods (+ \NUANCE\ modelling), while in the calculations with \GENIE3, we dealt with the true CCQE events.
  In other words, the definitions of the CCQE events are not fully identical in the MiniBooNE analysis and in the \GENIE3\
  simulation, due to different treatments of the FSI effects.

  (iv) Estimations of the $\chi^2$ values and normalization factors \Nbestfit\ may generally depend on the choice
  of the normalization uncertainties, $\delta$, which cannot always be unambiguously fixed. We tested however that
  for all models under consideration, the $\chi_{\mathcal{N}}^2$ values decrease very slowly as $\delta$ increases 
  and the factors \Nbestfit\ become almost independent of it as $\delta\gtrsim0.1$.
  Therefore possible indetermination of $\delta$
  does not significantly influence the subsequent comparisons and conclusions.
  
  (v) Since the \NUANCE\ generator takes into account the interference  between the amplitudes of the resonances
  with the same spin and orbital momentum (within the RS model),
  it turns out (thanks to the procedure described above) to be automatically accounted for in our estimates of
  the CCQE-like backgrounds.
  
  (vi) For transforming the $\overline{\nu}_\mu$CH$_2$ cross sections to the $\overline{\nu}_\mu{}^{12}$C ones,
  the MiniBooNE analysis assumes an effective axial mass of $1.02$~GeV for the quasi-free hydrogen scattering component.
  The $21$~MeV uncertainty is applied according to Refs.~\cite{Bernard:2001rs,Bodek:2007vi}. We did not try to correct
  the hydrogen component subtraction by accounting our best-fit value of $M_A$ and other inputs, since the expected effect
  is very small in comparison with the above-mentioned uncertainties.
 

  \subsubsection{Total CCQE and CCQE-like cross sections}
  \label{sec:MiniBooNE_TCS}

\begin{figure*}[htb]
  \centering
  \includegraphics[width=\textwidth]
  {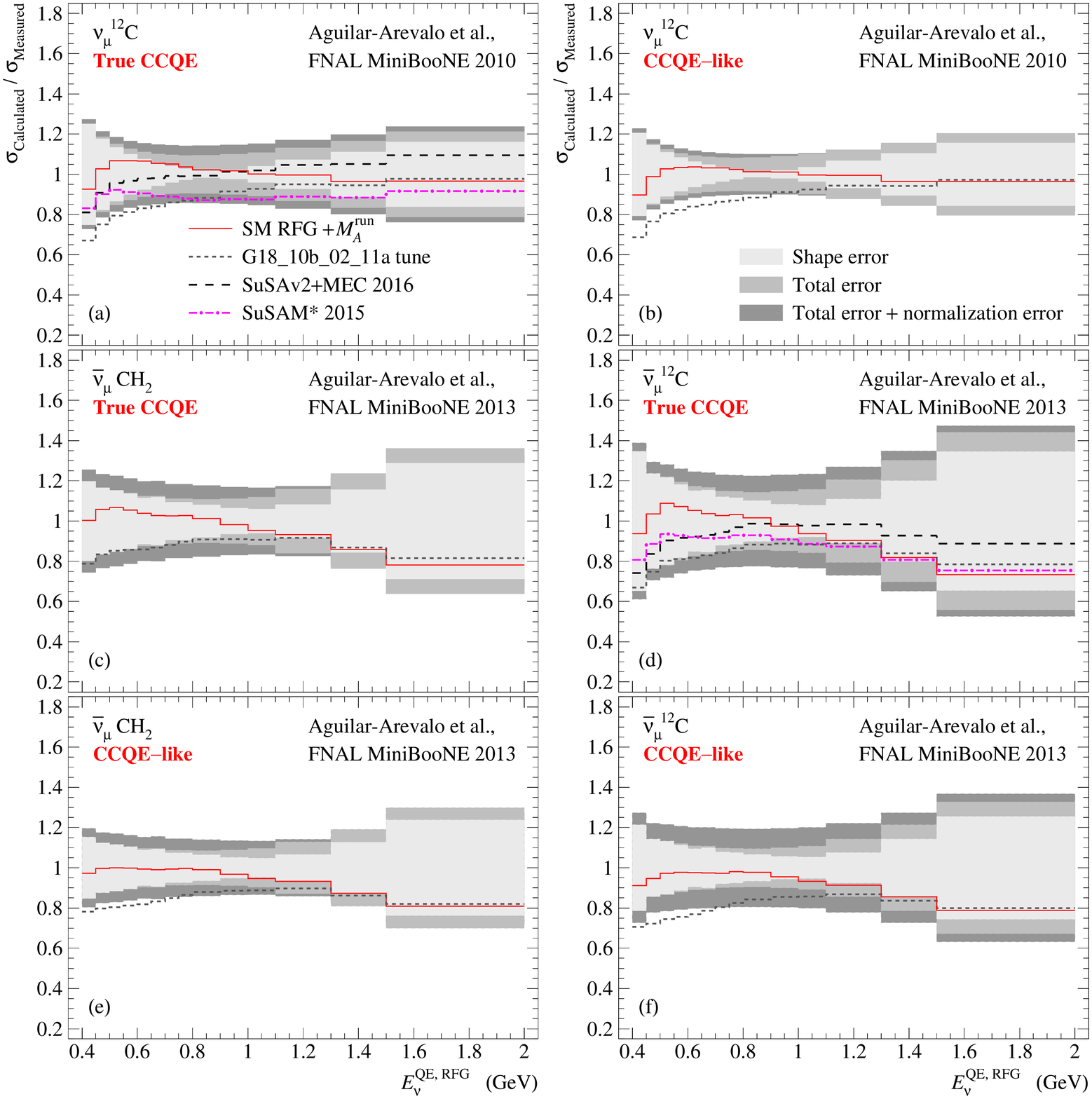}
  \caption{(Color online)
           Ratios of the total CCQE and CCQE-like cross sections measured by
           MiniBooNE \cite{Aguilar-Arevalo:2010zc,Aguilar-Arevalo:2013dva} to those predicted by several models.
           Shown are the ratios for the $\nu_\mu{}^{12}$C (a, b), $\overline{\nu}_\mu$CH$_2$ (c, e), and
           $\overline{\nu}_\mu{}^{12}$C (d, f) cross sections; all are plotted as functions of ``quasielastic''
           (anti)neutrino energy, $E_\nu^{\text{QE, RFG}}$, reconstructed using the RFG model.
           The light gray, gray and deep gray bands indicate, respectively,
           the shape errors, total (statistical and systematic) errors except the normalization ones, and
           the full errors which include the normalization uncertainties whose values are given in Tables
           \protect\ref{Tab:MiniBooNE_Neutrino}--\protect\ref{Tab:MiniBooNE_Antineutrino-CCQE-like}).
           Histograms represent the predictions of \SMRFG_MArun\ (\hA2018), \GENIEb\ tune,
           \SuSAv2MEC\ \cite{Megias:2016fjk}, and \SuSAM\
 		   \protect\cite{Amaro:2015zjaWithErratum,RuizSimo:2020}.
           Quantitative comparison between the models and data is given in the mentioned tables.
          }
  \label{Fig:sQESCC_Arevalo_MiniBooNE10_13_EXPvsTHE_101.3.31.301.6k_2_BBBA25}
  \end{figure*}

Figure~\ref{Fig:sQESCC_Arevalo_MiniBooNE10_13_EXPvsTHE_101.3.31.301.6k_2_BBBA25} shows the ratios
of the predicted and measured total cross sections for $\nu_{\mu}$ CCQE and CCQE-like scattering
on carbon and, for $\overline{\nu}_\mu$, CCQE and CCQE-like scattering on carbon and MiniBooNE's mineral oil.
The histograms representing the predictions of the \SMRFG_MArun\ (with \hA2018\ FSI), \GENIEb\,
\SuSAv2MEC\ \cite{Megias:2016fjk}, and
\SuSAM\ \cite{Amaro:2015zjaWithErratum} are plotted as functions of $E_\nu^{\text{QE, RFG}}$,
the $\nu_\mu/\overline\nu_\mu$ energy reconstructed using the SM RFG model.
In all calculations, we ignore minor (atlhough sometimes not entirely insignificant) differences in the reconstructed energies.
due to differences between the RFG parameters adopted in the MiniBooNE \NUANCE\ and \GENIE3 input settings.
The meaning of the shaded bands is explained in the legend and caption to the figure.
With reference to Fig.~\ref{Fig:sQESCC_Arevalo_MiniBooNE10_13_EXPvsTHE_101.3.31.301.6k_2_BBBA25} and
Tables \protect\ref{Tab:MiniBooNE_Neutrino}--\protect\ref{Tab:MiniBooNE_Antineutrino-CCQE-like}, it can be concluded
that the \SMRFGpMArun\ and the two superscaling models under consideration satisfactory describe both the absolute
CCQE cross sections and $\nu_{\mu}/\overline{\nu}_\mu$ energy shapes well within the shape errors. 

The \SuSAv2MEC\ model requires small or no renormalization ($|\mathcal{N}-1|\approx0.03$ for $\nu_\mu$ and $\approx 0.3$
for $\overline{\nu}_\mu$). The agreement of the model with the CCQE-like $\nu_\mu{}^{12}$C, $\overline{\nu}_\mu$CH$_2$,
and $\overline{\nu}_\mu{}^{12}$C data is exceptionally good below $\sim1$~GeV, the predictions only slightly deviate
from the mean measured values of the cross sections (always being within the shape errors) at higher energies.
The renormalization is also ever not needed or inessential ($|\mathcal{N}-1|\ll\delta$). 

The \G18\ tunes predictions are close to the data (and to those of the \SMRFG_MArun\ model) at
$E_\nu^{\text{QE, RFG}}\gtrsim1$~GeV but at lower energies, they exhibit poorer consistency with the MiniBooNE
CCQE and, to a greater degree, with the CCQE-like $\nu_\mu{}^{12}$C, $\overline{\nu}_\mu$CH$_2$, and
$\overline{\nu}_\mu{}^{12}$C data.
The agreement can be significantly improved by a renormalization, but with 
$\rat\equiv|\mathcal{N}-1|/\delta\approx1.2$ ($0.9$) and $\rat=1.4-1.5$ ($1.3$) for, respectively,
CCQE $\nu_\mu$ ($\overline\nu_\mu$) and CCQE-like $\nu_\mu$ ($\overline\nu_\mu$) data samples.
In other words, the \G18\ tunes well describe the shapes but not the absolute values of the cross sections.
The difference in the FSI models does not essentially disturb the predictions.

The relevant predictions of the two superscaling models are only available for the CCQE samples on carbon.
It is seen that both models are in good agreement with the data.
The agreement with the \SuSAM\ model can be further improved by the renormalization with
$\rat\approx1.2$ and $0.5$ for, respectively, $\nu_\mu$ and $\overline\nu_\mu$ data samples
(see Tables \ref{Tab:MiniBooNE_Neutrino} and \ref{Tab:MiniBooNE_Antineutrino-CCQE}).

It is instructive to compare our calculations with the total $\nu_\mu{}^{12}\text{C}\to\mu^-X$ cross section
measured by MiniBooNE~\cite{Aguilar-Arevalo:2018ylq} at precisely fixed $\nu_\mu$ energy of $236$~MeV, -- 
the world's-first known-energy, weak-interaction-only probe of the neutrino-nucleus interaction.
The reported cross section is
\[
  \sigma_{\text{tot}}^{\text{exp}} = \left(2.7\pm0.9\pm0.8\right)\times10^{-39}~\text{cm}^2/\text{neutron},
\]
where the first error represents the total ``rate+shape'' uncertainty and the second comes from the uncertainty
on the initial $K^+\to\mu^+\nu_\mu$ decay-at-rest neutrino flux. This result should be compared with 
\[
\dfrac{\sigma_{\text{tot}}^{\text{MC}}}{10^{-39}~\text{cm}^2/\text{neutron}} =
\left\{ 
\begin{aligned}
2.07 & \enskip (\text{\SuSAM}),             \\
2.05 & \enskip (\text{\SMRFG_MArun}),       \\
1.95 & \enskip (\text{\SuSAv2MEC}),         \\
1.62 & \enskip (\text{\G18\ tunes}),        \\
1.30 & \enskip (\text{\NuWro}),
\end{aligned}
\right.
\]
where the prediction of the \NuWro\ neutrino event generator \cite{Juszczak:2009qa,Golan:2012wx} is borrowed from
Ref.~\cite{Aguilar-Arevalo:2018ylq}. 
The theoretical predictions are very weakly sensitive to the FSI model since the neutrino energy is below the inelastic threshold.
The monoenergetic point is included into the joint MiniBooNE neutrino CCQE and CCQE-like datasets presented in Table~\ref{Tab:MiniBooNE_Neutrino}.
A visual comparison with several model predictions is shown in Fig.~\ref{Fig:sQESCC_mn_n_C+ma_p_C_101.3.31.301.6k_2_BBBA25_1}
(see \ref{sec:TotalCCQE}).

  \begin{figure*}[!hbt]
  \centering
  \includegraphics[width=\textwidth]
  {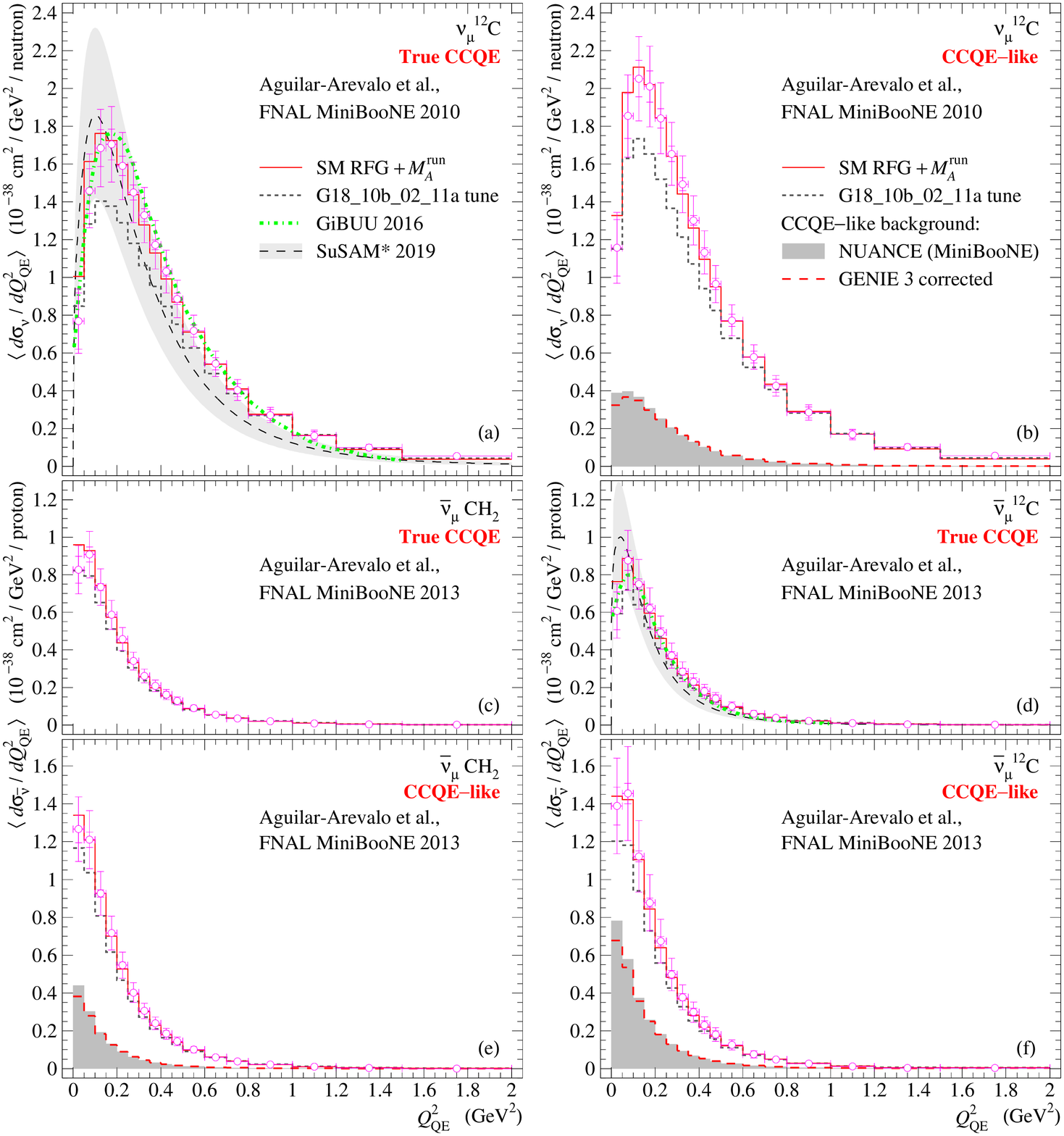}
  \caption{(Color online)
           Flux-weighted differential cross sections as functions of $Q_{\text{QE}}^2$ for the
           true CCQE and CCQE-like neutrino and antineutrino scattering from mineral oil and carbon,
           as measured by the MiniBooNE detector~\cite{Aguilar-Arevalo:2010zc,Aguilar-Arevalo:2013dva}.
           The inner and outer vertical error bars indicate the total errors without and with the
           overall normalization uncertainties.
           The data were not involved into the global fit.
           The solid and dashed histograms represent predictions of the \SMRFG_MArun\ model with \hA2018\ FSI
           contribution and \GENIEb\ tune.
           The dash-dotted curves in panels (a) and (d) show the results of the \GiBUU\ model~\cite{Gallmeister:2016dnq}
           and the dashed curves are the results of the \SuSAM\ model~\cite{RuizSimo:2018kdl,RuizSimo:2020}.
           The light gray bands around the dashed curves in panels (a) and (d) correspond to the uncertainties
           in the \SuSAM\ model input parameters.
           The shaded gray histograms in panels (b), (e), and (f) represent the \NUANCE\ simulated CCQE-like 
           backgrounds as reported in Refs.~\cite{Aguilar-Arevalo:2010zc,Aguilar-Arevalo:2013dva}.
           The long-dashed histograms in these panels represent the same backgrounds after applying the correction
           factors \protect\eqref{CCQE-likeCorrectionFactor} computed with \GENIE3\ using the \SMRFG_MArun\ (\hA2018)
           model setting.
           Corresponding $\chi^2/\ndf$ values are listed in Tables
           \protect\ref{Tab:MiniBooNE_Neutrino}--\protect\ref{Tab:MiniBooNE_Antineutrino-CCQE-like}.
          }
  \label{Fig:dsQES_L_CC_dQ2_Arevalo_MiniBooNE10_13_101.3.31.301.6k_2_BBBA25_1}
  \end{figure*}


\begin{table*}[!htb]
\caption{The values of $\chi^2_{\text{st}}/\ndf$, $\chi^2_1/\ndf$, $\chi^2_{\mathcal{N}}/(\ndf-1)$, and $\mathcal{N}$,
         evaluated with several models for the total, single-differential, and double-differential CCQE and CCQE-like
         cross sections for $\nu_\mu$ scattering from carbon as measured by MiniBooNE \cite{Aguilar-Arevalo:2010zc}.
         Respective numbers are shown before (CCQE) and after (CCQE-like) slashes.
         The results for the \GiBUU, \SuSAv2MEC, and \SuSAM\ models are only available for the CCQE cross sections.
         The relevant data and selected model predictions are shown in Figs.\
         \protect\ref{Fig:sQESCC_Arevalo_MiniBooNE10_13_EXPvsTHE_101.3.31.301.6k_2_BBBA25}~(a, b),  
         \protect\ref{Fig:dsQES_L_CC_dQ2_Arevalo_MiniBooNE10_13_101.3.31.301.6k_2_BBBA25_1}~(a, b), 
		 and
         \protect\ref{Fig:d2sQESCC_dEkdcosT_mn_n_CH_2.06}                                           
         (see also Fig.~\protect\ref{Fig:sQESCC_mn_n_C+ma_p_C_101.3.31.301.6k_2_BBBA25_1}~(a)).
         The meaning of the listed quantities is explained by Eqs.\ \protect\eqref{chi2_st}--\protect\eqref{chi2_mathcalN}
         (implying in this instance only diagonal covariances).
         The ndf (second column) represents the number of experimental bins for each calculation.
         The relative normalization uncertainty $\delta=10.7\%/9.8\%$ for the CCQE/CCQE-like cross sections
         is taken the same for all data subsets.
         Also shown the $\chi^2$s and normalization factors, $\mathcal{N}$, for the joint MiniBooNE neutrino CCQE and CCQE-like datasets
         including the measurement of $\nu_\mu{}^{12}\text{C}\to\mu^-X$ cross section at $E_\nu=236$~MeV reported in Ref.~\cite{Aguilar-Arevalo:2018ylq}.
         Here and in the tables below, three digits in mantissa of $\mathcal{N}$s are displayed to distinguish differences caused by the two FSI models.
        }
\begin{tabular*}{\textwidth}{@{\extracolsep\fill}lcllll}                                           \hline\noalign{\smallskip}
  \MC{1}{c}{Model}                 & ndf & \CN{\text{st}} & \CN{1}    & \CNN      & \Nbestfit\  \\ \noalign{\smallskip}\hline
  \mc{6}{c}{$\sigma_{\text{tot}}^{\text{QE}}$}  
  \SMRFG_MArun\ (\hA2018)          &  14 & 0.07/0.05      & 0.13/0.08 & 0.07/0.08 & 0.972/0.990 \\ 
  \WithMono\                       &  15 & 0.09/0.07      & 0.14/0.10 & 0.10/0.10 & 0.973/0.991 \\
  \SMRFG_MArun\ (\hN2018)          &  14 & 0.07/0.06      & 0.13/0.09 & 0.07/0.09 & 0.972/0.993 \\
  \WithMono\                       &  15 & 0.09/0.07      & 0.14/0.11 & 0.10/0.11 & 0.973/0.994 \\
  \GENIEa\                         &  14 & 0.78/0.99      & 1.39/1.88 & 0.44/0.45 & 1.129/1.139 \\
  \WithMono\                       &  15 & 0.78/0.98      & 1.35/1.81 & 0.44/0.45 & 1.131/1.140 \\
  \GENIEb\                         &  14 & 0.78/1.04      & 1.39/1.96 & 0.44/0.45 & 1.129/1.143 \\
  \WithMono\                       &  15 & 0.79/1.03      & 1.36/1.88 & 0.44/0.45 & 1.131/1.144 \\
  \SuSAv2MEC\ 2016                 &  14 & 0.08           & 0.11      & 0.12      & 1.004       \\
  \SuSAM\ 2019                     &  14 & 0.46           & 0.89      & 0.12      & 1.113       \\
  \mc{6}{c}{$d\sigma/dQ_{\text{QE}}^2$} 
  \SMRFG_MArun\ (\hA2018)          &  17 & 0.23/0.17      & 0.81/0.75 & 0.68/0.56 & 1.017/1.017 \\
  \SMRFG_MArun\ (\hN2018)          &  17 & 0.23/0.19      & 0.74/0.87 & 0.68/0.54 & 1.013/1.022 \\
  \GENIEa\                         &  17 & 1.43/1.58      & 19.3/22.0 & 1.66/1.31 & 1.203/1.195 \\
  \GENIEb\                         &  17 & 1.43/1.75      & 19.3/24.0 & 1.68/1.32 & 1.203/1.206 \\
  \GiBUU\,2016                     &  16 & 0.63           & 3.19      & 1.25      & 0.948       \\
  \SuSAM\,2019                     &  17 & 4.08           & 28.1      & 12.0      & 1.198       \\
  \mc{6}{c}{$d^2\sigma/dE_{\mu}d\cos\theta_\mu$} 
  \SMRFG_MArun\ (\hA2018)          & 137 & 0.20/0.21      & 0.27/0.26 & 0.25/0.26 & 0.980/1.000 \\
  \hspaceFS                        & 105 & 0.20/0.23      & 0.26/0.29 & 0.27/0.29 &             \\
  \hspaceBS                        &  32 & 0.18/0.13      & 0.28/0.16 & 0.19/0.16 &             \\ \noalign{\smallskip}
  \SMRFG_MArun\ (\hN2018)          & 137 & 0.20/0.23      & 0.27/0.28 & 0.25/0.28 & 0.977/1.008 \\
  \hspaceFS                        & 105 & 0.20/0.26      & 0.25/0.32 & 0.26/0.31 &             \\
  \hspaceBS                        &  32 & 0.19/0.14      & 0.31/0.17 & 0.19/0.18 &             \\ \noalign{\smallskip}
  \GENIEa\                         & 137 & 0.61/0.61      & 0.97/1.05 & 0.77/0.56 & 1.080/1.117 \\
  \hspaceFS                        & 105 & 0.55/0.63      & 0.94/1.12 & 0.55/0.40 &             \\
  \hspaceBS                        &  32 & 0.78/0.50      & 1.02/0.74 & 1.45/1.02 &             \\ \noalign{\smallskip}
  \GENIEb\                         & 137 & 0.61/0.64      & 0.97/1.12 & 0.77/0.51 & 1.079/1.131 \\
  \hspaceFS                        & 105 & 0.55/0.69      & 0.94/1.21 & 0.56/0.37 &             \\
  \hspaceBS                        &  32 & 0.77/0.45      & 1.02/0.71 & 1.44/0.88 &             \\ \noalign{\smallskip}
  \GiBUU\,2016                     & 137 & 3.52           & 3.82      & 3.53      & 0.916       \\
  \hspaceFS                        & 105 & 0.49           & 0.60      & 0.82      &             \\
  \hspaceBS                        &  32 & 13.4           & 14.2      & 12.3      &             \\ \noalign{\smallskip}
  \SuSAv2MEC\ 2016                 & 137 & 1.05           & 1.69      & 1.46      & 1.086       \\
  \hspaceFS                        & 105 & 0.81           & 1.15      & 1.19      &             \\
  \hspaceBS                        &  32 & 1.78           & 3.38      & 2.30      &             \\ \noalign{\smallskip}
  \SuSAM\,2019                     & 137 & 3.08           & 4.72      & 3.71      & 1.208       \\
  \hspaceFS                        & 105 & 1.78           & 2.71      & 1.99      &             \\
  \hspaceBS                        &  32 & 7.24           & 11.1      & 9.10      &             \\ \noalign{\smallskip}
  \mc{6}{c}{Joint dataset including monoenergetic point} 
  \SMRFG_MArun\ (\hA2018)          & 169 &                & 0.31/0.30 & 0.31/0.28 & 1.002/1.011 \\
  \SMRFG_MArun\ (\hN2018)          & 169 &                & 0.30/0.33 & 0.31/0.30 & 0.999/1.016 \\
  \GENIEa\                         & 169 &                & 2.84/3.32 & 1.04/0.83 & 1.158/1.169 \\
  \GENIEb\                         & 169 &                & 2.84/3.60 & 1.05/0.79 & 1.158/1.181 \\  \noalign{\smallskip}\hline
\end{tabular*}
\label{Tab:MiniBooNE_Neutrino}
\end{table*}


\begin{table*}[!htb]
\centering
\caption{Same as in Table~\protect\ref{Tab:MiniBooNE_Neutrino} but for the
         CCQE $\overline\nu_{\mu}$CH$_2$ and $\overline\nu_{\mu}{}^{12}$C cross sections \cite{Aguilar-Arevalo:2013dva};
         respective numbers are shown before (CH$_2$) and after (${}^{12}$C) slashes.
         Here and below, the abbreviation ``\GENIEab'' means that the listed $\chi^2$s and \Nbestfit\,s are the same
         (within the accepted rounding-off conventions) for the \GENIEa\ and \GENIEb\ tunes.
         The results for the \GiBUU, \SuSAv2MEC, and \SuSAM\ models are only available for the CCQE cross sections on carbon.
         The relevant data and selected model predictions are shown in Figs.
         \protect\ref{Fig:sQESCC_Arevalo_MiniBooNE10_13_EXPvsTHE_101.3.31.301.6k_2_BBBA25}~(c, d),        
         \protect\ref{Fig:dsQES_L_CC_dQ2_Arevalo_MiniBooNE10_13_101.3.31.301.6k_2_BBBA25_1}~(c, d),       
		 and
         \protect\ref{Fig:d2sQESCC_dEkdcosT_ma_p_CH_2.06}                                                 
         (see also Fig.~\protect\ref{Fig:sQESCC_mn_n_C+ma_p_C_101.3.31.301.6k_2_BBBA25_1}~(b)).
         The normalization uncertainties, $\delta$, for each data subset are shown in parentheses.
        }
\begin{tabular*}{\textwidth}{@{\extracolsep\fill}lcrrrr}                                               \hline\noalign{\smallskip}
  \MC{1}{c}{Model}                 &   ndf   & \CN{\text{st}} & \CN{1}    & \CNN      & \Nbestfit\  \\ \noalign{\smallskip}\hline
  \mc{6}{c}{$\sigma_{\text{tot}}^{\text{QE}}$ ($\delta=13.0\%/17.4$\%)}
  \SMRFG_MArun\ (\hA2018)          &   14    & 0.08/0.08      & 0.13/0.12 & 0.13/0.13 & 0.991/0.994 \\
  \SMRFG_MArun\ (\hN2018)          &   14    & 0.09/0.08      & 0.13/0.12 & 0.14/0.13 & 0.991/0.993 \\
  \GENIEab\                        &   14    & 0.43/0.41      & 0.82/0.75 & 0.14/0.16 & 1.122/1.161 \\
  \SuSAv2MEC\,2016                 &   14    &      0.10      &      0.15 &      0.09 &       1.050 \\
  \SuSAM\,2019                     &   14    &      0.16      &      0.29 &      0.07 &       1.093 \\
  \mc{6}{c}{$d\sigma/dQ_{\text{QE}}^2$ ($\delta=12.9\%/17.2$\%)}
  \SMRFG_MArun\ (\hA2018)          &   17    & 0.12/0.13      & 0.46/0.45 & 0.47/0.37 & 1.007/1.026 \\
  \SMRFG_MArun\ (\hN2018)          &   17    & 0.12/0.13      & 0.46/0.45 & 0.46/0.37 & 1.007/1.025 \\
  \GENIEa\                         &   17    & 0.48/0.40      & 4.85/3.91 & 0.55/0.37 & 1.128/1.169 \\
  \GENIEb\                         &   17    & 0.48/0.40      & 4.84/3.90 & 0.55/0.37 & 1.128/1.169 \\
  \GiBUU\,2016 ($\T=1$)            &   14    &      0.54      &      1.59 &      0.95 &       1.062 \\
  \SuSAM\,2019                     &   17    &      1.85      &      9.05 &      5.57 &       1.179 \\
  \mc{6}{c}{$d^2\sigma/dE_{\mu}d\cos\theta_\mu$ ($\delta=13.0\%/17.2$\%)}
  \SMRFG_MArun\ (\hAN2018)         &  78/75  & 0.39/0.39      & 0.56/0.57 & 0.55/0.57 & 0.991/0.992 \\ 
  \hspaceFS\                       &  70/67  & 0.34/0.34      & 0.52/0.53 & 0.52/0.53 &             \\
  \hspaceBS\                       &    8    & 0.80/0.82      & 0.84/0.89 & 0.80/0.85 &             \\ \noalign{\smallskip}
  \GENIEab\                        &  78/75  & 0.45/0.38      & 0.68/0.62 & 0.47/0.37 & 1.077/1.110 \\
  \hspaceFS\                       &  70/67  & 0.44/0.37      & 0.68/0.62 & 0.45/0.35 &             \\
  \hspaceBS\                       &    8    & 0.49/0.44      & 0.53/0.51 & 0.55/0.46 &             \\ \noalign{\smallskip}
  \GiBUU\,2016 ($\T=1$)            &  78/75  &      0.44      &      0.73 &      0.41 &      1.127  \\
  \hspaceFS\                       &  70/67  &      0.42      &      0.72 &      0.39 &             \\
  \hspaceBS\                       &    8    &      0.58      &      0.68 &      0.51 &             \\ \noalign{\smallskip}
  \SuSAv2MEC\,2016                 &  78/75  &      0.57      &      1.00 &      0.83 &      1.088  \\
  \hspaceFS\                       &  70/67  &      0.56      &      1.01 &      0.84 &             \\
  \hspaceBS\                       &    8    &      0.60      &      0.70 &      0.60 &             \\ \noalign{\smallskip}
  \SuSAM\,2019                     &  78/75  &      1.40      &      2.51 &      1.60 &      1.237  \\
  \hspaceFS\                       &  70/67  &      1.38      &      2.55 &      1.58 &             \\
  \hspaceBS\                       &    8    &      1.38      &      1.58 &      1.30 &             \\ \noalign{\smallskip}
  \mc{6}{c}{Joint dataset ($\delta_{\text{eff}}=13.0\%/17.3$\%)}
  \SMRFG_MArun\ (\hA2018)          & 109/106 &                & 0.48/0.49 & 0.49/0.49 & 1.000/1.011 \\
  \SMRFG_MArun\ (\hN2018)          & 109/106 &                & 0.48/0.49 & 0.49/0.49 & 1.000/1.010 \\
  \GENIEa\                         & 109/106 &                & 1.34/1.15 & 0.46/0.36 & 1.111/1.149 \\
  \GENIEb\                         & 109/106 &                & 1.34/1.15 & 0.46/0.35 & 1.111/1.149 \\ \noalign{\smallskip}\hline
\end{tabular*}
\label{Tab:MiniBooNE_Antineutrino-CCQE}
\end{table*}


\begin{table*}[!htb]
\centering
\caption{Same as in Table~\protect\ref{Tab:MiniBooNE_Antineutrino-CCQE} but for the CCQE-like 
         $\overline\nu_{\mu}$CH$_2$ and $\overline\nu_{\mu}{}^{12}$C MiniBooNE data subsets \cite{Aguilar-Arevalo:2013dva};
         respective numbers are shown before (CH$_2$) and after (${}^{12}$C) slashes.
         The relevant data and selected model predictions are shown in Figs.
         \protect\ref{Fig:sQESCC_Arevalo_MiniBooNE10_13_EXPvsTHE_101.3.31.301.6k_2_BBBA25}~(e, f),         
         \protect\ref{Fig:dsQES_L_CC_dQ2_Arevalo_MiniBooNE10_13_101.3.31.301.6k_2_BBBA25_1}~(e, f),        
		 and
         \protect\ref{Fig:d2sQELCC_dEkdcosT_ma_p_CH_2.06}.                                                 
		 The normalization uncertainties, $\delta$, for each data subset are shown in parentheses.
        }
\begin{tabular*}{\textwidth}{@{\extracolsep\fill}lccccc}                                               \hline\noalign{\smallskip}
  \MC{1}{c}{Model}                 &  ndf    & \CN{\text{st}} & \CN{1}    & \CNN      & \Nbestfit\  \\ \noalign{\smallskip}\hline
  \mc{6}{c}{$\sigma_{\text{tot}}^{\text{QE}}$ ($\delta=12.4\%/16.9$\%)}
   \SMRFG_MArun\ (\hA2018)         &   14    & 0.07/0.09      & 0.12/0.19 & 0.09/0.08 & 1.019/1.040 \\
   \SMRFG_MArun\ (\hN2018)         &   14    & 0.08/0.10      & 0.13/0.21 & 0.10/0.10 & 1.021/1.044 \\
   \GENIEa\                        &   14    & 0.88/0.90      & 2.11/2.70 & 0.29/0.37 & 1.164/1.223 \\
   \GENIEb\                        &   14    & 0.90/0.92      & 2.14/2.76 & 0.30/0.37 & 1.166/1.227 \\
   \mc{6}{c}{$d\sigma/dQ_{\text{QE}}^2$ ($\delta=12.3\%/16.7$\%)}
   \SMRFG_MArun\ (\hA2018)         &   17    & 0.06/0.05      & 0.32/0.40 & 0.25/0.15 & 1.013/1.028 \\
   \SMRFG_MArun\ (\hN2018)         &   17    & 0.06/0.06      & 0.41/0.62 & 0.26/0.16 & 1.018/1.038 \\
   \GENIEa\                        &   17    & 0.62/0.51      & 8.91/9.33 & 0.52/0.40 & 1.147/1.189 \\
   \GENIEb\                        &   17    & 0.66/0.56      & 9.70/10.4 & 0.56/0.43 & 1.154/1.203 \\
  \mc{6}{c}{$d^2\sigma/dE_{\mu}d\cos\theta_\mu$ ($\delta=12.4\%/16.7$\%)}
   \SMRFG_MArun\ (\hA2018)         &  78/75  & 0.36/0.33      & 0.52/0.56 & 0.52/0.54 & 1.003/1.011 \\
   \hspaceFS                       &  70/67  & 0.31/0.28      & 0.48/0.50 & 0.48/0.49 &             \\
   \hspaceBS                       &    8    & 0.75/0.77      & 0.79/0.83 & 0.80/0.88 &             \\ \noalign{\smallskip}
   \SMRFG_MArun\ (\hN2018)         &  78/75  & 0.36/0.34      & 0.53/0.57 & 0.53/0.55 & 1.007/1.019 \\
   \hspaceFS                       &  70/67  & 0.31/0.28      & 0.49/0.53 & 0.48/0.49 &             \\
   \hspaceBS                       &    8    & 0.76/0.78      & 0.80/0.84 & 0.82/0.92 &             \\ \noalign{\smallskip}
   \GENIEa\                        &  78/75  & 0.55/0.50      & 1.03/1.25 & 0.48/0.38 & 1.105/1.147 \\
   \hspaceFS                       &  70/67  & 0.55/0.50      & 1.07/1.30 & 0.45/0.36 &             \\
   \hspaceBS                       &    8    & 0.48/0.44      & 0.52/0.52 & 0.54/0.45 &             \\ \noalign{\smallskip}
   \GENIEb\                        &  78/75  & 0.57/0.53      & 1.09/1.36 & 0.49/0.39 & 1.110/1.157 \\
   \hspaceFS                       &  70/67  & 0.57/0.54      & 1.13/1.43 & 0.47/0.37 &             \\
   \hspaceBS                       &    8    & 0.48/0.45      & 0.52/0.52 & 0.54/0.45 &             \\ \noalign{\smallskip}
   \mc{6}{c}{Joint dataset ($\delta_{\text{eff}}=12.3\%/16.8$\%)}
   \SMRFG_MArun\ (\hA2018)         & 109/106 &                & 0.43/0.47 & 0.42/0.42 & 1.010/1.023 \\
   \SMRFG_MArun\ (\hN2018)         & 109/106 &                & 0.46/0.52 & 0.43/0.43 & 1.015/1.032 \\
   \GENIEa\                        & 109/106 &                & 2.38/2.72 & 0.48/0.40 & 1.136/1.178 \\
   \GENIEb\                        & 109/106 &                & 2.55/2.98 & 0.49/0.41 & 1.142/1.189 \\ \noalign{\smallskip}\hline
\end{tabular*}
\label{Tab:MiniBooNE_Antineutrino-CCQE-like}
\end{table*}


  \subsubsection{Single differential cross sections}
  \label{sec:MiniBooNE_SDCS}

  Figure \ref{Fig:dsQES_L_CC_dQ2_Arevalo_MiniBooNE10_13_101.3.31.301.6k_2_BBBA25_1}
  shows a comparison of the measured and calculated flux-folded CCQE and CCQE-like differential cross sections,
  $d\sigma/dQ_{\text{QE}}^2$, on carbon (for the $\nu_\mu$ case) and CCQE and CCQE-like cross sections
  on carbon and mineral oil (for the $\overline\nu_\mu$ case). The calculations were done in several models discussed above.
  The quantitative comparison is given in Tables~\ref{Tab:MiniBooNE_Neutrino}--\protect\ref{Tab:MiniBooNE_Antineutrino-CCQE-like}.
  To clarify the picture, the figure also shows (shaded histograms in panels (b), (e), and (f)) the CCQE-like background
  contributions estimated with the \NUANCE\ MC neutrino event generator which is used by the MiniBooNE Collaboration for
  reconstructing the CCQE cross sections from the CCQE-like datasets~\cite{Aguilar-Arevalo:2010zc,Aguilar-Arevalo:2013dva}.
  These backgrounds are compared against those are re-weighted with the factors \eqref{CCQE-likeCorrectionFactor} calculated
  with \GENIE3\ using the \SMRFGpMArun\ model setting, as described in Sec. \ref{sec:MArunGENIEsetting}, and \hA2018 FSI
  (long-dashed histograms); the correction factors for the four models are shown in
  Fig.~\protect\ref{Fig:dsQELCC_dQ2_BSc_R+S_Arevalo_MiniBooNE10_13_101.3.31.301.6k_2_BBBA25}
  for the $\nu_\mu{}^{12}$C and $\overline\nu_\mu{}^{12}$C scatterings.

  One can recognize that the \NUANCE\ and \GENIE3\ simulated background contributions are noticeably different
  in both shape and magnitude.
  The main reason of the differences is in the input parameters for pion production symultions, different descriptions
  of the SIS region and FSI models used in the two neutrino event generators.
  However, the differences themselves are relatively small in magnitude compared to the main contributions to the CCQE-like
  cross sections and thus the \SMRFG_MArun\ model well reproduces both CCQE and CCQE-like cross sections, adequately simulating
  the ``beyond RFG'' nuclear effects.

  
  It is seen from Fig.~\ref{Fig:dsQES_L_CC_dQ2_Arevalo_MiniBooNE10_13_101.3.31.301.6k_2_BBBA25_1} and
  Tables \protect\ref{Tab:MiniBooNE_Neutrino}--\protect\ref{Tab:MiniBooNE_Antineutrino-CCQE-like} that the \SMRFGpMArun\ model
  is in full accord with the $\nu_\mu$ and $\overline\nu_\mu$ CCQE and CCQE-like data within the full errors (all values of
  $\chi^2_{\text{st}}/\ndf$ are small) and almost no renormalization is needed ($\rat<0.2$ for all cross sections).
  Moreover, it is in agreement with almost all the data within the errors which do not include the normalization uncertainty.
  It is remarkably, that the \GENIE3\ corrections to the CCQE-like backgrounds
  (see Fig.~\ref{Fig:dsQELCC_dQ2_BSc_R+S_Arevalo_MiniBooNE10_13_101.3.31.301.6k_2_BBBA25})
  work in the right direction, i.e.\ they improve agreement with the CCQE-like data.
  At the low-$Q^2$ region, the effect is mainly due to a decrease in the single-pion production cross sections caused
  by account of the muon mass in the KLN-BS model; and at high $Q^2$s, it is due to a cut-off in $W$, which increases
  the DIS contribution in the SIS region.
 
  The \G18\ tunes substantially underestimate the CCQE and CCQE-like cross sections at low $Q^2$s.
  The resulting $\chi_1^2$ values are incredibly large
  (see Tables~\ref{Tab:MiniBooNE_Neutrino}--\protect\ref{Tab:MiniBooNE_Antineutrino-CCQE-like})
  but it must be taken into account that the statistical errors in this dataset are notably small.
  In contrast to the \SMRFG_MArun\ model, the \G18\ corrections to the \NUANCE\ simulated inelastic backgrounds
  either do not improve (for $\nu_\mu$) or even worsen (for $\overline\nu_\mu$) the agreement with the CCQE-like data,
  as a result of underestimating the inelastic contributions in the RES and SIS regions (see Sec.\ \ref{sec:GENIE3tune}).
  The description of the data (for both $\nu_\mu$ and $\overline\nu_\mu$) can be improved but at the cost
  of large normalization factors: $\rat\approx1.9$ ($1.0$) and $\rat=2.0-2.1$ ($1.1-1.2$) for, respectively,
  CCQE $\nu_\mu$ ($\overline\nu_\mu$) and CCQE-like $\nu_\mu$ ($\overline\nu_\mu$) data samples.
  As for the \SMRFG_MArun\ model, the differences due to distinctions in the two \GENIE3\ FSI models
  are expectedly small, although not entirely insignificant.
  For both \SMRFG_MArun\ model and \G18\ tunes, the agreement with the CCQE-like data on carbon is
  a little bit better than that on CH$_2$, but this is due mainly to lesser sistematic errors in the CH$_2$ data sample,
  which is less model dependent.


  
  The \GiBUU\ model ($\T=1$) \cite{Gallmeister:2016dnq} is in satisfactory agreement with the CCQE $\nu_{\mu}{}^{12}$C and
  $\overline\nu_{\mu}{}^{12}$C data at $Q_{\text{QE}}^2\lesssim1$~GeV$^2$ but underestimates the high-$Q^2$ tails
  of the differential cross sections. However a small renormalization ($-5.2$\% for $\nu_{\mu}$ and $+6.2$\% for $\overline\nu_{\mu}$)
  noticeably improves the agreement. At the same time, it should be pointed out that the authors of Ref.~\cite{Gallmeister:2016dnq}
  do not provide calculations for the highest-$Q^2$ bins.

  The \SuSAM\ model \cite{RuizSimo:2018kdl,RuizSimo:2020} rather poorly describes the CCQE cross section shapes
  and this cannot be substantially corrected by a renormalization.
  In Fig.~\ref{Fig:dsQES_L_CC_dQ2_Arevalo_MiniBooNE10_13_101.3.31.301.6k_2_BBBA25_1} we display the confidence bands,
  delineating the uncertainties arising due to variations in the input parameters of the \SuSAM\ model derived from
  the extensive global fit to electron scattering data \cite{Amaro:2018xdi}; in fact, the bands represent minimax
  over the $1\sigma$ uncertainties of the parameters defining the form of the scaling function.
  Let us remind that the $\chi^2$ values listed in the tables do not take into account these uncertainties.
  Accounting for them will certainly improve the formal consistency with the data.
  We also recall that the model was not tuned to neutrino data.

  \subsubsection{Double-differential cross sections}
  \label{sec:MiniBooNE_DDCS}


  Figure \ref{Fig:d2sQESCC_dEkdcosT_mn_n_CH_2.06} shows the MiniBooNE $\nu_\mu$ data for the flux-folded
  CCQE and CCQE-like sections $d^2\sigma_{\nu}/dE_{\mu}d\cos\theta_{\mu}$ on carbon.
  Figures \ref{Fig:d2sQESCC_dEkdcosT_ma_p_CH_2.06}--\ref{Fig:d2sQELCC_dEkdcosT_ma_p_CH_2.06}
  show the MiniBooNE $\overline\nu_\mu$ \cite{Aguilar-Arevalo:2013dva} data for the flux-folded CCQE and CCQE-like
  double-differential cross sections 
  on mineral oil and pure carbon.
  The cross sections are plotted as slices at fixed bins of $\cos\theta_{\mu}$ vs.\ $E_{\mu}$,
  where $\theta_{\mu}$ is the muon scattering angle and $E_{\mu}$ is the muon kinetic energy.
  In several panels of Figs.\ \ref{Fig:d2sQESCC_dEkdcosT_ma_p_CH_2.06} and \ref{Fig:d2sQELCC_dEkdcosT_ma_p_CH_2.06}, the experimental data
  and relevant model predictions are rescaled for easier comparison of the cross section shapes displayed in the adjacent panels.
  Only the CCQE data shown in Figs.~\ref{Fig:d2sQESCC_dEkdcosT_mn_n_CH_2.06} and \ref{Fig:d2sQESCC_dEkdcosT_ma_p_CH_2.06}
  were involved into the global fit of the running axial mass.
  The data are compared with several model predictions; no normalization has been applied.
  In order not to overload the figures, not all models under consideration are presented in the figures.
  In particular, we display only one version of the FSI corrections for the \SMRFG_MArun\ model and \G18\ tune
  (the differences due to the two FSI versions are usually small).
  

  \def\Vspace{3.0mm}       
  \def\W{0.975\textwidth}  

  \begin{figure*}[!htb]
  \centering
  \includegraphics[width=\W]
  {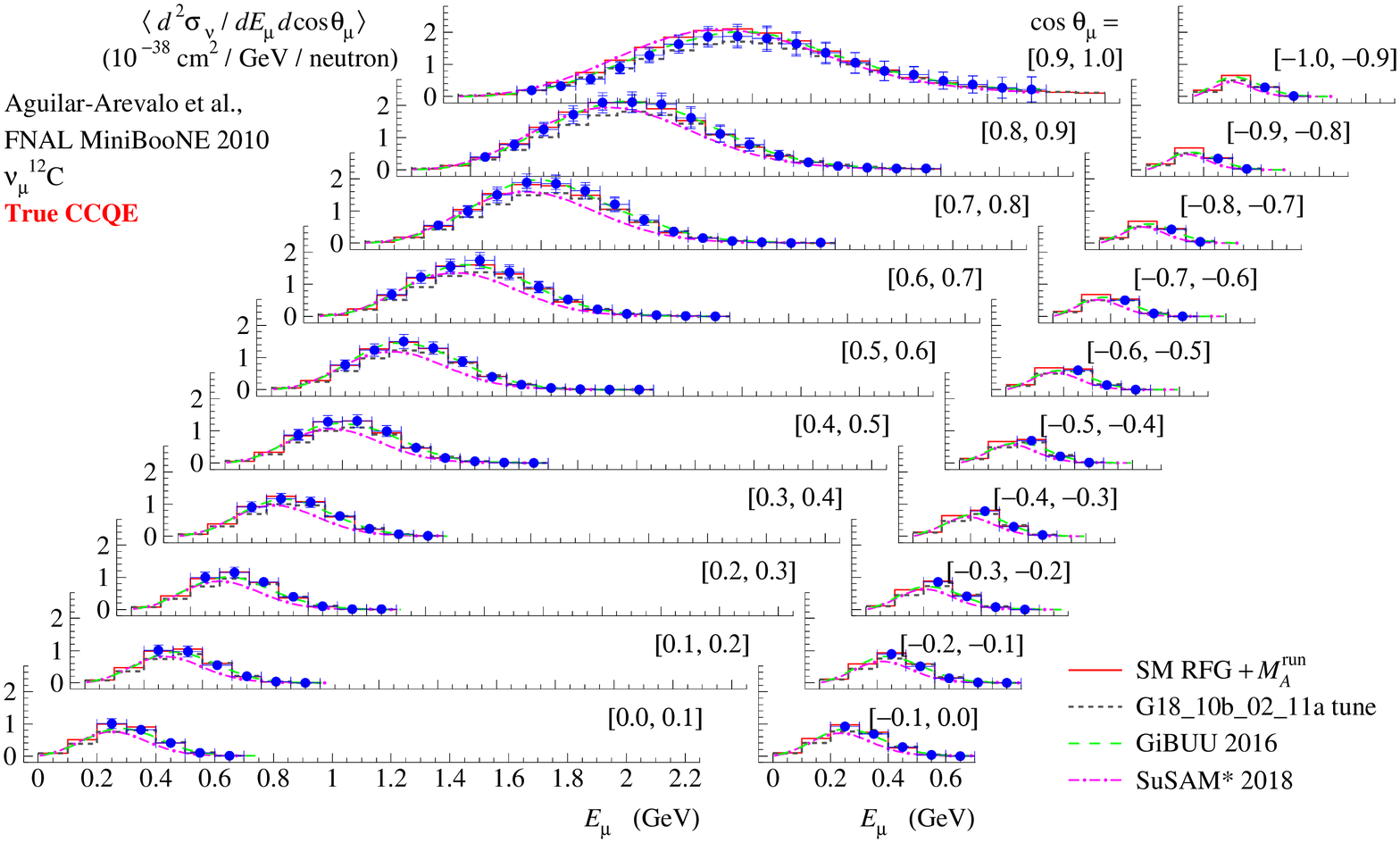}  \\[\Vspace]
  \includegraphics[width=\W]
  {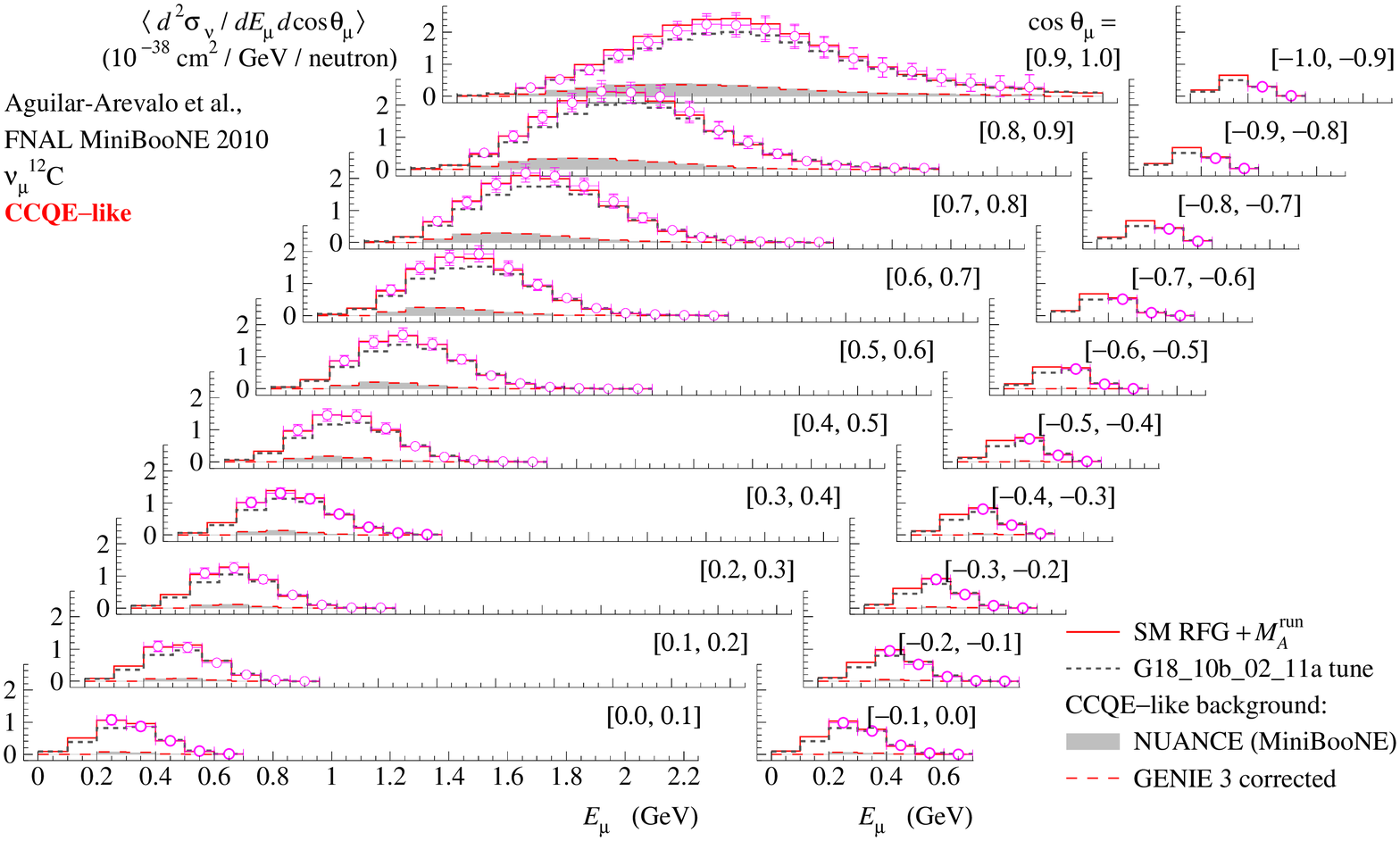}
  \caption{(Color online)
           Flux-weighted double-differential cross sections, $d^2\sigma_{\nu}/dE_{\mu}d\cos\theta_{\mu}$,
           for the true CCQE (twenty top panels) and CCQE-like (twenty bottom panels) $\nu_{\mu}$ scattering
		   from carbon as measured with MiniBooNE \cite{Aguilar-Arevalo:2010zc}.
           The cross sections are displayed as functions of the muon kinetic energy, $E_\mu$,
           for several intervals of the cosine of the muon scattering angle, $\theta_\mu$ (shown in square brackets).
           The inner and outer vertical error bars indicate, respectively, the total errors without and with
           the normalization uncertainty ($\delta=10.7$\% for CCQE and $9.8$\% for CCQE-like data).
           Histograms represent predictions of the \SMRFG_MArun\ model (with \hA2018\ FSI) and \GENIEb\ tune.
           Smooth curves represent the \GiBUU~\cite{Gallmeister:2016dnq} and \SuSAM~\cite{RuizSimo:2018kdl}
           model predictions.
		   The shaded gray and long-dashed histograms show, respectively, the CCQE-like background contributions
           reported in Ref.~\cite{Aguilar-Arevalo:2010zc} and the same after applying the correction factors
           \protect\eqref{CCQE-likeCorrectionFactor}.
           Corresponding $\chi^2/\ndf$ values are listed in Table~\protect\ref{Tab:MiniBooNE_Neutrino}.
          }
  \label{Fig:d2sQESCC_dEkdcosT_mn_n_CH_2.06}
  \end{figure*}

  \begin{figure*}[!htb]
  \centering
  \includegraphics[width=\W]
  {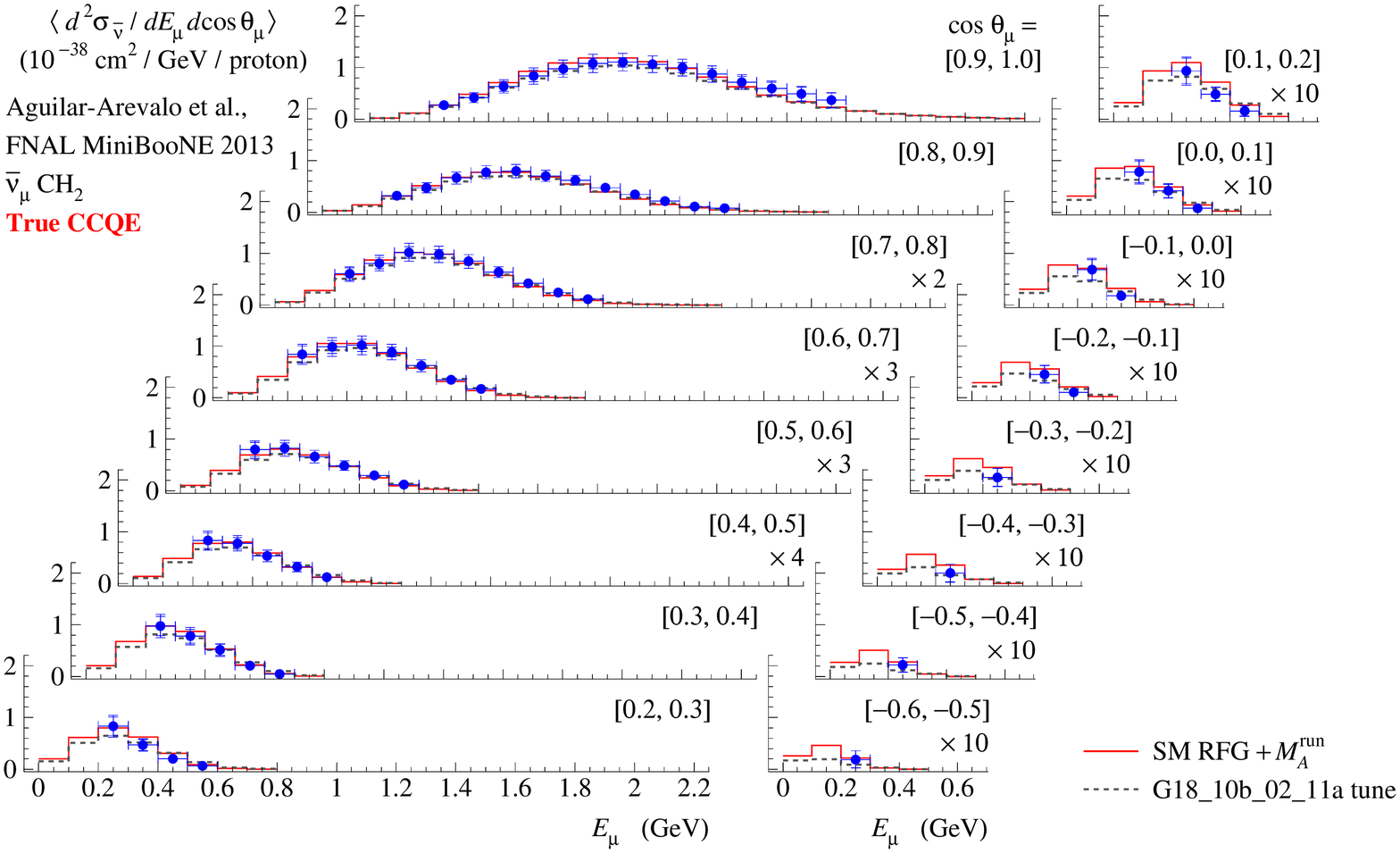}  \\[\Vspace]
  \includegraphics[width=\W]
  {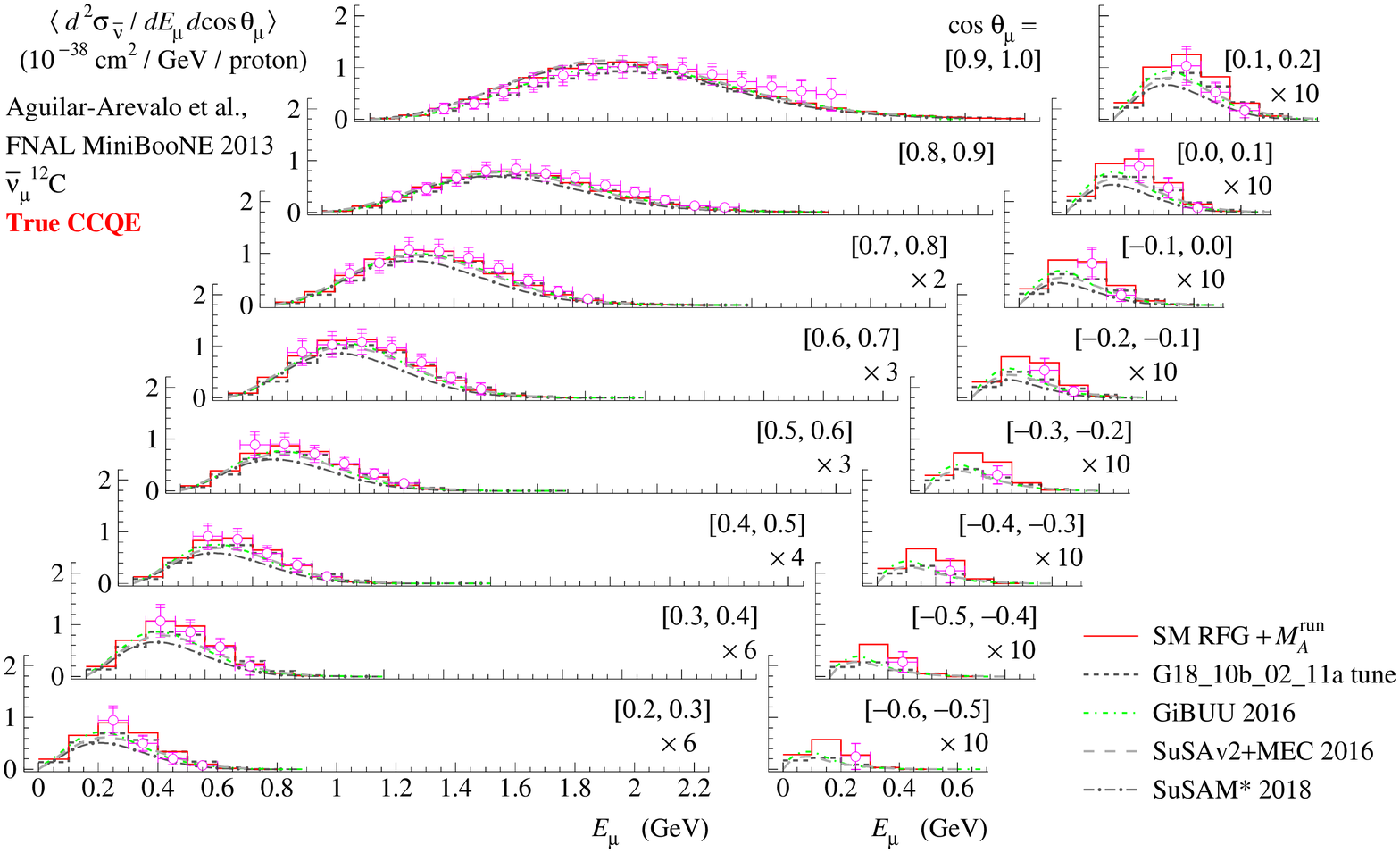}
  \caption{(Color online)
           Flux-weighted double-differential cross sections, $d^2\sigma_{\overline{\nu}}/dE_{\mu}d\cos\theta_{\mu}$,
           for the true CCQE $\overline{\nu}_{\mu}$ scattering from mineral oil (sixteen top panels) and pure carbon
           (sixteen bottom panels) as measured by MiniBooNE \cite{Aguilar-Arevalo:2013dva}.
           The notation is the same as in Fig.~\protect\ref{Fig:d2sQESCC_dEkdcosT_mn_n_CH_2.06}.
           The inner and outer vertical error bars indicate, respectively, the total errors without and with
           the normalization uncertainty ($\delta=13$\% for mineral oil and $17.2$\% for carbon).
           To aid the visualization, the data and histograms in some panels are multiplied by the factors
           indicated in the legends.
           The histograms represent predictions of the \SMRFG_MArun\ model (with \hA2018\ FSI) and \GENIEb\ tune.
           Smooth curves in bottom panels represent predictions of \GiBUU~\cite{Gallmeister:2016dnq} and of
		   the two superscaling models -- \SuSAv2MEC~\cite{Megias:2016fjk} and \SuSAM~\cite{RuizSimo:2018kdl}.
           Corresponding $\chi^2/\ndf$ values are listed in Table~\protect\ref{Tab:MiniBooNE_Antineutrino-CCQE}.
           }
  \label{Fig:d2sQESCC_dEkdcosT_ma_p_CH_2.06}
  \end{figure*}

  \begin{figure*}[!htb]
  \centering
  \includegraphics[width=\W]
  {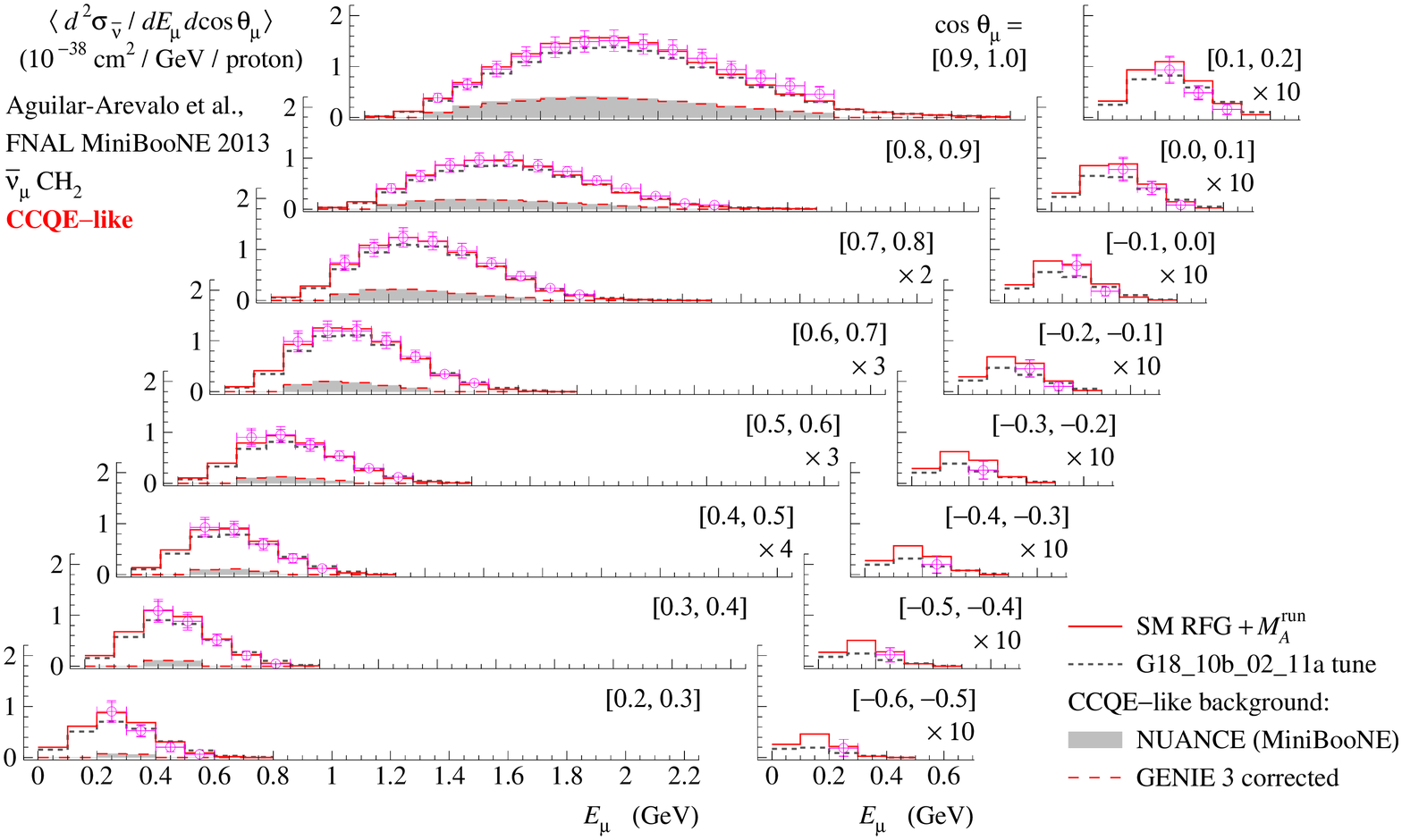} \\[\Vspace]
  \includegraphics[width=\W]
  {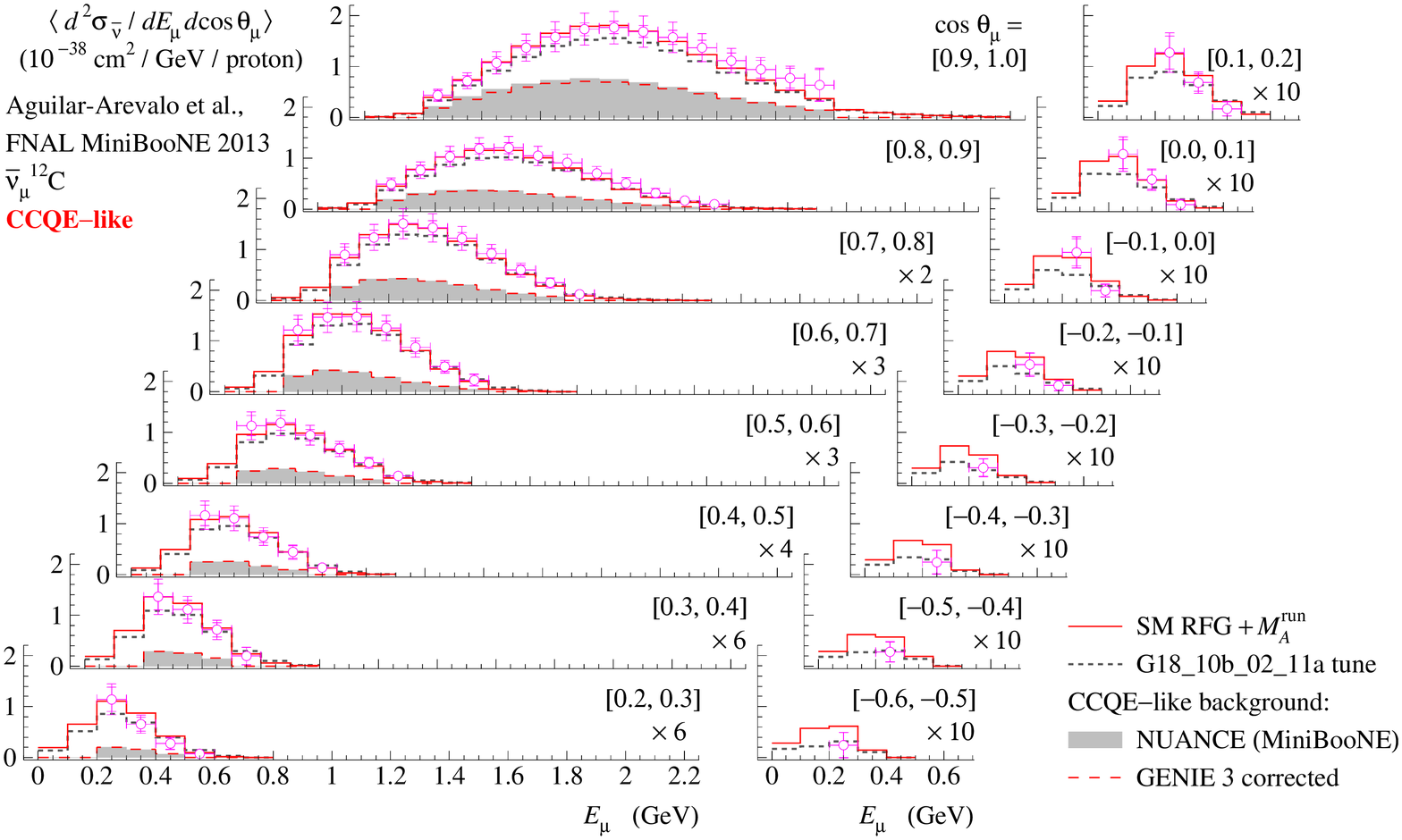}
  \caption{(Color online)
           Flux-weighted double-differential cross sections, $d^2\sigma_{\overline{\nu}}/dE_{\mu}d\cos\theta_{\mu}$,
           for the CCQE-like $\overline{\nu}_{\mu}$ scattering from mineral oil (sixteen top panels) and pure carbon
           (sixteen bottom panels) as measured by MiniBooNE \cite{Aguilar-Arevalo:2013dva}.
           The solid and dashed histograms represent predictions of the \SMRFG_MArun\ model (with \hA2018\ FSI)
           and \GENIEb\ tune, respectively.
           The shaded gray and long-dashed histograms represent, respectively, the CCQE-like background contributions
           reported in Ref.~\cite{Aguilar-Arevalo:2013dva} and the same after applying the correction factors
           \protect\eqref{CCQE-likeCorrectionFactor} calculated using the \SMRFG_MArun\ (\hA2018) model setting
           (see Fig.~\protect\ref{Fig:d2sQELCC_dEkdcosT_BSc_R+S_Arevalo_MiniBooNE10_13_101.3.31.301.6k_2_BBBA25_1_3D}~(b)).
           The inner and outer vertical error bars indicate, respectively, the total errors without and with
           the normalization uncertainty ($\delta=12.4$\% for mineral oil and $16.7$\% for carbon).
           Other designations have the same meaning as in Fig.~\protect\ref{Fig:d2sQESCC_dEkdcosT_ma_p_CH_2.06}.
           Corresponding $\chi^2/\ndf$ values are listed in Table~\protect\ref{Tab:MiniBooNE_Antineutrino-CCQE-like}.
          }
  \label{Fig:d2sQELCC_dEkdcosT_ma_p_CH_2.06}
  \end{figure*}

  In Figs.\ \ref{Fig:d2sQESCC_dEkdcosT_mn_n_CH_2.06},
            \ref{Fig:d2sQESCC_dEkdcosT_ma_p_CH_2.06}, and
            \ref{Fig:d2sQELCC_dEkdcosT_ma_p_CH_2.06},
  we show the CCQE-like backgrounds modeled with the \NUANCE\ neutrino event generator \cite{Aguilar-Arevalo:2010zc,Aguilar-Arevalo:2013dva}
  (shaded histograms) and those corrected using \GENIE3\ with the \SMRFG_MArun\ (\hA2018) model settings (long-dashed histograms); 
  the backgrounds that are visually indistinguishable from zero are not displayed to avoid cluttering the figures.
  The corresponding correction factors are presented in Fig.\ \ref{Fig:d2sQELCC_dEkdcosT_BSc_R+S_Arevalo_MiniBooNE10_13_101.3.31.301.6k_2_BBBA25_1_3D}.
  Tables \ref{Tab:MiniBooNE_Neutrino}--\ref{Tab:MiniBooNE_Antineutrino-CCQE-like} include detailed lists of $\chi^2$s for each model
  under examination, evaluated both for the entire kinematic range and for the forward and backward scattering subregions.
  
  It can be seen from the Figures and Tables \ref{Tab:MiniBooNE_Neutrino}--\ref{Tab:MiniBooNE_Antineutrino-CCQE-like} that
  the \SMRFG_MArun\ model provides very good agreement with the CCQE and CCQE-like data in all kinematic regions
  for both $\nu_\mu$ and $\overline\nu_\mu$ datasets. The model practically does not require renormalization ($\rat\lesssim0.2$).
  The two \G18\ tunes also provide reasonable agreement with the data but slightly underestimate the measured cross sections in certain kinematic domains.
  Since there is no specific difficulties related to the forward/backward scattering, consistency with the data can partly be improved
  by an overall renormalization with $\rat\approx0.6-1.2$ which is still acceptable within the reported experimental normalization
  uncertainty. 
  
  The \GiBUU\ and \SuSAv2MEC\ models both show good agreement with the CCQE $\nu_{\mu}{}^{12}$C data for the essential part of
  the forward scattering hemisphere (see Fig.~\ref{Fig:d2sQESCC_dEkdcosT_mn_n_CH_2.06}) and with the CCQE $\overline\nu_{\mu}{}^{12}$C data
  in the whole kinematic range (see Fig.~\ref{Fig:d2sQELCC_dEkdcosT_ma_p_CH_2.06}).
  In certain kinematic domains, the \GiBUU\ $\chi^2$ values for the $\overline\nu_{\mu}{}^{12}$C cross sections are similar
  to or even better than (after acceptable renormalization) those for the \SMRFGpMArun\ model.
  However, both \GiBUU\ and \SuSAv2MEC\  exhibit difficulties in reproducing the $\nu_{\mu}{}^{12}$C cross section shapes in the
  backward scattering hemisphere and an overall renormalization cannot resolve the conflict with the MiniBooNE data.
  The unexpectedly large $\chi^2$ values for the CCQE $\overline\nu_{\mu}{}^{12}$C double-differential cross section predicted by
  the \GiBUU\, in comparison with, e.g., \SuSAM\ model (see Table~\ref{Tab:MiniBooNE_Neutrino})
  arise because of systematic bias of the \GiBUU\ prediction from several data points at highest muon energies and
  at the backward scattering angles, visually indistinguishable in Fig.~\ref{Fig:d2sQESCC_dEkdcosT_mn_n_CH_2.06}.
  
  The \SuSAM\ model offers only a qualitative description of the CCQE cross section shapes (see Figs.~\ref{Fig:d2sQESCC_dEkdcosT_mn_n_CH_2.06}
  and \ref{Fig:d2sQESCC_dEkdcosT_ma_p_CH_2.06}), which cannot be substantially corrected by a renormalization (see Table~\ref{Tab:MiniBooNE_Neutrino}).
  We must, however, remind that the \SuSAM\ model operates with a set of adjustable parameters which can be further tuned
  to improve accordance with the MiniBooNE data. 
  

  \subsubsection{Joint datasets}
  \label{sec:MiniBooNE_JD}

  The joint $\nu_\mu{}^{12}$C, $\overline{\nu}_\mu$CH$_2$, and $\overline{\nu}_\mu{}^{12}$C datasets
  presented at the bottom of Tables \ref{Tab:MiniBooNE_Neutrino}--\ref{Tab:MiniBooNE_Antineutrino-CCQE-like}
  include the MiniBooNE data on the CCQE and CCQE-like total, single-differential, and double-differential cross sections.
  The analysis shows that
  \begin{itemize}
  \item[(a)] all \emph{six} MiniBooNE data subsamples are well consistent with each other and
  \item[(b)] the \SMRFG_MArun\ model very well describes all these subsamples almost irrespective of the FSI model.
  \end{itemize}
  It is essential that only a relatively small part of the full MiniBooNE dataset (true CCQE double-differential cross sections)
  has been involved into the fit of \MArun. This shows good predictive power of the model within the MiniBooNE energy range.
  On the other hand, even this part of the MiniBooNE data has significant statistical weight in the full experimental
  dataset used in the global fit. This is due to large number of the data-points and relatively small total errors,
  excluding the normalization uncertainties, which, however, do not substantially affect the above conclusions since the
  \SMRFG_MArun\ model requires almost no normalization to the data.
  Thus, the MiniBooNE dataset critically influences the fitting parameters $M_0$ and, even to a greater extent, $E_0$.
  The \G18\ tunes reasonably describe the MiniBooNE cross section shapes but claims essential renormalization
  (sometimes larger than the data normalization uncertainty) in order to fit the absolute values of the cross sections. 



  \subsection{T2K ND280}
  \label{sec:T2K}

  In this section, we discuss a comparison of the \SMRFGpMArun\ model and several
  others theoretical predictions with the recent data on pionless interactions
  from T2K's off-axis fully magnetized ND280 near detector on the J-PARC site \cite{Abe:2011ks},
  obtained using two detector targets: water based \cite{Abe:2017rfw} and one
  composed of plastic scintillator (C$_8$H$_8$) and metal \cite{Abe:2018pwo}.
  In our calculations, the cross sections predicted for both T2K ND280 experiments
  are averaged over the updated $\nu_\mu$ flux \cite{Abe:2015awa}
  in the full simulated neutrino energy range to about $30$ GeV.
  While the narrow energy spectrum of the beam is centered around $600$ MeV and has
  the mean energy of about $870$ MeV, the high-energy tail marginally ($\lesssim1$\%)
  contributes to the inelastic backgrounds.

  \subsubsection{Water target}
  \label{sec:T2K_water}

  The T2K ND280 experiment with the H$_2$O target selected the CCQE-like events
  without pions in the final state (so-called ``CC$0\pi$'' events).
  It is expected that more than one nucleons may be ejected out of the nucleus due to intranuclear 
  and multinucleon interactions.
  The number of post-FSI nucleons, however, are not fully controlled in the experiment.
  Our simulations reproduce all essential features of the experiment, in particularly, the CC$0\pi$
  events with any number of secondary nucleons are considered as the signal. 

  \begin{figure*}[!htb]
  \centering
  \includegraphics[width=\textwidth]
  {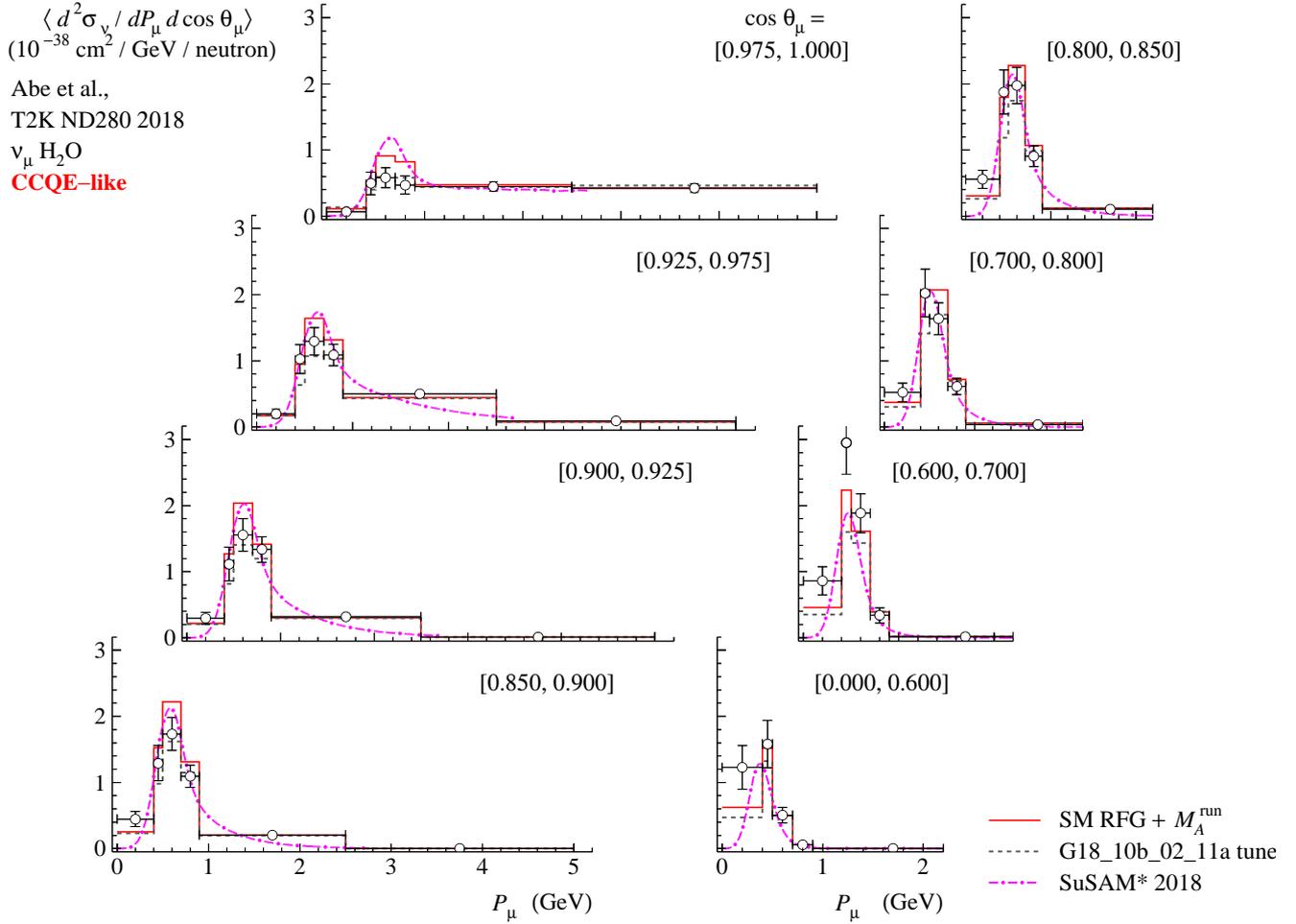}
  \caption{(Color online)
           Flux-weighted double-differential cross sections
           for the CCQE-like $\nu_{\mu}$ scattering from water target
           as measured by T2K ND280 \cite{Abe:2017rfw} and
           plotted vs.\ muon momentum, $P_\mu$, for several intervals $\cos\theta_\mu$ 
           (shown in square brackets).
           The vertical error bars represent the total errors
           including the normalization uncertainty of $8.76$\%.
           Histograms represent the \SMRFG_MArun\ model and \GENIEb\ tune,
           curves show prediction of the \SuSAM\,2018 model \cite{RuizSimo:2018kdl}
           obtained with no account for the FSI effects.
           For more details, see Table \protect\ref{Tab:T2K_H2O-2d_Neutrino-CCQE-like-standard+Normalizedlognorm}.
  }
  \label{Fig:T2K2018ddH2O}         
  \end{figure*}

    \begin{table*}[!htb]
  \centering
  \caption{The values of
           $\chi^2_{\text{st}}/\ndf$,
           $\chi^2_1/\ndf$, $\chi^2_{\mathcal{N}}/(\ndf-1)$, and
           $\mathcal{N}$,
           calculated for the CCQE-like double-differential cross section $d^2\sigma/dP_{\mu}d\cos\theta_\mu$,
           for neutrino scattering from water as measured by T2K ND280 \cite{Abe:2017rfw} (see Fig.~\protect\ref{Fig:T2K2018ddH2O}).
           The last column shows the ratios of the measured and predicted reduced flux-averaged total cross section
           $\sigma=\sigma_{\nu_{\mu}\text{H}_2\text{O}}^{\text{CC}0\pi}$.
           All calculations are performed with four \GENIE3\ models and by using the full covariance matrix with $\ndf=45$.
           Result of the \SuSAM\,2018 model is borrowed from Ref.~\cite{RuizSimo:2018kdl};
           the calculation was made without taking into account the FSI effects and for incomplete dataset ($\ndf=40$).
          }
  \begin{tabular*}{\textwidth}{@{\extracolsep\fill}lcccccc}                                                                      \hline\noalign{\smallskip}
  \MC{1}{c}{Model}          & \CN{\text{st}} & \CN{1} & \CNN & \Nbestfit\ & \CN{\text{log}}
                                                      & $\dfrac{\sigma_{\text{exp}}}{\sigma_{\text{MC}}}$ \\ \noalign{\smallskip}\hline\noalign{\smallskip}
  \SMRFG_MArun\ (\hA2018)   & 1.26           & 1.28   & 1.14 & 0.888      & 1.32        & $1.20 \pm 0.17$ \\
  \SMRFG_MArun\ (\hN2018)   & 1.29           & 1.30   & 1.16 & 0.888      & 1.32        & $1.20 \pm 0.17$ \\
  \GENIEa\                  & 0.85           & 1.01   & 1.02 & 1.033      & 1.31        & $1.41 \pm 0.20$ \\
  \GENIEb\                  & 0.85           & 1.01   & 1.02 & 1.034      & 1.31        & $1.41 \pm 0.20$ \\
  \SuSAM\,2018 (no FSI)     & 1.56           & 1.89   & 1.89 & 0.941      & 1.92        & $1.48 \pm 0.21$ \\ \noalign{\smallskip}\hline
  \end{tabular*}
  \label{Tab:T2K_H2O-2d_Neutrino-CCQE-like-standard+Normalizedlognorm}
  \end{table*}

  \begin{figure*}[!htb]
  \centering
  \includegraphics[width=\textwidth]
  {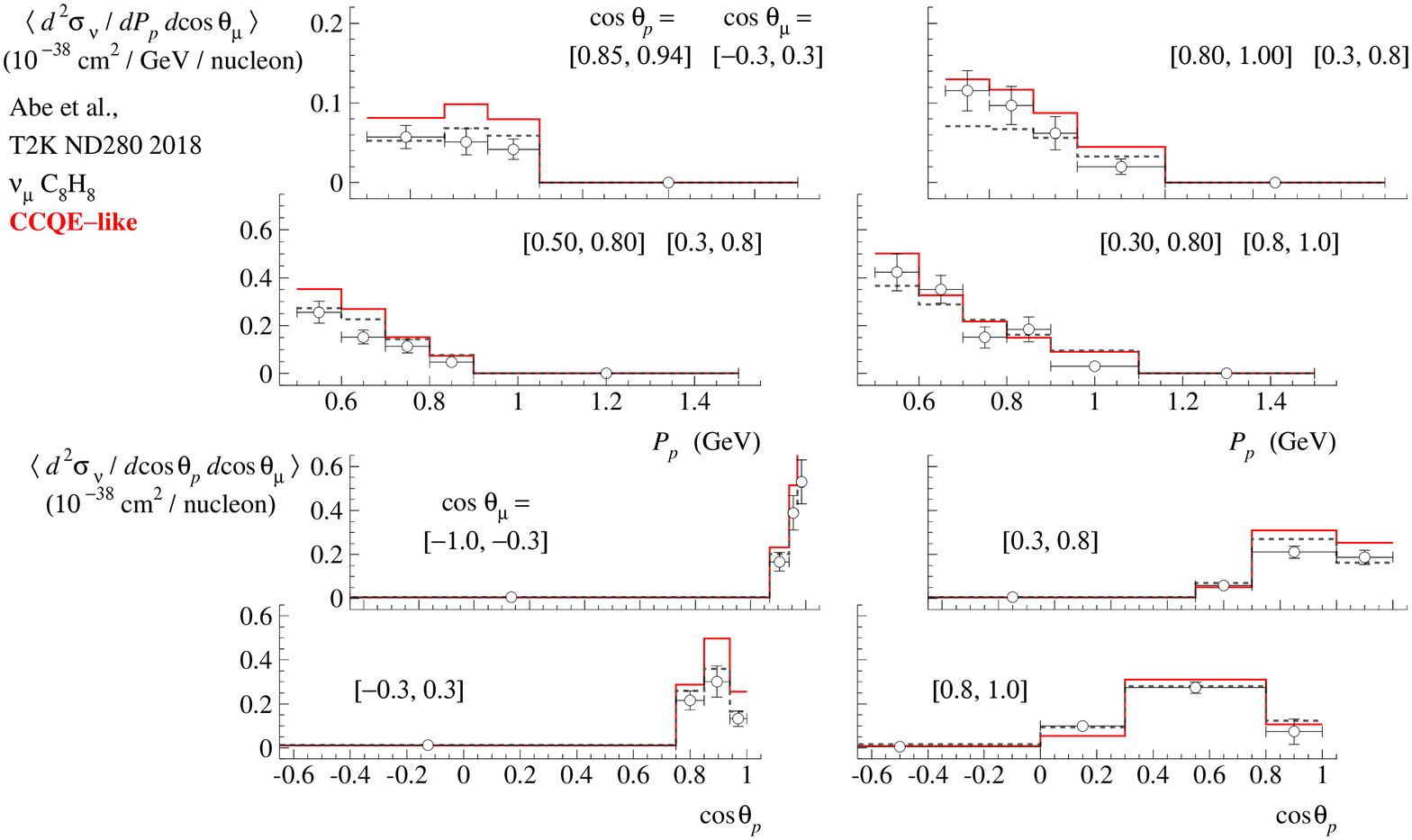} \\[3.0mm]
  \includegraphics[width=\textwidth]
  {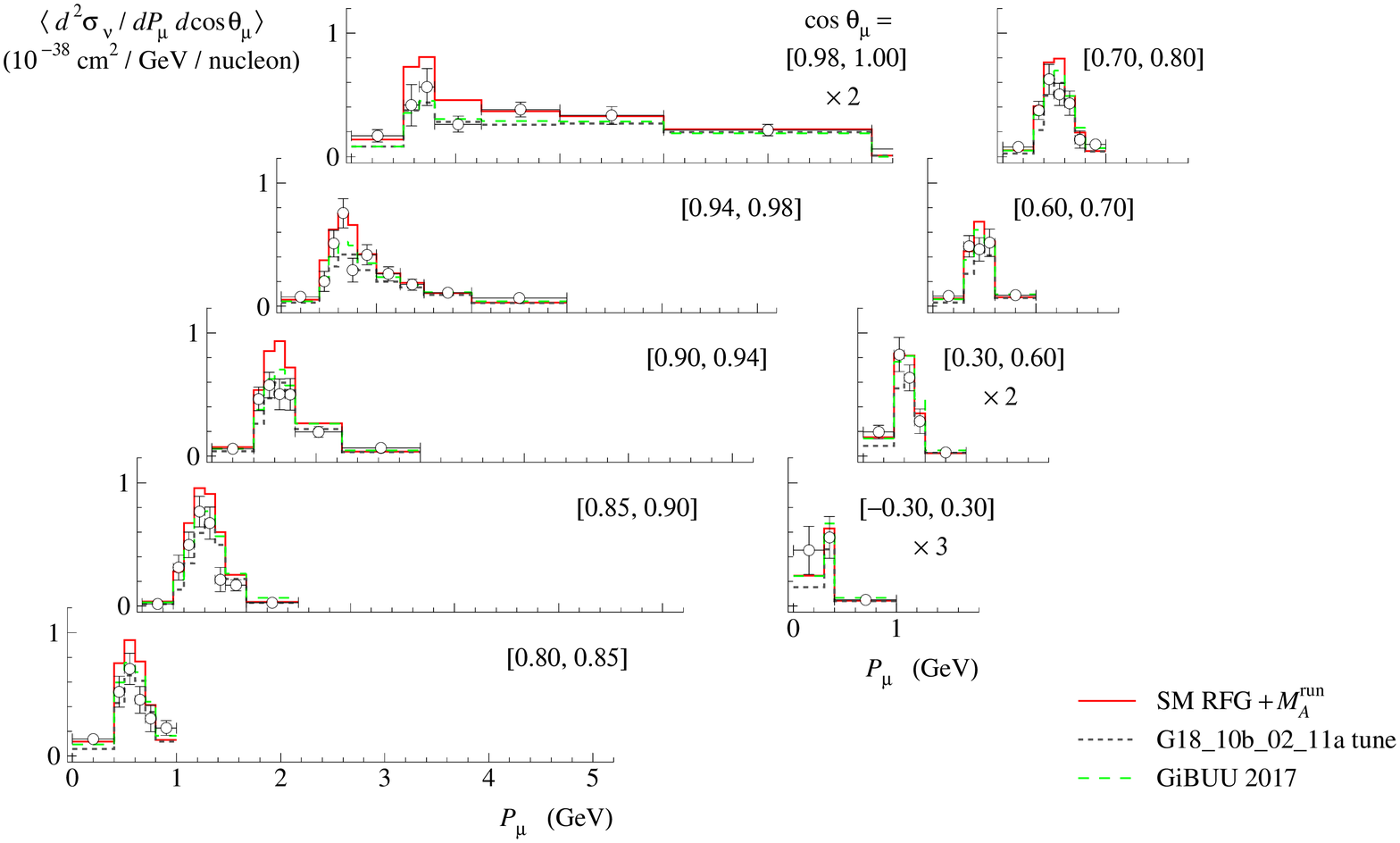}
  \caption{(Color online)
           Flux-weighted double-differential cross sections for the CCQE-like
           $\nu_{\mu}$ scattering from hydrocarbon target as measured by T2K ND280 \cite{Abe:2018pwo}
           and plotted as function of proton momentum, $P_p$, for several intervals of the cosines
           of the proton and muon scattering angles, $\theta_p$ and $\theta_\mu$ (four top panels);
           as function of $\cos\theta_p$ for several intervals of $\cos\theta_\mu$ (four middle panels); and
           as function of muon momentum, $P_\mu$, for several intervals of $\cos\theta_\mu$ (nine bottom panels).
           All intervals of fixed angular variables are shown in square brackets.
           The vertical error bars represent the total errors including the normalization uncertainty of $8.5$\%.
           Histograms represent the \SMRFG_MArun\ (\hA2018) model, \GENIEb\ tune, and \GiBUU\,2017 ($\T=0$) \cite{Dolan:2018sbb}.
           For more details, see Table \protect\ref{Tab:T2K_C8H8-2d_Neutrino-CCQE-like-standard+lognorm}.
          }
  \label{Fig:T2K2018ddC8H8}
  \end{figure*}

  Figure \ref{Fig:T2K2018ddH2O} shows a comparison of the model predictions with the flux-folded double-differential
  cross section for the CCQE-like $\nu_{\mu}$ scattering from the water target.
  Table \ref{Tab:T2K_H2O-2d_Neutrino-CCQE-like-standard+Normalizedlognorm} collects the values of 
  $\chi^2_{\text{st}}/\ndf$,
  $\chi^2_1/\ndf$,
  $\chi^2_{\mathcal{N}}/(\ndf-1)$, and of the normalization factor,
  $\mathcal{N}$.
  These values are obtained using the detailed contributions to the covariance matrix from all sources of
  uncertainties, provided by the authors. The normalization uncertainty, $\delta$, is taken to be $8.76$\%.
  It is seen that all the models under consideration are broadly consistent with the data.
  Formally, the best agreement occurs for the two \G18\ tunes; they as well require the least renormalization ($\rat\approx3.9$\%).
  The FSI effects simulated with the \hA2018\ and \hN2018\ models are practically indistinguishable.
  The $\chi^2$ values for the \SMRFG_MArun\ model (also very weakly dependent of the FSI effect modeling versions)
  are worse but entirely satisfactory and can be somewhat improved by an essential renormalization ($\rat\approx1.3$\%).
  The cross sections predicted by the \SuSAM\,2018 model are taken from Ref.\ \cite{RuizSimo:2018kdl}.
  The model does not take into account the FSI effects and this is, probably, one of the reasons of
  considerably high $\chi^2$s.
  We emphasize that Ref.\ \cite{RuizSimo:2018kdl} does not provide predictions for the muon momenta above $3$ GeV.
  So, the statistical analysis for this model is made with the reduced dataset.

  The last column in Table \ref{Tab:T2K_H2O-2d_Neutrino-CCQE-like-standard+Normalizedlognorm} shows the ratios
  of the measured and predicted flux-averaged total cross sections $\sigma=\sigma_{\nu_{\mu}\text{H}_2\text{O}}^{\text{CC}0\pi}$
  in the restricted region of the phase space. The experimental result of Ref.\ \cite{Abe:2017rfw} is
  \begin{multline*}
  \sigma_{\text{exp}} = \left(0.95 \pm 0.08_{\text{stat}}
                                   \pm 0.06_{\text{detector syst}} \right. \\
                        \left.     \pm 0.04_{\text{model syst}}
                                   \pm 0.08_{\text{flux}} \right) \times 10^{-38}~\text{cm}^2/\text{neutron}.
  \end{multline*}
  All the models under consideration predict essentially lower cross sections, namely
  \[\sigma_{\text{GENIE}}  =  0.79~(0.66)\times 10^{-38}~\text{cm}^2/\text{neutron} \]
  for \SMRFG_MArun\ (\G18); the predictions of these models are almost insensitive to the FSI model.
  A rough estimation made with the \SuSAM\,2018 model yields 
  \[\sigma_{\text{SuSAM}^*} \simeq 0.64  \times 10^{-38}~\text{cm}^2/\text{neutron}.\]
  For comparison, the \NEUT5\ neutrino generator (default in the present T2K analyses) predicts \cite{Abe:2017rfw}
  \[\sigma_{\text{NEUT}}   =    0.66     \times 10^{-38}~\text{cm}^2/\text{neutron} \]
  (the same as for the \G18\ tunes). According to Ref.\ \cite{Abe:2017rfw}, the tension is primarily (but not only)
  due to discrepancies between the data and MC symulations in the large-angle regions, which cover an essential
  part of the reduced phase space.
  This is, at least in part, likewise true for other models under examination and thus indicates some disagreements
  between the T2K and MiniBooNE measurements. 
  
  \subsubsection{Hydrocarbon target}
  \label{sec:T2K_hydrocarbon}

\begin{figure}[!htb]
  \centering
  \includegraphics[width=\linewidth]
  {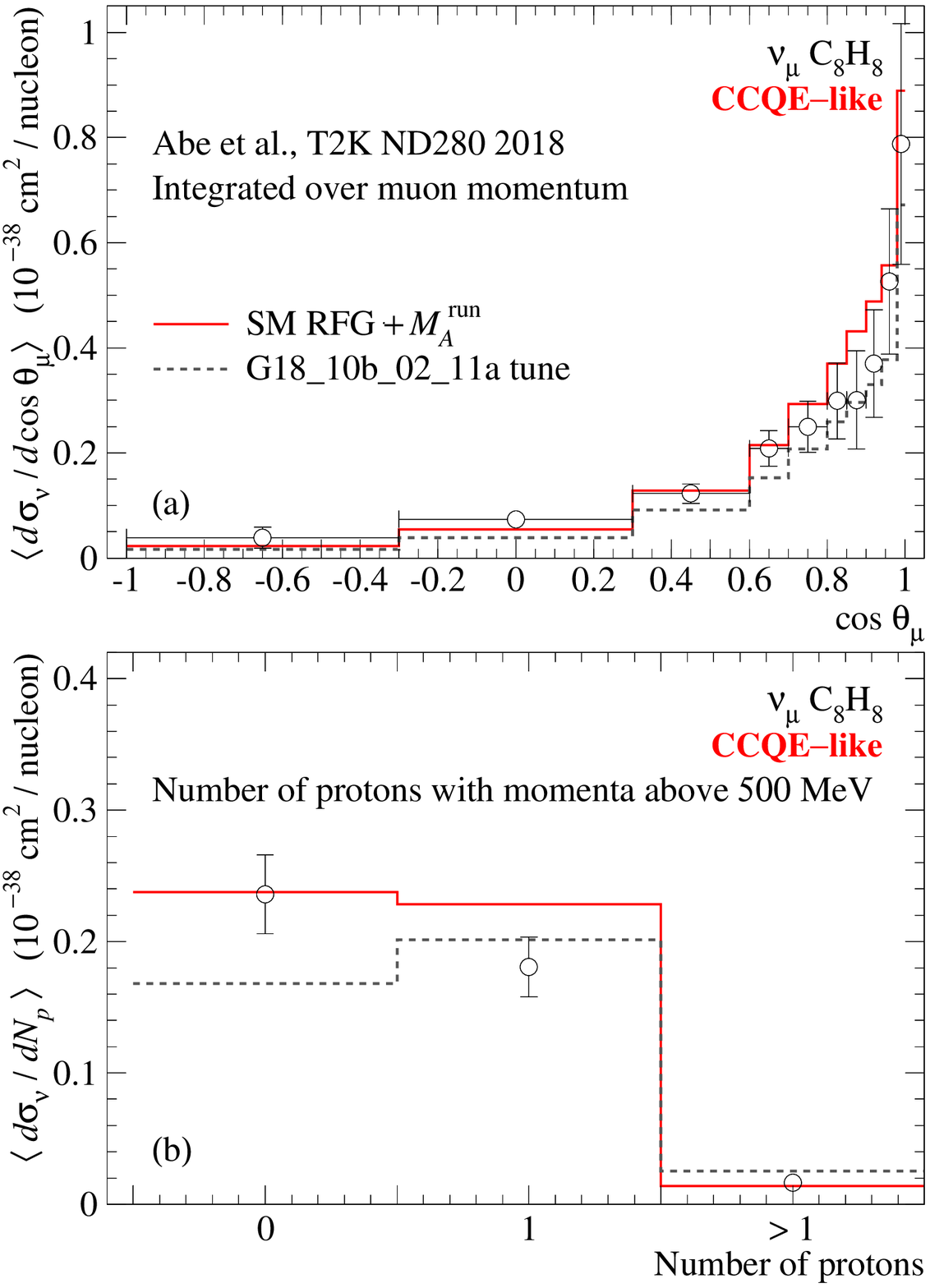}
  \caption{(Color online)
           Differential cross section vs.\ $\cos\theta_\mu$ integrated over $P_\mu$ (a)
           and distribution in the number of secondary protons with momenta $>500$~MeV$/c$ (b).
           The vertical error bars represent the total errors including the normalization uncertainty.
           The T2K ND280 data points are from Ref.\ \cite{Abe:2018pwo} (C$_8$H$_8$).
           Histograms represent the \SMRFG_MArun\ (\hA2018) model and \GENIEb\ tune.
           Following the prescription of the T2K Collaboration, only the left bin in panel (a) and the right
           bin in panel (b) are included into the dataset used in calculations of $\chi^2$s listed in
           Table \protect\ref{Tab:T2K_C8H8-2d_Neutrino-CCQE-like-standard+lognorm}.
          }
  \label{Fig:dsQELCC_dcosT+dsQELCC_dNp_ND28018_2_BBBA25_1}
  \end{figure}

  Figure \ref{Fig:T2K2018ddC8H8} shows the flux-weighted CCQE-like double-diffe\-rential cross sections
  of three types, plotted as functions of the leptonic variables, proton momentum, $P_p$, and cosine of the proton scattering angle,
  $\cos\theta_p$; the data are from the same experimental sample as shown in Fig.\ \ref{Fig:dsQELCC_dcosT+dsQELCC_dNp_ND28018_2_BBBA25_1}.
  We transformed the original data presented by the authors of experiment as the single-differential cross sections
  $d\sigma_\nu/dP_\mu$,
  $d\sigma_\nu/dP_p$, and
  $d\sigma_\nu/d\cos\theta_p$
  to the double-differential ones; 
  this is done for the convenience of comparing the data with each other and with similar data from \MINERvA\ and MiniBooNE.
  For similar reasons, the cross sections calculated as function of $P_\mu$ at lowest values
  of $|\cos\theta_\mu|$ are multiplied by the scale factors shown in the legends.
  
  The T2K ND280 experiment with the hydrocarbon target also selected CC0$\pi$ events, 
  but classified them by number of final state protons with momenta above $500$ MeV$/c$.
  The distribution of these events, $d\sigma_\nu/dN_p$, in the number of secondary protons, $N_p$,
  is shown in the bottom panel of Fig.\ \ref{Fig:dsQELCC_dcosT+dsQELCC_dNp_ND28018_2_BBBA25_1}.
  It is seen that the number of the final-state protons is in agreement with the \SMRFG_MArun\ model
  prediction for $N_p=0$ and $N_p\ge2$, but for the case $N_p=1$ the \GENIEa\ tune works better.
  The figure, as well, shows (see top panel) the differential CCQE-like cross section $d\sigma_\nu/d\cos\theta_\mu$
  integrated over the muon momentum, $P_\mu$. The \SMRFGpMArun\ model with \hA2018 FSI and \G18\ tune predict
  similar shapes (both are in conformity with the data), but somewhat different absolute values.
  Looking at this plot, it can hardly be said that one model works better than the other in describing
  the muon angular distribution.

  Let us now consider the comparison of the T2K data with predictions of the \SMRFG_MArun\ (with \hA2018 FSI) model,
  \GENIEb\ tune, and \GiBUU\,2017 ($\T=0$) \cite{Dolan:2018sbb}.
  A visual comparison of the measured and predicted cross sections shows that the agreement between
  all model predictions and the data is generally unsatisfactory.
  Considering that this experiment (in contrast to the T2K experiment with water target and similarly to the \MINERvA\
  experiment discussed below) does not provide the covariance matrix responsible for the flux uncertainties,
  one cannot unambiguously define the matrix $\widetilde{\mathbf{W}}$ (see Eq.\ \eqref{chi2_N}) and thus it is
  difficult to properly define the normalization factors without loss of information on the bin-by-bin correlations.
  We mention in passing that the $\chi^2$s are very slowly sensitive to the global normalization uncertainty.
  So to quantify the comparison more definitely, here and in the succeeding discussion,
  we use both the standard \eqref{chi2_st} and log-normal \eqref{chi2_log} least-squares criteria.

  Table \ref{Tab:T2K_C8H8-2d_Neutrino-CCQE-like-standard+lognorm} collects the values of $\chi^2_{\text{st}}/$ndf
  and $\chi^2_{\text{log}}/$ndf estimated for the full T2K dataset and for the subset containing only
  the differential cross sections measured in terms of the leptonic variables.
    \begin{table}[!htb]
  \centering
  \caption{Standard and log-normal $\chi^2/\ndf$ values calculated for the neutrino CC0$\pi$
           datasets on the flux-weighted differential cross sections on hydrocarbon target from 
           T2K ND280 experiment~\cite{Abe:2018pwo}
           (see Figs.\ \protect\ref{Fig:T2K2018ddC8H8}, \protect\ref{Fig:dsQELCC_dcosT+dsQELCC_dNp_ND28018_2_BBBA25_1}).
           Calculations are done for several models using the full covariance matrix for the full T2K dataset ($\ndf=93$)
           and for a subset of the data dependent only on the leptonic variables ($\ndf=60$).
           Following to Ref.~\cite{Abe:2018pwo}, only a part of the T2K data shown in Figs.~\protect\ref{Fig:T2K2018ddC8H8}
           and \protect\ref{Fig:dsQELCC_dcosT+dsQELCC_dNp_ND28018_2_BBBA25_1} is included into the analysis.
           The number in parentheses for the T2K-tuned \NEUT5\ model (incorporating ``LFG+RPA'' model with $1p1h$ and $2p2h$
           prediction by Nieves \emph{et al.}~\cite{Nieves:2011yp}) is taken from Ref.~\cite{Abe:2018pwo}.
          }
  \begin{tabular*}{\linewidth}{@{\extracolsep\fill}lcccc}                                              \hline\noalign{\smallskip}
                              & \MC{2}{c}{Full}                  & \MC{2}{c}{Leptonic}              \\
                              & \MC{2}{c}{dataset}               & \MC{2}{c}{variables}             \\ \noalign{\smallskip}\cline{2-5}\noalign{\smallskip}
  \MC{1}{l}{Model}            & \CN{\text{st}} & \CN{\text{log}} & \CN{\text{st}} & \CN{\text{log}} \\ \noalign{\smallskip}\hline\noalign{\smallskip}
  \SMRFG_MArun\ (\hA2018)     &  6.27          & 4.61            & 3.19           & 2.25            \\
  \SMRFG_MArun\ (\hN2018)     &  5.77          & 4.45            & 3.39           & 2.93            \\
  \GENIEa\                    &  4.19          & 4.46            & 2.17           & 2.87            \\
  \GENIEb\                    &  4.22          & 4.51            & 2.18           & 2.90            \\
  T2K-tuned \NEUT5\           & (3.99)         & 4.63            & 2.90           & 3.77            \\  
  \GiBUU\,2017 ($\T=0$)       &  ---           & ---             & 2.83           & 2.45            \\
  \GiBUU\,2017 ($\T=1$)       &  ---           & ---             & 3.90           & 2.75            \\ \noalign{\smallskip}\hline
  \end{tabular*}
  \label{Tab:T2K_C8H8-2d_Neutrino-CCQE-like-standard+lognorm}
  \end{table}
  Similar quantities estimated for the cross sections $d\sigma_\nu/dN_p$ are listed in
  Table \ref{Tab:T2K_C8H8-2d_Neutrino-CCQE-like-standard+lognorm-N}.
  It is seen from the Tables that none of the models, including the default one from the T2K-tuned
  MC neutrino event generator \NEUT5\ \cite{Hayato:2002sd,Hayato:2009zz}
  (whose predictions are not shown in Fig.\ \ref{Fig:dsQELCC_dcosT+dsQELCC_dNp_ND28018_2_BBBA25_1}),
  can accurately describe the T2K C$_8$H$_8$ data. 
  It should be recorded here that \NEUT5\ uses similar theoretical models for CCQE, $2p2h$, resonance pion
  production, coherent scattering, etc., as \G18\ tunes, but implementation differs in many details
  (e.g., in RPA corrections, $W$ cutoff between the RES and DIS regions) and in values of the input parameters.
  The same is true for FSI; for example, \NEUT5\ uses impulse approximation for the nucleon FSI and
  the model by Salcedo \emph{et al.} \cite{Salcedo:1987md} for pions which includes nuclear medium effects
  (cf.\ Sec.~\ref{sec:FSI}). 
  For the leptonic data subset, the \GENIEa\ tune provides the lowest $\chi^2_{\text{st}}$,
  while the \SMRFG_MArun\ model with \hA2018 FSI gives the lowest $\chi^2_{\text{log}}$;
  the \GiBUU\,2017 ($\T=0$) predictions yield intermediate values of these criteria.
  It is worthy of note that the \G18\ tunes are less sensitive to the FSI model version than the \SMRFG_MArun\ model.
  The greatest disagreement occurs for the full T2K C$_8$H$_8$ dataset where the correlations drastically increase
  $\chi^2$s for all models.
  A detailed comparison of several other models with the data has already been done in Ref.\ \cite{Abe:2018pwo}.
  We only note that all these models also result in unacceptably large standard least-squares values for
  the full dataset: $\chi_{\text{st}}^2/\ndf=4-6.2$. 
    \begin{table}[!htb]
  \centering
  \caption{Standard and log-normal $\chi^2/\ndf$ values calculated for the 
           distribution in the number of secondary protons with momenta above 500 MeV/c
           as measured by  T2K ND280~\cite{Abe:2018pwo}
           (see Fig.~\protect\ref{Fig:dsQELCC_dcosT+dsQELCC_dNp_ND28018_2_BBBA25_1}\;(b)).
           The \NuWro\ calculation is taken from Ref.~\cite{Abe:2018pwo}.
           }
  \begin{tabular*}{\linewidth}{@{\extracolsep\fill}lcc}             \hline\noalign{\smallskip}
  \MC{1}{l}{Model}            & \CN{\text{st}} & \CN{\text{log}} \\ \noalign{\smallskip}\hline\noalign{\smallskip}
  \SMRFG_MArun\ (\hA2018)     &  2.76          & 2.22            \\
  \SMRFG_MArun\ (\hN2018)     &  4.42          & 3.55            \\
  \GENIEa\                    &  7.29          & 8.30            \\
  \GENIEb\                    &  7.56          & 8.57            \\
  \NuWro\,2019                &  2.07          & 2.77            \\ \noalign{\smallskip}\hline
  \end{tabular*}
  \label{Tab:T2K_C8H8-2d_Neutrino-CCQE-like-standard+lognorm-N}
  \end{table}

  As is argued in Ref.\ \cite{Abe:2018pwo}, this analysis should be treated with caution. 
  In particular, such $\chi_{\text{st}}^2$ statistics can suffer from so-called Peelle's Pertinent Puzzle \cite{Peelle:87}
  (see Ref.~\cite{Bonus:2020yrd} for a possible remedy to deal with this problem).
  Recall that the \SMRFG_MArun\ models and \G18\ tunes satisfactory describe the T2K data on water target exposed
  to the same $\nu_\mu$ beam.
  Considering large statistical and systematic uncertainties, the relatively small expected differences between
  the $\nu_\mu$ scattering on oxygen and carbon are not very important. 
  Thus, it can be assumed that the main distinctions between the two T2K measurements are in different event selection
  criteria and data-processing methods.
  Moreover, the predicted cross sections calculated as functions of muon variables agree with the corresponding
  data subset much better than ones for the data subset which includes the final-state proton variables.
  One may therefore expect that the enormous disagreement between the model predictions and the full T2K C$_8$H$_8$
  dataset is likely partially, if not mainly, caused by the FSI problem common to all neutrino generators.
  \subsection{\MINERvA}
  \label{sec:MINERvA}

  \begin{figure*}[bht]                    
  \includegraphics[width=\linewidth]
  {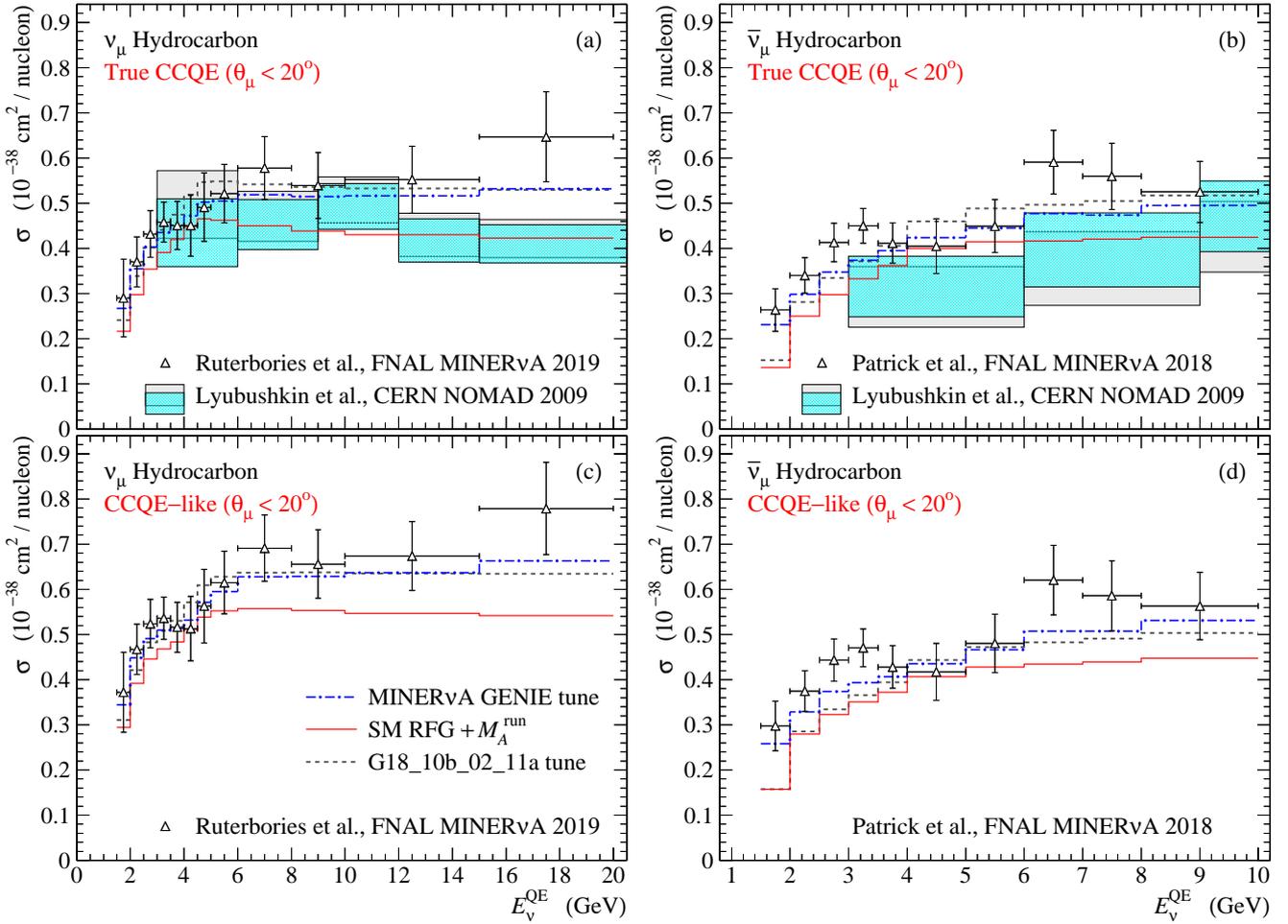}
  \caption{(Color online)
           Total cross sections vs.\ $E_\nu^{\text{QE}}$ for the true CCQE (a, b) and CCQE-like (c, d) $\nu_{\mu}$
           and $\overline\nu_{\mu}$ scattering from hydrocarbon as measured by \MINERvA\ \cite{Ruterbories:2018gub,Patrick:2018gvi}.
           Gray rectangles in the background represent the NOMAD data for carbon target~\cite{Lyubushkin:2008pe} and
           translucent rectangles are the result of a conversion of the NOMAD data to the \MINERvA\
           target composition and kinematic cuts by applying the \SMRFG_MArun\ model with \hA2018\ FSI.
           This conversions can only be thought of as an approximation for a qualitative comparison of the NOMAD
           and \MINERvA\ results in the area of their intersection.
           The vertical error bars and heights of rectangles represent the total errors including the normalization uncertainties.
           Histograms represent the \SMRFG_MArun\ model, \GENIEb\ tune, and the \MINERvA-tuned \GENIEG\ v1 model from
           Refs.~\cite{Ruterbories:2018gub,Patrick:2018gvi}.
           For more details, see Table~\protect\ref{Tab:MINERvA-1d_tot_Neutrino-CCQE+CCQE-like-standard+lognorm}.
          }
  \label{Fig:sQESCC_Ruterbories_Patrick_MINERvA18-19}
  \end{figure*}

  \begin{figure*}[!htb]
  \includegraphics[width=\textwidth]
  {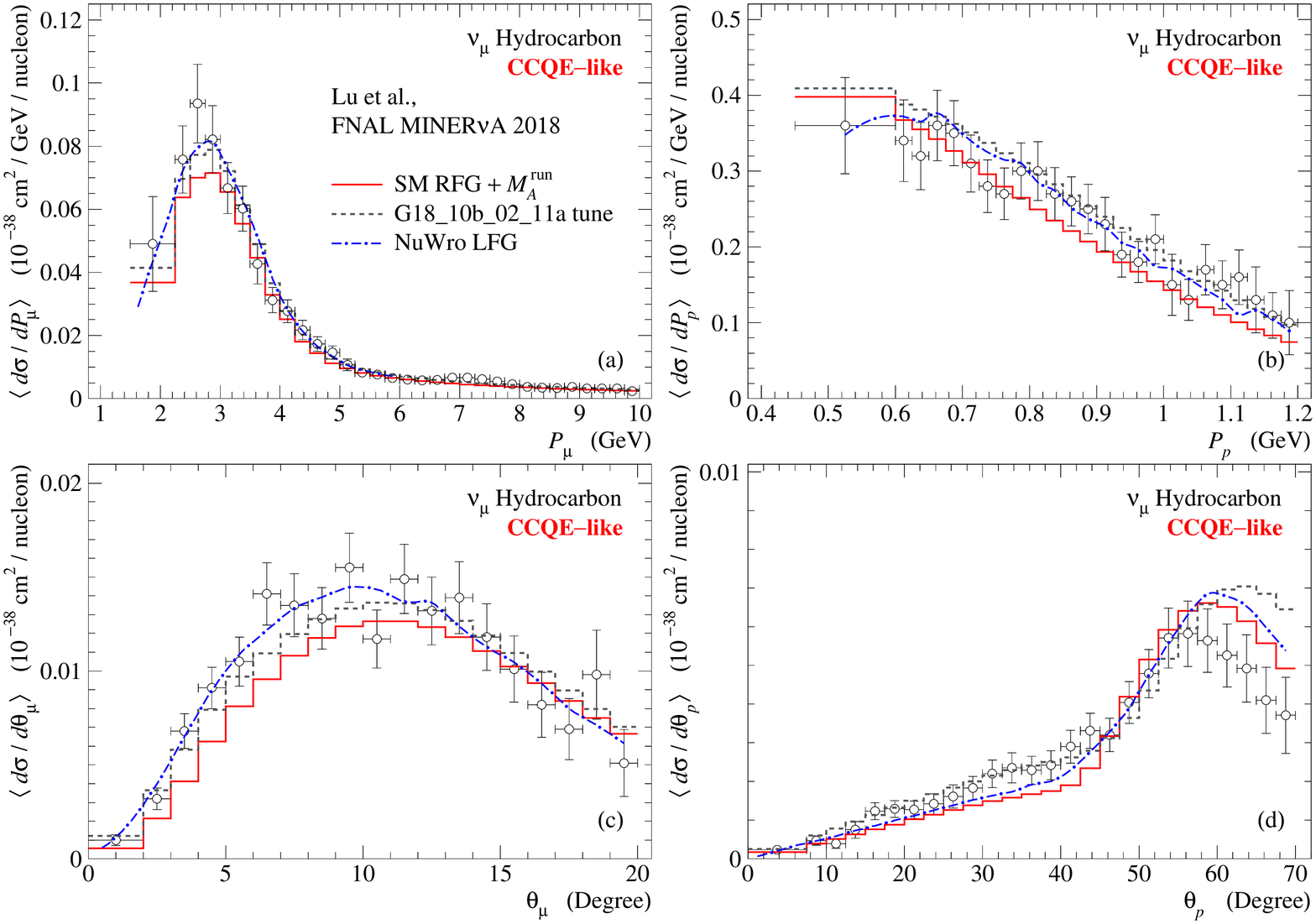}
  \caption{(Color online)
           Flux-weighted differential cross sections
           CCQE-like $\nu_{\mu}$ scattering from hydrocarbon
           as measured by \MINERvA\ \cite{Lu:2018stk}
           and plotted as functions of
           muon momentum, $P_\mu$ (a),
           proton momentum $P_p$ (b),
           muon scattering angle $\theta_\mu$ (c), and
           proton scattering angle $\theta_p$ (d).
           Vertical error bars represent the total errors including the normalization uncertainty of $7.5$\%.
           Histograms represent the \SMRFG_MArun\ model and \GENIEb\ tune.
           Dot-dash curves are calculated with the \NuWro\ generator using the LFG model.
           Both the data and the \NuWro\ predictions are borrowed from ancillary files to Ref.~\protect\cite{Lu:2015tcr}.
           For more details, see Table~\protect\ref{Tab:MINERvA-1d_Neutrino-CCQE-like}.
          }
  \label{Fig:MINERvA2018dd}
  \end{figure*}

  In this section, we discuss in detail the comparison between the experimental data of \MINERvA\ 
  \cite{Lu:2018stk, Ruterbories:2018gub, Patrick:2018gvi} and predictions from a set of different theoretical models.
  The results of the statistical analysis are presented in
  Figs.~\ref{Fig:sQESCC_Ruterbories_Patrick_MINERvA18-19}--\ref{Fig:MINERvA2018dQ2_E_QE} and
  Tables~\ref{Tab:MINERvA-1d_tot_Neutrino-CCQE+CCQE-like-standard+lognorm}--\ref{Tab:MINERvA-2d_Q2Antineutrino-CCQE+CCQE-like-standard+lognorm}.
  In all our calculations, we use the recent and most precise \emph{a priori} prediction of the NuMI low-energy flux
  \cite{Aliaga:2016oazWithErratum} 
  based on a simulation that has been modified to reproduce thin and thick target
  measurements of meson and nucleon production as well as measurements of meson and nucleon absorption cross sections.
  Although, for brevity, in the following we are talking about $\nu_\mu$ and $\overline\nu_{\mu}$ scattering from hydrocarbon target,
  in our simulations we actually accounted for the full chemical composition of the \MINERvA\ detector, which is a mix of
  $88.51$\% carbon,
  $ 8.18$\% hydrogen,
  $ 2.5$\% oxygen,
  $ 0.47$\% titanium,
  $ 0.2$\% chlorine,
  $ 0.07$\% aluminum, and
  $ 0.07$\% silicon~\cite{Ruterbories:2018gub}.

  The signal in the recent \MINERvA\ measurements of the CCQE-like $\nu_\mu/\overline\nu_\mu$ interactions
  is usually defined \cite{Lu:2018stk,Ruterbories:2018gub,Patrick:2018gvi} as an event which have post-FSI
  final states with one muon of angle $\theta_\mu<20^{\circ}$ with respect to the $\nu_\mu/\overline\nu_\mu$
  beam when exiting the nucleus, no mesons and heavy or excited baryons, any number of photons with energy
  $\le10$~MeV, and any number of protons or neutrons for incident neutrino, or with any number of protons having
  kinetic energy, $T_p$, below 120~MeV for incident antineutrino.
  More specific kinematic constraints will be mentioned when necessary.

  \subsubsection{Total CCQE and CCQE-like cross sections}
  \label{sec:MINERvA_TCS}

  In Fig.~\ref{Fig:sQESCC_Ruterbories_Patrick_MINERvA18-19}, we show the comparison of the \MINERvA\ data 
  \cite{Patrick:2018gvi,Ruterbories:2018gub} with three model predictions for the total CCQE (panels (a, b))
  and CCQE-like (panels (c, d)) total neutrino and antineutrino interaction cross sections on hydrocarbon
  as functions of $E_\nu^{\text{QE}}$. 
  Along with the \SMRFG_MArun\ model with \hA2018\ FSI and \GENIEb\ tune, we also examine a \MINERvA-tuned \GENIEG\,2.8.4\
  model (see Fig.\ \ref{Fig:sQESCC_Ruterbories_Patrick_MINERvA18-19}), which incorporates RPA and tuned $2p2h$ and
  is used as default in the \MINERvA\ analysis for extracting the cross sections.
  The \MINERvA\ measurements fill the gap between the modern low-energy (MiniBooNE, SciBooNE, T2K) and high-energy (NOMAD, LAr-TPC) 
  data and it is instructive to look at their intersection.
  For this purpose, panels (a, b) also show the NOMAD data~\cite{Lyubushkin:2008pe}.
  
  A comparison of the \MINERvA\ data with other recent measurements is shown in
  Fig.\ \ref{Fig:sQESCC_mn_n_C+ma_p_C_101.3.31.301.6k_2_BBBA25_1} (see \ref{sec:GoldData}).
  As is explained in Ref.~\cite{Patrick:2018gvi}, the \MINERvA\ measurements under consideration do not in fact yield
  exactly the total CCQE-like cross sections, $\sigma_{\text{tot}}$, but rather a single-differenti\-al projection of the
  double-differential cross sections, i.e.\ a well-defined approximation to $\sigma_{\text{tot}}$.
  This is in part because the RFG-based quantity $E_\nu^{\text{QE}}$ does not of course match the true (anti)neutrino energy
  $E_\nu^{\text{True}}$ (and it is not applicable to the contribution from the CCQE $\overline\nu_{\mu}$ scattering off hydrogen).
  However, as mentioned earlier, the quantities $E_\nu^{\text{QE}}$ and $E_\nu^{\text{True}}$ are closely correlated.
  Considering all this, we have performed accurate simulations with the kinematic restrictions reproducing the \MINERvA\ definition
  of $\sigma(E_\nu^{\text{QE}})$ as truly as possible.

  A direct quantitative comparison of the \MINERvA\ and NOMAD data is complicated by the kinematic cuts used in the \MINERvA\ analysis
  and by the difference in the chemical compositions of the two detector targets. The cuts considerably reduces the total CCQE
  cross section at low energies while becomes almost inessential above $5-6$~GeV; the difference in the 
  composition has a small effect at all energies.
  To qualitatively compare the \MINERvA\ and NOMAD data, we converted the latter by multiplying 
  bin-by-bin the original NOMAD data on a carbon-reach target (shown by gray rectangles in panels (a, b)
  of Fig.~\ref{Fig:sQESCC_Ruterbories_Patrick_MINERvA18-19}) by the factor 
  \begin{equation}
  \label{NOMAD2MINERvAconversionFactor}
  \sigma^{\text{MINERvA}}(E_\nu^{\text{QE}})/\sigma^{\text{NOMAD}}(E_\nu^{\text{QE}}),
  \end{equation}
  where $\sigma^{\text{MINERvA}}$ is the total CCQE cross section calculated for the \MINERvA\
  detector target using the aforementioned cuts (the main of which is the cut on the muon scattering angle)
  and $\sigma^{\text{NOMAD}}$ is the same cross section on carbon calculated for the
  full phase space. Both calculations are done using the \SMRFG_MArun\ model with \hA2018\ FSI and
  the result is shown by translucent rectangles.
  Note that the conversion factor \eqref{NOMAD2MINERvAconversionFactor} decreases the NOMAD neutrino cross section
  and increases the antineutrino one, although the effects due to the differences in the detector targets
  and kinematic cuts partially compensate each other.
  Despite the roughness and model dependence of such a conversion, a tension between the \MINERvA\ and NOMAD data
  is plainly visible both in normalization and in shape.
  The FSI model applied to the data calculation has little or no effect on this tension.
  Similar qualitative comparison of the \MINERvA\ and properly rescaled  MiniBooNE total CCQE cross sections
  indicates no significant tension between these data within the errors.
  \begin{table}[!htb]
\centering
\caption{The values of the standard and log-normal $\chi^2/\ndf$ calculated for the \MINERvA\
         neutrino \cite{Ruterbories:2018gub} CCQE and CCQE-like datasets
         on the total cross sections $\sigma(E_\nu^{\text{QE})})$
         (see Fig. \protect\ref{Fig:sQESCC_Ruterbories_Patrick_MINERvA18-19}).
         Calculations are done for several models using the full covariance matrices with $\ndf=12$.
         }        
\begin{tabular*}{\linewidth}{@{\extracolsep\fill}lcccc}                                                \hline\noalign{\smallskip}
                              & \MC{2}{c}{CCQE}                  & \MC{2}{c}{CCQE-like}             \\ \noalign{\smallskip}\cline{2-5}\noalign{\smallskip}
  \MC{1}{l}{Model}            & \CN{\text{st}} & \CN{\text{log}} & \CN{\text{st}} & \CN{\text{log}} \\ \noalign{\smallskip}\hline\noalign{\smallskip}
  \SMRFG_MArun\ (\hA2018)     & 1.27           & 1.89            & 1.32           & 1.44            \\
  \SMRFG_MArun\ (\hN2018)     & 1.28           & 1.91            & 1.41           & 1.63            \\
  \GENIEa\                    & 1.35           & 1.46            & 1.40           & 1.22            \\
  \GENIEb\                    & 1.39           & 1.50            & 1.41           & 1.31            \\
  \MINERvA-tuned \GENIEG\ v1  & 0.77           & 0.89            & 0.73           & 0.65            \\ \noalign{\smallskip}\hline
\end{tabular*}
\label{Tab:MINERvA-1d_tot_Neutrino-CCQE+CCQE-like-standard+lognorm}
\end{table}
  
  Figure \ref{Fig:sQESCC_Ruterbories_Patrick_MINERvA18-19} demonstrates (with minor reservations) reasonable agreement
  between the \MINERvA\ data and both \GENIEb\ tune and \MINERvA-tuned \GENIEG\ v1 model \cite{Patrick:2018gvi,Ruterbories:2018gub},
  but also a clear conflict of the data with the \SMRFG_MArun\ model, which significantly underestimates 
  the measured cross sections in both neutrino and antineutrino cases.
  Renormalization of the data does not essentially improve the situation due to the inconsistency presented also in the shapes.
  The fact that \SMRFG_MArun\ model describes the NOMAD (participated in the adjusting of the \MArun\ parameters) and fails
  in description of the \MINERvA\ data in the region of their intersection indicates a contradiction between these datasets.
  To quantify these findings, it is necessary to take into account strong correlations between the \MINERvA\ data points.
  This is not possible for the antineutrino case due to the unavailability of the covariance matrix for this dataset.
  For the neutrino case, the comparison is shown in Table~\ref{Tab:MINERvA-1d_tot_Neutrino-CCQE+CCQE-like-standard+lognorm}
  which collects the standard and log-normal $\chi^2/\ndf$ values calculated with the full covariance matrix.
  It is seen that the tension is partially softened after accounting for the correlations. 
  All four \GENIE3\ models listed in the Table have comparable $\chi^2$s indicating a small tension rather than contradiction
  with the \MINERvA\ results.
  Consistency between the standard and log-normal $\chi^2$s indicates the reliability of both statistical tests.
  The variations due to the FSI models are comparatively small and not statistically significant to make 
  a choice between these models.
  The \MINERvA-tuned \GENIEG\ v1 model \cite{Patrick:2018gvi,Ruterbories:2018gub} achieves (``by definition'')
  the best agreement with the data.
  This model (adjusted to the \MINERvA\ data) is a modified version of the \GENIEG\,2.8.4 default model,
  which in particular, reduces the standard, $1p1h$, CCQE cross section (by applying the RPA corrections), enhances the $2p2h$
  contribution (by about $53$\%, when integrated over the whole phase space), and decreases the non-resonant pion production
  (by about $43$\%).
  Just like the \GENIE3\ tunes \GENIEa\ and \GENIEb, both $2p2h$ and RPA contributions in the \MINERvA-tuned v1 model
  are the parts of the Valencia model
  \cite{Gil:1997bm, Gil:1997jg, Nieves:2004wxWithErratum}.
  Thus, we can conclude that the distinctions between the models are due mainly to different re-weighting 
  of the $2p2h$ and RPA parts.

  \subsubsection{Single-differential cross sections}
  \label{sec:MINERvA_SDCS}


    \begin{table*}[!htb]
  \centering
 \caption{The values of $\chi_{\text{st}}^2/\ndf$ and $\chi_{\text{log}}^2/\ndf$ (shown after slashes) calculated
           for the four \MINERvA\ CCQE-like datasets displayed in Fig.~\protect\ref{Fig:MINERvA2018dd}.
           Calculations are done with six models using the full covariance matrices.
           The columns ``a'', ``b'', ``c'', and ``d'' correspond to the panels with the same labels in Fig.~\ref{Fig:MINERvA2018dd};
           the last column shows the $\chi_{\text{st}}^2/\ndf$ and $\chi_{\text{log}}^2/\ndf$ evaluated for the full data set
           involving all four data subsets (assuming no correlations between them). The \ndf\ values are given in parentheses (second row).
           Asterisks mark the \NuWro\ calculations made using local Fermi gas (LFG) and spectral function (SF) models,
           and covering incomplete data subsets (\ndf=16 and 86 for, respectively, the second and last columns).
           The NuWro predictions are taken from the ancillary files of Ref.~\protect\cite{Lu:2015tcr}.
          }
  \begin{tabular*}{\textwidth}{@{\extracolsep\fill}lccccc}                                                              \hline\noalign{\smallskip}
                          &     a          &     b     &     c     &     d     &  Full dataset   \\
  \MC{1}{l}{Model}        &    (32)        &    (19)   &    (25)   &    (26)   &     (102)       \\ \noalign{\smallskip}\hline\noalign{\smallskip}
  \SMRFG_MArun\ (\hA2018) &  0.89/1.08     & 2.05/2.52 & 1.23/1.33 & 2.36/2.48 &   1.56/1.77     \\
  \SMRFG_MArun\ (\hN2018) &  0.92/1.12     & 1.84/2.19 & 1.47/1.54 & 2.41/2.62 &   1.61/1.80     \\
  \GENIEa\                &  1.07/1.02     & 1.93/1.77 & 1.27/0.99 & 3.01/1.68 &   1.77/1.32     \\
  \GENIEb\                &  1.06/1.01     & 1.85/1.64 & 1.50/1.13 & 2.99/1.70 &   1.81/1.33     \\
  \NuWro\ (LFG)           & ~1.29/1.19$^*$ & 1.21/1.15 & 0.99/1.01 & 1.99/1.81 & ~~1.40/1.31$^*$ \\
  \NuWro\ (SF)            & ~1.03/1.21$^*$ & 1.13/1.25 & 0.84/1.00 & 1.29/1.21 & ~~1.08/1.16$^*$ \\ \noalign{\smallskip}\hline
  \end{tabular*}
  \label{Tab:MINERvA-1d_Neutrino-CCQE-like}
  \end{table*}
  \begin{table}[!htb]
\centering
\caption{The values of the standard and log-normal $\chi^2/\ndf$ calculated for the \MINERvA\
         neutrino \cite{Ruterbories:2018gub} CCQE and CCQE-like datasets
         on the flux-weighted differential cross sections $d\sigma/dQ^2$
         (see Fig. \protect\ref{Fig:dsQES_L_CC_dQ2_QE_E_QE_MINERvA_020_MINERvA18_19}).
         Calculations are done for several models using the full covariance matrices with $\ndf=16$.
         }        
\begin{tabular*}{\linewidth}{@{\extracolsep\fill}lcccc}                                                \hline\noalign{\smallskip}
                              & \MC{2}{c}{CCQE}                  & \MC{2}{c}{CCQE-like}             \\ \noalign{\smallskip}\cline{2-5}\noalign{\smallskip}
  \MC{1}{l}{Model}            & \CN{\text{st}} & \CN{\text{log}} & \CN{\text{st}} & \CN{\text{log}} \\ \noalign{\smallskip}\hline\noalign{\smallskip}
  \SMRFG_MArun\ (\hA2018)     & 1.75           & 1.95            & 1.31           & 1.46            \\
  \SMRFG_MArun\ (\hN2018)     & 1.73           & 1.92            & 1.11           & 1.47            \\
  \GENIEa\                    & 2.43           & 2.77            & 2.12           & 2.72            \\
  \GENIEb\                    & 2.43           & 2.77            & 2.21           & 2.99            \\
  \MINERvA-tuned \GENIEG\ v1  & 2.90           & 2.08            & 4.02           & 3.08            \\ \noalign{\smallskip}\hline
\end{tabular*}
\label{Tab:MINERvA-1d_Q2_Neutrino-CCQE+CCQE-like-standard+lognorm}
\end{table}
    \begin{figure*}[!htb]
  \includegraphics[width=\linewidth]
  {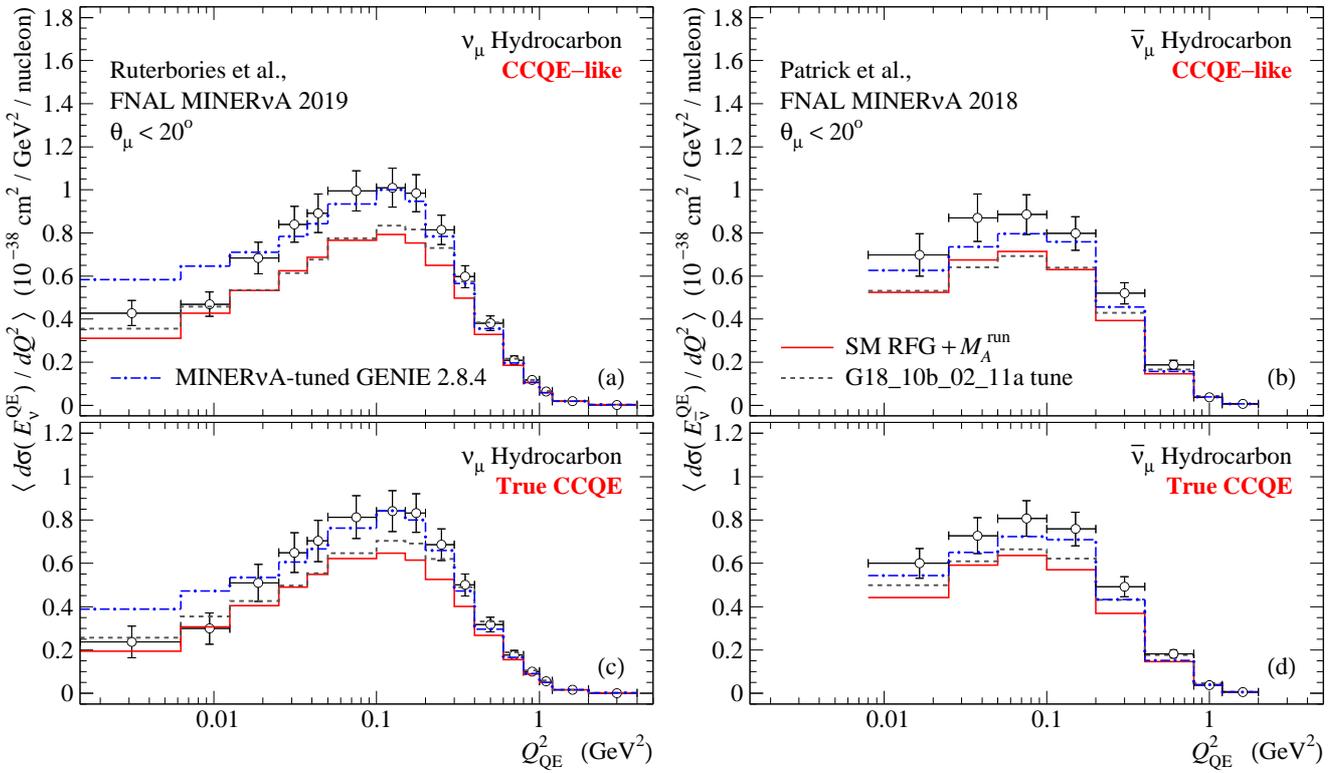}
  \caption{(Color online)
            Flux-weighted differential cross sections $d\sigma/dQ^2$ vs.\ $Q_{\text{QE}}^2$
            for the CCQE-like (a, b) and true CCQE (c, d) $\nu_{\mu}$ and
            $\overline\nu_{\mu}$ scattering from hydrocarbon as measured by \MINERvA\
            \cite{Patrick:2018gvi,Ruterbories:2018gub}. Vertical error bars represent
            the total errors including the normalization uncertainty of $7.5$\%.
            Histograms and curves represent the \SMRFG_MArun\ model and \GENIEb\ tune predictions.
            Also shown the \MINERvA-tuned \GENIEG\ v1 model from Refs.~\cite{Patrick:2018gvi,Ruterbories:2018gub}.
            For more details, see Table~\protect\ref{Tab:MINERvA-1d_Q2_Neutrino-CCQE+CCQE-like-standard+lognorm}.
           }
  \label{Fig:dsQES_L_CC_dQ2_QE_E_QE_MINERvA_020_MINERvA18_19}
  \end{figure*}
  
  A comparison of the flux-weighted differential CCQE-like $\nu_{\mu}$ cross sections
  $d\sigma/dP_\mu$, $d\sigma/dP_p$, $d\sigma/d\theta_\mu$, and $d\sigma/d\theta_p$ measured by \MINERvA\
  \cite{Lu:2018stk} with the respective preditions of three models, \SMRFG_MArun, \GENIEb, and \NuWro\ LFG,
  is shown in Fig.\ \ref{Fig:MINERvA2018dd}. 
  Here $P_\mu$ and $\theta_\mu$ ($P_p$ and $\theta_p$) are, respectively, the muon (proton) momentum
  and scattering angle.
  The additional requirements for the \MINERvA\ signal definition include
  $1.50~\text{GeV} < P_\mu < 10~\text{GeV}$,  $\theta_\mu < 20^{\circ}$,
  $0.45~\text{GeV} < P_p   < 1.2~\text{GeV}$, $\theta_p   < 70^{\circ}$;
  there must be at least one final-state proton satisfying the above conditions.
  At first glance, the solid histogram corresponding to the \SMRFG_MArun\ model shown in panel (a) describes 
  the experimental data ($d\sigma/dP_\mu$) at small muon momenta worse than two other models.
  For other panels, it is more difficult to tell by eye which model best describes the experimental data.
  Therefore, it is better to refer to Table~\protect\ref{Tab:MINERvA-1d_Neutrino-CCQE-like},
  in which we present the standard and log-normal $\chi^2/\ndf$ values for these and three other models.
  All calculations take into account the full correlation matrices.
  As one can see from the table, the \NuWro\ SF model has the least $\chi_{\text{st}}^2/\ndf$ for all
  data subsets except those shown in panel (a).
  The \NuWro\ LFG model demonstrates a little worse results, but in general better than the other models everywhere,
  except the case shown in panel (a), where just the \SMRFG_MArun\ model shows the best $\chi_{\text{st}}^2/\ndf$ value. 
  This example clearly demonstrates the discrepancy between the visual assessment of the quality of the data
  description and a calculation that takes into account the play of correlations.
  Comparing the values of $\chi_{\text{st}}^2/\ndf$ for the full dataset presented in Fig.~\ref{Fig:MINERvA2018dd}
  (last column in Table~\protect\ref{Tab:MINERvA-1d_Neutrino-CCQE-like}) one finds that the \NuWro\ results
  obtained with the LFG and SF models show the least $\chi_{\text{st}}^2/\ndf$ values, the \G18\ tunes give
  the worst ones, and the $\chi_{\text{st}}^2/\ndf$ values for the \SMRFG_MArun\ model lie between them.
  What concerns the $\chi_{\text{log}}^2/\ndf$ values for the full dataset, here again the two \NuWro\ versions
  show the best result, but the worst result in this case is shown by the \SMRFG_MArun\ models.
  In all cases, the description is slowly sensitive to the FSI model.

  The flux-weighted  cross sections $d\sigma/dQ_{\text{QE}}$ for the $\nu_{\mu}$ and $\overline\nu_{\mu}$
  scattering from hydrocarbon along with few model predictions are shown in
  Fig.~\ref{Fig:dsQES_L_CC_dQ2_QE_E_QE_MINERvA_020_MINERvA18_19}, where the two top (bottom) panels display
  the CCQE-like (CCQE) datasets.
  The $\chi_{\text{st}}^2/\ndf$ and $\chi_{\text{log}}^2/\ndf$ values for several models
  are listed in Table~\ref{Tab:MINERvA-1d_Q2_Neutrino-CCQE+CCQE-like-standard+lognorm} (for the neutrino case only).
  Although by eye the \SMRFG_MArun\ model describes the data worse than the \G18\ tune and \MINERvA-tuned \GENIEG,
  the quantitative comparison in Table \ref{Tab:MINERvA-1d_Q2_Neutrino-CCQE+CCQE-like-standard+lognorm} shows a completely
  opposite picture, also illustrating the unpredictable effect of correlations.
 
  A comparison of the flux-weighted differential cross sections $d\sigma/dp_T$ and $d\sigma/dp_L$
  (where $p_T$ and $p_L$ are the muon transverse and longitudinal momenta) for the true CCQE and CCQE-like
  $\nu_\mu$ and $\overline{\nu}_\mu$ scattering from hydrocarbon \cite{Ruterbories:2018gub,Patrick:2018gvi}
  shows that the \G18\ tune and \MINERvA-tuned \GENIEG\ both give a reasonable description of the data
  while the \SMRFG_MArun\ model underestimates the \MINERvA\ cross sections. 
  We do not discuss here the corresponding $\chi^2$s since these datasets represent just the one-dimensional slices
  from the double-differential cross sections shown below in Figs.\ \ref{Fig:MINERvA2018ddA}--\ref{Fig:MINERvA2018dQ2_E_QE}
  and the covariance matrices for these slices are not provided by the authors.

  \subsubsection{Double-differential cross sections}
  \label{sec:MINERvA_DDCS}
 

  \def\Vspace{3.0mm}           
  \def\W{\textwidth}

  \begin{figure*}[!htb]
  \includegraphics[width=\W]
  {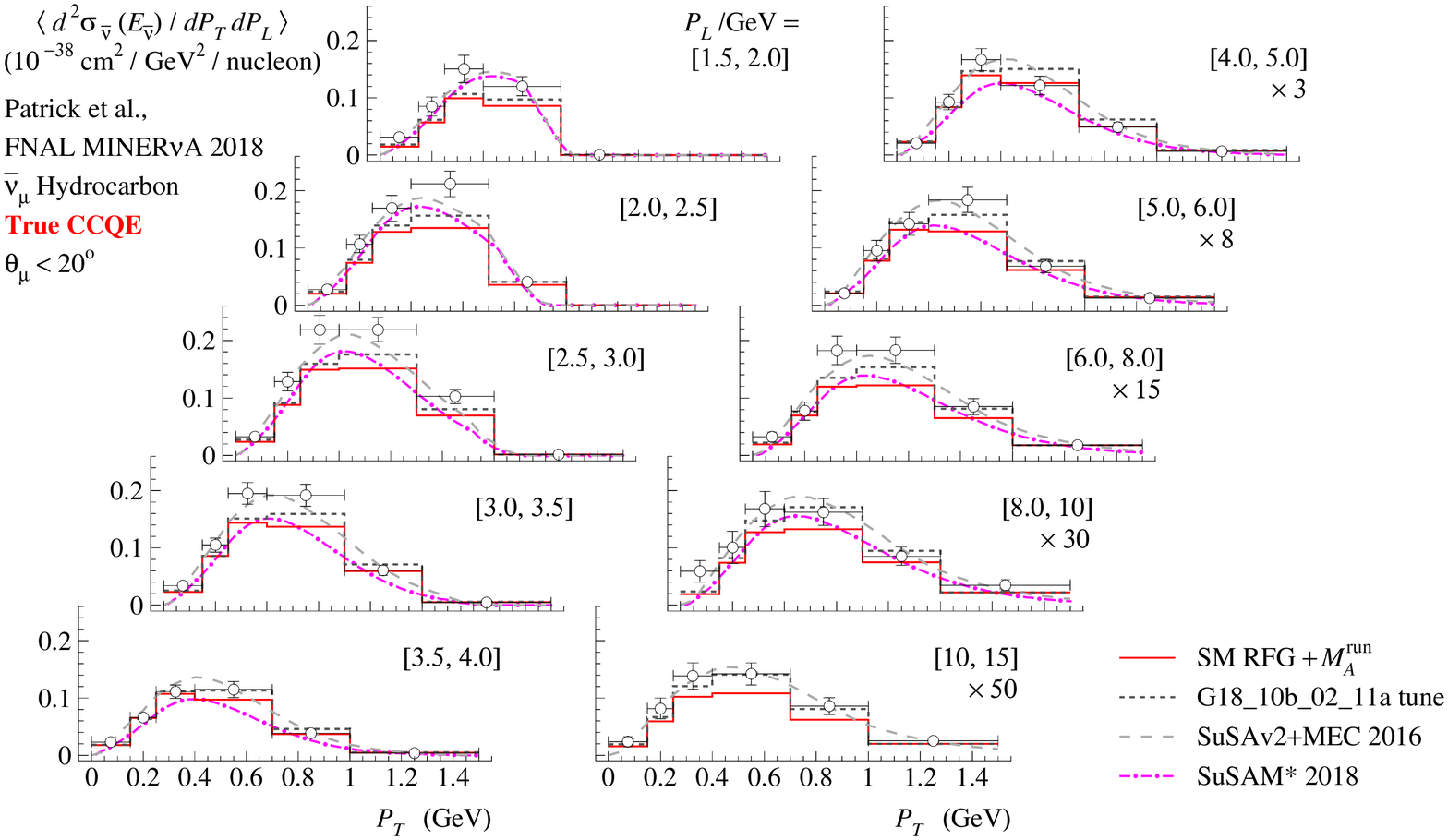} \\[\Vspace]
  \includegraphics[width=\W]
  {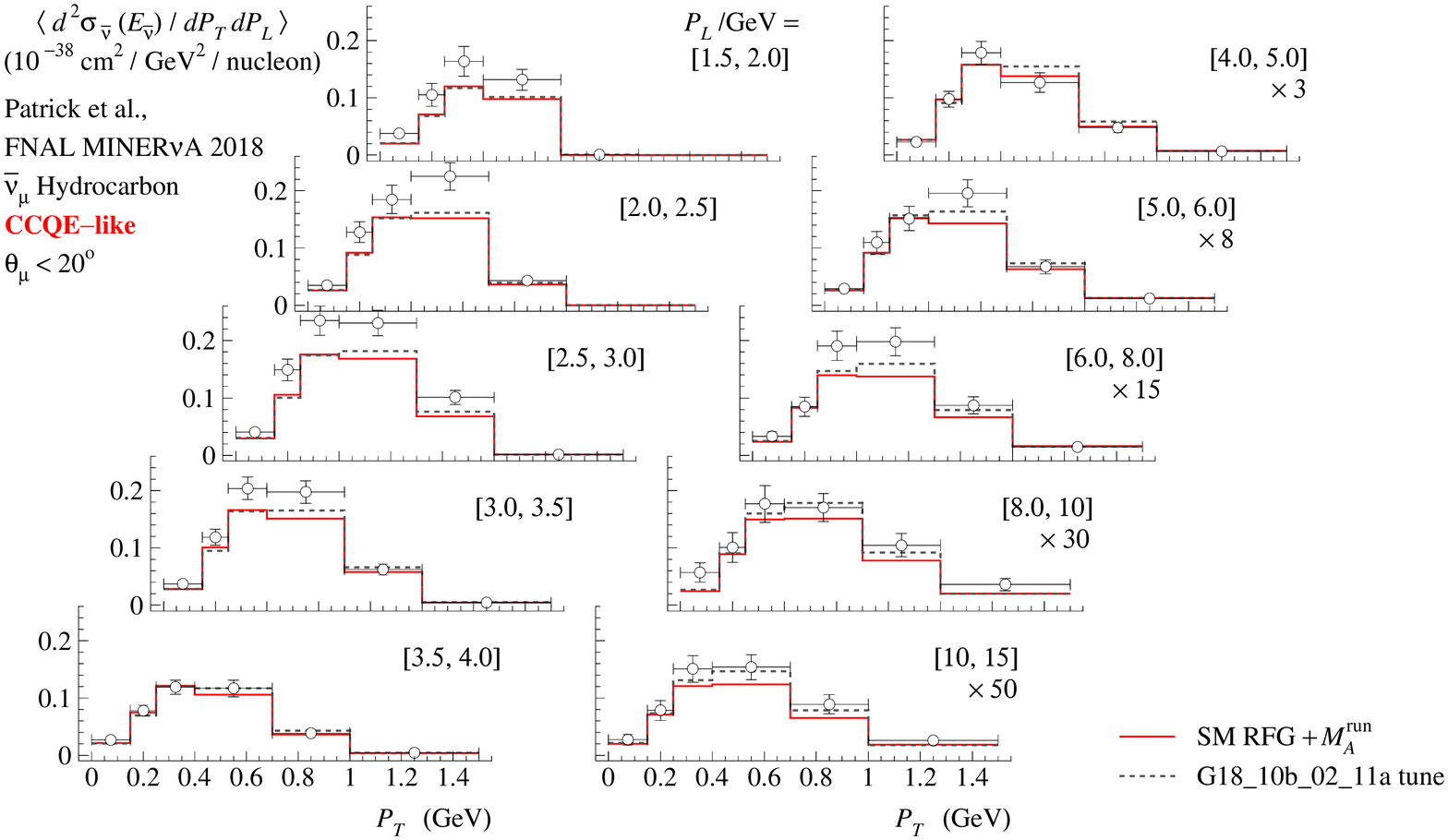}
  \caption{(Color online)
           Flux-weighted double-differential cross sections,  $d^2\sigma/dP_TdP_L$, for the
           CCQE (ten top panels) and CCQE-like (ten bottom panels) $\overline{\nu}_{\mu}$ scattering
           from hydrocarbon as measured by \MINERvA\ \cite{Patrick:2018gvi} and plotted vs.\ transverse
		   muon momentum, $P_T$, for several intervals of the longitudinal momentum, $P_L$
		   (shown in square brackets).
           Vertical error bars represent the total errors including the normalization uncertainty of $7.5$\%.
           Histograms represent the \SMRFG_MArun\ model and \GENIEb\ tune,
           curves show \SuSAv2MEC\ 2016~\cite{Megias:2016fjk} and
           \SuSAM\ 2018~\cite{RuizSimo:2018kdl} models.
           See Table~\protect\ref{Tab:MINERvA-2d_Neutrino+Antineutrino-CCQE+CCQE-like-standard+lognorm}.
          }
  \label{Fig:MINERvA2018ddA}
  \end{figure*}

  \begin{figure*}[!htb]
  \includegraphics[width=\W]
  {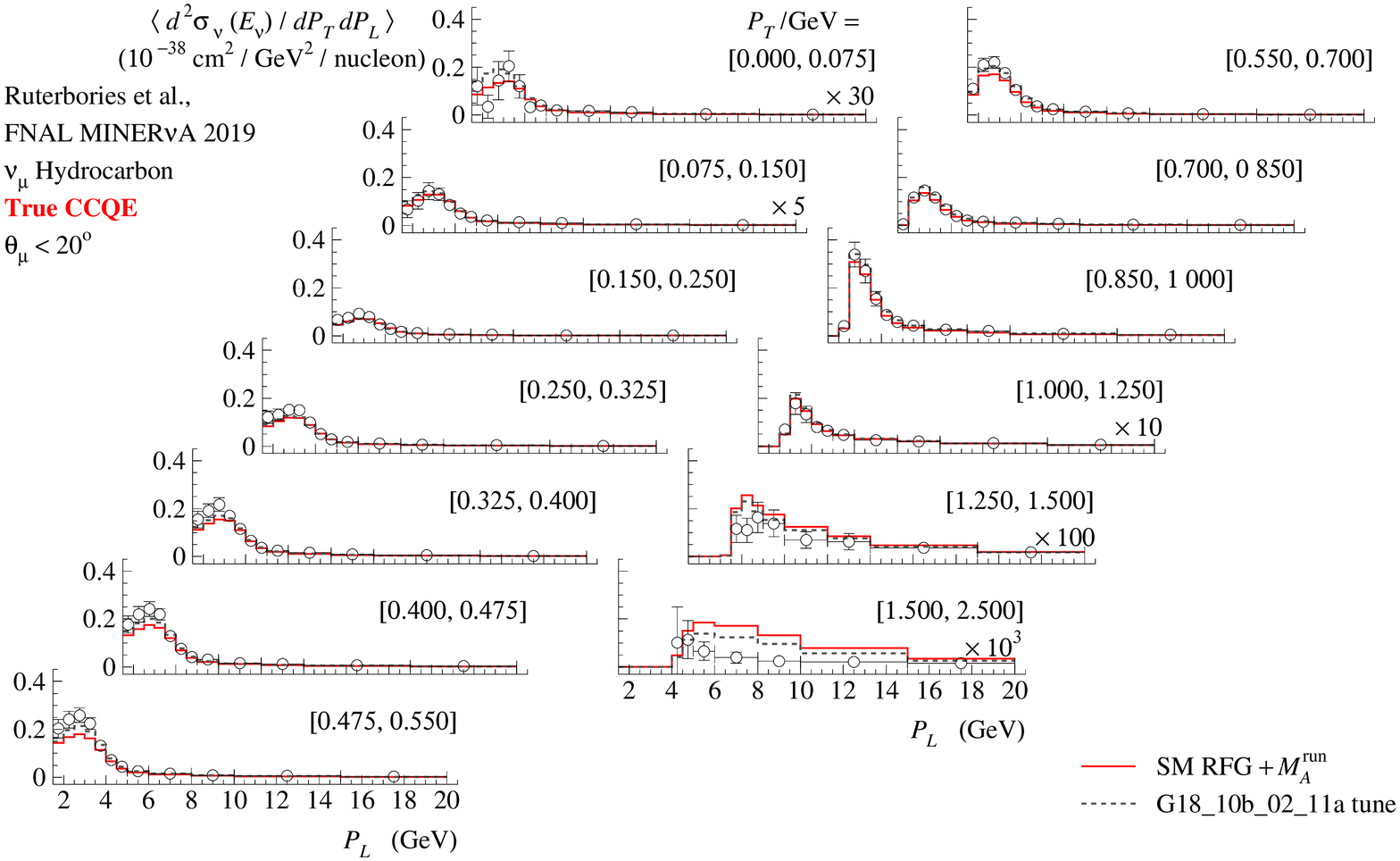}  \\[\Vspace]
  \includegraphics[width=\W]
  {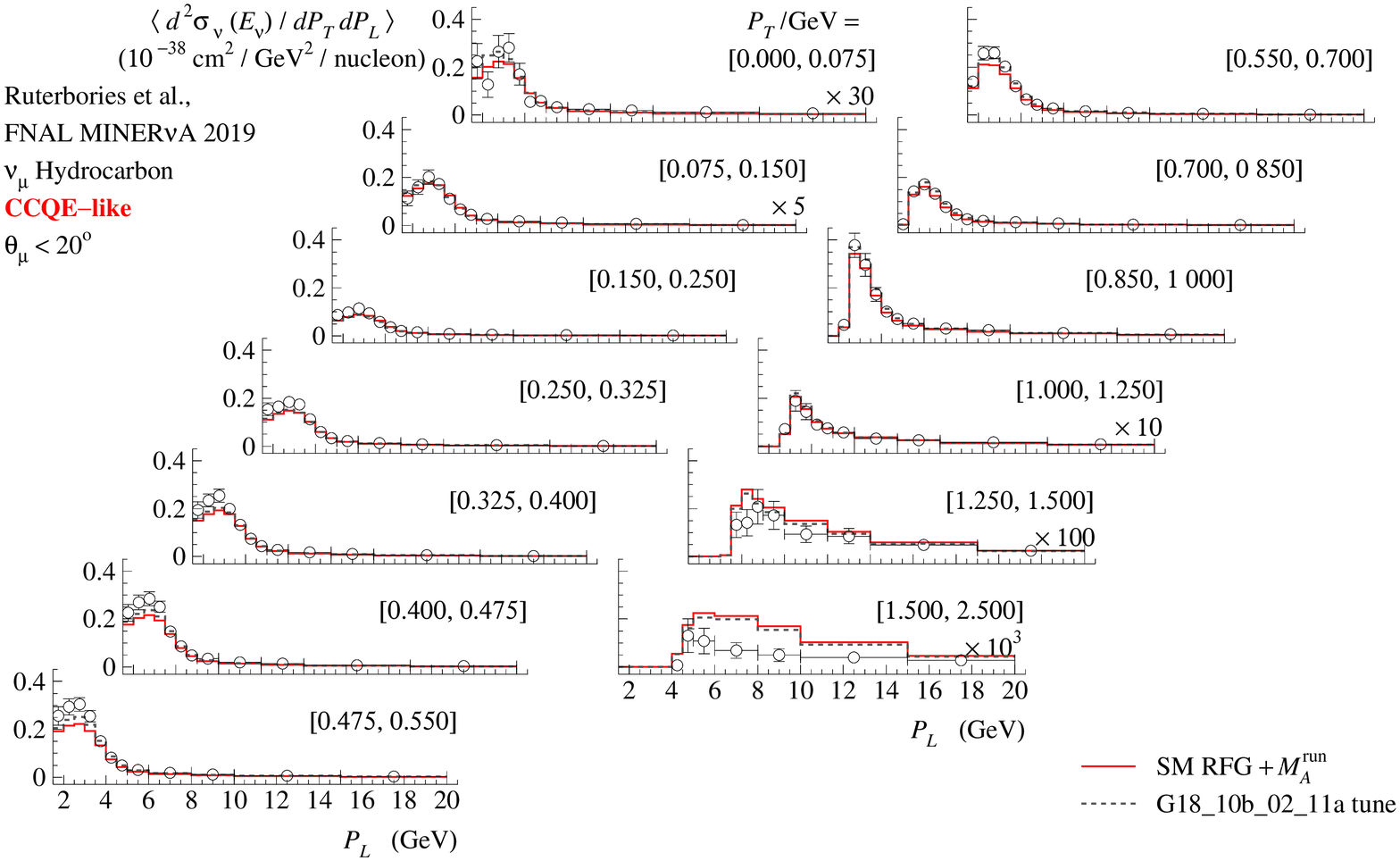}
  \caption{(Color online)
           Flux-weighted double-differential cross sections, $d^2\sigma/dP_TdP_L$, for the
           CCQE (thirteen top panels) and CCQE-like (thirteen bottom panels) $\nu_{\mu}$ scattering
           from hydrocarbon as measured by \MINERvA\ \cite{Ruterbories:2018gub} and
           plotted vs.\ transverse muon momentum, $P_T$,
           for several intervals of the longitudinal momentum, $P_L$ (shown in square brackets).
           Vertical error bars represent the total errors including the normalization uncertainty of $7.5$\%.
           Histograms represent the \SMRFG_MArun\ model and \GENIEb\ tune.
           For more details, see Table~\protect\ref{Tab:MINERvA-2d_Neutrino+Antineutrino-CCQE+CCQE-like-standard+lognorm}.
          }
  \label{Fig:MINERvA2018ddN-CCQE}
  \end{figure*}

  \begin{figure*}[!htb]
  \includegraphics[width=\W]
  {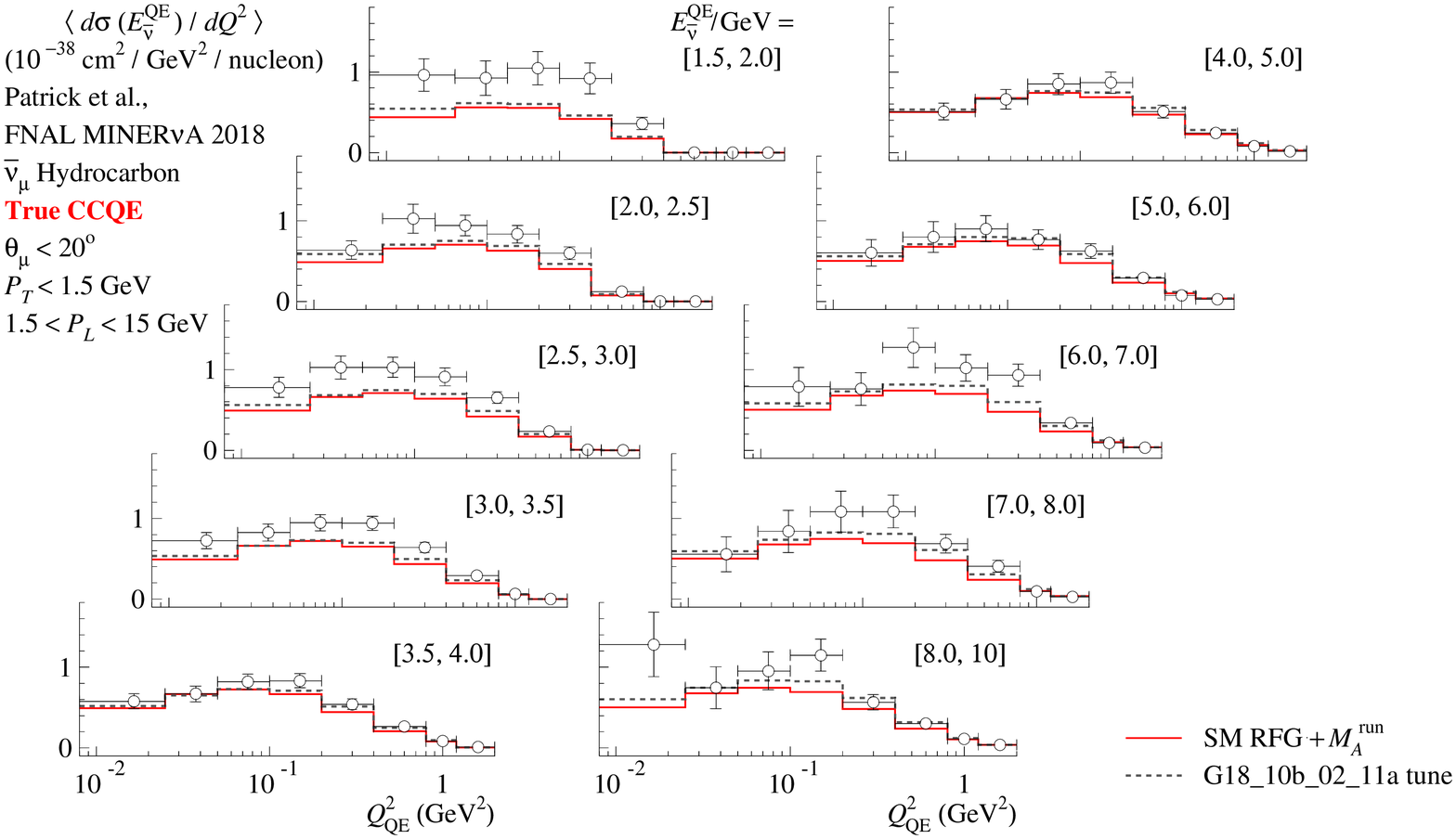} \\[\Vspace]
  \includegraphics[width=\W]
  {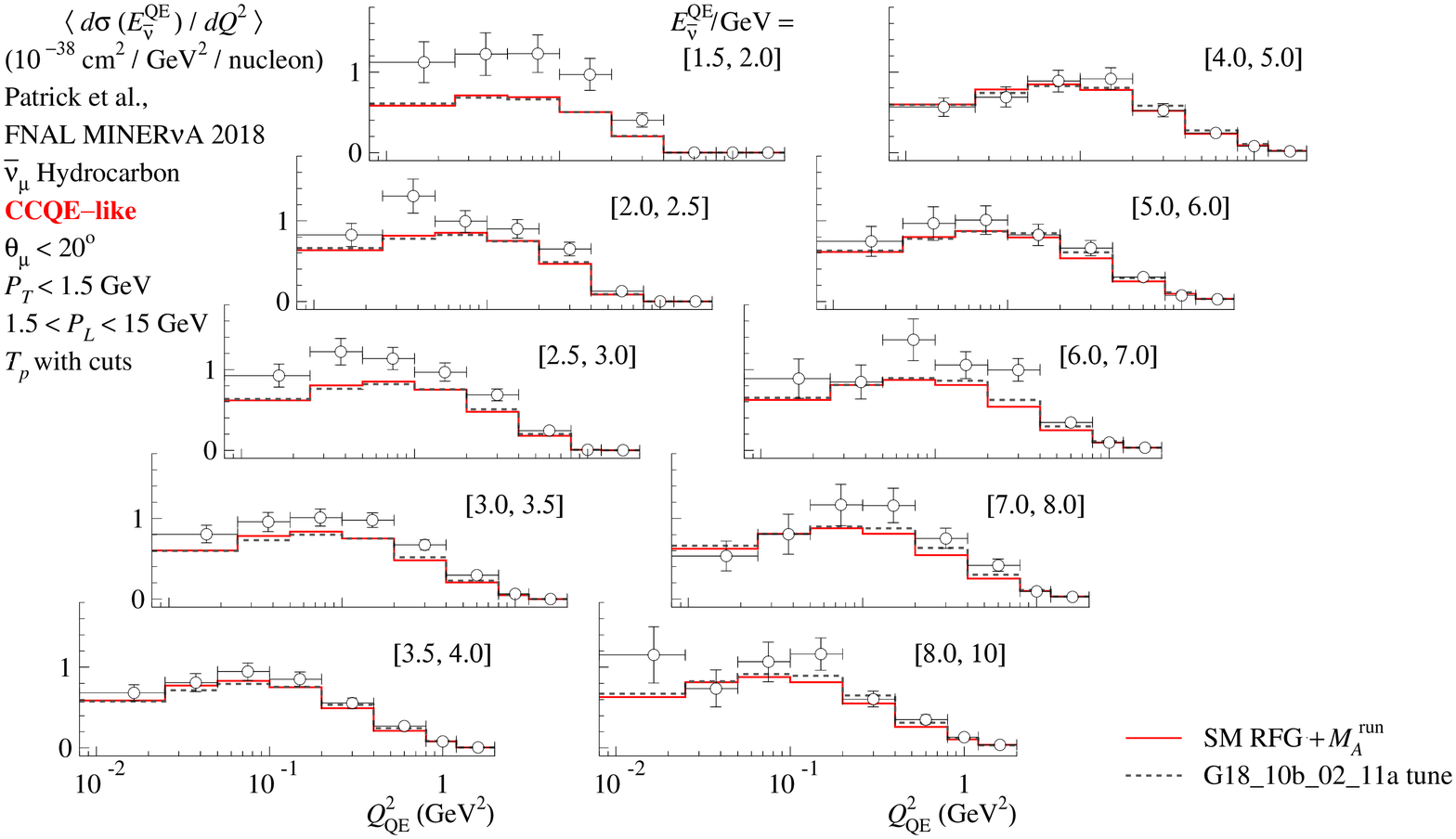}
  \caption{(Color online)
           Flux-weighted double-differential cross sections for the CCQE (ten top panels) and
           CCQE-like (ten bottom panels) $\overline{\nu}_{\mu}$ scattering
           from hydrocarbon as measured by \MINERvA~\cite{Patrick:2018gvi} and plotted as functions of $Q^2_{\text{QE}}$
           for several intervals of the antineutrino energy $E^{\text{QE}}_{\overline\nu_\mu}$ (shown in square brackets) 
           The vertical error bars represent the total errors including the normalization uncertainty of $7.5$\%.
           Histograms represent the \SMRFG_MArun\ model and \GENIEb\ tune.
           For more details, see Table~\protect\ref{Tab:MINERvA-2d_Q2Antineutrino-CCQE+CCQE-like-standard+lognorm}.
          }
  \label{Fig:MINERvA2018dQ2_E_QE}
  \end{figure*}

    \begin{table}[!htb]
  \centering
  \caption{The values of the standard and log-normal $\chi^2/\ndf$
           calculated for the \MINERvA\ neutrino \cite{Ruterbories:2018gub} and
           antineutrino \cite{Patrick:2018gvi} CCQE and CCQE-like datasets
           on the flux-weighted double-differential cross sections $d^2\sigma/dP_TdP_L$
           (see Fig.\ \protect\ref{Fig:MINERvA2018ddA}).
           Calculations are done for several models using the full covariance matrices.
          }
  \begin{tabular*}{\linewidth}{@{\extracolsep\fill}lcccc}                                                                       \hline\noalign{\smallskip}
                              & \MC{2}{c}{CCQE}                  & \MC{2}{c}{CCQE-like}             \\ \noalign{\smallskip}\cline{2-5}\noalign{\smallskip}
  \MC{1}{l}{Model}            & \CN{\text{st}} & \CN{\text{log}} & \CN{\text{st}} & \CN{\text{log}} \\ \noalign{\smallskip}\hline\noalign{\medskip}
                                \MC{5}{c}{{\it Neutrino dataset} (ndf=144)}                         \\ \noalign{\medskip}
  \SMRFG_MArun\ (\hA2018)     & 1.66           & 1.53            & 1.73           & 1.81            \\
  \SMRFG_MArun\ (\hN2018)     & 1.65           & 1.52            & 1.49           & 1.86            \\
  \GENIEa\                    & 1.43           & 1.20            & 1.48           & 1.32            \\
  \GENIEb\                    & 1.45           & 1.22            & 1.26           & 1.34            \\
  \MINERvA-tuned \GENIEG\     & 1.51           & 1.04            & 1.77           & 1.41            \\ \noalign{\medskip}
                                \MC{5}{c}{{\it Antineutrino dataset} (ndf=58)}                      \\ \noalign{\medskip}
  \SMRFG_MArun\ (\hA2018)     & 1.34           & 2.22            & 1.49           & 2.06            \\
  \SMRFG_MArun\ (\hN2018)     & 1.34           & 2.11            & 1.49           & 2.14            \\
  \GENIEa\                    & 1.46           & 1.89            & 1.44           & 1.74            \\
  \GENIEb\                    & 1.44           & 1.85            & 1.40           & 1.72            \\
  \SuSAv2MEC\ 2018            & 2.13           & 2.02            & ---            & ---             \\
  \SuSAM\ 2018                & 1.51           & 2.64            & ---            & ---             \\
  \MINERvA-tuned \GENIEG\     & 1.58           & 1.85            & 1.40           & 1.70            \\ \noalign{\smallskip}\hline
  \end{tabular*}
  \label{Tab:MINERvA-2d_Neutrino+Antineutrino-CCQE+CCQE-like-standard+lognorm}
  \end{table}
  \begin{table}[!htb]
\centering
\caption{The values of the standard and log-normal $\chi^2/\ndf$ calculated for the \MINERvA\
         antineutrino CCQE and CCQE-like datasets on the flux-weighted double-differential cross sections
         $d^2\sigma/dQ_{\text{QE}}^2dE_\nu^{\text{QE}}$ (see Fig.\ \protect\ref{Fig:MINERvA2018dQ2_E_QE}
		 and $d^2\sigma/dQ_{\text{QE}}^2dE_\nu^{\text{True}}$ (not illustrated) \cite{Patrick:2018gvi}.
          Calculations are done for several models using a simplified (block-diagonal) covariance matrices with 74 and 76 degrees of freedom 
         for the cross sections represented in terms of $E_\nu^{\text{QE}}$ and $E_\nu^{\text{True}}$, respectively.
         The reasons for the modification of the covariance matrices are explained in the main text.
        }        
\begin{tabular*}{\linewidth}{@{\extracolsep\fill}lcccc}                                                                         \hline\noalign{\smallskip}
                              & \MC{2}{c}{$E_\nu^{\text{QE}}$}   & \MC{2}{c}{$E_\nu^{\text{True}}$} \\ \noalign{\smallskip}\cline{2-5}\noalign{\smallskip}
  \MC{1}{l}{Model}            & \CN{\text{st}} & \CN{\text{log}} & \CN{\text{st}} & \CN{\text{log}} \\ \noalign{\smallskip}\hline\noalign{\medskip}
                                \MC{5}{c}{{\it CCQE dataset}}                                       \\ \noalign{\smallskip}
  \SMRFG_MArun\ (\hA2018)     & 0.58           & 0.53            & 0.58           & 0.54            \\
  \SMRFG_MArun\ (\hN2018)     & 0.60           & 0.55            & 0.60           & 0.55            \\
  \GENIEa\                    & 0.56           & 0.37            & 0.27           & 0.35            \\
  \GENIEb\                    & 0.56           & 0.37            & 0.29           & 0.49            \\
  \SuSAM\ 2018                & 0.05           & 0.21            & ---            & ---             \\ \noalign{\medskip}
                                \MC{5}{c}{{\it CCQE-like dataset}}                                  \\ \noalign{\medskip}
  \SMRFG_MArun\ (\hA2018)     & 0.54           & 0.86            & 0.58           & 0.86            \\
  \SMRFG_MArun\ (\hN2018)     & 0.62           & 0.96            & 0.66           & 0.94            \\
  \GENIEa\                    & 0.61           & 0.86            & 0.48           & 0.97            \\
  \GENIEb\                    & 0.67           & 0.93            & 0.54           & 1.14            \\ \noalign{\smallskip}\hline
\end{tabular*}
\label{Tab:MINERvA-2d_Q2Antineutrino-CCQE+CCQE-like-standard+lognorm}
\end{table}
  
  The \MINERvA\ results for antineutrino scattering on hydrocarbon~\cite{Patrick:2018gvi}
  with several predictions are shown in Fig.~\ref{Fig:MINERvA2018ddA}.
  The CCQE and CCQE-like flux-weighted double-differenti\-al cross sections $d^2\sigma/dP_TdP_L$ are presented
  at ten top and ten bottom panels, respectively, for the fixed intervals of the muon longitudinal momentum $P_L$
  (shown in square brackets) as functions of the muon transverse momentum $P_T$;
  the data are compared with the four  models as listed in the legend.
  Similar plots for the neutrino CCQE and CCQE-like datasets~\cite{Ruterbories:2018gub} are presented in
  Fig.\ \ref{Fig:MINERvA2018ddN-CCQE} as functions of $P_L$ for the fixed intervals of $P_T$;
  the data are compared with the two models.
  The signal definition for CCQE-like events in this case is post-FSI CC-events with $\theta_\mu<20^\circ$,
  any numbers of nucleons and low-energy photons (with energies below $10$~MeV) and, as above,
  without mesons and heavy or excited baryons.
  In the case of the $\nu_\mu/\overline\nu_\mu$ true CCQE signal, only the cuts on leptonic variables are essential.
  
   The values of the redused standard and log-normal $\chi^2$s the models presented in
  are summarized in Table~\protect\ref{Tab:MINERvA-2d_Neutrino+Antineutrino-CCQE+CCQE-like-standard+lognorm}.
  The \SMRFG_MArun\ model and \G18 tunes are presented for the two FSI versions.
  The table also includes the corresponding numbers for the \MINERvA-tuned \GENIEG\,2.8.4\ model, which was used in the \MINERvA\
  experiment for extracting the cross sections. 
  The calculations were carried out using the full covariance matrices.
  Note that our estimation of $\chi_{\text{st}}^2/\ndf$ for the \SuSAv2MEC\ model ($\overline\nu_\mu$ case, CCQE)
  differs noticeably from author's result of $1.79$ \cite{Megias:2018ujz}.
  Overall, the two \G18\ tunes provide the best description of the CCQE and CCQE-like data for both $\overline\nu_\mu$
  and $\nu_\mu$ cases; this can be seen both from the figures and from the comparison of the standard and log-normal $\chi$s
  in Table~\protect\ref{Tab:MINERvA-2d_Neutrino+Antineutrino-CCQE+CCQE-like-standard+lognorm}.
  As for other models, then  the \SMRFG_MArun\ shows better agreement in terms of $\chi_{\text{st}}^2/\ndf$ for the $\overline\nu_\mu$ CCQE case, 
  and \MINERvA-tuned \GENIEG\ v1 has the lowest values of $\chi_{\text{st}}^2/\ndf$
  for $\nu_\mu$ CCQE and of $\chi_{\text{log}}^2/\ndf$ for the $\overline\nu_\mu$ CCQE-like case.
  
  The CCQE and CCQE-like flux-weighted $\overline\nu_\mu$ double-dif\-fe\-ren\-tial cross sections
  of the same data as shown in Fig.~\ref{Fig:MINERvA2018ddA}
  with the same cuts~\cite{Patrick:2018gvi} but for other kinematic variables
  are plotted as functions of $Q^2_{\text{QE}}$ for several intervals of the fixed energy
  $E^{\text{QE}}_{\overline\nu_\mu}$ (Fig.\ \ref{Fig:MINERvA2018dQ2_E_QE}).
  (the energy bins are shown in square brackets).
  We do not present a comparison with analogous cross sections for the fixed true energy intervals, because 
  they are very similar to those in Fig.\ \ref{Fig:MINERvA2018dQ2_E_QE}.
  The values $E^{\text{QE}}_{\overline\nu_\mu}$ and $Q^2_{\text{QE}}$ are defined by Eqs.\ \eqref{eq:QE-quantities}
  with $E_b=30$~MeV.
  The quantitative comparison 
  is given in Table~\ref{Tab:MINERvA-2d_Q2Antineutrino-CCQE+CCQE-like-standard+lognorm}.
  An important remark should be made regarding these data.
  The covariance matrices reported by the \MINERvA\ Collaboration for the $d^2\sigma/dQ^2_{\text{QE}}dE^{\text{QE}}_{\overline\nu_\mu}$
  and $d^2\sigma/dQ^2_{\text{QE}}dE^{\text{True}}_{\overline\nu_\mu}$ cross sections are not positive-definite
  \footnote{In all our calculations, we automatically check the symmetry, non-singularity, and positive definiteness of
            all covariance/correlation matrices involved into the analysis.}.
  Detailed investigation revealed that the incorrect covariances are located in the matrix elements that take into account
  the correlations between the $Q^2_{\text{QE}}$ bins at fixed antineutrino energy bins (for both $E^{\text{QE}}_{\overline\nu_\mu}$
  and $E^{\text{True}}_{\overline\nu_\mu}$ cases), namely -- in the elements outside the eight main-diagonal blocks.
  So we decided to neglect the doubtful correlations and simplify the standard and log-normal covariance matrices
  by putting the elements of the corresponding off-diagonal submatrices equal to zero. The ``scrubbed'' covariance matrices
  are therefore the partitioned matrices containing only six $10\times10$ and two $8\times8$ main-diagonal blocks.
  These matrices are positive-definite, while account only a part of correlations.
  Therefore, the estimations listed in Table~\ref{Tab:MINERvA-2d_Q2Antineutrino-CCQE+CCQE-like-standard+lognorm} should be treated with caution.
  Keeping this in mind, 
  we may conclude that all the models satisfactory describe this data subset.
  It is seen that the \SuSAM\ model has the least standard and log-normal $\chi^2$s for the CCQE cross section
  $d^2\sigma/dQ_{\text{QE}}^2dE_\nu^{\text{QE}}$.
  As for other models, the \G18\ tunes seem better for the CCQE dataset, but for CCQE-like data it is hard to pick out the best one.

  \section{Conclusions}
  \label{sec:Conclusions}

  In this paper, we suggest the phenomenological notion of running (energy-dependent)
  axial-vector mass, $M^\text{run}_A(E_\nu)$, as a flexible tool for description of the
  quasielastic interactions of neutrinos and antineutrinos with nuclei within the
  framework of the Smith-Moniz RFG model.
  This intention was inspired by the heuristic fact that the effective dipole axial-vector mass 
  of the nucleon extracted (within the RFG model) in several recent experiments on
  the CCQE and CCQE-like $\nu_\mu$ and $\overline\nu_\mu$ interactions with carbon-rich targets
  increases with decreasing the mean energy of the $\nu_\mu$ and $\overline\nu_\mu$ beams
  (see Fig.~\ref{Fig:MA_QES_REVIEW_shortlist}).

  The function $M_A^\text{run}(E_\nu)$ is defined by only two adjustable parameters,
  $M_0$ and $E_0$, independent of $Z$ for $Z\geq6$. 
  The best-fit values of the parameters were obtained from a global statistical analysis
  of all available self-consistent CCQE and CCQE-like data for substantial variety
  of nuclear targets and $\nu/\overline\nu$ energy spectra (see Sec.\ \ref{sec:StatisticalAnalysis}).
  It is important that the best-fit value of $M_0$ is in very good agreement with the axial mass
  value extracted from the deuterium data as well as with the results of the former statistical
  analyses \cite{Bodek:2007ym,Kuzmin:2007kr}.
  The parameter $M_0$ can be therefore treated as the current (dipole) axial-vector mass of the nucleon.

  The \SMRFG_MArun\ model has compared with several competing models and extensively
  tested on large amounts of recent CCQE and CCQE-like data from the experiments MiniBooNE, T2K, and \MINERvA. 
  In most cases, the model describes the data with a reasonable and in some cases (MiniBooNE)
  with remarkable accuracy. 
  The biggest disagreement is with the T2K C$_8$H$_8$ distributions over the final-state
  proton variables \cite{Abe:2018pwo} which, however, provide the problem for the rival models as well,
  even after applying the renormalization procedure (see Sec.\ \ref{sec:Comparison}).
  It is thought that the discrepancies are at least in part due to the incompleteness
  of the FSI models implemented into the modern neutrino event generators and perhaps with
  certain difficulties in the analysis of the post-FSI protons in the T2K near detectors.
  In general, it can be concluded that there is no single contemporary model that could
  satisfactorily describe all the current data.
  Moreover, there are indications of some inconsistencies among the modern data,
  namely, between low-energy data from T2K ND280 (hydrocarbon) and MiniBooNE and between
  the higher-energy data from \MINERvA\ and NOMAD. 
  
  The best-fit values of $M_0$ and $E_0$ are somewhat sensitive to variations
  of the input parameters of the SM RFG model (Fermi momenta, separation energies) and
  of the models for the nucleon electromagnetic form factors.
  However the fit can almost automatically be repeated with the modifications
  of the RFG model (e.g., Bodek--Ritchie RFG), or its extensions (SF, LFG, CFG, etc.),
  as well as with the more advanced nuclear models.
  A more sophisticated parametrization of the function $M_A^\text{run}(E_\nu)$
  seems to be unreasonable for the present-day level of accuracy of the CCQE and CCQE-like
  data but may be needed in the future.
  Individual parametrizations for different nuclei or nuclear groups are also unreasonable today,
  but mainly because the currently available dataset for the inorganic heavy nuclear targets
  is not sufficiently accurate and self-consistent.

  There is no statistically significant difference between the \MArun\ parameters extracted separately
  from the $\nu_\mu$ and $\overline{\nu}_\mu$ data, but there is a faint hint on possible difference
  (larger $M_0$ and smaller $E_0$ for $\overline{\nu}_\mu$ interactions).
  In any case, the available $\overline{\nu}_\mu$ dataset is not yet sufficient for a more definite statement.
  To draw more robust conclusions it is desirable to compare the \SMRFG_MArun\ model predictions with
  the very new high-statistics measurements of
  \MINERvA\ \cite{Carneiro:2019jds} (broad-spectrum $\nu_\mu$ beam peaking around $6$~GeV) and 
  T2K ND280/INGRID \cite{Abe:2020jbf} (off-axis $\nu_\mu$ and $\overline\nu_\mu$), 
                   \cite{Abe:2020uub} (off-axis $\nu_\mu$),
                   \cite{Abe:2019iyx} (on-axis, $\nu_{\mu}$), and 
                   \cite{Abe:2020iop} (combined, $\overline{\nu}_{\mu}$ and $\overline{\nu}_{\mu}+\nu_{\mu}$)
  on hydrocarbon and water targets.
  For a further tune of the model, it would be instructive to add the modern data into the global fit. 

\begin{acknowledgements}
  We thank to
  Jose Amaro,                                                                   
  Stephen Dolan,                                                                
  Ulrich Mosel,  and                                                            
  Ignacio Luis Ruiz Sim\'{o}                                                    
  for providing us with the scripts and/or tabulated results of calculations based on their models.
  We are very grateful to
  \fbox{Samoil Bilenky,}                                                        
  Arie Bodek,                                                                   
  Askhat Gazizov,                                                               
  Kendall Mahn,                                                                 
  Dmitry Naumov,                                                                
  Alexander Olshevskiy,                                                         
  Olga Petrova,                                                                 
  Oleg Samoylov,                                                                
  Victor Shamanov,                                                              
  Oleg Teryaev,                                                                 
  and our colleagues from the \GENIEG\ Collaboration,
  for helpful discussions and critical comments.
  We appreciate the assistance of JINR cloud team with cloud infrastructure particularly
  Nikita Balashov,                                                              
  Alexander Baranov, and                                                        
  Nikolay Kutovskiy.                                                            
  Research of I.\,K.\ has been supported by the Russian Science Foundation grant No.~18-12-00271.
\end{acknowledgements}

  \appendix

\def\W{0.970\linewidth}

\section{Including ``mixed'' data into $\chi^2$ sum}
\label{Method_adds}

The expression \eqref{Blobel_fixed} for $\chi^2$ becomes a bit more complicated when some data 
represent a combination (for example, sum, as in the IHEP–ITEP experiment \cite{Belikov:1985mw})
of quantities measured in experiments with neutrino and antineutrino beams.
Here we consider the simple case when one has to analyze the three {uncorrelated} datasets --
$\mathbf{E}_1$, $\mathbf{E}_2$, and $\mathbf{E}_3=\mathbf{E}_1'+\mathbf{E}_2'$, where the pairs
$(\mathbf{E}_1,\mathbf{E}_1')$ and $(\mathbf{E}_2,\mathbf{E}_2')$ have the common normalization
uncertainties. Let us construct the sum for minimization in this case:
\begin{align*}
\chi_{\text{mix}}^2
= &\  \sum_{k=1,2}\left\{\sum_{j \in G_k}\frac{\left[N_kT_{kj}(\boldsymbol{\lambda})-E_{kj}\right]^2}{\sigma_{kj}^2}
     +\frac{(N_k-1)^2}{\delta_k^2}\right\}  \\
  &\ +\sum_{j \in G_3}\frac{\left[N_1T_{1j}'(\boldsymbol{\lambda})+
                                  N_2T_{2j}'(\boldsymbol{\lambda})-E_{3j}\right]^2}{\sigma_{3j}^2}
\end{align*}
(the notation is similar to that in the main text). From the minimization conditions
\[
\frac{\partial\chi_{\text{mix}}^2}{\partial N_1}=\frac{\partial\chi_{\text{mix}}^2}{\partial N_2}=0 
\]
we obtain the expressions for the normalization factors:
\[
N_1 = \frac{C_1A_{22}-C_2A_{12}}{D}, \quad
N_2 = \frac{C_2A_{11}-C_1A_{12}}{D},
\]
where
\begin{align*}
A_{12} = &\ \sum_{j \in G_3}\frac{T_{1j}'  (\boldsymbol{\lambda})T_{2j}'(\boldsymbol{\lambda})}{\sigma_{3j}^2}, \\
A_{kk} = &\ \sum_{j \in G_k}\frac{T_{kj}^2 (\boldsymbol{\lambda})}{\sigma_{kj}^2}
           +\sum_{j \in G_3}\frac{T_{kj}'^2(\boldsymbol{\lambda})}{\sigma_{3j}^2}+\frac{1}{\delta_k^2},         \\
C_k    = &\ \sum_{j \in G_k}\frac{T_{kj}   (\boldsymbol{\lambda})E_{kj}}{\sigma_{kj}^2}
           +\sum_{j \in G_3}\frac{T_{kj}'  (\boldsymbol{\lambda})E_{3j}}{\sigma_{3j}^2}+\frac{1}{\delta_k^2},   \\
D      = &\  A_{11}A_{22}-A_{12}^2, \enskip\text{and}\enskip k=1,2.
\end{align*}

  \section{Comparison with earlier data}
  \label{sec:GoldData}

  In this appendix, we present an extended comparison of the \SMRFG_MArun\ model
  predictions with the accelerator data on CCQE total and differential cross
  sections and event distributions in $Q^2$, published before 2010. 
  We also briefly discuss some technical details regarding our statistical analysis,
  which were omitted from the main text.
  Since the modern experiments operate mainly with carbon targets, the expected
  predictive power of the model for other nuclear targets is based only on the
  earlier (``golden'') dataset used in the global fit of \MArun.
  Below, these data are shown by filled symbols and  all others by open symbols.
  For completeness sake, we also show comparison with the partially outdated but
  not disapproved data.
  This compilation is not only of historical interest; it also demonstrates a level
  of consistency (not always satisfactory) between the data subsamples, as well as
  stability and conformity of the early and later measurements. Besides, it may be
  particularly useful to display the remaining data gaps of all kinds
  (nuclear targets, kinematic ranges) to be filled in future experiments.
  In all figures, the calculations performed with the best-fit values of the
  parameters $M_0$ and $E_0$ \eqref{Eqn:ME_QES_default} are shown as solid curves,
  the inner and outer shaded bands correspond, respectively, to $1\sigma$ and
  $2\sigma$ deviations from the best-fit. No renormalization is applied.
  In most cases, the \SMRFG_MArun\ model satisfactory describes the earlier CCQE data
  in the wide energy and $Q^2$ ranges, including multi-GeV (anti)neutrino energies
  out of sight of the current accelerator experiments.
  The remaining problems with description of the earlier data are mainly related to 
  imperfections of the previous analyses of the inelastic backgrounds, FSI effects,
  and, for non-deuterium targets and at low energies, to the poorly controlled
  systematic uncertainty in the discrimination between the CCQE and CCQE-like data samples.

  \subsection{Total CCQE cross sections}
  \label{sec:TotalCCQE}

  \subsubsection{NOMAD}
  \label{sec:NOMAD}

\begin{figure}[htb]
  \includegraphics[width=\linewidth]
  {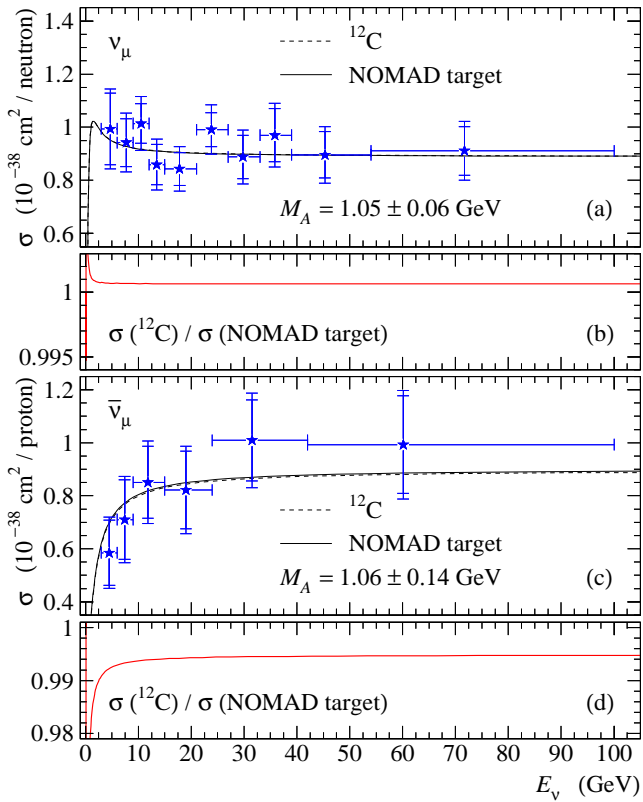}
  \caption{(Color online)
           Total CCQE $\nu_\mu$ and $\overline{\nu}_\mu$ cross sections measured
           in the NOMAD experiment~\cite{Lyubushkin:2008pe} in comparison with
           predicted cross sections per bounded nucleons for pure carbon target
           and NOMAD composite target (panels (a) and (c)).
           The outer and inner error bars indicate total errors and those with
           subtracted uncertainties due to the (anti)neutrino flux normalization,
           respectively.
           Ratios of the cross sections for carbon and NOMAD target are shown
           in panels (b) and (d).
           All calculations are done using the fixed values of the axial masses
           obtained by the authors of experiment as shown in the legends.
          }
  \label{Fig:sQESCC_Lyubushkin_NOMAD09}
  \end{figure}

  Considering that the asymptotic behavior of the total CCQE cross sections in the \SMRFG_MArun\ model
  is largely determined by the data of the NOMAD experiment~\cite{Lyubushkin:2008pe}
  (by adjusting the model parameters to the data), let us dwell on the latter in more detail.
  Figure \ref{Fig:sQESCC_Lyubushkin_NOMAD09} displays the total CCQE $\nu_\mu$ and $\overline{\nu}_\mu$
  cross sections measured in the NOMAD experiment~\cite{Lyubushkin:2008pe} in comparison with the model
  predictions for the pure carbon target and for the real NOMAD composite target.
  Also shown the ratios of the cross sections calculated for the carbon target and NOMAD target.
  According to Ref.\ \cite{Anfreville:2001zi} the NOMAD drift chamber consists 
  (by weight) of
  $64.30$\% carbon,
  $ 5.14$\% hydrogen,
  $22.13$\% oxygen,
  $ 5.92$\% nitrogen,
  $ 0.30$\% chlorine,
  $ 1.71$\% aluminum,
  $ 0.27$\% silicon,
  $ 0.19$\% argon, and
  $ 0.03$\% copper.
  Calculations show that the $\nu_\mu$ ($\overline{\nu}_\mu$) cross sections in the multi-GeV region 
  are fully (almost) insensitive to the admixtures to the main carbon component and thus the complexity
  of the target composition does not provide additional uncertainties.
  For simplicity, the calculations are done with the current axial masses $M_A$
  (rather than \MArun). It is evident that this simplification does not affect the above conclusion.
  It should be also recorded that the NOMAD results for the cross sections measurement and axial mass
  extraction are obtained using an extended Gari-Kr\"umpelmann model of the nucleon electromagnetic
  form factors, ``GKex(05)'' \cite{Lomon:2006xb}.
  The $M_A$ extractions performed with the BBBA(07) form factor model \cite{Bodek:2007ym,Bodek:2007vi}  
  increase the axial mass value by about $0.8$\% and $0.9$\% for, respectively, $\nu_\mu$ and
  $\overline{\nu}_\mu$ induced reactions.
  It is pertinent to note here that this is a typical uncertainty in global analyses of $M_A$ due to
  the current uncertainty in our knowledge of electromagnetic form factors.

  \begin{figure*}[htb]
  \centering
  \includegraphics[width=\W]
  {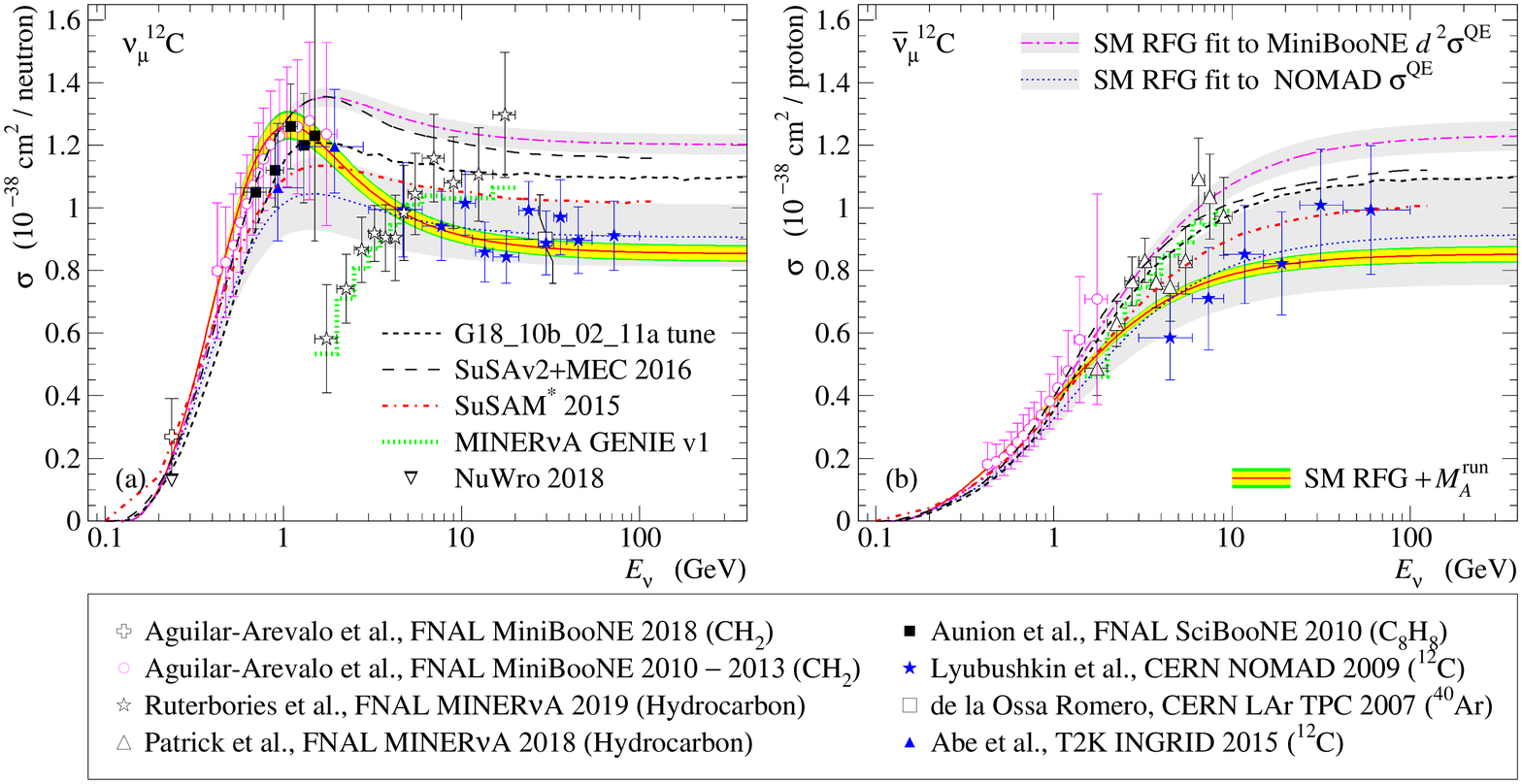}
  \caption{(Color online)
            Total CCQE $\nu_{\mu}$ (a) and $\overline{\nu}_{\mu}$ (b)
            cross sections per interacting nucleon for carbon reach targets and argon,
            as measured in
            MiniBooNE \cite{Aguilar-Arevalo:2010zc,Aguilar-Arevalo:2013dva,Aguilar-Arevalo:2018ylq},
            SciBooNE \cite{AlcarazAunion:2009ku,Aunion:2010zz},
            LAr-TPC \cite{MartinezdelaOssaRomero:2007oxj},
            NOMAD \cite{Lyubushkin:2008pe},
            T2K INGRID \cite{Abe:2014iza}, and
            \MINERvA\ \cite{Ruterbories:2018gub,Patrick:2018gvi}.
            The latter are not true total CCQE cross sections, but differential cross ones
            (vs.\ $E_\nu^{\text{QE}}$) extracted after applying several rigid kinematic cuts, see text;
            and the dotted histograms represent the respective predictions of the \MINERvA\ \GENIEG\ tune v1
            \cite{Ruterbories:2018gub,Patrick:2018gvi}.
            The vertical error bars represent the total experimental errors including the 
            flux normalization uncertainties.
            The data points marked with open symbols were not included into the statistical analysis.
            The solid curves and narrow inner/outer bands correspond to the best-fit values of $M_0$ and $E_0$
            \protect\eqref{Eqn:ME_QES_default} and $1\sigma/2\sigma$ deviations from the best fit.
            The dashed-dotted and dotted curves represent, respectively, the SM RFG predictions with the values
            of the current axial mass, \MARFG, obtained from individual fits to the MiniBooNE double-differential
            CCQE dataset and to the NOMAD CCQE dataset for $\sigma_{\text{tot}}^{\text{QE}}$.
            The gray bands indicate the $1\sigma$ uncertainty of these fits.
            Other curves represent the predictions by the two superscaling models and \GENIEb\ tune.
            Also shown the \NuWro\ prediction for the total $\nu_{\mu}{}^{12}$C cross section at $E_\nu=236$~MeV
            (borrowed from Ref.\ \cite{Aguilar-Arevalo:2018ylq}).
           }
  \label{Fig:sQESCC_mn_n_C+ma_p_C_101.3.31.301.6k_2_BBBA25_1}
  \end{figure*}


  The new improved analysis of the NOMAD dataset is presented in Ref.\ \cite{Petti:2013}.
  The official results of the NOMAD Collaboration~\cite{Lyubushkin:2008pe} were based on
  the $\nu_\mu$ CCQE candidates selected from $14,021$ single- and two-tracks events
  with total selection efficiency and purity of about $34$\% and $50$\%, respectively.
  The $\overline{\nu}_\mu$ CCQE candidates were selected from $2,237$ single-track events only,
  with efficiency and purity of about $64$\% and $38$\%, respectively.
  The new (yet unpublished) result on the CCQE cross sections is based on the analysis 
  of the full kinematic range. It, in particular, uses the $[0,2\pi]$ range for the muon azimuthal
  scattering angle, $\phi_\mu$, while the previous analysis has been limited by the condition
  $0\leq\phi_\mu\leq\pi$ (this means that the proton track must lie in the bottom hemisphere).
  Also the new analysis includes a larger fiducial volume, modified selection criteria for
  the CCQE events, better understanding of the sources of systematics errors for event reconstruction,
  improved calibration of background events, and model-independent study of nuclear effects
  and FSI from a comparison of single- and two-tracks events.
  The improved analysis of high-purity samples with tighter kinematic cuts obtained
  $16,800$ total    two-track candidates with efficiency (purity) of $25$\% ($57$\%) and
  $18,600$ total single-track candidates with efficiency (purity) of $29$\% ($57$\%).
  As a result, the average total CCQE cross section in the neutrino energy range of $2.5-100$ GeV
  was found to be
  \begin{equation*}
  \overline\sigma_\nu=\left(0.914\pm 0.013_{\text{stat}}\!
                                 \pm 0.038_{\text{syst}}\right)\times10^{-38}~\text{cm}^2/\text{neutron}.
  \end{equation*}
  It should be compared with the earlier NOMAD result for the neutrino and antineutrino cross sections
  averaged over the energy range of $3-100$ GeV:
  \begin{align*}
  \overline\sigma_\nu              = &\
  \left(0.92 \pm 0.02_{\text{stat}} \pm 0.06_{\text{syst}}\right) \times 10^{-38}~\text{cm}^2/\text{neutron}, \\
  \overline\sigma_{\overline{\nu}} = &\
  \left(0.81 \pm 0.05_{\text{stat}} \pm 0.09_{\text{syst}}\right) \times 10^{-38}~\text{cm}^2/\text{proton}.
  \end{align*}
  As is seen, the new and previous results for the $\nu_\mu$ cross section are in very good agreement,
  while remaining in tension with the \MINERvA\ measurement in the area of intersection.
  Needless to say, the cross sections averaged over a wide energy interval
  (about two orders of magnitude for NOMAD) are rather rough characteristic, especially for the
  $\overline{\nu}_\mu$ cross section shape.
  Anyway the data can be compared with the \SMRFG_MArun\ predictions
%
%
%
  \begin{align*}
  \overline\sigma_\nu              = &\
  0.880^{+0.022(0.026)}_{-0.021(0.025)} \times 10^{-38}~\text{cm}^2/\text{neutron}, \\
  \overline\sigma_{\overline{\nu}} = &\
  0.824^{+0.020(0.024)}_{-0.019(0.023)} \times 10^{-38}~\text{cm}^2/\text{proton}
  \end{align*}
  for the $\nu_\mu$ and $\overline{\nu}_\mu$ reactions, respectively.
  Here the errors correspond $1\sigma$ ($2\sigma$) deviations from the best-fit values of $M_0$ and $E_0$
  obtained in the global fit (narrow shaded bands in Fig.\ \ref{Fig:sQESCC_mn_n_C+ma_p_C_101.3.31.301.6k_2_BBBA25_1}).
  The averaging was made over the energy range of $2.5-100$ GeV. 
  Taking into account the normalization factor of about $1.036$ and $1.016$ for, respectively,
  $\nu_\mu$ and $\overline{\nu}_\mu$ (see Table~\ref{Tab:N_test}), we may conclude that the \SMRFG_MArun\ model
  is consistent with both earlier and new NOMAD data.


  \begin{table*}[htb]
\centering
\caption{Best-fit values of axial mass, \MARFG, and normalization factors, $\mathcal{N}$,
         derived from the NOMAD data on CCQE total cross sections \cite{Lyubushkin:2008pe}  
         and MiniBooNE data on CCQE double-differential cross sections
         \cite{Aguilar-Arevalo:2010zc,Aguilar-Arevalo:2013dva}                              
         using the SM RFG model.
         }        
\begin{tabular*}{\linewidth}{@{\extracolsep\fill}lcccc}                                   \hline\noalign{\smallskip}
  Dataset                        & \ndf & \MARFG\ (GeV)   & $\mathcal{N}$     & \CNNt  \\ \noalign{\smallskip}\hline\noalign{\smallskip}
  NOMAD ($\nu_\mu$)              &  10  & $1.07 \pm 0.11$ & $1.000 \pm 0.103$ & $0.68$ \\
  NOMAD ($\overline\nu_\mu$)     &   6  & $1.08 \pm 0.19$ & $0.992 \pm 0.137$ & $0.55$ \\
  MiniBooNE ($\nu_\mu$)          & 137  & $1.35 \pm 0.03$ & $1.003 \pm 0.013$ & $0.26$ \\
  MiniBooNE ($\overline\nu_\mu$) &  75  & $1.38 \pm 0.04$ & $1.016 \pm 0.021$ & $0.48$ \\ \noalign{\smallskip}\hline
\end{tabular*}
\label{Tab:RFG}
\end{table*}


  \begin{figure*}[htb]
  \centering
  \includegraphics[width=\W]
  {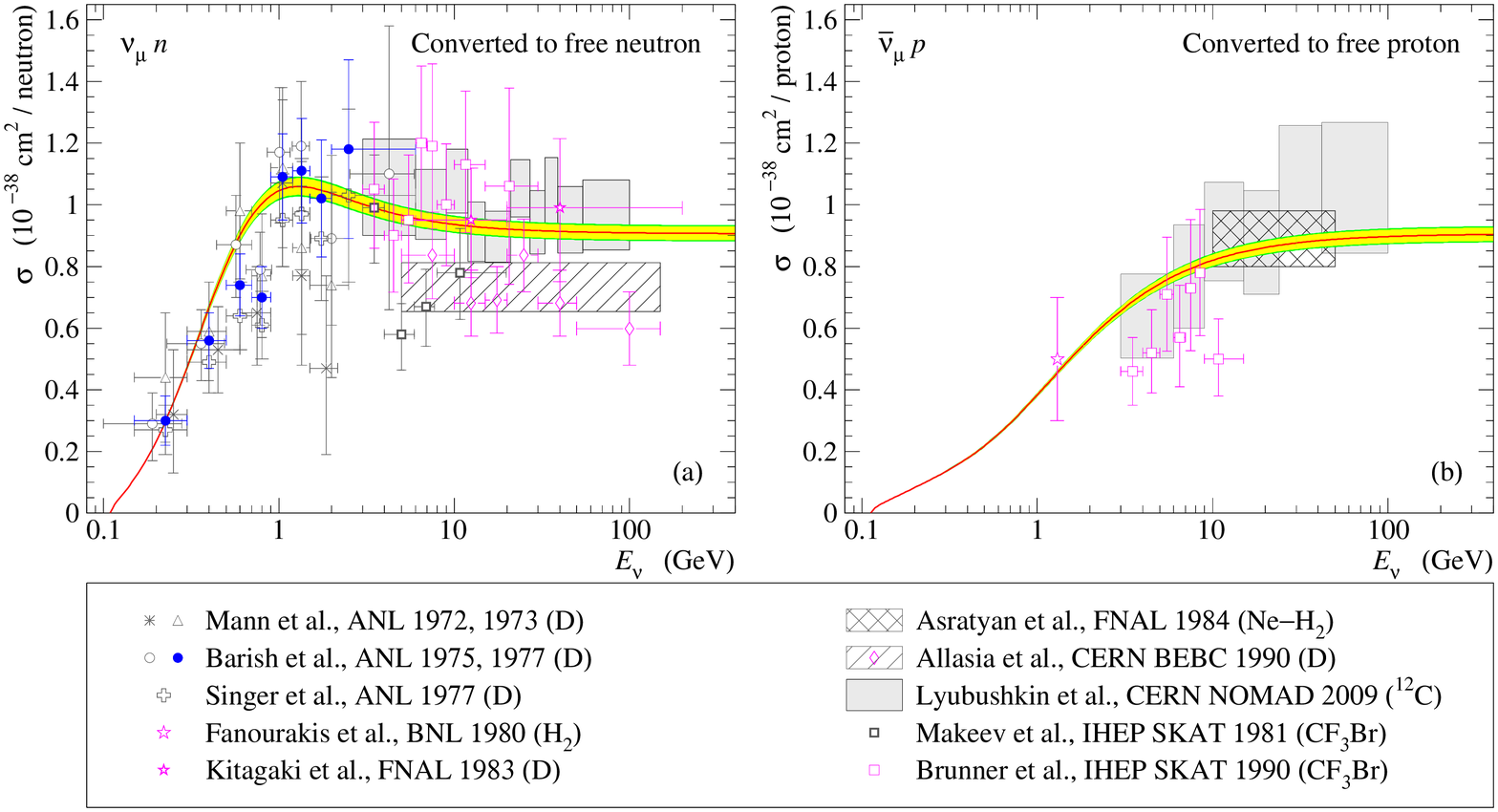}
  \caption{(Color online)
           Total CCQE cross sections for $\nu_{\mu}$ and $\overline{\nu}_{\mu}$
           scattering on free neutrons (a) and protons (b) extracted from the
           measurements on
           deuterium
           (ANL 1972~\cite{Mann:1972},
                1973 \cite{Mann:1973pr},
                1975~\cite{Barish:1976dz,Perkins:1975bj},
                1977~\cite{Singer:1977rs,Barish:1977qk},
           CERN BEBC 1990~\cite{Allasia:1990uy}, and
           FNAL 1983~\cite{Kitagaki:1983px}),
           hydrogen
           (BNL 1980~\cite{Fanourakis:1980si}),
           neon-hydrogen mixture
           (FNAL 1984~\cite{Asratyan:1984gh}),
           heavy freon (IHEP SKAT 1981~\cite{Makeev:1981em} and 1990~\cite{Brunner:1989kw}), and
           carbon with small admixtures (NOMAD 2009~\cite{Lyubushkin:2008pe}).
           The conversions to the free nucleon targets have been performed by the authors of the experiments.
           The vertical error bars and heights of rectangles represent the total errors
           including the flux normalization and conversion uncertainties.
           The solid curves and two shaded confidence bands around them correspond to
           calculations made with the best-fit values of $M_A=M_0$ and $1\sigma$ and $2\sigma$ deviations
           from the best-fit curves, respectively.
          }
  \label{Fig:sQESCC_mn_n_D+ma_p_H_2_101.3.31.301.6k_2_BBBA25}
  \end{figure*}
    
  \begin{figure*}[htb]
  \centering
  \includegraphics[width=\W]
  {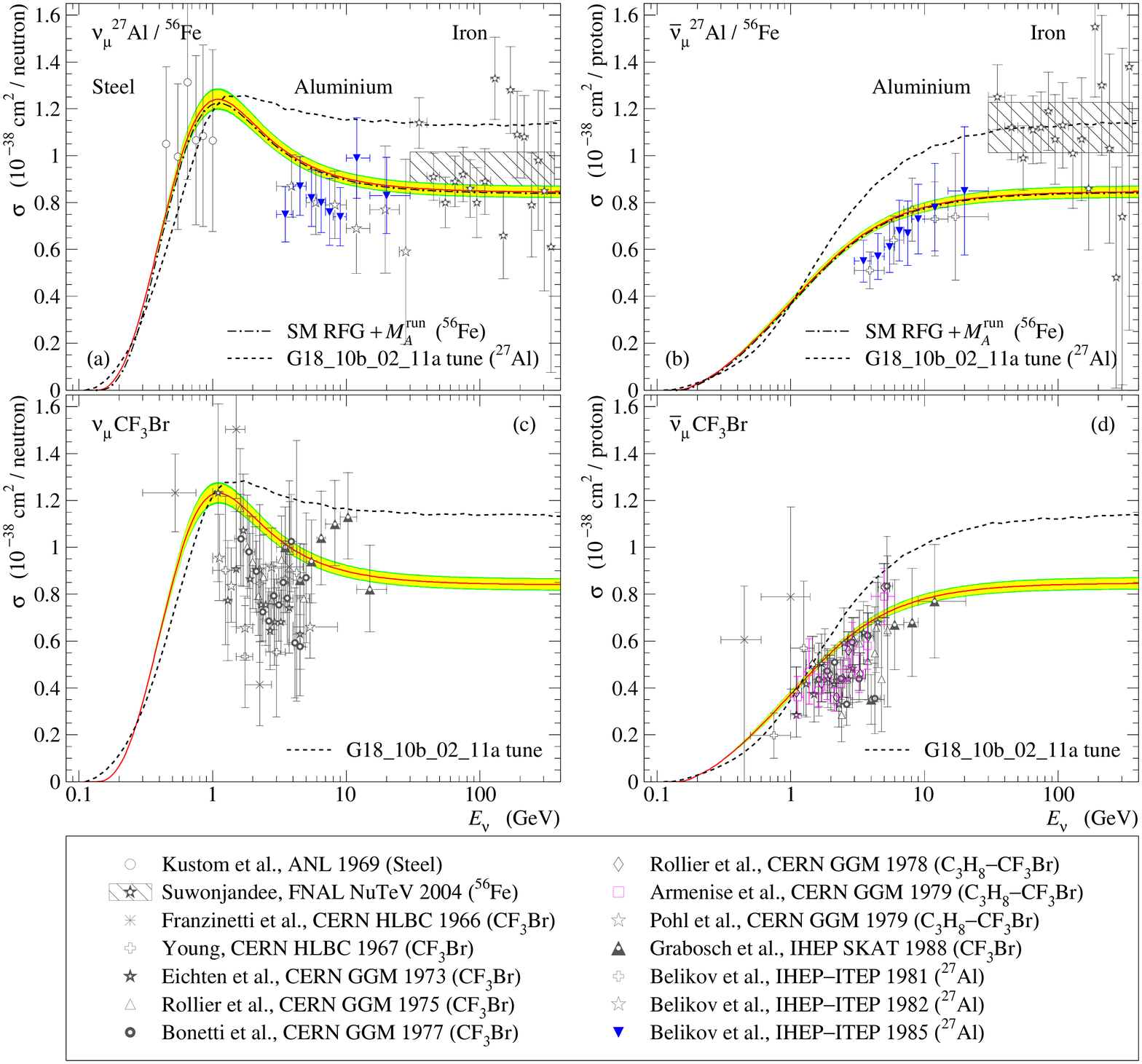} 
  \caption{(Color online)
           Total CCQE cross sections for $\nu_{\mu}$ and $\overline{\nu}_{\mu}$
           scattering on bound nucleons measured in experiments on
           steel (ANL 1969~\cite{Kustom:1969dh}),
           aluminium (IHEP--ITEP 1981~\cite{Belikov:1981fq},
                                 1982~\cite{Belikov:1981ut},
                                 1985~\cite{Belikov:1983kg,Belikov:1985mw}),
           iron (FNAL NuTeV~2004~\cite{Suwonjandee:2004aw}),
           trifluorobromomethane (CERN HLBC 1966~\cite{Franzinetti:1965},
                                            1967~\cite{Young:1967ud},
                                        GGM 1973~\cite{Eichten:1973cs},
                                            1975~\cite{Rollier:1975qr},
                                            1977~\cite{Bonetti:1977cs},
                                  IHEP SKAT 1988~\cite{Grabosch:1988js}, and
           propane-freon mixture (CERN  GGM 1978~\cite{Rollier:1978kr},
                                            1979~\cite{Armenise:1979zg,Pohl:1979zm}).
           The vertical error bars represent the total errors including the flux normalization and
           conversion uncertainties.
           Solid curves represent the \SMRFG_MArun\ model predictions;
           shaded confidence bands around these curves represent $1\sigma$ and $2\sigma$ deviations
           from the best-fitted values of the \MArun\ parameters.
           The short-dashed curves represent the cross sections calculated with the \GENIEb\ tune.
           The dashed-dotted curves in panels (a) and (b) represent the cross sections
           calculated for iron target.
          }
  \label{Fig:sQESCC_101.3.31.301.6k_2_BBBA25_NT}
  \end{figure*}

  Figure \ref{Fig:sQESCC_mn_n_C+ma_p_C_101.3.31.301.6k_2_BBBA25_1} displays the total cross sections (per interacting nucleon)
  for the CCQE $\nu_{\mu}$ and $\overline{\nu}_{\mu}$ scattering on carbon-rich detector targets for a wide energy range.
  Shown are the data of MiniBooNE \cite{Aguilar-Arevalo:2010zc,Aguilar-Arevalo:2013dva,Aguilar-Arevalo:2018ylq},
  SciBooNE \cite{AlcarazAunion:2009ku,Aunion:2010zz},
  NOMAD \cite{Lyubushkin:2008pe}, and
  T2K INGRID \cite{Abe:2014iza}.
  Also shown are the LAr-TPC data point \cite{MartinezdelaOssaRomero:2007oxj} (liquid argon) and
  the data points from \MINERvA\ \cite{Ruterbories:2018gub,Patrick:2018gvi} (hydrocarbon).
  The \MINERvA\ result has been discussed at length in Sec.~\ref{sec:MINERvA_TCS}. 
  Here we only remind that it was obtained using cuts on the muon-momentum components and
  muon scattering angle ($\theta_\mu<20^\circ$).
  According to our estimations, these cuts considerably reduce the total CCQE cross section below
  $\sim4$ GeV -- up to about $60$\% in the lowest energy bin. So, these data provide only the lower
  limits for the total CCQE cross section.
  The estimations also suggest that the MiniBooNE and \MINERvA\ CCQE data are consistent within the errors,
  but there is an essential tension between the \MINERvA\ and NOMAD data.
  The experimental data are compared with the predictions of 
  \SMRFG_MArun, \GENIEb, \SuSAv2MEC\ \cite{Megias:2016fjk},
  \SuSAM\ \cite{Amaro:2015zjaWithErratum}, and
  \NuWro\ neutrino generator~\cite{Juszczak:2009qa,Golan:2012wx}
  (the latter number is taken from Ref.\ \cite{Aguilar-Arevalo:2018ylq}).
  All calculations are done for pure carbon.
  For contrasting purposes, we also show the result of the standard SM RFG calculations
  made with the best-fit values of the current axial mass, \MARFG,
  extracted separately from the MiniBooNE double-differential CCQE cross section data for CH$_{2}$ target
  (shown in Figs.\ \ref{Fig:d2sQESCC_dEkdcosT_mn_n_CH_2.06} and \ref{Fig:d2sQESCC_dEkdcosT_ma_p_CH_2.06},
  Sec.\ \ref{sec:MiniBooNE_DDCS}) and from the NOMAD total CCQE cross section data
  (plotted in Fig.\ \ref{Fig:sQESCC_mn_n_C+ma_p_C_101.3.31.301.6k_2_BBBA25_1}).
  These fits were performed separately for $\nu_\mu$ and $\overline{\nu}_\mu$ data subsets.
  The gray bands around the curves represent the $1\sigma$ uncertainties due to the errors in determination
  of \MARFG\ and normalization factors $\mathcal{N}$.
  The corresponding parameters are listed in Table \ref{Tab:RFG}. 


  It is not news that the RFG models cannot describe the total CCQE cross sections at both low
  (MiniBooNE, SciBooNE, T2K) and high (NOMAD, LAr-TPC) energies, but it is less well known that
  the same is also true for more sophisticated recent models, namely,
  the models that account for the MEC contribution are broadly consistent with the low-energy data
  and with the high-energy \MINERvA\ data, but are in conflict with the NOMAD and LAr-TPC measurements,
  as well as with essentially all earlier high-energy data for other nuclear targets 
  (e.g., IHEP--ITEP 1985~\cite{Belikov:1983kg,Belikov:1985mw}).
  On the contrary, the \SMRFG_MArun\ model, being tuned to these data, does not fit the \MINERvA\ measurements.
  We will return to this problem later.
  MiniBooNE's measurement of monoenergetic CC $\nu_\mu$ interactions \cite{Aguilar-Arevalo:2018ylq}
  (point at $236$~MeV in panel (a), see also Sec.~\ref{sec:MiniBooNE_TCS}) is of special interest.
  Recall that formally it represents the total $\nu_\mu{}^{12}$C CC cross section,
  but since the neutrino energy is below the inelastic threshold,
  it is less model-dependent in comparison to the higher-energy data.

  \subsubsection{Earlier data}
  \label{sec:EarlierData}

  Figures \ref{Fig:sQESCC_mn_n_D+ma_p_H_2_101.3.31.301.6k_2_BBBA25} and
  \ref{Fig:sQESCC_101.3.31.301.6k_2_BBBA25_NT} show comparison of calculations
  with the early experimental data on the total cross sections for the CCQE $\nu_{\mu}$ and
  $\overline{\nu}_{\mu}$ scattering on nucleons bound in various nuclei.
  Figure \ref{Fig:sQESCC_mn_n_D+ma_p_H_2_101.3.31.301.6k_2_BBBA25} displays
  the cross sections on free neutrons and protons extracted from the ANL and CERN measurements on
  deuterium performed in ANL \cite{Mann:1972,Mann:1973pr,Barish:1976dz,Singer:1977rs,Barish:1977qk}
  (see also Ref.\ \cite{Perkins:1975bj}),
  and CERN \cite{Allasia:1990uy},
  hydrogen (BNL) \cite{Fanourakis:1980si},
  neon-hydrogen mixture (FNAL) \cite{Asratyan:1984gh}, 
  freon (IHEP) \cite{Makeev:1981em,Brunner:1989kw}, and
  CERN NOMAD drift chambers used as an active target and made of nearly isoscalar material
  ($N : Z \simeq 47.56\% : 52.43\%$, $Z/A \simeq 0.52$) \cite{Lyubushkin:2008pe}.
  The extraction procedures for the three latter datasets are obviously more ambiguous than
  for the deuterium data; they, in particular, did not take into consideration the MEC contributions.
  So, taking into account other systematic uncertainties, these data can be considered as qualitative
  upper limits for the CCQE cross sections on free nucleons.
  It is seen, however, that the NOMAD neutrino data are in agreement with the deuterium data from ANL
  (in the overlap region) and FNAL (in the multi-GeV region), but are considerably in excess of the
  CERN BEBC result \cite{Allasia:1990uy}.
  We may therefore again conclude that the NOMAD and essentially all other data shown in
  Fig.\ \ref{Fig:sQESCC_mn_n_D+ma_p_H_2_101.3.31.301.6k_2_BBBA25} do not support the significant MEC
  contribution needed according the recent \MINERvA\ measurements discussed in Sec.~\ref{sec:MINERvA_TCS}.
  The figure also demonstrates the extreme poverty of the experimental data on the CCQE $\overline{\nu}p$
  cross section: in fact, currently we have the only data point from the only direct measurement
  (BNL)~\cite{Fanourakis:1980si} performed more than forty ears ago.
  From all published data of the ANL 12-ft bubble chamber experiment E412, only the ANL 1977 \cite{Barish:1977qk}
  data are included into the global fits of $M_A^{\text{D}}$ and \MArun. The authors of the experiment
  evaluated the flux normalization uncertainty to be $\pm12$\%. However in Ref.\ \cite{Singer:1977rs}
  it has been reconsidered to be $\pm15$\% in the energy range $0.5-1.5$ GeV and $\pm25$\% elsewhere.
  In the fits, we use this more conservative uncertainty.
  In the prior ANL publication \cite{Mann:1972,Mann:1973pr,Barish:1976dz,Perkins:1975bj} the corresponding
  uncertainty was not reported and thus the errors in Fig.\ \ref{Fig:sQESCC_mn_n_D+ma_p_H_2_101.3.31.301.6k_2_BBBA25}
  are reproduced as is in the citing articles.

  Figure \ref{Fig:sQESCC_101.3.31.301.6k_2_BBBA25_NT} displays the total CCQE $\nu_{\mu}$ and $\overline{\nu}_{\mu}$
  cross sections on bound nucleons measured in experiments on
  steel (ANL) \cite{Kustom:1969dh},
  iron (FNAL NuTeV) \cite{Suwonjandee:2004aw},
  aluminium (IHEP--ITEP) \cite{Belikov:1981fq,Belikov:1981ut,Belikov:1983kg,Belikov:1985mw},
  freon (CERN, IHEP) \cite{Franzinetti:1965,Young:1967ud,Eichten:1973cs,Rollier:1975qr,Bonetti:1977cs,Grabosch:1988js}, and
  propane-freon mixtures (CERN GGM) \cite{Rollier:1978kr,Armenise:1979zg,Pohl:1979zm}.
  These data are compared with the \SMRFG_MArun\ model and \GENIEb\ predictions.
  Only the most recent IHEP--ITEP data from Refs.\ \cite{Belikov:1983kg,Belikov:1985mw} are
  included into the global fits of the running axial mass parameters.
  The experimental data measured with aluminium and iron targets are combined in single panel, because the nuclear effects
  for these targets are not essentially different and the predicted cross sections for iron are only slightly lower
  than for aluminum.
  It can be seen that the \G18\ tune predicts the cross sections that are significantly higher than most of the data
  above $1-2$ GeV.
  The only exception provide the high-energy data obtained in the NuTeV experiment~\cite{Suwonjandee:2004aw}.
  The data were obtained at record high energies of neutrinos and antineutrinos.  
  This, in particular, means that the CCQE contribution to the total number of events is very small compared to
  the inelastic contributions, and it is very difficult to unambiguously distinguish the quasielastic events,
  especially considering that the NuTeV experiment was carried out with an inactive iron target.
  For this reason, NuTeV data have a very large scatter and large statistical and systematic errors.
  We do not include these data in the global fit, following Bodek's recommendation \cite{Bodek:2008},
  and considering that these results were never approved by the NuTeV collaboration.  
  
  \subsection{Differential cross sections and distributions}
  \label{sec:DifferentialCCQE}

  \begin{figure*}[htb]
  \centering
  \includegraphics[width=\W]
  {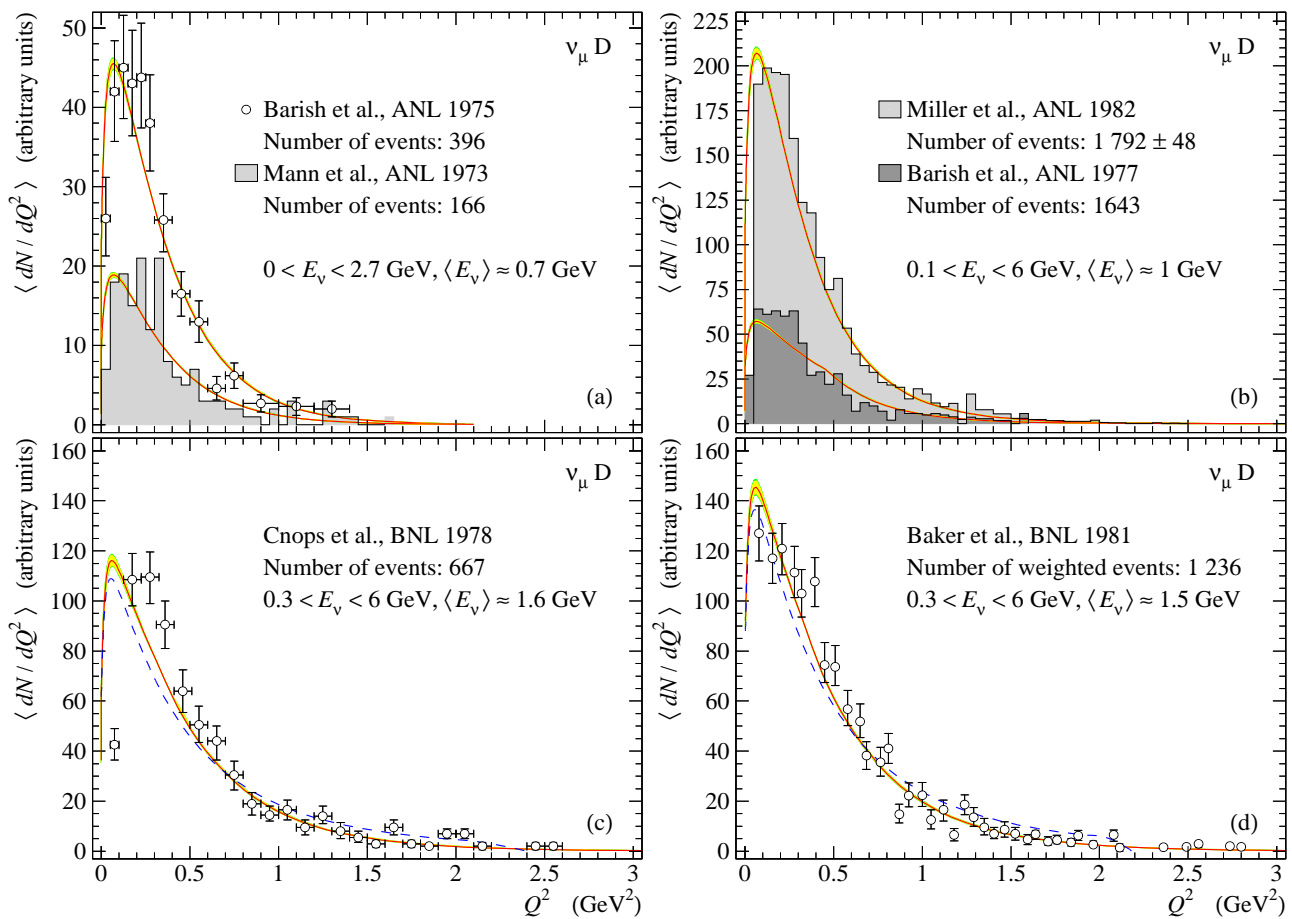}
  \caption{(Color online)
           Flux-weighted $Q^2$ distributions for $\nu_\mu$ CCQE reaction 
           measured with ANL and BNL deuterium-filled bubble chambers.
           Vertical error bars represent the total experimental errors including
           the $\nu_\mu$ flux normalization uncertainties.
           The predicted distributions are normalized to the data.
           Bubble chamber in the experiments ANL 1973 \cite{Mann:1973pr} (open circles)
           and ANL 1975 \cite{Perkins:1975bj} (light gray histogram) 
           were exposed to the $\nu_{\mu}$ beam described in Ref.\ \cite{Barish:1976dz}
           (panel (a)).
           The bubble chambers in the experiments ANL 1977 \cite{Barish:1977qk} (dark gray histogram)
           and ANL 1982 \cite{Miller:1982qi} (light gray histogram)
           were exposed to the $\nu_{\mu}$ beam described in Ref.\ \cite{Miller:1982qi}
           (panel (b)).
           The chambers in the BNL 1978 and 1981 experiments were exposed to the
           $\nu_{\mu}$ beam described in Ref.\ \cite{Baker:1981su} (panels (c) and (d)).
           Dashed curves in panels (c) and (d) represent the $Q^2$ distributions calculated
           at the mean neutrino energies (see text for explanation).  
          }
  \label{Fig:dNQESCC_dQ2_ANL-BNL_101.3.31.301.6k_2_BBBA25}
  \end{figure*}

  \begin{figure*}[htb]
  \centering
  \includegraphics[width=\W]
  {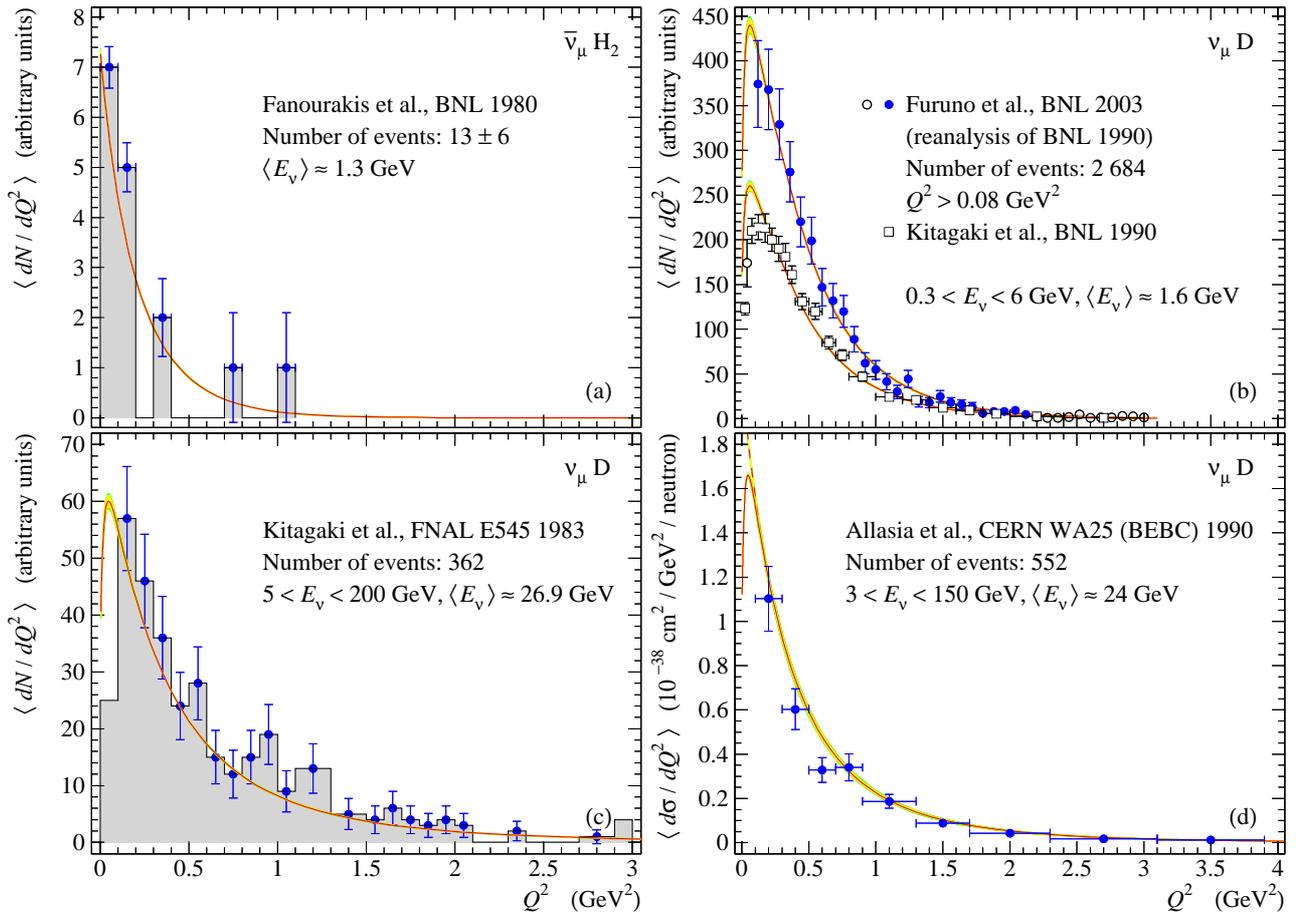}
  \caption{(Color online)
           Flux-weighted $Q^2$ distributions and cross section measured in the
           BNL, FNAL, and CERN experiments.
           Panel (a) shows distribution for $\overline{\nu}_\mu$ CCQE reaction 
           measured with the BNL 7-ft hydrogen-filled bubble chamber exposed to the
           $\overline{\nu}_\mu$ broad-band, horn-focused beam from the $30$ GeV proton
           beam at the Brookhaven Alternating Gradient Synchrotron (AGS) \cite{Fanourakis:1980si}.
           The curve is computed using the mean antineutrino energy of $1.3$ GeV.
           Panel (b) shows the $Q^2$ distributions for $\nu_\mu$ CCQE reaction
           measured in the BNL experiment with the 7-ft deuterium-filled bubble chamber
           exposed to the BNL wide-band neutrino beam at AGS \cite{Kitagaki:1990vs} (open squares).
           Also shown (filled circles) the result of a later reanalysis of the BNL 1990
           data sample described in Ref.\ \cite{Furuno:2003ng}.
           The highest $Q^2$ data points shown by open circles are excluded from
           the global fits since the ordinates of these points could not be reliably
           digitized from the available versions of the paper~\cite{Furuno:2003ng}.
           The very first point from that dataset (shown by open circle) is also
           excluded from the fit because it spoils the correlation matrix.
           Panel (c) shows the $Q^2$ distribution for $\nu_\mu$ CCQE reaction measured with
           the Fermilab 15-ft deuterium-filled bubble chamber~\cite{Kitagaki:1983px}.
           The curve is computed at the mean antineutrino energy of $26.9$ GeV.
           The points and histograms in panels (a) and (c) represent the same datasets and
           vertical error bars in these panels are purely statistical;
           the error bars shown in panels (b) and (d) represent the total experimental
           errors including the neutrino flux normalization uncertainties.
           Panel (d) shows the flux-weighted differential cross section (per neutron)
           for $\nu_\mu$ CCQE reaction as measured in the CERN WA25 experiment with the 
           Big European Bubble Chamber (BEBC) filled with deuterium and exposed to
           the high-energy $\nu_\mu$ beam at the CERN Super Proton Synchrotron (SPS) \cite{Allasia:1990uy}
           Solid and dashed lines in this panel represent the distributions calculated for
           the neutron bounded in deuterium and for the bare neutron, respectively.
           In our calculation, the experimental $\nu_{\mu}$ energy spectrum was borrowed from
           Ref.\ \cite{Allasia:1983dq}.
          }
  \label{Fig:dNQESCC_dQ2_BNL+FNAL_dsQESCC_dQ2_Allasia_BEBC90_101.3.31.301.6k_2_BBBA25}
  \end{figure*}

  \begin{figure*}[htb]
  \centering
  \includegraphics[width=\W]
  {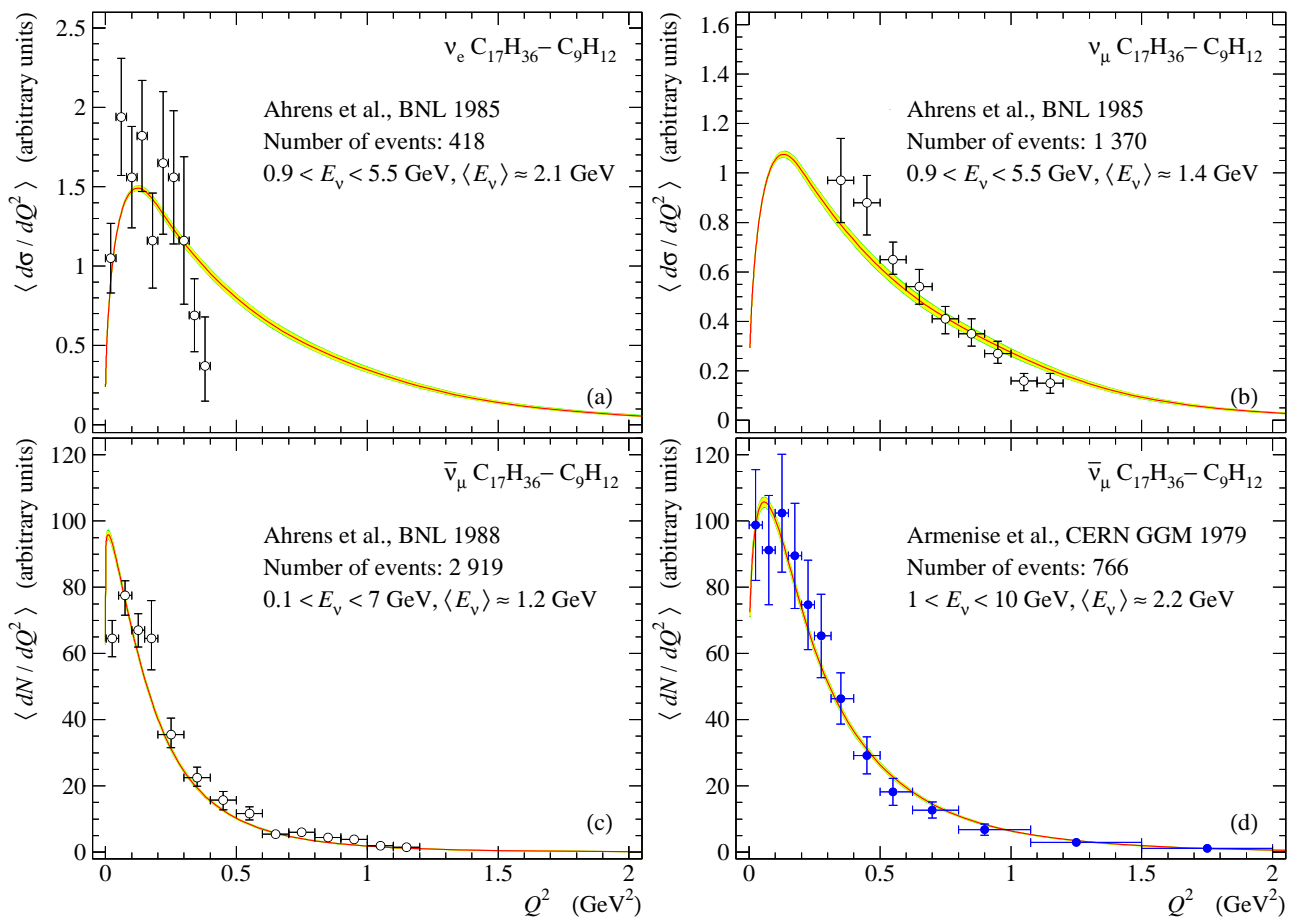}
  \caption{(Color online)
           Flux-weighted differential cross sections per neutron and
           $Q^2$ distributions for $\nu_e$ (panel (a)), $\nu_\mu$ (panel (b))
           and $\overline{\nu}_{\mu}$ CCQE reactions
           measured at BNL \cite{Ahrens:1984gp,Ahrens:1988rr} and 
           CERN \cite{Armenise:1979zg}.
           Vertical error bars represent the total experimental errors including
           the $\nu/\overline{\nu}$ flux normalization uncertainties.
           The BNL detector was filled with a liquid scintillator
           and exposed to the Brookhaven wide-band $\nu_e/\nu_\mu$ beams
           at AGS \cite{Ahrens:1986ke}.
           The CERN data are obtained with the bubble chamber Gargamelle filled
           with light propane--freon mixture (87 mole per cent of propane)
           and exposed to the CERN-PS $\overline\nu_{\mu}$ beam \cite{Armenise:1979zg}.
           All the data are given in arbitrary units
           and the predicted cross sections are normalized to the data.
           }
  \label{Fig:dsQESCC_dQ2_BNL_dNQESCC_dQ2_BNL+Armenise_GGM79_101.3.31.301.6k_2_BBBA25}
  \end{figure*}

  \begin{figure}[!htb]
  \centering
  \includegraphics[width=\linewidth]
  {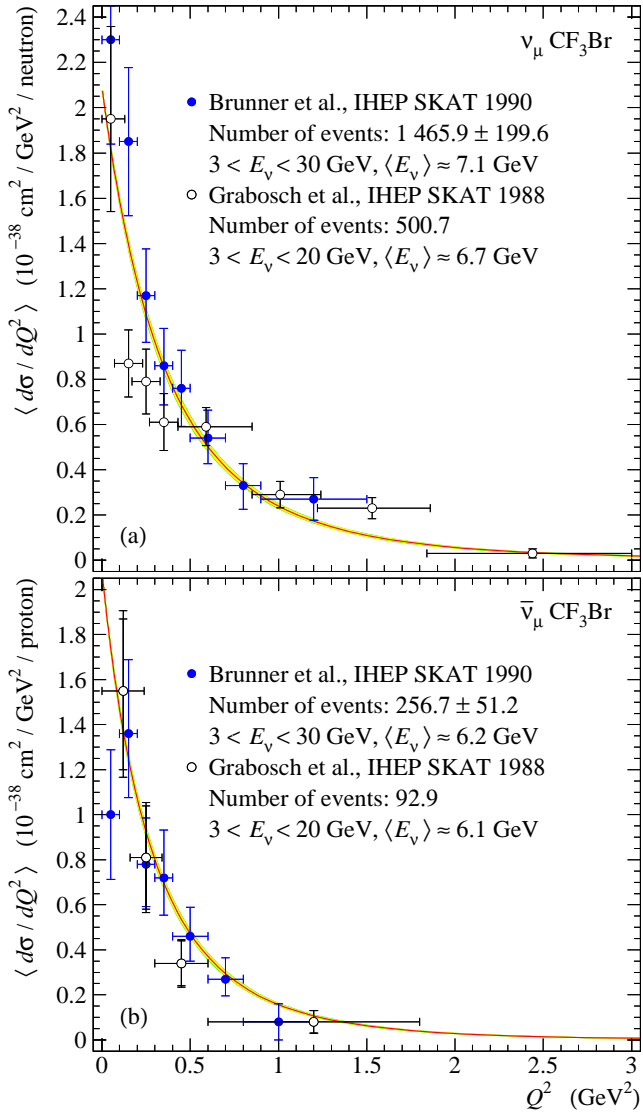}
  \caption{(Color online)
           Flux-weighted differential cross sections per neutron or proton
           for CCQE $\nu_\mu$ (panels (a)), $\overline{\nu}_\mu$ (panels (b))
           reactions measured with the freon-filled bubble chamber SKAT exposed to
           the U70 broad-band $\nu_{\mu}$ and $\overline\nu_{\mu}$ beams of the
           Serpukhov proton synchrotron (PS)
           \cite{Grabosch:1988js,Brunner:1989kw,Ammosov:1992ak}
           (see also Ref.\ \cite{Grabosch:1986nu} for the preceding analyses
           of the of the SKAT experiment).
           Experimental $\nu_{\mu}$ and $\overline\nu_{\mu}$ energy spectra
           are borrowed from Ref.\ \cite{Ammosov:1992ak}.
           The systematic error includes the uncertainties due to the
           cross section normalization and nuclear Monte Carlo.
          }
  \label{Fig:dsQESCC_dQ2_Grabosch_SKAT88+Brunner_SKAT90_101.3.31.301.6k_2_BBBA25}
  \end{figure}

  \begin{figure}[htb]
  \includegraphics[width=\W]{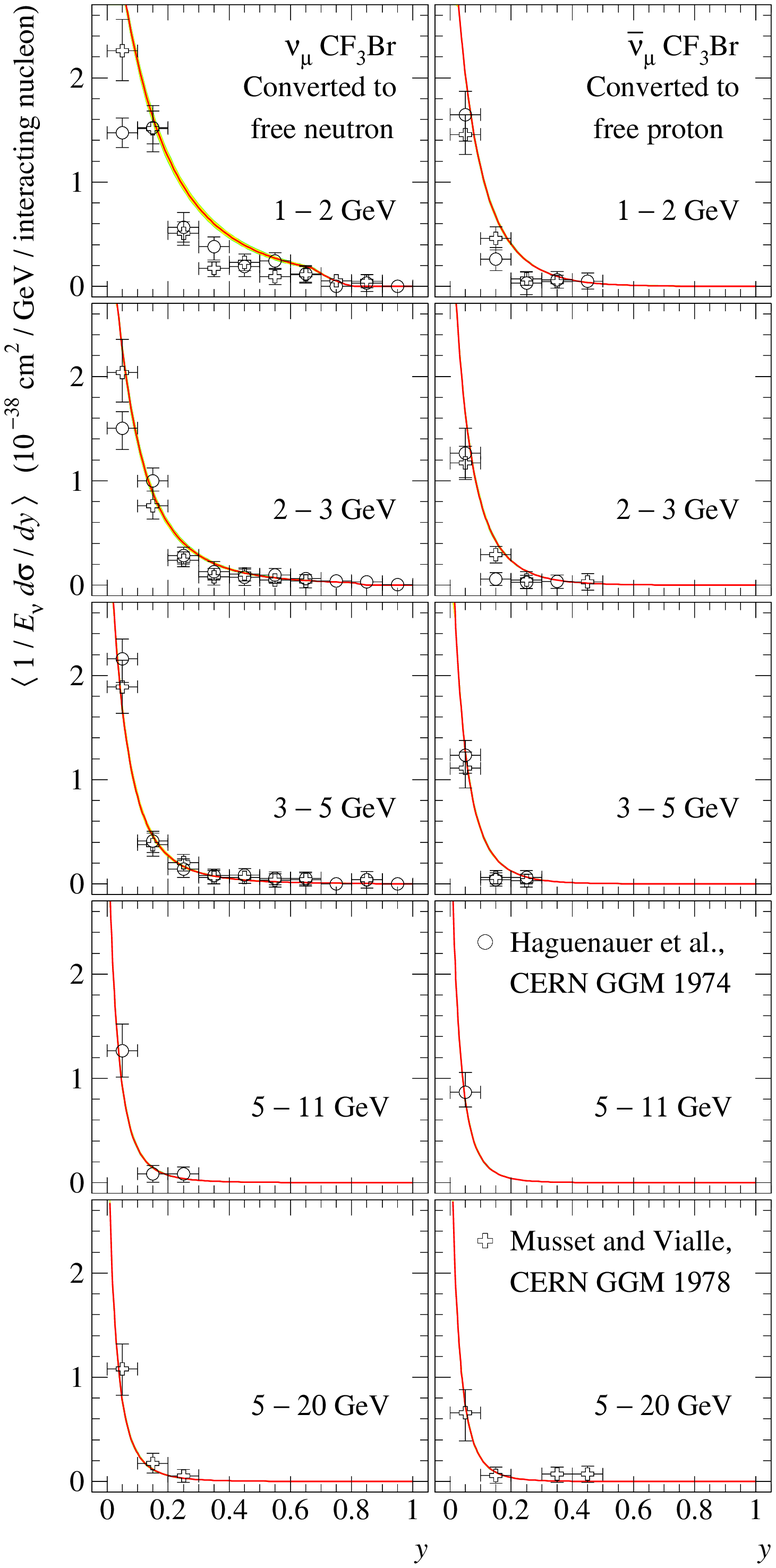}
  \caption{(Color online)
           Flux-weighted differential cross sections 
           divided by energy for $\nu_{\mu}n$ and $\overline{\nu}_{\mu}p$
           CCQE reactions, measured with the heavy-liquid freon-filled bubble chamber Gargamelle
           and exposed to the wide-band CERN-PS $\nu_{\mu}$ and $\overline\nu_{\mu}$ beams.
           The data from Ref.\ \cite{Haguenauer:74} ($E_{\nu}<11$ GeV) and \cite{Musset:1978gf}
           (all energy intervals) represent two analyses of the same data sample.
           The measured cross sections were converted to free nucleon target
           by the authors of the experiment.
           The quoted error bars include uncertainties in the $\nu_{\mu}/\overline\nu_{\mu}$
           flux and in nuclear Monte Carlo.
          }
  \label{Fig:dsQESCC_dy_101.3.31.301.6k_2_BBBA25}
  \end{figure}

  \begin{figure*}[htb]
  \centering
  \includegraphics[width=\W]
  {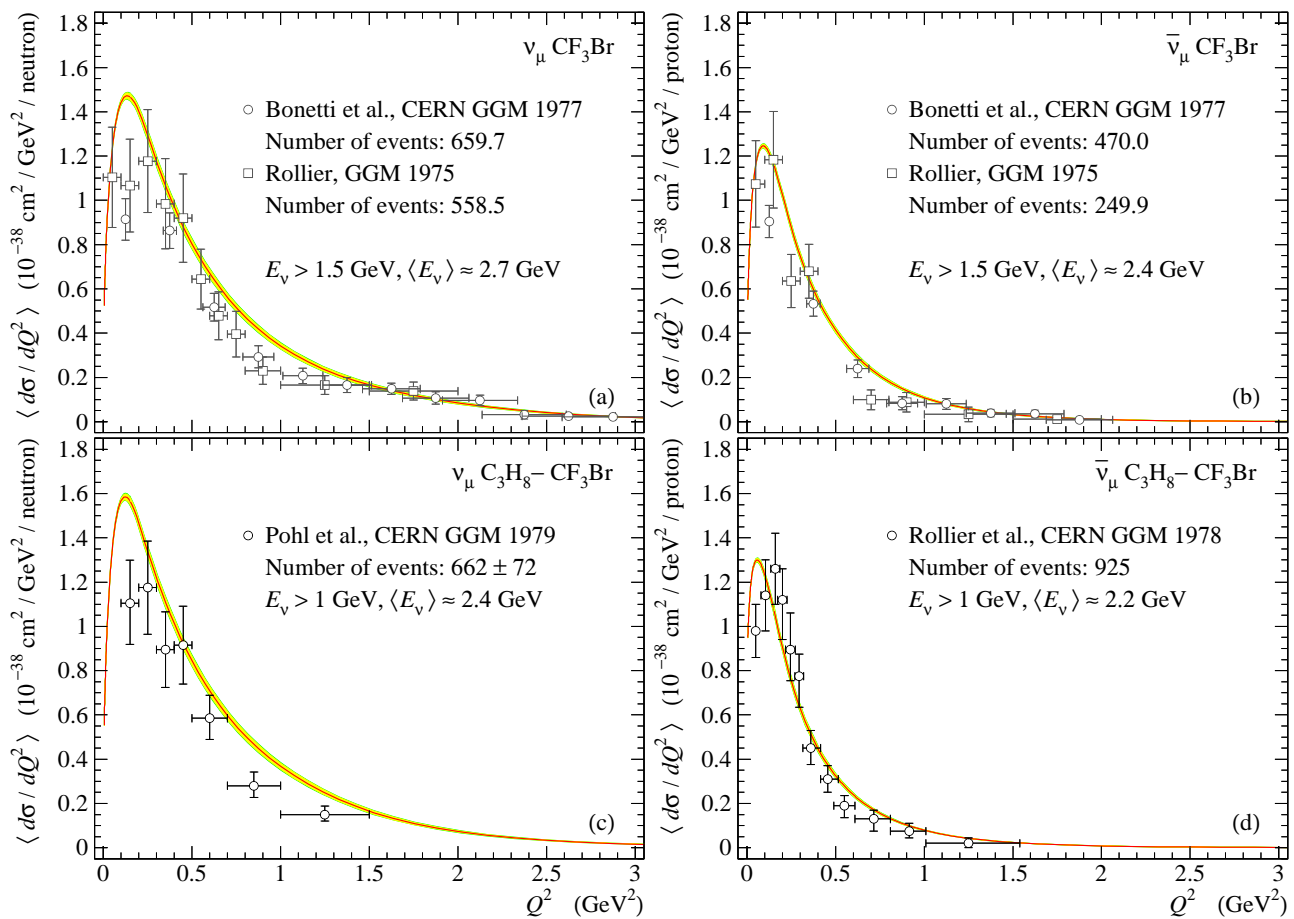}
  \caption{(Color online)
           Flux-weighted differential cross sections for
           $\nu_\mu$ (panel (a)) and $\overline\nu_\mu$ (panel (b)) CCQE reactions
           measured with the heavy-liquid bubble chamber Gargamelle
           filled with heavy freon and exposed to the CERN-PS $\nu_{\mu}$ and 
           $\overline\nu_{\mu}$ beams \cite{Bonetti:1977cs,Musset:1978gf}.
           Obsolete data measured with CERN GGM 1975 \cite{Rollier:1975qr,Perkins:1975bj}
           experiment are also shown for completeness.
           Flux-weighted differential cross sections for
           $\nu_\mu$ (panel (c)) and $\overline{\nu}_\mu$ (panel (d))  reactions
           measured with the bubble chamber Gargamelle filled with light
           propane--freon mixture and exposed to the CERN-PS beams \cite{Pohl:1979zm,Rollier:1978kr}.
           The fluxes are borrowed from Ref.\ \cite{Pohl:1979zm,Armenise:1979zg}.
           The vertical error bars represent the total experimental errors
           including the $\nu_\mu$ or $\overline{\nu}_\mu$ flux normalization uncertainties.
          }
  \label{Fig:dsQESCC_dQ2_Bonetti_GGM77_Rollier_GGM78_Pohl_GGM79_101.3.31.301.6k_2_BBBA25}
  \end{figure*}

  \begin{figure*}[htb!]
  \includegraphics[width=\W]
  {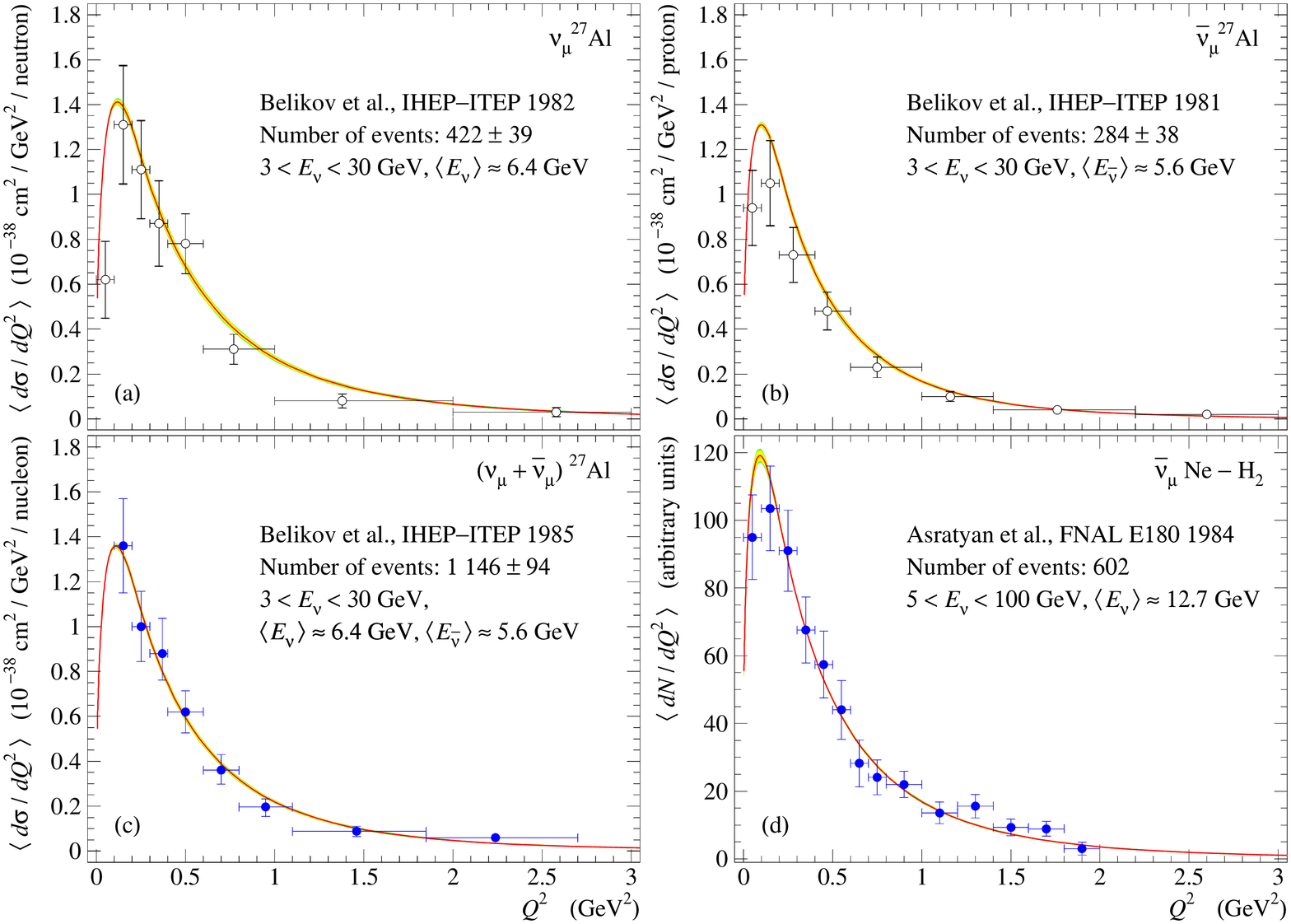}
  \caption{(Color online)
           Flux-weighted differential cross sections per interacting nucleon
           for $\nu_\mu$ (panel (a)) and $\overline\nu_\mu$ (panel (b)) CCQE reactions,
           and semi-sum of the $\nu_\mu$ and $\overline\nu_\mu$ cross sections (panel (c))
           as  measured in the IHEP--ITEP experiment using a spark chamber detector
           with aluminum filters exposed to the U70 broad-band $\nu_{\mu}$ and
           $\overline\nu_{\mu}$ beams of the Serpukhov PS
           \cite{Belikov:1981fq,Belikov:1981ut,Belikov:1985mw}.
           The error bars in panels (a)--(c) include the uncertainties due to
           the flux normalization and nuclear Monte Carlo.
           Panel (d) shows the $Q^2$ distribution per proton for $\overline\nu_\mu$ CCQE
           interactions, as measured in the FNAL E180 experiment with the 15-ft bubble chamber
           filled with heavy neon--hydrogen mixture \cite{Asratyan:1984gh,Asratyan:1984ir}.
           The vertical error bars represent the total experimental errors
           including the $\overline{\nu}_\mu$ flux normalization uncertainties.
          }
  \label{Fig:dNQESCC_dQ2_Asratyan_FNAL84_dsQESCC_dQ2_Belikov_IHEP-ITEP81,82,85_101.3.31.301.6k_2_BBBA25}
  \end{figure*}
 

  Figures 
  \ref{Fig:dNQESCC_dQ2_ANL-BNL_101.3.31.301.6k_2_BBBA25}--\ref{Fig:dsQESCC_dy_101.3.31.301.6k_2_BBBA25}
  show the differential cross sections $d\sigma/dQ^2$ and $d\sigma/dy$, and also the 
  $Q^2$ distributions predicted with the \SMRFGpMArun\ model in comparison with large experimental
  $\nu_\mu$ and $\overline{\nu}_\mu$ datasets and very meager $\nu_e$ dataset for the CCQE scattering
  on bound nucleons.
  Only a part of these data is presented in Fig.\ \ref{Fig:dsQESCC_dQ2_EXPvsTHE_2_101.1.30.301.8a_101.3.31.301.6k_BBBA25}
  The data under consideration were obtained using single-element detector targets consist of
  hydrogen     \cite{Fanourakis:1980si},
  deuterium    \cite{Mann:1973pr,Perkins:1975bj,Barish:1977qk,Miller:1982qi,Cnops:1978zi,Baker:1981su,%
                     Kitagaki:1990vs,Kitagaki:1986ct,Furuno:2003ng,Kitagaki:1983px,Allasia:1983dq,Allasia:1990uy},
  aluminum     \cite{Belikov:1981fq,Belikov:1981ut,Belikov:1985mw}, and more complex 
  targets such as 
  neon-hydrogen mixture
                \cite{Asratyan:1984gh,Asratyan:1984ir},
  propane       \cite{Budagov:1969bg},
  freon         \cite{Franzinetti:1965,Young:1967ud,Block:1964,Bonetti:1977cs,Musset:1978gf,Rollier:1975qr,%
                      Perkins:1975bj,Grabosch:1988js,Ammosov:1992ak,Haguenauer:74,Sciulli:1974we,%
                      Deden:1974uw,Baltay:1994he,Rollier:1975qr}, and
  various propane-freon mixtures
                \cite{Rollier:1978kr,Pohl:1979zm,Ahrens:1984gp,Ahrens:1988rr,Armenise:1979zg}.
  The cross sections and $Q^2$ distributions shown in Figs.\ 
  \ref{Fig:dNQESCC_dQ2_ANL-BNL_101.3.31.301.6k_2_BBBA25}--%
  \ref{Fig:dsQESCC_dQ2_Grabosch_SKAT88+Brunner_SKAT90_101.3.31.301.6k_2_BBBA25}
  are grouped by detector targets, the data in Fig.\
  \ref{Fig:dsQESCC_dQ2_Bonetti_GGM77_Rollier_GGM78_Pohl_GGM79_101.3.31.301.6k_2_BBBA25} 
  collect the CERN GGM experiments with freon and propane-freone targets,
  Fig.~\ref{Fig:dNQESCC_dQ2_Asratyan_FNAL84_dsQESCC_dQ2_Belikov_IHEP-ITEP81,82,85_101.3.31.301.6k_2_BBBA25}
  shows the data on IHEP-ITEP measurements on aluminum target and FNAL measurement on neon-hydrogent target. 
  The legends in the figures list experiment name, reaction type, target, number of events involved into
  the analysis of the given measured quantity, (anti)neutrino energy range, and estimated mean energy.
  The area normalized $Q^2$ distributions are shown in arbitrary units.
 
  The ``golden data'' 
  are from the following experiments:
   BNL 1980 \cite{Fanourakis:1980si},
       1990 \cite{Kitagaki:1990vs,Furuno:2003ng,Kitagaki:1986ct},
  FNAL 1983 \cite{Kitagaki:1983px},
  CERN BEBC 1990 \cite{Allasia:1990uy}
  (see panels (a), (b), and (d) of Fig.\ \ref{Fig:dNQESCC_dQ2_BNL+FNAL_dsQESCC_dQ2_Allasia_BEBC90_101.3.31.301.6k_2_BBBA25}),
  FNAL 1984 \cite{Asratyan:1984gh} (see panel (d) of Fig.\ \ref{Fig:dNQESCC_dQ2_Asratyan_FNAL84_dsQESCC_dQ2_Belikov_IHEP-ITEP81,82,85_101.3.31.301.6k_2_BBBA25}),
  CERN GGM 1979 \cite{Armenise:1979zg} 
  (panel (d) of Fig.\ \ref{Fig:dsQESCC_dQ2_BNL_dNQESCC_dQ2_BNL+Armenise_GGM79_101.3.31.301.6k_2_BBBA25}),
  IHEP SKAT 1990 \cite{Brunner:1989kw}
  (panels (b) and (d) of Fig.\ \ref{Fig:dsQESCC_dQ2_Grabosch_SKAT88+Brunner_SKAT90_101.3.31.301.6k_2_BBBA25}),
  IHEP-ITEP 1985 \cite{Belikov:1985mw}
  (panel (c) of Fig.\ \ref{Fig:dNQESCC_dQ2_Asratyan_FNAL84_dsQESCC_dQ2_Belikov_IHEP-ITEP81,82,85_101.3.31.301.6k_2_BBBA25}).
  The data presented in other figures are not involved into the fits and are collected here for comparison only.


  Figures~\ref{Fig:dNQESCC_dQ2_ANL-BNL_101.3.31.301.6k_2_BBBA25} and 
  \ref{Fig:dNQESCC_dQ2_BNL+FNAL_dsQESCC_dQ2_Allasia_BEBC90_101.3.31.301.6k_2_BBBA25}
  show the flux-weighted $Q^2$ distribution for $\nu_\mu$ CCQE reaction
  measured by ANL, BNL, and FNAL detectors with deuterium targets.
  We estimate the flux uncertainty in the ANL experiments to be $\pm15$\%
  except for the highest energies, where the lack of measurements of $K^+$ production
  forces us to assign uncertainty to $\pm25$\%.
  Analysis of the BNL experiment \cite{Cnops:1978zi} is based on 569,000 neutrino
  pictures taken in the BNL 7-fit bubble chamber filled with D$_2$ and an additional 204,000
  pictures in H$_2$ and thus represented as study of ``neutrino-deuterium'' reactions.
  In panel (c) of Fig.\ \ref{Fig:dNQESCC_dQ2_ANL-BNL_101.3.31.301.6k_2_BBBA25} 
  we show the predicted $Q^2$ distribution calculated for the $\nu_\mu$D reaction
  with no $\nu_\mu$H$_2$ contribution.
  It should be recorded that the neutrino spectra from the BNL wide-band $\nu_{\mu}$ beam
  used for the processing of the data by the BNL 1978 experiment \cite{Cnops:1978zi} is not known.
  For our calculations we use the $\nu_\mu$ spectrum from Ref.\ \cite{Baker:1981su}.
  The flux averaging of the predicted $Q^2$ distribution for the BNL 1981 experiment \cite{Baker:1981su}
  is more definite because the neutrino energy distribution is explicitly provided by the authors.
  For methodological purposes, in panels (c) and (d) of Fig.\ \ref{Fig:dNQESCC_dQ2_ANL-BNL_101.3.31.301.6k_2_BBBA25}
  we show the distributions $dN/dQ^2$ calculated at the mean neutrino beam energy
  ($\langle{E_\nu}\rangle=1.6$ and $1.5$ GeV for, respectively, panels (c) and (d)).
  It is clearly seen that such calculations are too rough: they do not reproduce
  the shapes of the measured $Q^2$ distributions and break off at the kinematic boundaries
  (of about $2.4$ and $2.2$ GeV$^2$ corresponding to the mentioned mean energies),
  much to the left of the real (experimental) endpoints of the distributions.
  This exercise also demonstrates the importance of knowing the neutrino energy spectrum for accurate analysis.
  It is probably the wrong spectrum that explains the contradiction between the data and
  the calculations shown in panel (c).
  
  Predicted $Q^2$ distributions in comparison with data 
  measured with BNL 1980 \cite{Fanourakis:1980si} and FNAL E545 \cite{Kitagaki:1983px}
  are calculated for an average antineutrino and neutrino energies
  because the flux used in the analysis are unknown.
  The average antineutrino energy in BNL 1980 experiment is estimated to be $1.3$ GeV,
  while the energy distribution in the antineutrino beam is peaked at about $1.1$ GeV.
  The spectrum was estimated from quasielastic events measured in a previous experiment
  with the same bubble chamber filled with deuterium (unpublished).
  Although the number of events measured in the BNL experiment is very small and determined with
  an error of about $50$\%, we include this data into the global fit, because this experiment
  provides a unique dataset for quasielastic antineutrino scattering on pure hydrogen.

  In the FNAL 1983 experiment \cite{Kitagaki:1983px}, the neutrino flux was obtained from 
  the analysis of the CCQE events using the predicted cross sections with $M_A=1.05$ GeV.
  In the energy range below $30$ GeV we use the tabulated data for the ``effective flux''
  and above $30$ we use the author's Monte Carlo simulation of the flux.
  By applying this composite flux, we reproduce the average energies at different energy intervals
  ($12.5$ GeV for $5<E_\nu<20$ GeV,
   $39.6$ GeV for $20<E_\nu<200$ GeV, and
   $23.7$ GeV for $5<E_\nu<200$ GeV).
  The expected uncertainties in the neutrino flux prediction differ significantly
  for different energies and grow form $\sim10$\% at $14.8$ GeV to $\sim60$\% at $200$ GeV.

  The BNL 1990 data \cite{Kitagaki:1990vs} shown in
  Fig.\ \ref{Fig:dNQESCC_dQ2_BNL+FNAL_dsQESCC_dQ2_Allasia_BEBC90_101.3.31.301.6k_2_BBBA25} (b)
  were collected in two periods of runs, in 1976--1977 and 1979--1980, when the total of
  $1.8\times10^6$ pictures were obtained. 
  The result published in Ref.\ \cite{Kitagaki:1990vs} has been reanalised later \cite{Furuno:2003ng}
  (see also Ref.\ \cite{Sakuda:2003qn}). In the improved analysis, the background to CCQE events
  has been reevaluated and, as a result, the number of the CCQE events has been increased.
  The final data sample in the CCQE channel obtained was $2,684$ and the estimated correction
  to this number (coming from many sources) was found to be $1.06\pm0.11$.
  In this experiment, the neutrino flux was not independently measured and thus the
  flux independent analysis of Ref.\ \cite{Furuno:2003ng} provided the model dependent cross sections.
  The $\langle{dN/dQ^2}\rangle$ data set of BNL 1990 \cite{Kitagaki:1990vs,Furuno:2003ng-proc}
  for neutrino reactions on deuterium are reliably measured under the condition
  $Q^2>0.08$ (GeV/$c$)$^2$ (we do not discuss the reasons, but refer to the earlier
  mentioned original papers). 
  The analysis of Ref.\ \cite{Furuno:2003ng} has not been published, but we prefer to use this
  more recent result in our statistical analysis. At the same time, we are forced to do some rejections.
  The lowest-$Q^2$ data point is excluded from our fits because it strongly contradicts
  to all model calculation of the $Q^2$ distribution. Probable explanation of this discrepancy
  is the effect of missing recoil proton at very low $Q^2$ discussed in the Ref.\ \cite{Kitagaki:1990vs}.
  Other reasons include a simplified description of the deuteron effects (e.g.\ neglecting the MEC contribution).
  Unfortunately, the author's tabulated data for the $Q^2$ distribution are not available.
  Therefore, we used the data taken from the figure in Ref.\ \cite{Furuno:2003ng};
  the digitization of the data points and, more importantly, associated error bars is extremely
  unreliable for $Q^2\gtrsim2$ GeV$^2$.
  Hence we show these data points for completeness, but exclude them from the global fit.
  It is seen that the high-$Q^2$ data points do not conflict with the predicted distribution. 
  

  Let us dwell a little more on the data of CERN BEBC WA25 experiment \cite{Allasia:1990uy}
  shown in panel (d) of Fig.\ \ref{Fig:dNQESCC_dQ2_BNL+FNAL_dsQESCC_dQ2_Allasia_BEBC90_101.3.31.301.6k_2_BBBA25}.
  In order to demonstrate the magnitude of nuclear effects for these measurements at relatively high energies
  ($\langle E_\nu \rangle \approx 24$ GeV), we show the predicted $Q^2$ distributions calculated
  for neutron bounded in deuterium nucleus and for the bare neutron.
  As evident from the comparison of the solid and dashed curves at the lowest $Q^2$ bin,
  the fit to the BEBC WA25 cross sections should be practically insensitive to the nuclear corrections.
  For the global fit, we slightly modified the dataset of the flux-weighted differential cross section,
  $\langle{d\sigma_\nu/dQ^2}\rangle$, for $\nu_\mu$ CCQE reaction per neutron.
  The full experimental dataset contains $9$ bins.
  It is obvious that the value obtained in the $Q^2$ bin $[0.5,0.7]$ GeV$^2$ is not
  consistent with the values obtained in the neighboring bins that yields unsatisfactory covariance matrix (from MINUIT)
  and artificial change in the fitted parameters. 
  To avoid this, we combined the data from the two bins $[0.5,0.7]$ and $[0.7,0.9]$ GeV$^2$ into
  the single one, $[0.5,0.9]$ GeV$^2$; the corresponding systematic error in the combined bin
  is estimated to be $\pm10$\%.
  It is assumed that the flux uncertainty of $8.6$\% is included into the quoted systematic error.


  Figure \ref{Fig:dsQESCC_dQ2_BNL_dNQESCC_dQ2_BNL+Armenise_GGM79_101.3.31.301.6k_2_BBBA25}
  shows the cross sections and $Q^2$ distributions measured in the BNL experiment
  \cite{Ahrens:1984gp,Ahrens:1988rr} with liquid scintillator consisting of 
  $\sim60$\% light mineral oil (C$_{17}$H$_{36}$),
  $\sim35$\% 1,2,4-trimethylbenzene (C$_9$H$_{12}$), and
  $\sim5$\% proprietary ingredients and has the H/C ratio of about $1.9$ \cite{Sulak:1975zz}.
  Therefore we use the simplified formula, $60$\% of C$_{17}$H$_{36}$ + 40\% of C$_9$H$_{12}$,
  to approximate its chemical composition.
  The uncertainty of the calculated $\nu_e$ and $\nu_\mu$ spectra
  from the Brookhaven Alternating Gradient Synchrotron is about $20$\% and the related
  errors for the measured cross sections is about $30$\%.
  In the BNL 1988 experiment were obtained two data sets of event samples
  so called main and stopping muon samples.
  Stopping muon sample was used to check the event selection of the main data sample
  in the low-$Q^2$ region and was selected by requiring a single long track stopping
  in the detector regardless of the existence of a secondary vertex due to a neutron
  interaction \cite{Ahrens:1988rr}.
  Panel (c) in Fig.\ \ref{Fig:dsQESCC_dQ2_BNL_dNQESCC_dQ2_BNL+Armenise_GGM79_101.3.31.301.6k_2_BBBA25}
  includes both the stopping muon data sample (points at $Q^2<0.2$ GeV$^2$ and main data sample
  (points at $Q^2>0.2~\text{GeV}^2$).

  Figure \ref{Fig:dsQESCC_dQ2_Grabosch_SKAT88+Brunner_SKAT90_101.3.31.301.6k_2_BBBA25}
  shows the experimental data on the $\nu_\mu$ and $\overline\nu_\mu$ differential cross sections
  obtained with the IHEP SKAT bubble chamber filled with heavy freon and converted to
  a free nucleon target by the authors of the experiment.
  We do not show the partially obsolete data of the IHEP SKAT 1981 \cite{Makeev:1981em} and
  1986 \cite{Grabosch:1986nu} obtained for the slightly different neutrino energy ranges
  in comparison with the posterior and final (full statistics) SKAT dataset \cite{Brunner:1989kw}.
  The earlier SKAT data are consistent with the final resultat within the uncertainties
  and taking into account the difference in the constraints on (anti)neutrino energy used
  in the analyses.
  The current axial mass parameter extracted from the measured total and differential
  cross sections was $1.06 \pm 0.15$~GeV for neutrino and $0.71 \pm 0.22$~GeV for antineutrino;
  the analysis used a simple dipole ansatz for the electromagnetic form factor.
  From Fig.\ \ref{Fig:dsQESCC_dQ2_Grabosch_SKAT88+Brunner_SKAT90_101.3.31.301.6k_2_BBBA25} 
  it can be seen that the SKAT data agree with the universal value of $M_A=M_0=1.008$~GeV
  obtained in our analysis. But it is not the case for total CCQE $\nu_\mu$ and $\overline\nu_\mu$
  cross sections (see Fig.\ \ref{Fig:sQESCC_mn_n_D+ma_p_H_2_101.3.31.301.6k_2_BBBA25}),
  where agreement with SKAT data is unsatisfactory.
  Note that this is a common situation for almost all data on the total CCQE cross sections
  measured using freon-field detectors.
  It is, of course, mainly related to the outdated analysis of nuclear effects in old experiments.
  

  Experimental data on the slopes of the differential cross sections
  measured in terms of inelasticity $y=1-E_\mu/E_\nu$ are employed for 
  additional test of our best-fit value of the axial mass.
  Figure \ref{Fig:dsQESCC_dy_101.3.31.301.6k_2_BBBA25} shows the flux-weighted slopes
  $E_\nu^{-1}d\sigma/dy$ for the $\nu_\mu$ and $\overline\nu_\mu$ CCQE reactions as measured
  using the CERN bubble chamber Gargamelle filled with CF$_3$Br \cite{Haguenauer:74,Musset:1978gf};
  the data were converted by the authors of experiment to those on free interacting nucleons,
  taking into account Fermi motion and Pauli suppression.
  The data from Refs.\ \cite{Haguenauer:74} and \cite{Musset:1978gf} represent two different
  analyses of the same full raw data sample; they partially overlap in the regions below $5$ GeV
  but at higher energies, the two analyses use different bins ($5-11$ and $5-20$ GeV). 
  The analyses are generally consistent with each other and mutually complementary,
  except the lowest-energy bin and at small inelasticities.
  Alternative analyses were reported in Refs.\ \cite{Sciulli:1974we,Deden:1974uw,Baltay:1994he}
  but they do not provide additional data for comparison.
  
  The predicted cross sections are consistent with the experimental data from four of the five
  narrow instrumental energy ranges of $2-3$, $3-5$, $5-11$, and $5-20$ GeV.
  It is also seen that the agreement is generally better for the later analysis \cite{Musset:1978gf}.
  However the predicted $\nu_\mu$ and $\overline\nu_\mu$ cross sections in the lowest-energy range
  ($1-2$ GeV) systematically overestimate the data. 
  The discrepancy can be partially explained by the outdated nuclear models (including FSI)
  used in the analyses of the GGM data;
  as we now know, the MEC contributions and correct FSI modelling are critical in the few-GeV energy region.
  We cannot recalculate the nuclear effects and use this data sample in the global fit of axial mass,
  because the required Monte Carlo simulation details are not available.
  
  Another source of the observed disagreement may be associated with the systematics of the
  (anti)neutrino flux normalization and energy spectra used in the GGM analysis.
  According to author's explanation, the uncertainties in the $\nu_\mu$ and $\overline\nu_\mu$ spectra
  are estimated to be $9$\% and $12$\% for the energy ranges between $2$ and $6$ GeV and above $6$ GeV,
  respectively. Below $2$ GeV, the $\nu_\mu/\overline\nu_\mu$ spectra cannot be measured or adjusted
  in the GGM experiment and therefore an extrapolation from the higher energies has been used,
  which is not very reliable.  
  Moreover, the $\nu_\mu/\overline\nu_\mu$ fluxes were estimated by using the observed number
  of elastic events and the expected (simulated) elastic cross section.
  The author's rough estimation of the number of inclusive events gives about $2,500$ and $1,000$
  $\nu_\mu$ and $\overline{\nu}_\mu$ events, respectively, but the number of elastic events separated
  from the full inclusive sample is uncertain.


  Figure \ref{Fig:dsQESCC_dQ2_Bonetti_GGM77_Rollier_GGM78_Pohl_GGM79_101.3.31.301.6k_2_BBBA25}
  shows the flux-weighted differential cross sections measured with the large heavy-liquid
  bubble chamber Gargamelle exposed to $\nu_\mu$ and $\overline\nu_\mu$ beams at CERN
  \cite{Bonetti:1977cs,Musset:1978gf,Rollier:1978kr,Pohl:1979zm}.
  The total error on the flux used in CERN GGM 1977 experiment
  ranges from $9$\% and to $\sim12$\% in the energy regions dominated by neutrinos from
  pion and kaon decays, respectively.
  Systematic errors of $10$\% in the experimental data of CERN GGM 1979 \cite{Pohl:1979zm}
  arise due to nuclear corrections (including FSI) and uncertainties in the shape and
  absolute normalization of the $\nu_\mu$ flux.
  The data of CERN GGM 1978 \cite{Rollier:1978kr} are rescaled to the cross sections
  per neutron and proton of freon nucleus.
  The total errors include $12$\% uncertainty of the (anti)neutrino fluxes.
  

  Panels (a)--(c) in Fig. \ref{Fig:dNQESCC_dQ2_Asratyan_FNAL84_dsQESCC_dQ2_Belikov_IHEP-ITEP81,82,85_101.3.31.301.6k_2_BBBA25}
  show the differential cross sections measured with the IHEP--ITEP experiment
  \cite{Belikov:1981fq,Belikov:1981ut,Belikov:1985mw}.
  The systematic errors in these measurement make up about $10$\% and arise mainly from the neutrino
  and antineutrino flux normalization ($3$\% -- record-low uncertainty) and errors in scanning ($3$\%)
  and trigger ($4$\%) efficiencies \cite{Belikov:1983kg}. 
  In the analysis, we only use the data from  Ref.\ \cite{Belikov:1985mw}, as they are based on much more extensive statistics.
  Panel (d) shows the $Q^2$ distribution for $\overline\nu_\mu$ induced CCQE scattering measured in the FNAL E180 experiment
  with the 15-ft bubble chamber filled with heavy neon--hydrogen mixture (64\% of neon atoms) and exposed to the FNAL wide-band
  $\overline\nu_{\mu}$ beam \cite{Asratyan:1984gh,Asratyan:1984ir}.
  Predicted distribution is calculated using the mean $\overline{\nu}_\mu$ energy of $12.7$ GeV.

 
\onecolumn

  \subsubsection{Normalization factors for selected datasets}
  \label{sec:SelectedDatasets}

The normalization factors and corresponding values of $\chi^2/\text{ndf}$ obtained from individual fits
to each experimental dataset discussed in Sect. 3 are listed in Table 13. The data types used for fitting are also shown. 

\hspace*{5mm}
{\footnotesize 
 \LTcapwidth=\textwidth                             
 \setlength\LTleft{0pt}
 \setlength\LTright{0pt}
 \def\e{=}
 \begin{longtable}{lcccr}
 \caption{\label{Tab:N_test} \small
          Results of individual fits of selected sets of experimental data (published before 2016)
          for the true CCQE (anti)neutrino-nucleus interactions.
          Shown are the data types, normalization factors, $\mathcal{N}$, with the estimated $1\sigma$ 
		  and $2\sigma$ uncertainties (the latters are in parentheses), and also
          the values of $\chi^2/\ndf$, where $\ndf=\max(\text{NP}-1,1)$ and NP is the number of the
 		  experimental data points (bins) in the corresponding dataset.
		  The differential cross sections and $Q^2$ distributions are spectrum averaged.
          The cross sections for hydrogen and deuterium targets as well as those obtained with other targets,
		  but converted by the authors into the data on bare nucleons, are calculated with the current axial mass,
		  $M_A=M_0$; in all other cases (marked with an asterisk), we use the running axial mass, \MArun,
		  with the parameters $M_0$ and $E_0$ obtained in the global fit (see Eq.~\protect\eqref{Eqn:ME_QES_default}).
         }                                                                                                                                                                                                                                                                 \\ \hline\noalign{\smallskip}
  Authors / Experiment / Date                                                                  & Ref.                                           & Data type                                                          & $\mathcal{N}$           &\MC{1}{c}{$\chi^2/\ndf$}   \\ \noalign{\smallskip}\hline\noalign{\smallskip}\endfirsthead
  \noalign{\smallskip}\hline\noalign{\smallskip}\endhead
  \MC{5}{c}{Deuterium, hydrogen, and nuclear targets recalculated to bare nucleons}                                                                                                                                                                                        \\ \noalign{\smallskip}\hline\noalign{\medskip}
  Mann~\emph{et al.}, ANL 1972                                                                 &       \cite{Mann:1972}                         &         $\sigma_\nu$                                               & $0.79 \pm 0.07\,(0.14)$ &       $5.92/  5 \e 1.18$  \\ \noalign{\medskip}
  Mann~\emph{et al.}, ANL 1973                                                                 &       \cite{Mann:1973pr}                       &         $\sigma_\nu$                                               & $0.94 \pm 0.07\,(0.14)$ &       $4.13/  6 \e 0.69$  \\ \noalign{\medskip}
  Barish~\emph{et al.}, ANL 1975                                                               &       \cite{Perkins:1975bj}                    &         $\sigma_\nu$                                               & $0.99 \pm 0.05\,(0.09)$ &       $7.86/  7 \e 1.12$  \\ \noalign{\medskip}
  Singer~\emph{et al.}, ANL 1977                                                               &       \cite{Singer:1977rs}                     &         $\sigma_\nu$                                               & $0.78 \pm 0.05\,(0.09)$ &       $8.43/  7 \e 1.20$  \\ \noalign{\medskip}
  Barish~\emph{et al.}, ANL 1977                                                               &       \cite{Barish:1977qk}                     &       ~{$\sigma_\nu$}$^*$                                          & $0.89 \pm 0.03\,(0.07)$ &       $16.0/  7 \e 2.28$  \\ \noalign{\medskip}
  Miller~\emph{et al.}, ANL 1982                                                               &       \cite{Miller:1982qi}                     &       ~{${dN_\nu/dQ^2}$}$^*$                                       & ---                     &       $73.9/ 38 \e 1.95$  \\ \noalign{\medskip}
  Fanourakis~\emph{et al.}, BNL 1980                                                           &       \cite{Fanourakis:1980si}                 &       ~{${dN_{\overline{\nu}}/dQ^2}$}$^*$                          & ---                     &       $1.37/  4 \e 0.34$  \\
                                                                                               &                                                &         $\sigma_{\overline{\nu}}$                                  & ---                     &       $            0.05$  \\ \noalign{\medskip}
  Kitagaki~\emph{et al.}, BNL 1990 (updated, 2003)                                             &       \cite{Kitagaki:1990vs,Furuno:2003ng-proc}&       ~{${dN_\nu/dQ^2}$}$^*$                                       & ---                     &       $21.5/ 36 \e 0.60$  \\ \noalign{\medskip}
  Kitagaki~\emph{et al.}, FNAL 1983                                                            &       \cite{Kitagaki:1983px}                   &       ~{${dN_\nu/dQ^2}$}$^*$                                       & ---                     &       $8.53/ 19 \e 0.45$  \\
                                                                                               &                                                &         $\sigma_{\nu}$                                             & $1.02 \pm 0.08\,(0.15)$ &       $0.18/  1 \e 0.18$  \\ \noalign{\medskip}
  Allasia~\emph{et al.}, CERN BEBC 1990                                                        &       \cite{Allasia:1990uy}                    &       ~{${d\sigma_\nu/dQ^2}$}$^*$                                  & $0.94 \pm 0.04\,(0.07)$ &       $4.77/  7 \e 0.68$  \\
                                                                                               &                                                &         $\sigma_{\nu}$                                             & $0.87 \pm 0.03\,(0.06)$ &       $15.1/  5 \e 3.02$  \\ \noalign{\smallskip}\hline\noalign{\medskip}
  \MC{5}{c}{Heavy targets converted to bare nucleons}                                                                                                                                                                                                                      \\ \noalign{\smallskip}\hline\noalign{\medskip}
  Lyubushkin~\emph{et al.}, CERN NOMAD 2009                                                    &       \cite{Lyubushkin:2008pe}                 &         $\sigma_\nu$                                               & $1.05 \pm 0.03\,(0.05)$ &       $7.49/  9 \e 0.83$  \\
                                                                                               &                                                &         $\sigma_{\overline{\nu}}$                                  & $1.04 \pm 0.06\,(0.11)$ &       $2.85/  5 \e 0.57$  \\ \noalign{\medskip}
  Brunner~\emph{et al.}, IHEP SKAT 1990                                                        &       \cite{Brunner:1989kw,Ammosov:1992ak}     &       ~{${d\sigma_\nu/dQ^2}$}$^*$                                  & $1.08 \pm 0.06\,(0.12)$ &       $3.89/  7 \e 0.56$  \\
                                                                                               &                                                &         $\sigma_\nu$                                               & $1.06 \pm 0.06\,(0.12)$ &       $3.14/  7 \e 0.45$  \\
                                                                                               &                                                &       ~{${d\sigma_{\overline{\nu}}/dQ^2}$}$^*$                     & $0.90 \pm 0.06\,(0.13)$ &       $7.31/  6 \e 1.22$  \\
                                                                                               &                                                &         $\sigma_{\overline{\nu}}$                                  & $0.82 \pm 0.06\,(0.11)$ &       $8.10/  6 \e 1.35$  \\ \noalign{\medskip}
  Makeev~\emph{et al.}, IHEP SKAT 1981                                                         &       \cite{Makeev:1981em}                     &         $\sigma_\nu$                                               & $0.74 \pm 0.06\,(0.11)$ &       $6.03/  3 \e 2.01$  \\ \noalign{\medskip}
  Grabosch~\emph{et al.}, IHEP SKAT 1988                                                       &       \cite{Grabosch:1988js}                   &         ${d\sigma_\nu/dQ^2}$                                       & $0.85 \pm 0.05\,(0.10)$ &       $37.2/  7 \e 5.32$  \\
                                                                                               &                                                &         $\sigma_\nu$                                               & $1.03 \pm 0.06\,(0.11)$ &       $4.30/  6 \e 0.72$  \\
                                                                                               &                                                &         ${d\sigma_{\overline{\nu}}/dQ^2}$                          & $0.85 \pm 0.10\,(0.20)$ &       $3.75/  3 \e 1.25$  \\
                                                                                               &                                                &         $\sigma_{\overline{\nu}}$                                  & $0.75 \pm 0.09\,(0.18)$ &       $5.40/  3 \e 1.80$  \\ \noalign{\smallskip}\hline\noalign{\medskip}
  \MC{5}{c}{Heavy targets}                                                                                                                                                                                                                                                 \\ \noalign{\smallskip}\hline\noalign{\medskip}
  Kustom~\emph{et al.}, ANL 1969                                                               &       \cite{Kustom:1969dh}                     &       ${d\sigma_\nu/dQ^2}$                                         & $0.60 \pm 0.03\,(0.06)$ &       $185/  13 \e 14.21$ \\
                                                                                               &                                                &       $\sigma_\nu$                                                 & $1.12 \pm 0.05\,(0.10)$ &       $12.4/  5 \e  2.47$ \\ \noalign{\medskip}
  Ahrens~\emph{et al.}, BNL 1985                                                               &       \cite{Ahrens:1984gp}                     &       ${d\sigma_{\nu_e}/dQ^2}$                                     & $0.92 \pm 0.08\,(0.15)$ &       $14.6/  9 \e 1.62$  \\
                                                                                               &                                                &       ${d\sigma_\nu/dQ^2}$                                         & $0.64 \pm 0.03\,(0.06)$ &       $15.1/  8 \e 1.89$  \\ \noalign{\medskip}
  Asratyan~\emph{et al.}, FNAL 1984                                                            &       \cite{Asratyan:1984gh}                   &     ~{${dN_{\overline{\nu}}/dQ^2}$}$^*$                            & ---                     &       $7.93/ 13 \e 0.61$  \\
                                                                                               &                                                &       $\sigma_{\overline{\nu}}$                                    & ---                     &       $            0.06$  \\ \noalign{\medskip}
  Suwonjandee~\emph{et al.}, FNAL NuTeV 2004                                                   &       \cite{Suwonjandee:2004aw}                &       $\sigma_\nu$                                                 & $1.10 \pm 0.04\,(0.08)$ &       $21.0/ 16 \e 1.31$  \\
                                                                                               &                                                &       $\sigma_{\overline{\nu}}$                                    & $1.29 \pm 0.04\,(0.08)$ &       $21.7/ 16 \e 1.36$  \\ \noalign{\medskip}
  Alcaraz-Aunion~\emph{et al.}, FNAL SciBooNE 2010                                             &       \cite{AlcarazAunion:2009ku}              &       $\sigma_\nu$                                                 & $0.95 \pm 0.03\,(0.06)$ &       $1.89/  4 \e 0.47$  \\ \noalign{\medskip}
  Aguilar-Arevalo~\emph{et al.}, FNAL MiniBooNE (CH$_2$) 2010                                  &       \cite{Aguilar-Arevalo:2010zc}            &     ~{${d^2\sigma_\nu/dE_{\mu}d\cos\theta_{\mu}}$}$^*$             & $1.00 \pm 0.01\,(0.02)$ &       $35.8/136 \e 0.26$  \\ \noalign{\smallskip}
  Aguilar-Arevalo~\emph{et al.}, FNAL MiniBooNE (CH$_2$) 2013                                  &       \cite{Aguilar-Arevalo:2013dva}           &     ~{${d^2\sigma_{\overline{\nu}}/dE_{\mu}d\cos\theta_\mu}$}$^*$  & $1.01 \pm 0.01\,(0.03)$ &       $31.9/ 77 \e 0.42$  \\ \noalign{\smallskip}
  \hspace*{84pt}                 FNAL MiniBooNE ($^{12}$C) 2013                                &       \cite{Aguilar-Arevalo:2013dva}           &       ${d^2\sigma_{\overline{\nu}}/dE_{\mu}d\cos\theta_\mu}$       & $1.01 \pm 0.01\,(0.03)$ &       $44.7/ 74 \e 0.61$  \\ \noalign{\medskip}
  Fiorentini~\emph{et al.}, FNAL \MINERvA\ 2016 ($\nu$, $\theta_\mu \leq 20^\circ$)            &\MR{2}{\cite{Fiorentini:2013ezn}}               &\MR{2}{${d\sigma_\nu/dQ^2}$}                                        & ---                     &       $11.24/ 8 \e 1.41$  \\ \noalign{\smallskip}
  \hspace*{62pt}            FNAL \MINERvA\ 2016 ($\nu$, no cut)                                &                                                &                                                                    & ---                     &       $12.15/ 8 \e 1.52$  \\ \noalign{\smallskip}
  Fields~\emph{et al.},     FNAL \MINERvA\ 2016 ($\overline\nu$, $\theta_\mu\leq20^\circ$)     &\MR{2}{\cite{Fields:2013zhk}}                   &\MR{2}{${d\sigma_{\overline{\nu}}/dQ^2}$}                           & ---                     &        $5.06/ 8 \e 0.63$  \\ \noalign{\smallskip}
  \hspace*{46pt}            FNAL \MINERvA\ 2016 ($\overline\nu$, no cut)                       &                                                &                                                                    & ---                     &        $5.62/ 8 \e 0.70$  \\ \noalign{\medskip}
  \hspace*{46pt}            FNAL \MINERvA\ 2016 ($\nu+\overline\nu$, $\theta_\mu\leq20^\circ$) &\MR{2}{\cite{Fiorentini:2013ezn,Fields:2013zhk}}&\MR{2}{${d\sigma_\nu/dQ^2}\,\&\,{d\sigma_{\overline{\nu}}/dQ^2}$}   & ---                     &       $19.66/16 \e 1.23$  \\ \noalign{\smallskip}
  \hspace*{46pt}            FNAL \MINERvA\ 2016 ($\nu+\overline\nu$, no cut)                   &                                                &                                                                    & ---                     &       $22.08/16 \e 1.38$  \\ \noalign{\medskip}
  Franzinetti,CERN HLBC 1966                                                                   &       \cite{Franzinetti:1965}                  &         $\sigma_\nu$                                               & $0.80 \pm 0.14\,(0.27)$ &       $10.8/  4 \e 2.70$  \\
                                                                                               &                                                &         $\sigma_{\overline{\nu}}$                                  & $1.07 \pm 0.29\,(0.57)$ &       $0.89/  1 \e 0.89$  \\ \noalign{\medskip}
  Young~\emph{et al.}, CERN HLBC 1967                                                          &       \cite{Young:1967ud}                      &         $\sigma_\nu$                                               & $0.64 \pm 0.07\,(0.15)$ &       $4.99/  4 \e 1.25$  \\
                                                                                               &                                                &         $\sigma_{\overline{\nu}}$                                  & $0.87 \pm 0.18\,(0.36)$ &       $1.51/  1 \e 1.51$  \\ \noalign{\medskip}
  Budagov~\emph{et al.}, CERN HLBC 1969                                                        &       \cite{Budagov:1969bg}                    &         $\sigma_{\nu}$                                             & $0.51 \pm 0.05\,(0.10)$ &       $18.2/  4 \e 4.55$  \\ \noalign{\medskip}
  Eichten~\emph{et al.}, CERN GGM 1973                                                         &       \cite{Eichten:1973cs}                    &         $\sigma_{\nu}$                                             & $0.74 \pm 0.04\,(0.08)$ &       $9.66/ 12 \e 0.81$  \\
                                                                                               &                                                &         $\sigma_{\overline{\nu}}$                                  & $0.83 \pm 0.05\,(0.10)$ &       $8.47/ 11 \e 0.77$  \\ \noalign{\medskip}
  Rollier~\emph{et al.}, CERN GGM 1975                                                         &       \cite{Rollier:1975qr,Perkins:1975bj}     &         $\sigma_{\nu}$                                             & $0.86 \pm 0.04\,(0.08)$ &       $6.44/  8 \e 0.81$  \\
                                                                                               &                                                &         $\sigma_{\overline{\nu}}$                                  & $0.85 \pm 0.05\,(0.10)$ &       $8.33/ 11 \e 0.76$  \\ \noalign{\medskip}
  Bonetti~\emph{et al.}, CERN GGM 1977                                                         &       \cite{Bonetti:1977cs}                    &         ${d\sigma_\nu/dQ^2}$                                       & $0.79 \pm 0.03\,(0.07)$ &       $18.8/ 11 \e 1.71$  \\
                                                                                               &                                                &         $\sigma_\nu$                                               & $0.80 \pm 0.04\,(0.08)$ &       $10.8/ 12 \e 0.90$  \\
                                                                                               &                                                &         ${d\sigma_{\overline{\nu}}/dQ^2}$                          & $0.82 \pm 0.04\,(0.08)$ &       $8.68/  7 \e 1.24$  \\
                                                                                               &                                                &         $\sigma_{\overline{\nu}}$                                  & $0.84 \pm 0.05\,(0.09)$ &       $12.6/  9 \e 1.40$  \\ \noalign{\medskip}
  Rollier~\emph{et al.}, CERN GGM 1978                                                         &       \cite{Rollier:1978kr}                    &         ${d\sigma_{\overline{\nu}}/dQ^2}$                          & $0.88 \pm 0.03\,(0.06)$ &       $29.3/ 11 \e 2.67$  \\
                                                                                               &                                                &         $\sigma_{\overline{\nu}}$                                  & $0.85 \pm 0.03\,(0.06)$ &       $23.2/ 10 \e 2.32$  \\ \noalign{\medskip}
  Pohl~\emph{et al.}, CERN GGM 1979                                                            &       \cite{Pohl:1979zm}                       &         ${d\sigma_\nu/dQ^2}$                                       & $0.79 \pm 0.04\,(0.08)$ &       $13.1/  6 \e 2.19$  \\
                                                                                               &                                                &         $\sigma_\nu$                                               & $0.77 \pm 0.04\,(0.08)$ &       $11.9/  6 \e 1.98$  \\ \noalign{\medskip}
  Armenise~\emph{et al.}, CERN GGM 1979                                                        &       \cite{Armenise:1979zg,Singh:1992dc}      &         $\sigma_{\overline{\nu}}$                                  & $0.78 \pm 0.03\,(0.06)$ &       $20.6/ 10 \e 2.06$  \\ \noalign{\medskip}
  De la Ossa Romero, CERN LAr-TPC 2007                                                         &       \cite{MartinezdelaOssaRomero:2007oxj}    &       ~{$\sigma_\nu$}$^*$                                          & ---                     &       $            0.09$  \\ \noalign{\medskip}
  Lyubushkin~\emph{et al.}, CERN NOMAD 2009                                                    &       \cite{Lyubushkin:2008pe}                 &       ~{$\sigma_\nu$}$^*$                                          & $1.04 \pm 0.03\,(0.05)$ &       $5.91/  9 \e 0.66$  \\
                                                                                               &                                                &       ~{$\sigma_{\overline{\nu}}$}$^*$                             & $1.02 \pm 0.06\,(0.11)$ &       $3.09/  5 \e 0.62$  \\ \noalign{\medskip}
  Belikov~\emph{et al.}, IHEP--ITEP 1981                                                       &       \cite{Belikov:1981fq}                    &         $\sigma_\nu$                                               & $0.90 \pm 0.06\,(0.12)$ &       $1.97/  5 \e 0.40$  \\
                                                                                               &                                                &         ${d\sigma_{\overline{\nu}}/dQ^2}$                          & $0.91 \pm 0.05\,(0.11)$ &       $2.07/  7 \e 0.30$  \\
                                                                                               &                                                &         $\sigma_{\overline{\nu}}$                                  & $0.90 \pm 0.06\,(0.11)$ &       $3.21/  5 \e 0.80$  \\ \noalign{\medskip}
  Belikov~\emph{et al.}, IHEP--ITEP 1982                                                       &       \cite{Belikov:1981ut}                    &         ${d\sigma_\nu/dQ^2}$                                       & $0.88 \pm 0.06\,(0.11)$ &       $10.2/  7 \e 1.46$  \\
                                                                                               &                                                &         $\sigma_\nu$                                               & $0.90 \pm 0.06\,(0.12)$ &       $1.78/  5 \e 0.36$  \\ \noalign{\medskip}
  Belikov~\emph{et al.}, IHEP--ITEP 1985                                                       &       \cite{Belikov:1983kg,Belikov:1985mw}     &        \MR{2}{{${d\sigma_{\nu+\overline{\nu}}/dQ^2}$}$^*$ \Bel}    & $0.98 \pm 0.11\,(0.18)$ &\MR{2}{$5.68/  6 \e 0.95$} \\
                                                                                               &                                                &                                                                    & $0.99 \pm 0.13\,(0.21)$ &                           \\
                                                                                               &                                                &       ~{$\sigma_\nu$}$^*$                                          & $0.87 \pm 0.04\,(0.07)$ &       $7.02/  7 \e 1.00$  \\
                                                                                               &                                                &       ~{$\sigma_{\overline{\nu}}$}$^*$                             & $0.90 \pm 0.05\,(0.09)$ &       $2.66/  7 \e 0.38$  \\ \noalign{\medskip}
  Abe~\emph{et al.}, T2K ND280 (off-axis) 2014                                                 &       \cite{Abe:2014iza}                       &         $\sigma_\nu$                                               & ---                     &       $9.57/  5 \e 1.91$  \\ \noalign{\medskip}
  Abe~\emph{et al.}, T2K INGRID (on-axis) 2015                                                 &       \cite{Abe:2015oar}                       &       ~{$\sigma_\nu$}$^*$                                          & $0.98 \pm 0.09\,(0.17)$ &       $0.55/  1 \e 0.55$  \\ \noalign{\medskip}
\noalign{\smallskip}\hline
\end{longtable}
}

\twocolumn

  \bibliographystyle{springer}
  \bibliography{references}

\end{document}